\documentclass[twocolumn,usenatbib,useAMS]{mnras}

\PassOptionsToPackage{dvipsnames,svgnames,x11names}{xcolor}
\usepackage{xparse,soul}

\makeatletter
\NewDocumentCommand{\sotwo}{O{red}O{black}+m}
    {%
        \begingroup
        \color{#1}%
        \setul{-.5ex}{.4pt}%
        \def\SOUL@uleverysyllable{%
            \rlap{%
                \color{#2}\the\SOUL@syllable
                \SOUL@setkern\SOUL@charkern}%
            \SOUL@ulunderline{%
                \phantom{\the\SOUL@syllable}}%
        }%
        \ul{#3}%
        \endgroup
    }
\makeatother

\usepackage{graphicx}	
\usepackage{amsmath}	
\usepackage{amssymb}	
\usepackage{multicol} 
\usepackage{multirow}
\usepackage{bm}	
\usepackage{pdflscape}
\usepackage{caption}
\usepackage{subfigure}
\usepackage{siunitx}
\usepackage{color}
\usepackage{lastpage}
\usepackage{xspace}
\usepackage{color}
\usepackage{booktabs}

\newcommand{\simba}{\mbox{{\sc Simba}}\xspace}
\newcommand{\colibre}{\mbox{{\sc Colibre}}\xspace}
\usepackage[T1]{fontenc}
\usepackage{ae,aecompl}

\usepackage{newtxtext,newtxmath}

\title[FLAMINGO secondary CMB anisotropies]{Self-consistent CMB Secondaries in the FLAMINGO Simulations}

\title[FLAMINGO secondary CMB anisotropies]{Self-consistent secondary cosmic microwave background anisotropies and extragalactic foregrounds in the FLAMINGO simulations}

\author[T. Yang et al.]{Tianyi Yang,$^{1}$\thanks{E-mail: t.yang@ljmu.ac.uk}
Ian G. McCarthy,$^{1}$\thanks{E-mail: i.g.mccarthy@ljmu.ac.uk}
Fiona McCarthy,$^{2,3}$
Boris Bolliet,$^{3,4}$
Jens Chluba,$^{5}$\newauthor
William Coulton,$^{2,3}$
John C. Helly,$^{6}$
Matthieu Schaller,$^{7,8}$
Joop Schaye$^{7}$
\\
$^{1}$Astrophysics Research Institute, Liverpool John Moores University, Liverpool, L3 5RF, UK\\
$^{2}$DAMTP, Centre for Mathematical Sciences, University of Cambridge, Wilberforce Road, Cambridge CB3 OWA, UK\\
$^{3}$Kavli Institute for Cosmology Cambridge, Madingley Road, Cambridge CB3 0HA, UK\\
$^{4}$Cavendish Astrophysics, University of Cambridge, Madingley Road, Cambridge CB3 0HA, UK\\
$^{5}$Jodrell Bank Centre for Astrophysics, Alan Turing Building, University of Manchester, Manchester M13 9PL\\
$^{6}$Institute for Computational Cosmology, Department of Physics, Durham University, South Road, Durham, DH1 3LE, United Kingdom\\
$^{7}$Leiden Observatory, Leiden University, PO Box 9513, 2300 RA Leiden, the Netherlands\\
$^{8}$Lorentz Institute for Theoretical Physics, Leiden University, PO box 9506, NL-2300 RA Leiden, the Netherlands}

\date{Accepted XXX. Received YYY; in original form ZZZ}

\pubyear{2025}

\begin{document}
\label{firstpage}
\pagerange{\pageref{firstpage}--\pageref{lastpage}}
\maketitle

\begin{abstract}
Secondary anisotropies in the cosmic microwave background (CMB) contain information that can be used to test both cosmological models and models of galaxy formation. 
Starting from lightcone-based HEALP\textsc{ix} maps and catalogues, we present a new set of mock CMB maps constructed in a self-consistent manner from the FLAMINGO suite of cosmological hydrodynamical simulations, including CMB lensing, thermal and kinetic Sunyaev-Zel’dovich effects, cosmic infrared background, radio point source and anisotropic screening maps. We show that these simulations reproduce a wide range of observational constraints.  We also compare our simulations with previous predictions based on dark matter-only simulations which generally model the secondary anisotropies independently from one another, concluding that our hydrodynamical simulation mocks perform at least as well as previous mocks in matching the observations whilst retaining self-consistency in the predictions of the different components. Using the model variations in FLAMINGO, we further explore how the signals depend on cosmology and feedback modelling, and we predict cross-correlations between some of the signals that differ significantly from those in previous mocks. 
The mock CMB maps should provide a valuable resource for exploring correlations between different secondary anisotropies and other large-scale structure tracers, and can be applied to forecasts for upcoming surveys.
\end{abstract}


\section{Introduction}\label{sec::intro}

The cosmic microwave background (CMB) is a relic radiation field from the early Universe. During the epoch of recombination, photons decoupled from matter and began to propagate freely, leaving imprints of temperature fluctuations at the surface of last scattering. These fluctuations give rise to the primary anisotropies of the CMB, which contain information about the initial conditions and key fundamental parameters of the Universe. Observations of these anisotropies provide strong support for the standard cosmological model, characterised by cold dark matter and a cosmological constant---the so-called $\Lambda$CDM model \citep[e.g.][]{Planck2020_cosmology, ACT_DR6_CIB_ref}.

As CMB photons travel through space following decoupling, they interact with intervening large-scale structure (LSS), which causes a range of secondary anisotropies in CMB observations. These include (but are not limited to): gravitational lensing of the CMB photons, which is sensitive to the total mass distribution along the line of sight; the Sunyaev–Zel’dovich (SZ) effects that probe the thermal energy density (thermal SZ, or tSZ) and the momentum (kinetic SZ, or kSZ) of the ionised gas; and so-called late-time anisotropic (or `patchy') screening due to Thomson scattering off ionised electrons in and/around haloes. In addition to CMB photons, other sources of emission are present at radio and microwave wavelengths that can act as contaminants for CMB measurements. These include the so-called Cosmic Infrared Background (CIB), which probes the cosmic evolution of dusty star-forming galaxies, and radio point sources that reflect compact non-thermal emission from active galactic nuclei and other extragalactic populations. As the CIB and radio sources are associated with galaxies and therefore also trace LSS, they are expected to (and indeed are observed to) spatially correlate with CMB secondary anisotropies.  Together, the primary CMB anisotropies and foregrounds, their secondary effects, and their cross-correlations with late-time tracers such as galaxies and weak gravitational lensing, provide powerful and complementary tools of cosmology and the astrophysical processes related to structure formation \citep[e.g.][]{P14_CIB, P16_tSZ_map, P16_CIB_tSZ, Hojjati_tSZ_weak_lensing, P18_radio, P18_lensing_ref, tSZ_tomo_Chiang, Fiona_CIB_CMB_lensing_fnl,Fiona_2024_CMBlensing_tSZ}.

Extracting a component of interest from CMB observations (e.g., the tSZ effect) is a challenging task owing to the finite sensitivity, resolution, and wavelength coverage of the observations coupled with uncertainties in the spectral characteristics of the CIB and radio sources (and their potential dependence on galaxy properties and their evolution) as well as uncertainties in the predicted spatial clustering of both these and the secondary CMB anisotropies and their evolution. To disentangle these contributions, several component separation algorithms have been developed, such as the \textsc{commander} method \citep{COMMANDER_ref}, which relies on a physical model of the sky, requiring both spatial and frequency-dependent templates for each component, and other more empirical methods such as the \textsc{ilc}, \textsc{nilc}, \textsc{gilc}, \textsc{sevem}, \textsc{smica}, and \textsc{milca}  algorithms \citep[see e.g.][and references therein]{MILCA_ref,CMB_separation_pipeline_paper}, which rely primarily on just the spectral dependence of the target signal.  Unsurprisingly, inaccurate assumptions employed in the component separation algorithms can result in biases in the extracted signals. For instance, \citet{CIB_deproj_ref3} demonstrated that the level of foreground leakage in the recovered tSZ power spectrum can vary significantly depending on the deprojection strategy used in the NILC algorithm. 

To interpret observations, realistic multi-component CMB simulations are essential for predicting how different astrophysical and cosmological signals are correlated. Such mocks are also an invaluable tool for testing how cleanly the signals can be separated from one another in realistic observing conditions, as well for testing routines for source detection (e.g., cluster finding). 

A number of previous studies have produced CMB secondary predictions based on the halo model formalism (e.g., \textsc{class\_sz}\footnote{\url{https://github.com/CLASS-SZ}}; \citealt{class_sz}) or dark matter-only cosmological simulations, which typically use parametric prescriptions for `painting on' observable components such as the Sunyaev-Zel'dovich effect.  We will mostly focus our comparison on the latter class of predictions, as they are more similar in spirit to our aims of producing full-sky maps that can be analysed in a way that is faithful to the observational analyses.

In a pioneering study producing the first CMB mocks based on dark matter-only simulations, \citet{Sehgal_2010} started with a moderate resolution 1000 $h^{-1}$ Mpc volume dark matter particle field, with haloes identified using a friends-of-friends (FoF) algorithm. The free parameters that govern the gas physics and CIB source counts were calibrated against existing external observations, including X-ray gas fractions and far-infrared measurements available at that time. Updated models using this framework have been used for Simons Observatory forecasting \citep[see][]{Sehgal_radio_SO_forecast}.

More recently, \citet{Websky_ref} produced the WebSky CMB simulations\footnote{WebSky maps are available at: \url{https://lambda.gsfc.nasa.gov/simulation/mocks_data.html}} (see also \citealt{Li_websky_radio} for associated radio catalogues construction) that use the mass-Peak Patch formalism \citep{mass_peak_patch_ref1,mass_peak_patch_ref2} to predict the growth of structure and the location of haloes.  While this formalism approximates non-linear evolution (as opposed to full cosmological simulations, which directly evaluate it), a major strength is its speed and ability to follow large simulation volumes that better sample rare, massive haloes. Baryons within haloes are assigned using analytical prescriptions calibrated against hydrodynamical simulations and observations, while the dynamics and matter distribution of the diffuse field component are modelled using second-order Lagrangian perturbation theory \citep[2LPT,][]{2LPT_ref}. 

More similar to the approach of \citet{Sehgal_2010}, the AGORA CMB simulations\footnote{AGORA maps are available at: \url{https://yomori.github.io/agora/index.html}} \citep{agora_ref} are subarcminute-resolution CMB sky maps based on the higher-resolution MultiDark \textit{Planck} 2 simulation \citep[MDPL2,][]{MDPL2_ref}. Extragalactic observables are assigned using observationally constrained analytical models, and the resulting mock signals have been validated against recent CMB observations from \textit{Planck} \citep{Planck_intro}, SPT-SZ \citep{SPT_intro}, and SPTpol \citep{SPTpol_intro}. 

While dark matter-only and halo model-based simulations have proven valuable for constructing CMB secondary anisotropy maps, they rely on simplified or empirically tuned models to assign baryonic properties and in general there is no physical association between the components (e.g., the CIB and tSZ effect) apart from the fact that their clustering is spatially correlated with haloes.  In contrast, hydrodynamical simulations offer a self-consistent treatment of gas physics, naturally capturing the interplay between complex processes such as radiative cooling, star formation, AGN feedback, diffuse gas flows, as well as their gravitational back reaction on the dark matter. Hydrodynamical simulations also have important uncertainties, since they cannot resolve all relevant scales or the predict from first principles the efficiencies of important processes such as feedback, thus requiring so-called subgrid models (see the discussion \citealt{FMG_SF_ref_15}).  Nevertheless, 
for a given feedback implementation the baryons (hot and cold gas and stars), dark matter, black holes, and neutrinos (if the latter are included) are followed in a self-consistent manner, which in general will result in correlations between the components. For example, relatively weak feedback might be expected to lead to higher gas fractions in groups and clusters but also to elevated star formation rates, potentially producing a much stronger cross-correlation between the CIB and the hot gas compared to a model where the feedback is considerably stronger.  Furthermore, one can study the sensitivity of the results to the subgrid modelling by systematically varying the key parameters of the models, as we do here (see also, e.g., \citealt{OWLS_ref, TNG_feedback_var_ref, camels_feedback_var,antilles,simba_feedback_var}).  
In addition, environmental processes such as ram pressure stripping of gas from satellites are also naturally captured, as are effects which are challenging to paint on dark matter-only simulations, including the effects of substructure, triaxiality, mis-centring, etc. on observable baryonic quantities (e.g., effects of triaxiality on the tSZ effect). 

Individual secondary observables have been studied within the setups of several hydrodynamical simulation suites, including the SZ effect \citep[e.g.][]{SZ_B12,SZ_TNG,SZ_cosmoowls,SZ_Magneticum,SZ_BAHAMAS,SZ_simba} and the infrared source reconstruction \citep[e.g.][]{C13,B17_FR,L21,C23}. However, to our knowledge, none of these simulations have been used for producing full-sky, multi-component CMB maps, which in some cases is due to their limited box sizes, which restricts the ability to sample the large-scale modes and rare massive objects that are important for reproducing the statistical properties of full-sky signals such as the tSZ effect. 

In this study, we present a set of full-sky CMB secondary anisotropy mock maps based on the FLAMINGO suite of cosmological hydrodynamical simulations \citep{FLAMINGO_ref_Joop,FLAMINGO_kugel}, where the signals are directly derived from the physical properties of matter, gas, and accreting black holes within the simulation volume. FLAMINGO features a large fiducial run with a comoving box size of 2.8 Gpc. In addition, the suite includes multiple 1 Gpc runs with systematically varied subgrid feedback models and cosmological parameters, allowing us to further explore how different baryonic and cosmological assumptions affect the predicted signals and their correlations. This setup provides, for the first time, a powerful framework to study the signal correlations and their model dependencies using large-volume hydrodynamical simulations, which could bridge the gap between theoretical predictions and forthcoming high-resolution CMB observations. 

This work presents a description of our methodology, simulated maps, and corresponding power spectra, along with initial comparisons to observational data and to previous mock studies based on dark matter-only simulations. In future work, we will extend this framework to evaluate the performance of component-separation algorithms and to conduct a wider range of comparisons with observations from current and forthcoming CMB experiments.

This paper is organised as follows. In Section \ref{sec::flamingo_sim}, we briefly summarise the setup of the FLAMINGO fiducial run and its model variants used for map construction. In Section \ref{sec::Map_Generation_and_Power_Spectrum_Modelling}, we detail the procedures and models adopted to construct the maps of different CMB secondaries. Section \ref{sec::Maps_and_power_spectra} presents the auto- and cross-power spectra of various quantities from our mock maps, along with their comparisons to other multi-component CMB simulations and observations. In Section \ref{sec::discussions}, we investigate the feedback and cosmology dependencies of these power spectra results. Finally, we conclude in Section~\ref{sec::conclusion}.

\section{FLAMINGO simulations}\label{sec::flamingo_sim}

We provide here a brief summary of the FLAMINGO simulations. A detailed description of the simulation is presented in \citet{FLAMINGO_ref_Joop}.

FLAMINGO is a suite of large-scale cosmological hydrodynamical simulations designed to study cosmology and LSS physics. The suite includes three different baryonic gas particle mass resolutions: a high-resolution run with $m_{\rm gas} = 1.3\times10^{8}~\rm M_{\odot}$ (referred to as m8), an intermediate-resolution run with $m_{\rm gas} = 1.1\times10^{9} ~\rm M_{\odot}$ (m9), and a low-resolution run with $8.6\times10^{9}~\rm M_{\odot}$ (m10). The flagship runs are the $(1 ~\rm Gpc)^{3}$ high-resolution run and the $(2.8 ~\rm Gpc)^{3}$ intermediate-resolution run (denoted as L1$\_$m8 and L2p8$\_$m9 respectively). The latter follows $2.8 \times 10^{11}$ particles, making it the largest hydrodynamical simulation evolved to $z=0$ at the time it was run\footnote{See the latest, larger Frontier-E simulation \citep{Frontier_E_sim}}. Most runs adopt a Dark Energy Survey Year Three \citep[DES Y3;][]{DES_Y3_Abbott} 3 $\times$ 2pt + All Ext. $\Lambda$CDM cosmology (see Table \ref{table_summary_cosmoinfo} for the list of parameters). This cosmology assumes a spatially flat universe and is based on a combination of constraints from three DES Y3 two-point correlation functions—cosmic shear, galaxy clustering, and galaxy–galaxy lensing—along with external data from baryon acoustic oscillations (BAO), redshift-space distortions, Type Ia supernovae, \textit{Planck} observations of the CMB (including CMB lensing), Big Bang nucleosynthesis, and local measurements of the Hubble constant. See \citet{DES_Y3_Abbott} for details.

The simulations were performed using \textsc{Swift} \citep{SWIFT_ref}, a fully open-source coupled cosmology, gravity, hydrody-
namics, and galaxy formation code. Short- and long-range gravitational forces are computed using a $4^{\rm th}$-order fast multipole method \citep{force_cal_ref1,force_cal_ref2,force_cal_ref3} and a particle-mesh method solved in Fourier space, respectively, following the force-splitting approach of \citet{force_cal_ref4}. The hydrodynamic equations are solved using the smoothed particle hydrodynamics (SPH) method \citep[for a review, see][]{SPH_review_price}, in particular the SPHENIX flavour of SPH \citep{Borrow_SPHENIX}, which was designed specifically for simulations of galaxy formation. The initial conditions are obtained from a modified version of monofonIC \citep{IC_ref1,IC_ref2}, and neutrinos are implemented with the $\delta f$ method \citep{FLAMINGO_neutrino}.

Unresolved physical processes are modelled using subgrid physics with various parameter choices. The simulation includes a wide range of physical processes, including radiative cooling and heating \citep{FMG_radiative_heating_cooling}, star formation and evolution \citep{FMG_SF_ref_08S,Wiersma_09b}, black hole growth \citep{Booth_schaye_09,Bahe_22}, feedback from young stars and supernovae \citep{FMG_SF_ref_22b,FMG_SF_ref_22a}, and AGN feedback \citep{Booth_schaye_09,Husko_jet}. A detailed description of the subgrid physics implementation is provided in \citet{FLAMINGO_ref_Joop}.

Four subgrid parameters—two related to stellar feedback, one to black hole growth, and one to AGN feedback—are calibrated to match the observed present-day galaxy stellar mass function (SMF, \citealt{Driver_SMF}) and the low-redshift gas mass fraction within $R_{\rm 500c}$ for galaxy groups and clusters \citep{Akino_fgas,FLAMINGO_kugel}. Machine learning-based emulators were used in the calibration \citep{FLAMINGO_kugel}.  Note that the emulators were used not only to calibrate to the observations as in the fiducial model, but also to produce different model variations as discussed below.

The goal of our study is to provide multicomponent full-sky CMB mock maps predicted from hydrodynamical simulations. We use the $(2.8 ~\rm Gpc)^{3}$ intermediate-resolution run as our primary model, referred to as the fiducial run. To further explore the feedback dependence of the correlation between different secondary anisotropies, we also make use of feedback model variants from a series of $(1 ~\rm Gpc)^{3}$ runs that vary the strength of stellar and/or AGN feedback. These runs share the same particle resolution as the large-box fiducial model, but apply shifts to the observed galaxy stellar masses (for $M_{\ast}$) or cluster gas fractions (for $f_{\rm gas}$) during calibration. In the strong stellar feedback model (denoted as $M_{\ast}-\sigma$), the observed stellar mass function (SMF) was shifted by \(-0.14\) dex lower in stellar mass. The AGN feedback variants (denoted as $f_{\rm gas}\pm N\sigma$) were calibrated to the observed gas mass fractions shifted by +2, -2, -4, -8 times the measurement error, respectively. An additional jets and strong jets model are also included (denoted as Jet and Jet$\_f_{\rm gas}-4\sigma$, respectively), where AGN feedback is implemented via kinetic kicks to gas particles within the black hole's SPH kernel, rather than through isotropic thermal energy injection. The jet model is calibrated against the same set of observational data as the fiducial thermal AGN feedback model. All feedback-variation runs are evolved under the fiducial DES Y3 cosmology.

The FLAMINGO simulations have previously been compared with a wide range of observations beyond the measurements used for calibration, including scaling relations between galaxy stellar mass and black hole mass, halo mass, quenched fractions, the cosmic star formation rate vs. redshift \citep{FLAMINGO_ref_Joop}, cluster X-ray and tSZ scaling relations and profiles \citep{FLAMINGO_ref_Joop,Braspenning2024}, the stacked kSZ effect profiles of massive galaxies \citep{Ian_kSZ_feedback_FLAMINGO,Siegel2026,Bigwood2026}, the auto and cross-power spectra of CMB lensing, cosmic shear, and the tSZ effect \citep{referee_ref_Ian_lowsigma8}, the X-ray--cosmic shear cross-power spectra \citep{McDonald2026}, among others.

To study the cosmological dependence of those signals, we also include two cosmology variants from the $(1 ~\rm Gpc)^{3}$ runs (see Table \ref{table_summary_cosmoinfo} for the list of parameters). One uses the \citet{Planck2020_cosmology} maximum likelihood cosmology, denoted as Planck, while the other takes the lensing cosmology from \citet[][denoted as LS8]{amon_lensing_cosmo}. This LS8 model has a lower amplitude of the power spectrum, $S_{8}=0.766$, compared with 0.815 and 0.833 for the fiducial and Planck runs, and provides a better match to some previous cosmic shear measurements, including from DES Y3 \citep{DES_Y3_Abbott}, HSC Y1 \citep{Aihara_HSC_Y1}, and KiDS-1000 \citep{Kuijken_KIDS}. All cosmology-variation runs are evolved using the same calibrated feedback model as the fiducial run.

\begin{table*}
\centering
\caption{Cosmological parameters adopted in different FLAMINGO simulations. Only the models used in this work are listed here (for the full table, please see \citealt{FLAMINGO_ref_Joop}). These include the dimensionless Hubble constant, $h$; the total matter density parameter, $\Omega_{\rm m}$; the dark energy density parameter, $\Omega_{\Lambda}$; the baryonic matter density parameter, $\Omega_{\rm b}$; the sum of the particle masses of the neutrino species, $\Sigma m_{\nu}c^{2}$; the amplitude of the
primordial matter power spectrum, $A_{\rm s}$; the power-law index of the primordial matter power spectrum, $n_{\rm s}$; the amplitude of the initial
power spectrum parametrized as the r.m.s. mass density fluctuation in spheres of radius 8 $h^{-1} ~\rm Mpc$ extrapolated to $z = 0$ using linear
theory, $\sigma_{8}$; the amplitude of the initial power spectrum parametrised as $S_{8}\equiv\sigma_{8}\sqrt{\Omega_{\rm m}/0.3}$; the neutrino matter density parameter, $\Omega_{\nu}\cong\Sigma m_{\nu}c^{2}/(93.14 ~h^{2}~\rm eV)$.  Note that the values of the Hubble and density parameters are given at $z= 0$.}
\begin{tabular}{c|cccccccccc} 
 \hline
 Model name & $h$  &$\Omega_{\rm m}$& $\Omega_{\rm \Lambda}$ & $\Omega_{\rm b}$  & $\Sigma m_{\nu}c^{2}$ & $A_{\rm s}$ & $n_{\rm s}$ &$\sigma_{8}$& $S_{8}$& $\Omega_{\nu}$\\
 \hline
 Fiducial & 0.681& 0.306& 0.694& 0.0486& 0.06 eV&$2.099\times10^{-9}$ &0.967 &0.807 & 0.815& $1.39\times10^{-3}$ \\ 
 Planck & 0.673& 0.316& 0.684& 0.0494& 0.06 eV&$2.101\times10^{-9}$ &0.966 &0.812 & 0.833& $1.42\times10^{-3}$\\
 LS8 & 0.682& 0.305& 0.695& 0.0473& 0.06 eV&$1.836\times10^{-9}$ &0.965 &0.760 & 0.766& $1.39\times10^{-3}$\\
 \hline
\end{tabular}
\label{table_summary_cosmoinfo}
\end{table*}

\section{Methodology}\label{sec::Map_Generation_and_Power_Spectrum_Modelling}

In this section, we describe our methodology for producing maps of the various CMB secondary anisotropies, based on the lightcone maps from the FLAMINGO simulations. Section \ref{ssec: lightcone_generation} provides a brief summary of the construction of lightcone maps and particle lightcone data in the simulation. In Section \ref{ssec::CMB_kappa_field}, we summarise the modelling of the stacked CMB lensing convergence field. Section \ref{ssec: SZ_effect} discusses the thermal and kinetic SZ effects, including the modelling of their relativistic corrections. Section \ref{ssec::patchy_screening} provides a brief introduction to the optical depth map construction, which is directly related to the patchy screening effect. Sections \ref{ssec::CIB} and \ref{ssec::Radio_Point_Sources} describe the construction of the CIB map and mock catalogues for radio point sources. In these sections, we also highlight some key differences between our models and those used in other mock CMB simulations, such as AGORA and WebSky. Finally, in Section \ref{ssec::Lensing_of_observables}, we outline how lensing effects are added to each simulated quantity. Relevant maps and their power spectra are discussed in Section \ref{sec::Maps_and_power_spectra}.

\subsection{HEALP\textsc{ix} lightcone maps and halo lightcone catalogues}\label{ssec: lightcone_generation}

Here we provide a brief summary of the lightcone data generation. A detailed description of the algorithm can be found in the appendix of \citet{FLAMINGO_ref_Joop}.

To produce the HEALP\textsc{ix} \citep{Healpix_ref} maps for each quantity, an observer's past lightcone is split into a set of concentric spherical shells, with a redshift interval of $\Delta z = 0.05$ from $z = 0$ to 3 and $\Delta z = 0.25$ for higher redshifts. Whenever a particle is found to have crossed the lightcone, we determine which shell it lies in at the time of crossing and accumulate the particle’s contributions to the HEALP\textsc{ix} maps for that shell. In this study, we mainly focused on the full-sky maps of total matter, Compton $y$, Doppler B (for the kSZ), optical depth (for anisotropic screening), star formation rate (for the CIB), and black holes (for radio point sources).

To produce the halo lightcone data, structure finding is performed in post-processing on the snapshot particle data using HBT-HERONS, which is a modified version of the HBT algorithm \citep{halo_lc_ref1,halo_lc_ref2,halo_lc_ref3}. We first read in the snapshot subhalo catalogue corresponding to a given snapshot, then load a spherical shell from the black hole particle lightcone that spans the redshift range halfway to the previous snapshot and halfway to the next. For each halo in the snapshot, we identify a black hole particle to use as a tracer of that halo. Whenever the tracer particle ID appears among the particles read from the black hole particle lightcone, we place a copy of the halo at that position. This process is then repeated for every snapshot to build the full halo lightcone catalogue.

In our study, the mock maps are integrated up to either $z = 4.5$ or $z = 3.0$, depending on the simulation box size. Such a redshift range is sufficient for CIB-related statistics (see Figure \ref{CIB_monopole}), which are the main focus of our study here. For the default fiducial $(2.8~\rm Gpc)^{3}$ run, we use full-sky particle lightcone HEALP\textsc{ix} maps per shell with $N_{\rm side}$=4096, extending out to $z = 4.5$. For the feedback model and cosmology variants in $(1~\rm Gpc)^{3}$ boxes, however, lightcone map outputs are only available up to $z = 3.0$. To account for any missing signal beyond $z = 3.0$ for the $(1~\rm Gpc)^{3}$ runs, we rescale their power spectra using a redshift evolution correction with the ratio $C_{\ell, ~\textrm{L2p8$\_$m9}~z < 4.5}/C_{\ell, ~\textrm{L2p8$\_$m9}~z < 3.0}$. This ensures a fair comparison with the results obtained from the fiducial $(2.8~\rm Gpc)^3$ run and that our summary statistics are fully converged in terms of redshift integration.  Note that, as only the fiducial feedback model in the fiducial cosmology was run in the large volume, the correction factor that we apply is assumed to be independent of cosmology and feedback.  In principle the correction factor ought to depend on both feedback and cosmology, but in practice we expect these dependencies to be small for the summary statistics that we examine. For radio sources, we integrate the halo lightcone data up to $z = 2.5$ only, which corresponds to the maximum redshift for which the adopted radio luminosity function is available.

We use the \textsc{NaMaster}\footnote{\url{https://namaster.readthedocs.io/en/latest/source/installation.html}} \citep{NaMaster_ref} package to compute the auto- and cross-power spectra for various statistics. We initially use a multipole moment binning (bandpower) of $\Delta \ell = 7$, and then smooth the resulting curves using a Savitzky–Golay filter of order 3 with a window size of 15, which is only for visualization purposes. We deconvolve the $N_{\rm side} = 4096$ pixel window function from the computed spectra using the \textit{pixwin} function in the HEALP\textsc{ix} package. To avoid repeating structures along the line of sight, the shells in the lightcone are rotated randomly every box-length interval, which preserves the line-of-sight correlations over the scales captured by the simulations. To preserve the correlations between the different signals (e.g., CIB and tSZ), the same rotations are applied to all quantities considered in this study.

\subsection{CMB lensing maps}\label{ssec::CMB_kappa_field}

CMB lensing is caused by the deflection of CMB photons when they travel through the LSS of the Universe. As these photons propagate from the last scattering surface to the observer, their paths are deflected by the intervening matter distribution, leading to small distortions in the observed CMB temperature and polarization images. This effect is characterised by a projected convergence field, $\kappa$, which can be used to infer the (CMB kernel-weighted) integrated mass distribution along the line of sight \citep[see][for a review]{CMB_weak_lensing_review}.

To produce the projected convergence map, we first compute the two-dimensional projected overdensity field $\delta(\chi, \bm{\theta})$ using the total mass \textsc{HEALPix} output lightcone maps from the FLAMINGO simulation. Here, $\chi$ denotes the comoving radial distance along the line of sight, and $\bm{\theta}$ represents the angular position on the sky. Therefore, each map corresponds to the projected overdensity within a thin shell at $\chi$. We then integrate these overdensity maps (shells) along the line of sight up to $z = 4.5$, weighting them by the CMB lensing kernel to yield the total CMB lensing convergence map. This is expressed as:
\begin{equation}\label{eqn::CMB_lensing_eqn}
\kappa_{\rm CMB}(\bm{\theta}) = \int_0^{\chi(z_{\rm max})} W^{\kappa_{\rm CMB}}(\chi)\delta(\chi,\bm{\theta}), \mathrm{d}\chi.
\end{equation}
In constructing the convergence field we adopt the Born approximation, i.e. the lensing kernel is integrated along unperturbed photon paths. Previous studies have shown that, for CMB lensing and other weak-lensing–dominated probes, corrections from full ray-tracing are typically at the percent level and therefore subdominant for the statistics considered in this work \citep[e.g.][]{referee_ref_BA_ref1,referee_ref_BA_ref2,referee_ref_BA_ref3,referee_ref_BA_ref4}. The CMB lensing kernel is written as:

\begin{equation}\label{eqn::CMB_lensing_kernel}
W^{\kappa_{\rm CMB}}(\chi) = \frac{3}{2} \left(\frac{H_0}{c}\right)^2 \Omega_{\rm m} \frac{\chi}{a(\chi)}\left(1 - \frac{\chi}{\chi_{\rm CMB}}\right),
\end{equation}
with $\chi_{\rm CMB}$ as the comoving distance to the last scattering surface (assumed to be at $z_{\rm CMB} = 1100$), and $a(\chi)$ the scale factor at comoving distance $\chi$. Although a non-negligible fraction of the $\kappa_{\rm CMB}$ signal might be expected from $z > 4.5$, given the broadness of the CMB lensing kernel, integration up to this redshift is sufficient for studying cross-correlations between the convergence field and the other fields we consider in this study (e.g., CIB, tSZ effect), which are the main summary statistics of interest in this study.

A detailed discussion of the $\kappa_{\rm CMB}$ auto power spectrum and the $\kappa_{\rm CMB}$–cosmic shear power spectra can be found in \citet{Ian_low_S8_FMG_paper}. In summary, the $\kappa_{\rm CMB}$ statistics predicted by the FLAMINGO simulation show good convergence with respect to variations in box size, resolution and feedback models. A \textit{Planck}-like cosmology yields excellent agreement between the predicted $\kappa_{\rm CMB}$ auto-power spectrum and measurements from \textit{Planck} \citep{P18_lensing_ref}, SPTpol \citep{SPTpol_lensing_ref}, and ACT DR6 \citep{ACTDR6_lensing_ref}. Under a \textit{Planck}-like cosmology, the predicted amplitude of the $\kappa_{\rm CMB}$–cosmic shear cross-power spectrum is marginally higher than that measured between KiDS-1000 and \textit{Planck} and ACT CMB lensing \citep{KIDS_Planck_ACT_shear_cross}, though this tension is not statistically significant. As we have previously considered the cross correlation between CMB lensing and cosmic shear and CMB lensing and Compton $y$ (see \citealt{FLAMINGO_ref_Joop}), we focus our discussion on lensing--CIB cross-correlation, as outlined in the following sections.

\subsection{Sunyaev-Zel'dovich effect maps}\label{ssec: SZ_effect}

The Sunyaev–Zel’dovich (SZ) effect arises from the interaction between CMB photons, emitted at the last scattering surface, and intervening free electrons \citep[][for a review]{SZ_effecr_ref1,SZ_effecr_ref2,tSZ_review_tony}. This effect can be primarily classified as the thermal SZ (tSZ) effect and the kinetic SZ (kSZ) effect.

The thermal SZ effect results from inverse Compton scattering of CMB photons off hot electrons around and within galaxy groups and clusters. Through this scattering process, photons gain energy, causing a distortion of the energy distribution compared to the original CMB radiation spectrum. This causes a decrease of CMB intensity at frequencies $\lesssim$ 217 GHz and an increase at higher frequencies. The distortion is expressed as
\begin{equation}\label{eqn::delta_tSZ}
    \frac{\Delta T_{\rm tSZ}}{T_{\rm CMB}} = yf(\nu),
\end{equation}
where the frequency dependence $f(\nu)$ is given by $x\textrm{coth}(x/2)-4$ with $x = h\nu/k_{\rm B}T_{\rm CMB}$. $T_{\rm CMB}$ is the CMB temperature and $k_{\rm B}$ is the Boltzmann constant. The magnitude of the distortion is characterised by the Compton parameter $y$, a dimensionless quantity that is related to the integral of gas pressure along the line of sight to the surface of last scattering:
\begin{equation}\label{eqn::y_tsz}
y=\frac{\sigma_{\rm T}}{m_{\rm e}c^2}\int P_{\rm e}\;dl \propto \int n_{\rm e} T_{\rm e}\;dl,   
\end{equation}
where $\sigma_{\rm T}$ is the cross-section for Thomson scattering and $m_{\rm e}c^2$ is the electron rest energy. $P_{\rm e}$ is the electron pressure calculated as $n_{\rm e}k_{\rm B}T_{\rm e}$, with $n_{\rm e}$ and $T_{\rm e}$ being the electron number density and temperature respectively.

The tSZ effect is linearly proportional to the product of gas density and temperature, and it is independent of redshift. These properties make it a powerful probe of the thermal state of the ionised gas over a wide range of spatial and mass scales, including not only massive clusters but also galaxy groups and diffuse components like the circumgalactic and intergalactic medium. Furthermore, the tSZ signal can be cross-correlated with other LSS tracers to provide joint constraints on the thermodynamic and baryonic evolution of the  Universe \citep[e.g.,][]{Vikram_2017_group_SZ,Tilman_KIDS_y_WL,Battaglia_2015_y_lensing,b_pe_constrain_most_precise}.

The kSZ effect, on the other hand, originates from the line-of-sight bulk motion of the ionised gas in the Universe, inducing a Doppler shift in the thermodynamic temperature of the CMB photons. Unlike the tSZ effect, the kSZ effect does not produce a spectral distortion but instead causes a frequency-independent shift in the CMB temperature. This shift is described by
\begin{equation}\label{eqn::kSZ_def}
    \frac{\Delta T_{\rm kSZ}}{T_{\rm CMB}} = -\frac{\sigma_{\rm T}}{c} \int_{\rm los}\frac{d\chi}{1+z} n_{\rm e}(\chi,z) (\bm{v_{\rm e,p}}\cdot\hat{\bm{n}}),
\end{equation}
where $\chi$ is the comoving distance to redshift $z$, $\bm{v_{\rm e,p}}\cdot\hat{\bm{n}}$ is line-of-sight part of the peculiar velocity for the free electron field, and where $\hat{\bm{n}}$ is the line-of-sight unit vector pointing away from the observer. 

All the above equations assume the SZ effect is induced by non-relativistic electrons. However, in galaxy clusters, where electron temperatures can exceed $T_{\rm e} \gtrsim 0.1~ \mathrm{keV}$, relativistic effects become relevant and should be considered \citep[e.g.][]{Challinor_rSZ}. As a result, the SZ spectral distortion becomes temperature, and velocity-dependent, and the standard frequency kernel $f(\nu)$ must be replaced with a more general relativistic kernel. In the case of the thermal SZ effect\footnote{Here, we focus only on the relativistic corrections to the tSZ effect. However, non-negligible corrections are also expected from the kSZ and the tSZ–kSZ cross terms \citep[see e.g.][]{Will_forecast_for_rSZ}. A full discussion of these corrections is left to future work.}, the relativistically corrected fluctuation is given by
\begin{equation}\label{eqn::delta_I_rtSZ}
    \Delta I_{\rm rtSZ} (\mathbf{\hat{n}}, \nu, T_{\rm e}(\mathbf{\hat{n}})) = f^{\rm rel}(\nu, T_{\rm e}(\mathbf{\hat{n}}))y(\mathbf{\hat{n}}),
\end{equation}
where $f^{\rm rel}(\nu, T_{\rm e}(\mathbf{\hat{n}}))$ is the relativistic frequency response function, which depends both on frequency and the local electron temperature. This correction introduces additional angular and spectral anisotropies in SZ maps, and has become an important direction of recent observational and theoretical studies aiming to better constrain the thermal properties of massive clusters \citep[see e.g.][]{rSZ_cluster_temp_ref1,Kay_2024_rSZ,rSZ_cluster_temp_ref2,rSZ_cluster_temp_ref3}.

In our simulations, Compton $y$ parameter maps of the tSZ effect are constructed for each lightcone. When a gas particle crosses the lightcone, we accumulate the following dimensionless quantity
\begin{equation}\label{eqn::FMG_y}
    y = \frac{\sigma_{\rm T}k_{\rm B}}{m_{\rm e}c^{2}}\frac{m_{\rm g}n_{\rm e}T_{\rm e}}{\Omega_{\rm pix}d^{2}_{\rm A}\rho}
\end{equation}
to the map, where $m_{\rm g}$ and $\rho$ are the mass and mass density of the gas particle. $\Omega_{\rm pix}$ is the solid angle of a HEALP\textsc{ix} pixel and $d_{\rm A}$ is the angular diameter distance to the observer. Since the gas particles have associated smoothing lengths, quantities derived from the gas are smoothed onto the HEALP\textsc{ix} maps. A detailed description of the smoothing scheme can be found in \citet{FLAMINGO_ref_Joop}. 

Similarly, smoothed lightcone maps of the dimensionless Doppler parameter $b$ for the kSZ effect are also provided. This is computed as
\begin{equation}\label{eqn::FMG_doppler_b_def}
    b=\frac{n_{\rm e}m_{\rm g}\sigma_{\rm T}v_{\rm r}}{\Omega_{\rm pix}d^{2}_{\rm A}\rho c},
\end{equation}
where $v_{\rm r}$ is the particle’s radial velocity relative to the observer. Under this definition, the mapping between the Doppler parameter is simply given by $\Delta T_{\rm kSZ}/T_{\rm CMB} = -b$, where $T_{\rm CMB}$ is taken as 2.73 K. The stacked SZ map is then produced by summing all HEALP\textsc{ix} lightcone maps from individual shells up to $z = 4.5$.

To account for relativistic corrections to the tSZ map, we construct the $y$-weighted electron temperature map, $\langle T_{\textrm{e}, y}(\mathbf{\hat{n}}) \rangle$, by dividing the stacked gas-temperature–Compton $y$ product map\footnote{This map is not available in the public repository and is regenerated using the Lightcone I/O code, which is available online at \url{https://github.com/jchelly/LightconeIO}} by the stacked $y$ map (e.g., \citealt{rSZ_cluster_temp_ref1,rSZ_cluster_temp_ref2,rSZ_cluster_temp_ref3}). The relativistic frequency response function, $f^{\rm rel}(\nu, \langle T_{\textrm{e}, y}(\mathbf{\hat{n}}) \rangle)$, is computed using the \texttt{COMBO} integration mode of the \textsc{SZpack}\footnote{\textsc{SZpack} is available at: \url{https://github.com/CMBSPEC/SZpack}.} \citep{SZpack_ref1,Szpack_ref2}. This code allows us to obtain precise SZ response functions for all electron temperatures.

\subsection{Anisotropic screening (optical depth $\tau$) maps}\label{ssec::patchy_screening}

The anisotropic, or `patchy', screening effect originates from Thomson scattering of CMB photons by free electrons along the line of sight between the last scattering surface and the observer. This effect provides a novel probe of ionized gas in and around haloes, as well as during the epoch of reionization. Patchy screening effectively acts as a spatially varying screen on the observed CMB sky, damping the primary CMB anisotropies and generating new polarization patterns. Its amplitude is generally much smaller than that of the tSZ and kSZ effects and it scales as
\begin{equation}\label{eqn::patchy_screening_eq}
    \Big(\frac{\Delta T}{T_{\rm CMB}}\Big)_{\rm screening} \approx \tau(\mathbf{\hat{n}})\Big(\frac{\Delta T}{T_{\rm CMB}}\Big)_{\rm primary},
\end{equation}
where $\tau(\mathbf{\hat{n}})$ is the line-of-sight optical depth, determined by the integrated free electron density in the direction $\mathbf{\hat{n}}$ on the sky ($\tau = \int a n_{\rm e}\sigma_{\rm T}~d\chi$).

When combined with the tSZ and kSZ effects, the patchy screening effect has the potential to provide a more complete picture of the electron distribution in and around haloes. Joint analyses can also help disentangle degeneracies between electron density, gas temperature, and galaxy velocities. Several observational attempts have been made to study patchy screening, often through cross-correlations of current high-resolution CMB surveys such as ACT with LSS surveys including unWISE, BOSS, and DESI \citep[e.g.][]{patchy_screening_will,patchy_screening_Boryana}. However, current measurements remain highly challenging, as they are significantly contaminated by other sources of anisotropy in the CMB. Future CMB surveys, such as the Simons Observatory (SO) and CMB-HD, with greater sensitivity and resolution, are expected to reduce these contaminants, which would open the possibility of robust detections of the patchy screening effect and its use as a new probe of gas thermodynamics \citep[e.g.][]{patchy_screening_Schutt,patchy_screening_kramer}.

The screening effect is proportional to the line-of-sight integrated optical depth $\tau(\mathbf{\hat{n}})$, which is simply as $\tau = \Sigma^{N_{\rm shell}}_{i} ~\sigma_{\rm T} ~\mathrm{\Delta\tau}(\mathbf{\hat{n}})$.  We provide a smoothed map of the optical depth, defined as the accumulated contribution per pixel of
\begin{equation}\label{eqn::DM_equation}
\Delta \mathrm{\tau} = \frac{n_{\rm e} m_{\rm g}}{\Omega_{\rm pix} d_{\rm A}^{2} \rho}.
\end{equation}
We generate stacked $\tau$ maps for different model variations in our repository. Meanwhile, a detailed analysis, including cross-correlations with other LSS tracers and the study of relevant statistics, is left to future work (Conley et al., in prep). 

\subsection{Cosmic infrared background maps}\label{ssec::CIB}

The cosmic infrared background (CIB) is primarily sourced by star-forming regions. Light emitted by young, massive stars in star-forming galaxies is absorbed by surrounding dust grains and then gets re-emitted at infrared wavelengths. Such emission builds up over time to form a generally unresolved background, which traces the cumulative infrared output of dusty galaxies across a wide range of redshift, $0<z\lesssim6$.

The CIB is a dominant component at high frequencies ($\nu \gtrsim$ 200 GHz), where it has been measured in detail by surveys such as \textit{IRIS} \citep{IRIS_CIB}, \textit{Herschel} \citep{Hershel_CIB_ref1,Hershel_CIB_ref2} and \textit{Planck} \citep{Pearly_CIB_results,P14_CIB}. At these frequencies, the CIB is often treated as a contaminant that must be carefully removed in order to isolate other signals of interest \citep[e.g.][]{CIB_deproj_ref1,CIB_deproj_ref2, CIB_deproj_ref3}. At lower frequencies, however, the CIB becomes subdominant and is therefore more poorly constrained. Nevertheless, CMB experiments detect and must include models of the emission properties of infrared sources \citep[e.g.][]{ACT_CIB_ref1,SPT_CIB_ref,SPTpol_CIB_ref,ACT_DR6_CIB_ref}. Beyond being a foreground contaminant to CMB studies, the CIB is an astrophysically rich signal in its own right. It directly traces the cosmic star formation rate density and is sensitive to the emission properties of dusty infrared galaxies. Because the CIB traces faint, unresolved galaxies, it offers a unique tool to study the star formation activities in low-mass or high-redshift systems \citep[e.g.][]{viero_CIB_galaxy_pop,SFR_CIB_M18,CIB_tomo_ref2,CIB_tomo_ref1}.  In addition, because it traces galaxies and galaxies trace LSS, the clustering of the CIB (and its cross-correlation with other tracers of LSS) will also be sensitive to the underlying cosmological model. 

However, modelling of the CIB signal is challenging because of significant uncertainties in how both the SFR and spectral energy distribution (SED) of dusty sources potentially vary with galaxy/halo mass, redshift, local environment, and so on. Previous studies based on dark matter-only simulations adopted the halo model formalism to describe the link between SFR (or sometimes dust mass) and galaxy/halo mass and its evolution (e.g., \citealt{Sehgal_2010,Websky_ref,agora_ref}), though the specific choices differed in detail.  Different studies also approached the SED modelling in different ways \citep[e.g.][]{Magdis_CIB_SED,Bethermin_CIB_SED,P14_CIB,P16_CIB_SED_template, M2020_CIB_kappa}. On the observational side, accurate separation of the CIB from Galactic dust foregrounds is non-trivial. In particular, the Galactic cirrus emission, which is often traced by low-$z$ HI surveys, must be subtracted carefully to avoid large-scale biases in the measured CIB signal \citep[e.g. see][]{L19_CIB,Fiona_CIB23}. In this regard, a detailed understanding of the CIB is crucial not only for studying galaxy evolution but also for precise CMB cosmology.

\subsubsection{Three-parameter model}\label{sssec::CIB_default_model}

Since the CIB is strongly correlated with the distribution of star-forming galaxies, for this work, we use the star formation rate (SFR) lightcone output maps from the simulations and convert them into mock CIB maps. We start with the conversion between the SFR and the bolometric luminosity of infrared sources, $L_{\rm bol, \rm IR}$, by assuming that the SFR is proportional to $L_{\rm bol, \rm IR}$ for a dusty galaxy \citep{LIR_SFR_relation}. For a \citet{Chabrier_2003_IMF} stellar initial mass function, as adopted by the FLAMINGO simulation suite, the conversion between $L_{\rm bol, \rm IR}$ and SFR is given by (also see e.g. \citealt{viero_CIB_galaxy_pop,Figueira_SFR_L_conversion})
\begin{equation}\label{eq::kennicutt_law}
\frac{L_{\rm bol, \rm IR}}{1\times10^{10} ~L_{\odot}} = \frac{\textrm{SFR}}{1 ~\rm M_{\odot} \textrm{yr}^{-1}}.
\end{equation}

As discussed earlier, what is observed is not the bolometric luminosity, but rather the luminosity within a specific bandpass. To convert $L_{\rm bol, \rm IR}$ into the IR luminosity at a given frequency, one needs to include the SED of infrared sources, $\Phi(\nu, T_{\rm dust})$. We follow \citet{P16_CIB_SED_template} and express the SED as 
\begin{equation}\label{eqn::SED_galaxies}
  \Phi(\nu, T_{\rm dust}, z) =
    \begin{cases}
      \Big[ \textrm{exp} \big( \frac{h\nu}{k_{\rm B} T_{\rm dust} (z)} \big) -1 \Big]^{-1} \nu^{\beta_{d}+3} & (\nu \leq \nu^{\prime}),\\
     \Big[ \textrm{exp} \big( \frac{h\nu^{\prime}}{k_{\rm B} T_{\rm dust}(z)} \big) -1 \Big]^{-1} \nu^{\prime \beta_{d}+3} \big( \frac{\nu}{\nu^{\prime}} \big)^{-\alpha_{d}} & (\nu > \nu^{\prime}),
    \end{cases}       
\end{equation}
which is modelled as a greybody radiation at low frequencies, with a power law transition at high frequencies. $\beta_{d}$ is a free parameter in the model that correlates with dust properties, and we model the dust temperature as a simple power law in redshift: $T_{\rm dust} = T_{0}(1+z)^{\alpha}$ \citep[as modelled and used in e.g.][]{Viero_13,Websky_ref,CIB_tomo_ref1,SZII_25}, since dust properties are not explicitly modelled in the FLAMINGO simulations. $T_0$ and $\alpha$ are also free parameters. In the regime considered in this work, $z \lesssim 4.5$ and $\nu < 1000$ GHz, the SED is expected to remain a greybody and the power-law cutoff component is not considered in the following calculations (e.g. the $\alpha_{d}$ defined in Equation \ref{eqn::SED_galaxies} is not used in Equation \ref{eqn::L_per_nu}, with the latter is simply an integral over a greybody). The luminosity at a specific frequency is then calculated as
\begin{equation}\label{eqn::L_per_nu}
    L_{\nu, \rm IR} = L_{\rm bol, \rm IR}(\textrm{SFR})\frac{\Phi(\nu, T_{\rm dust})}{\int d\nu\Phi(\nu, T_{\rm dust})}.
\end{equation}
Since we will fit our model to data from the \textit{Planck} survey, we further convolve the SED with a detector bandpass function in the modelling\footnote{The \texttt{HFI-RIMO-R3.00.fits} file available at \url{https://irsa.ipac.caltech.edu/data/Planck/release_3/ancillary-data/HFI_Products.html}}, which is defined as the normalised spectral transmission profile of a particular \textit{Planck} channel. The flux is then calculated based on the source's luminosity, radial comoving distance, and redshift:
\begin{equation}\label{eqn::CIB_flux}
    S_{\nu, \rm IR} = \frac{L_{\nu(1+z), \rm IR}}{4\pi \chi^{2} (1+z)},
\end{equation} 
with the luminosity is calculated in the rest frame. As shown in Equation \ref{eqn::L_per_nu}, we shift both the SED and the bandpass function to the rest frame. 

In summary, the free parameters in our CIB model are: $\beta_{d}$, $T_{0}$, and the redshift dependence of the dust temperature, $\alpha$. We fit the model to the auto-power spectra measured by \citet{L19_CIB} at 353, 545, and 857 GHz. In principle, while varying different sets of parameters to determine the best fit (which we do via Monte Carlo Markov chain optimisation), we need to recalculate the power spectrum for a new set of CIB HEALP\textsc{ix} lightcone maps at each redshift, which can be computationally expensive. To reduce this cost during the fitting procedure, we take advantage of the fact that the SFR maps are independent of the modelled CIB parameters. As a result, any CIB model-dependent terms can be factored out as constant normalisation terms in the final power spectrum calculation. Therefore, for each lightcone shell, we first compute the SFR power spectrum independently of the fitting routine. The modelled stacked $C_{\ell, ~\rm CIB}$ can then be expressed as:
\begin{equation}\label{eqn::C_ell_CIB_eqn}
    C^{\rm stacked}_{\ell, ~\rm CIB} = \Sigma^{N}_{i=1} A^{2}_{i}C^{i,i}_{\ell, ~\rm SFR} + 2\Sigma_{i>j} A_{i}A_{j}C^{i,j}_{\ell, ~\rm SFR},
\end{equation}
where the stacking of lightcone shells is performed up to $z = 4.5$. Here, $C^{i,i}_{\ell, ~\rm SFR}$ denotes the SFR auto-power spectrum for each lightcone shell. Because the redshift intervals of our output maps are relatively narrow at low $z$ ($\Delta z = 0.05$), while diffuse star-forming structures can span a broader redshift range, we also include the cross-shell term $C^{i,j}_{\ell, ~\rm SFR}$ of the SFR power spectrum in the calculation. The factor $A{i}$ encapsulates all the free parameters in our CIB model and is defined from Equations \ref{eq::kennicutt_law} to \ref{eqn::CIB_flux}.

In the fitting, a simple Gaussian likelihood is assumed, with
\begin{equation}\label{eqn::lnlike}
    \ln \mathcal{L}(\Vec{d}|\Vec{\theta}) = -\frac{1}{2} \big[\Vec{d}-\Vec{m}(\Vec{\theta}) \big]^{T} \Sigma^{-1} \big[\Vec{d}-\Vec{m}(\Vec{\theta}) \big]^{T},
\end{equation}
where $\Vec{\theta}=[\beta_{d}, T_{0}, \alpha]$. $\Vec{d}$ is the data vector and $\Vec{m}$ are the modelled powers at each frequency band. The covariance matrix is estimated using a simple Gaussian approach. In the absence of sky masks, this is given by 
\begin{equation}\label{eqn::cov_matrix}
    \Sigma = \delta_{\ell \ell^{\prime}}\frac{1}{(2\ell+1)}\Big[C_{\alpha \gamma}(\ell)C_{\beta \delta}(\ell)+C_{\alpha \delta}(\ell)C_{\beta \gamma}(\ell) \Big],
\end{equation}
where a Kronecker delta $\delta_{\ell \ell^{\prime}}$ indicates that the covariance is diagonal in this simple Gaussian limit. The subscripts $[\alpha, \gamma, \beta, \delta]$ correspond to one of the frequencies 353, 545, or 857 GHz. The $C_{\ell}$s in the covariance matrix are taken as the auto-spectra from the \citet{L19_CIB} CIB maps.  We note that the power spectra in \citet{L19_CIB} are measured from CIB maps where individual bright IR point sources have been masked.  In our standard fitting methodology we do not include any masking of the simulations, however we show in Appendix \ref{appen:D} that applying a realistic flux cut on the simulations has only a very small impact on the predicted power spectra and does not change any of our main results or conclusions.

\begin{figure}
\includegraphics[width=\columnwidth]{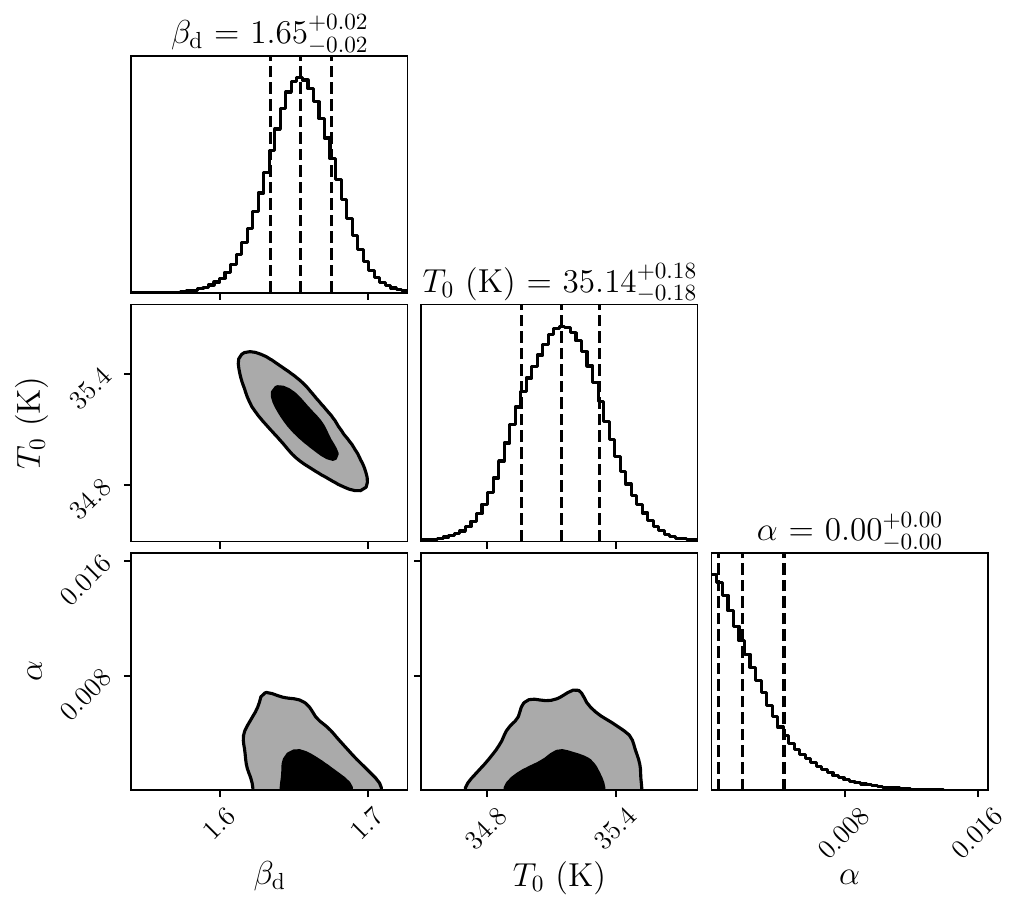}
   \caption{Constraints on the $\beta_{d}$, $T_{0}$, and $\alpha$ parameters of the SED of the CIB model, obtained by fitting to the 353/545/857 GHz auto-power spectrum measurements from \citet{L19_CIB}.}
\label{lensed_CIB_mcmc_L2800N5040_three_params}
\end{figure}

Figure \ref{lensed_CIB_mcmc_L2800N5040_three_params} shows the best-fitting values of $\beta_{d}$, $T_{0}$, and $\alpha$, which are used to generate the mock maps and compute the power spectrum. We find $\beta_{d} = 1.65 \pm 0.02$ and $T_{0} = 35.14 \pm 0.18$ K for the fiducial FLAMINGO model (we discuss how these parameters may vary as a function of feedback and cosmology in Section \ref{sssec::feedback_dependencies_CIB_SED_varied}), where the values are quite close to those adopted in the \citet{P16_CIB_SED_template} study. As indicated by the $\alpha$ parameter, our best-fitting model prefers weak to no redshift evolution of the dust temperature. 

\subsubsection{Extended models}

In the three-parameter model discussed above, the bolometric IR luminosity is specified by the SFR only and the mapping is precisely specified using the \citet{LIR_SFR_relation} relation.  However, in reality this relation has uncertainties and may depend on other properties such as dust mass and galaxy stellar mass. To examine how sensitive our results are to these possible effects, we explore two alternative cases. For the first case, we allow the constant of proportionality in the \citet{LIR_SFR_relation} relation to vary:
\begin{equation}\label{eqn::kenicutt_law_propor_varied}
\frac{L_{\rm bol, \rm IR}}{1\times10^{10+N_{\rm s}} ~L_{\odot}} = \frac{\textrm{SFR}}{1 ~\rm M_{\odot} \textrm{yr}^{-1}},
\end{equation}

In the second case, we adopt an alternative parametric form for modelling the infrared flux density, motivated by e.g. \citet{C13, L21, C23, K25}. Analogous to those studies, we introduce a non-linear dependence on both SFR and dust mass in the computation of the infrared luminosity:
\begin{equation}\label{eqn::new_L_IR_eq}
    L_{\rm bol, \rm IR} = 10^{10+a}L_{\odot}\Big(\frac{\rm SFR}{100~\rm M_{\odot}\textrm{yr}^{-1}}\Big)^{b}\Big(\frac{M_{\rm dust}}{10^{8}~\rm M_{\odot}}\Big)^{c}.
\end{equation}
Due to strong parameter degeneracies in this model, we fix the indices $b$ and $c$ to 0.5, since they do not vary significantly across different simulation setups \citep[e.g., see Table 2 in][]{K25}. For the $M_{\rm dust}$ calculation, we assume a fixed dust-to-metal mass ratio (DTM) of 0.4 (e.g. see measurements from \citealt{De_Cia_13_DTM,De_Cia_16_DTM,Wiseman_17_DTM} and predictions from \citealt{Qi_19_DTM_simba})\footnote{Note that the exact choice of DTM is inconsequential, since the uncertainty of this constant factor is absorbed into the free amplitude in Equation \ref{eqn::new_L_IR_eq}.}. Since metal mass lightcone maps are not available for the fiducial $(2.8~\rm Gpc)^{3}$ run, we construct our own HEALP\textsc{ix} maps using halo lightcone outputs from the fiducial $(1~\rm Gpc)^{3}$ run up to $z = 3.0$. We then apply a normalisation to account for the redshift evolution of the curve to $z = 4.5$, as described in Section \ref{ssec: lightcone_generation}. While computing the metal mass per halo, we consider either only the cold star-forming gas, or all gas within a 3D aperture of 30 kpc. Note that for both of these models, we keep the rest of the CIB modelling, such as the SED of infrared galaxies, unchanged.

In the following discussion, the CIB model with [$\beta_{d}$, $T_{0}$, $\alpha$] as free parameters is treated as our default case (denoted the \textit{three-parameter} fit case). The corresponding best-fitting CIB power spectra are shown later in Figure \ref{CIB_CIB_stats}. The version with a varied proportionality is referred to as the \textit{four-parameter} fit case, and the model with both SFR and $M_{\rm dust}$ dependencies in $L_{\rm bol, IR}$ is denoted as the \textit{extended} model (with three free parameters: $\beta_{d}$, $T_{0}$ and the amplitude $a$ as defined in Equation \ref{eqn::new_L_IR_eq}). The results of these alternative models are provided and discussed in Appendix \ref{appen:A}. 
In general, the resulting curves are similar in all three cases, although the best-fitting parameters vary across models. 

It is worth noting that more complexity can also be introduced into the SED template, for example when modelling the dust temperature and its redshift evolution. In this work, we adopt a simple power law relation; however, one could in principle trace the dust properties directly in hydrodynamical simulations if explicit dust modelling is included \citep[e.g. in \simba and \colibre, see ][]{simba_sim_ref,COLIBRE_ref}, or model them as functions of galaxy properties which could be fitted to the data. Since our simplified model already provides a good fit to the data we compare to, we leave such analyses to future work.

\subsection{Radio point sources}\label{ssec::Radio_Point_Sources}

Radio point sources make a non-negligible contribution to the CMB anisotropy signal at low frequencies, typically for $\nu \lesssim 150~\rm GHz$.  Individually detected bright sources are often masked out prior to component separation. If such sources are strongly correlated with the signal of interest, however, there is a risk that simply masking these radio sources will also mask the main contribution to the signal of interest. For example, if most massive local clusters have bright radio sources that are masked, the contribution of these clusters to the measured tSZ power spectrum may be underestimated, depending on the angular size of the mask and what compensation (if any) is made for the excluded area. Aside from individually detected bright sources, we also expect there to be many more undetected radio sources that could contribute significantly to the measurements and this contribution therefore needs to be modelled.

In the simulations, black holes are characterised solely by their masses and accretion rates (and positions and velocities).  As there is no explicit modelling of the accretion disk or of the scales near it, the simulations cannot predict which black holes ought to correspond to radio-loud AGN (as opposed to radiatively efficient, radio-quiet quasars, for example).  We therefore must identify potential radio AGN based on properties that are followed and then assign each radio AGN an appropriate luminosity.  

To identify potential radio AGN sources, we use the ratio of the bolometric luminosity to the derived Eddington luminosity for each black hole. This is motivated based on both theoretical and observational studies \citep[e.g. see][]{Best_Heckman_12,Fabian_2012,Heckman_Best_14}, which show that radiatively-inefficient radio-loud AGN are much more likely to have low Eddington ratios, typically below a few percent.  In more detail, for each black hole with a subgrid mass greater than $10^{7} ~\rm M_{\odot}$, we compute its bolometric luminosity using $L_{\rm bol} = \epsilon_{r} (1-\epsilon_{f})\dot M_{\rm BH}c^{2}$, where $\dot M_{\rm BH}$ is the accretion rate of each black hole particle, $\epsilon_{r}$ is the fraction of accreted rest-mass energy converted into radiation and $(1-\epsilon_{f})$ accounts for the fraction of radiated energy not coupled to feedback\footnote{Note that the inclusion of $(1-\epsilon_{f})$ factor implies that radiation that is coupled to the gas (i.e., as feedback) is not observable. Its inclusion may not be appropriate for comparisons to observations that have been explicitly corrected for intrinsic absorption.  We include the factor, but its inclusion has no practical consequences for our results as it does not change the rank ordering of black hole luminosities, and our results are also insensitive to small variations in the Eddington ratio cut.}. The FLAMINGO simulations set $\epsilon_{r}$ and $\epsilon_{f}$ as 0.1 and 0.15, respectively.  We apply the aforementioned $10^{7} ~\rm M_{\odot}$ mass cut to include only physically meaningful black holes in the analysis, which is far from the black hole seed mass, i.e. the mass scale at which it reaches the black hole–halo mass relation (see Fig. 1 in \citealt{Emiliy_paper}).  We explore the sensitivity of our results to the specific Eddington ratio cut applied. Specifically, we consider three cuts in Eddington ratio, $\lambda_{\rm Edd} \equiv L_{\rm bol}/L_{\rm Edd}$: $\lambda_{\rm Edd}<10^{-2}, 10^{-3} ~\textrm{and}~10^{-6}$.

After applying an Eddington ratio cut, we proceed to assign radio luminosities to each selected black hole.  As radio point source statistics are poorly constrained at high frequency bands in CMB surveys due to their low number density, our approach is to assign radio fluxes to each selected black hole based on deep, well-measured luminosity functions at low frequencies observed with LOFAR. These fluxes are then extrapolated to higher frequencies using a simple power law spectral model (as expected for synchrotron radiation), with free parameters calibrated against the source number count curves from other CMB surveys. 

\begin{figure}
\includegraphics[width=\columnwidth]{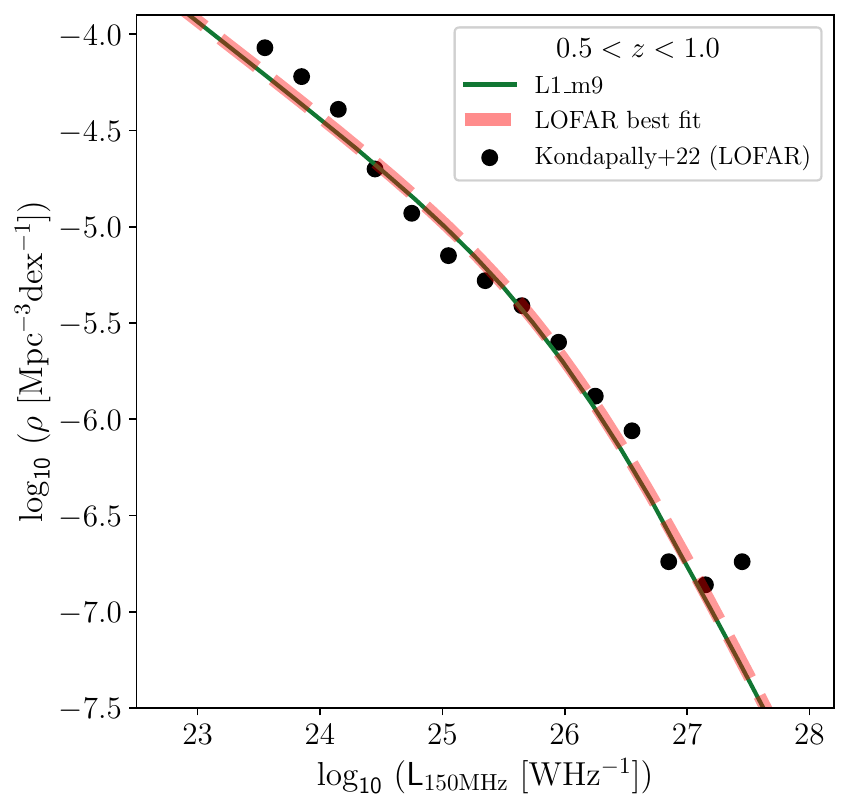}
   \caption{The 150 MHz radio luminosity function (RLF) reconstructed from the black hole particle lightcone in the fiducial $(1~\mathrm{Gpc})^{3}$ run. This is an example showing the case for a black hole selection of $\lambda_{\rm Edd} < 10^{-2}$ over the redshift range $0.5 < z < 1.0$. The observed RLF from the LOFAR survey within the same redshift interval (black points), as well as its best-fitting parametric model (red thick dashed line), are overplotted for comparison \citep[see][]{LOFAR_RLF}.  Abundance matching is used to map from the bolometric luminosity function of black holes in the simulation to the observed RLF (see text).}
\label{AM_RLF_example_0_point_01_case}
\end{figure}

More specifically, for each of the $\lambda_{\rm Edd}$ cases, we rank order the selected black holes based on their $L_{\rm bol}$ values\footnote{We note that the FLAMINGO simulations do not reproduce the observed evolution of the quasar luminosity function particularly well \citep[see][]{QLF_shortage_of_bright_quasar}, suggesting that the bolometric luminosities of the simulated BHs may not be realistic.  However, this should not be an issue for our purposes since only the luminosity \textit{ranking} matters in the abundance matching when assigning radio luminosities to the simulated BH samples.}.  We then re-assign the radio luminosity of each black hole by abundance matching to the radio luminosity function (RLF) of low-excitation radio galaxies (LERG) measured at 150 MHz from the LOFAR survey \citep{LOFAR_RLF}. We select these measurements because they are less uncertain (when compared to the RLF of high-excitation radio galaxies), and the parametric fits as a function of redshift (up to $z = 2.5$) are well calibrated against observations\footnote{It is worth noting that it is now possible to explore the radio properties of AGNs at higher redshifts \citep[e.g. at $z\sim5-6$ from the James Webb Space Telescope (JWST) AGN samples,][]{JWST_radio_properties}. We will investigate the cross-correlation between CMB secondary anisotropies and radio emission at higher redshifts in future work.}.  Abundance matching is done by computing the cumulative number density as a function of luminosity from the simulation and comparing it to the observed cumulative number density curve at 150 MHz, ranked by luminosity. An example of the reconstructed RLF from our simulation is shown in Figure \ref{AM_RLF_example_0_point_01_case}. The fluxes are then computed as 
    \begin{equation}\label{eqn::fluc_radio_cal}
        S_{\rm 150~MHz} = \frac{L_{150 ~\rm MHz}}{4\pi D^{2}_{L}(z)}(1+z)^{1+\alpha},
    \end{equation}
    with $\alpha$ assumed to be $-0.7$ \citep[as adopted in e.g.][and references therein]{LOFAR_RLF}.

\begin{figure*}
    \begin{minipage}[b]{1.0\textwidth}
        \centering
        \includegraphics[width=0.9\linewidth]{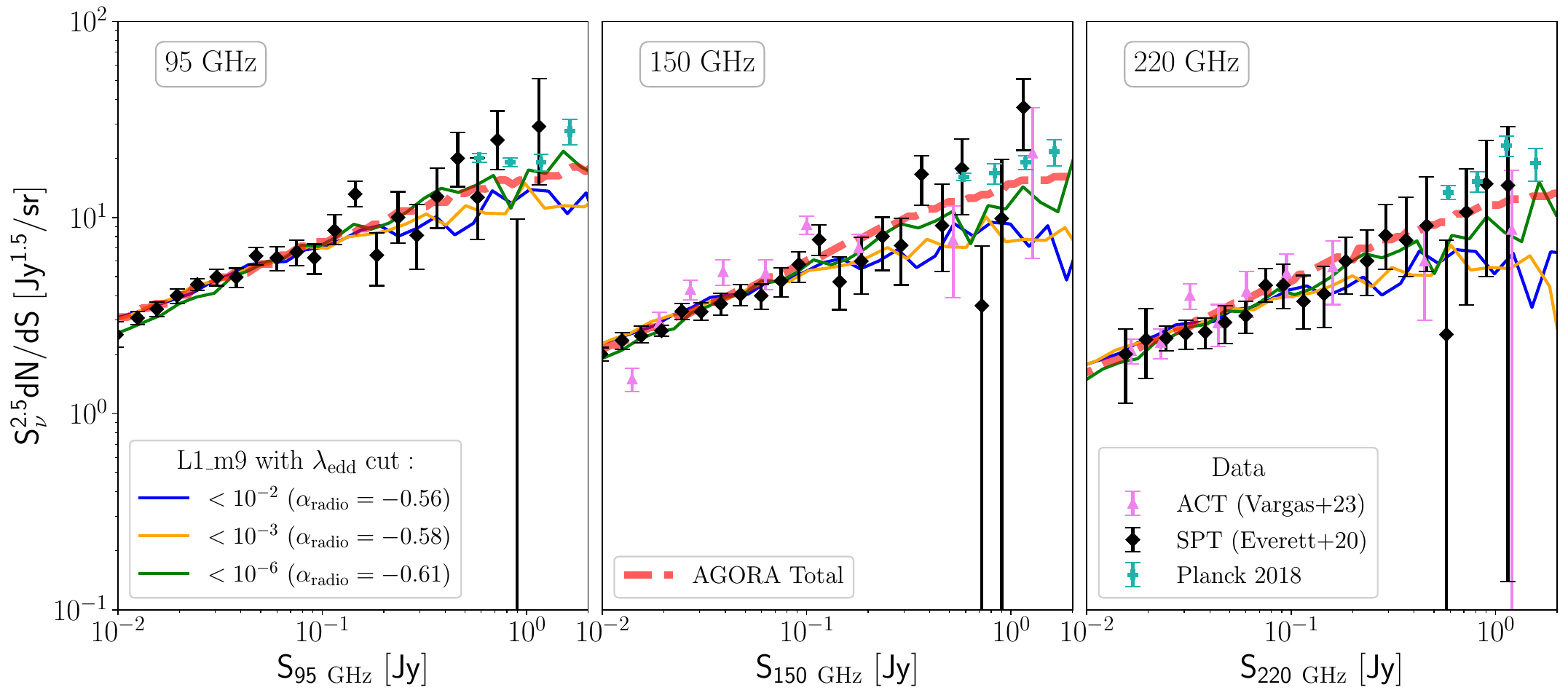}
    \end{minipage}%
    \vspace{+0.1cm}
\caption{Comparison of the differential source number counts between observations and simulations. The blue, orange, and green solid curves are the best-fitting source count models for the three black hole selections considered in this study (corresponding to different Eddington rate cuts), with the best-fitting $\alpha_{\rm radio}$ values obtained by fitting to the measured SPT data at three frequencies \citep{Everett_SPT} (black diamonds) in the flux range $S_{\nu} = 10^{-2}$–$10^{-1}$ Jy. Other observations from the ACT \citep{Vargas_ACT} and \textit{Planck} \citep{P18_radio} are shown as violet triangles and light green crosses respectively. Results from the AGORA simulation are shown as a red thick dashed line for comparison.  Overall, the simulations reproduce the number counts well.}
\label{radio_source_num_count}
\end{figure*}

From the 150 MHz fluxes we extend to other CMB frequencies using a simple power law, modelled as $S_{\nu}=S_{\rm 150~MHz}~ (\nu/150 ~\rm MHz)^{\alpha_{\rm radio}}$, where $\alpha_{\rm radio}$ is obtained by jointly fitting our curve to the measured differential source number counts from the SPT survey at 95, 150, and 220 GHz \citep{Everett_SPT}. Specifically, we fit to the range $10^{-2}$–$10^{-1}~\rm{Jy}$, since the shape of the bright end depends strongly on the particular $\lambda_{\rm Edd}$ cut and is also affected by large uncertainties in the measured data. Figure \ref{radio_source_num_count} shows a comparison between our recovered differential source number counts, $S^{2.5} dN/dS$, at the three frequencies and the SPT measurements. We also include the measurements from ACT and \textit{Planck}, as well as the predictions from the AGORA simulation, in the same figure. We find good agreement between our recovered curves and the other measurements over the fitted range. We find that the best-fitting $\alpha_{\rm radio}$ values are not very sensitive to the particular frequency we fit to.  For example, the best-fitting values are -0.56, -0.58 and -0.61 when fitting independently to the 95, 150, and 220 GHz SPT measurements, respectively, as opposed to our default method of jointly fitting all three data sets. 

\subsection{Lensing of observables}\label{ssec::Lensing_of_observables}

Large-scale structure not only acts as a lens for the primary CMB, it will also lens any CMB secondary sources more distant than the lens.  To account for the lensing effects on each mock CMB secondary anisotropy map, we use \textsc {pixell}\footnote{\url{https://github.com/simonsobs/pixell}} to deflect the background maps. Observables in the lightcone are lensed shell by shell, using the integrated $\kappa$-map up to the shell of interest. For radio point sources, we add the lensing effect to each object before projecting them onto 2D maps. The lensed flux density at the undeflected positions is given by $S^{\rm lens} = |\mu| S^{\rm unlensed}$, where $\mu$ is the magnification map. This map is constructed based on 
\begin{equation}
    \mu = \frac{1}{(1-\kappa)^{2}-|\gamma|^{2}},
\end{equation}
where we use the relationship \citep[see e.g.][]{kappa_phi_gamma_alm_conversion}
\begin{align*} 
\kappa_{\ell} &= \frac{\ell(\ell+1)}{2}\Phi_{\ell}, \\ 
\gamma_{\ell} &=\frac{1}{2}\sqrt{\frac{(\ell+2)!}{(\ell-2)!}}\Phi_{\ell}
\end{align*}
to convert the convergence $\kappa$ map into the shear $\gamma$ map. 

\begin{figure}
\includegraphics[width=\columnwidth]{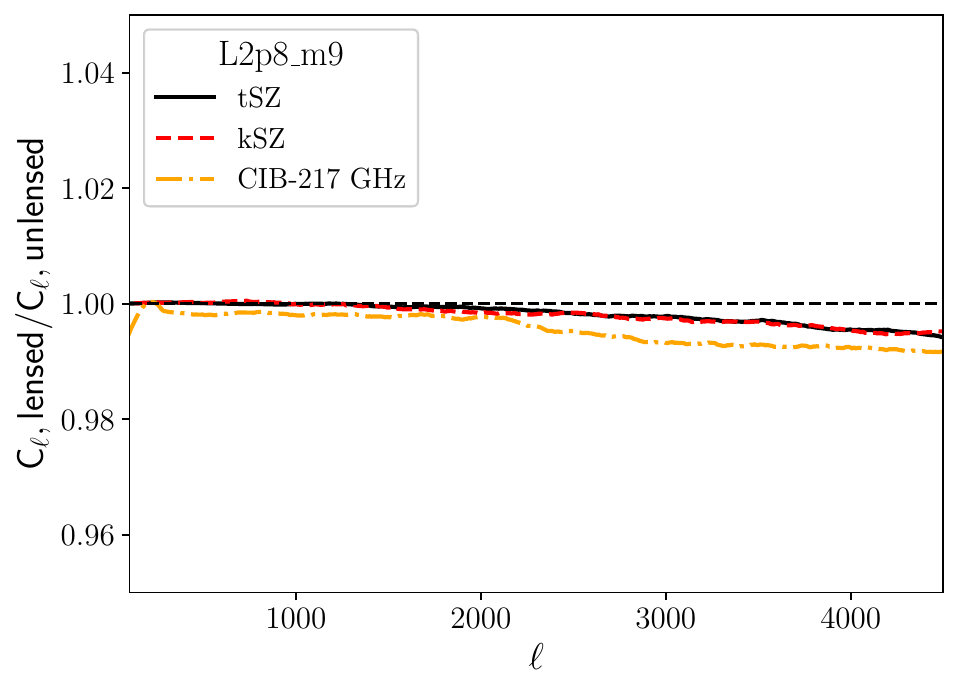}
   \caption{Comparison of the lensed and unlensed auto-power spectra of different LSS tracers from the fiducial $(2.8~\rm Gpc)^3$ run. All curves are obtained by averaging over 8 independent lightcones. A 1-$2\%$ suppression of power at small scales is expected from foreground lensing effects.}
\label{Cly_lensed_unlensed_comp}
\end{figure}

Figure \ref{Cly_lensed_unlensed_comp} shows an example of the impact of this lensing effect on the resulting power spectrum analysis, where we show the tSZ, kSZ and 217 GHz CIB auto-power spectrum as an illustration. In general, the lensing-induced modifications are small, where a suppression of power by around 1-$2\%$ is expected at small scales. The lensing effect on the CIB map at 217 GHz is slightly stronger than that on the SZ effect, due to its broader redshift kernel which has more overlap with the CMB lensing kernel. Although this is a relatively small effect, it may become important for future high-precision cosmological analyses.

\section{Maps and power spectra}\label{sec::Maps_and_power_spectra}

In this section, we present the mock maps and the resulting power spectra predicted by the FLAMINGO simulations, and we compare them with other multicomponent CMB simulations and observational data. 

\subsection{tSZ and kSZ auto-spectra}\label{ssec:tSZ_kSZ_auto}

\begin{figure*}
    \begin{minipage}[b]{1.0\textwidth}
        \centering
        \includegraphics[width=0.3\linewidth]{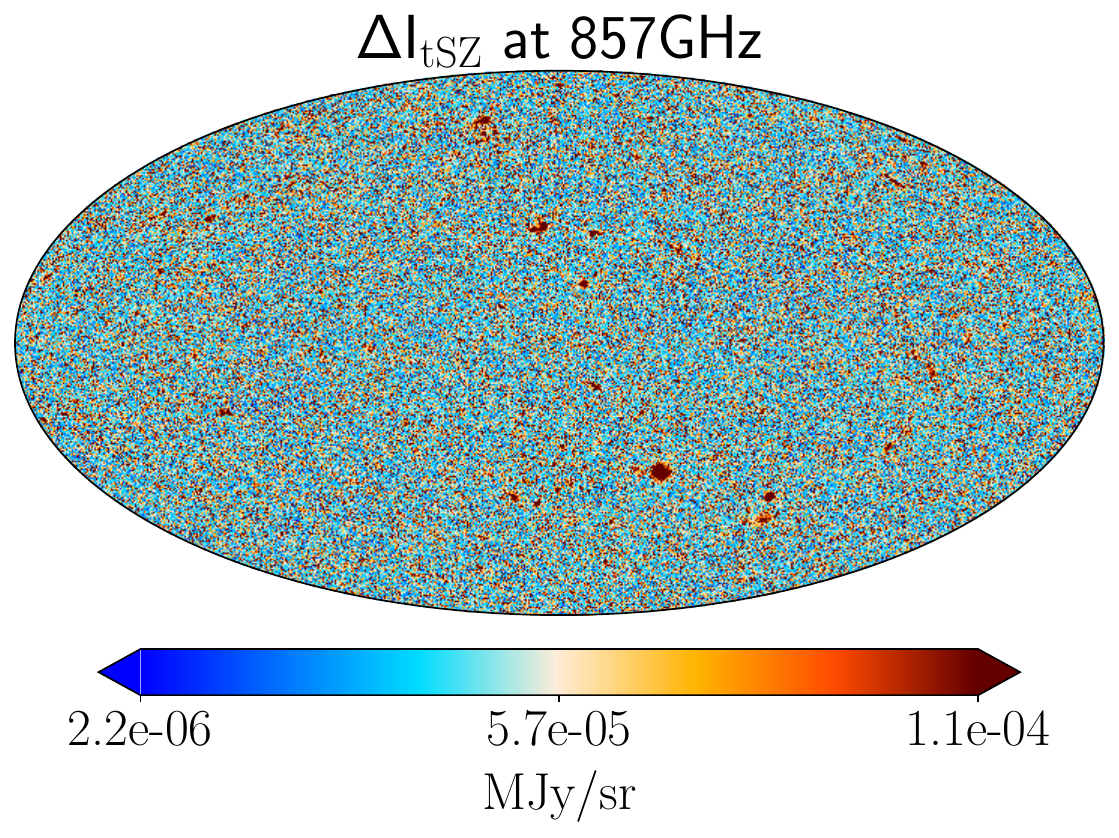}
        \includegraphics[width=0.3\linewidth]{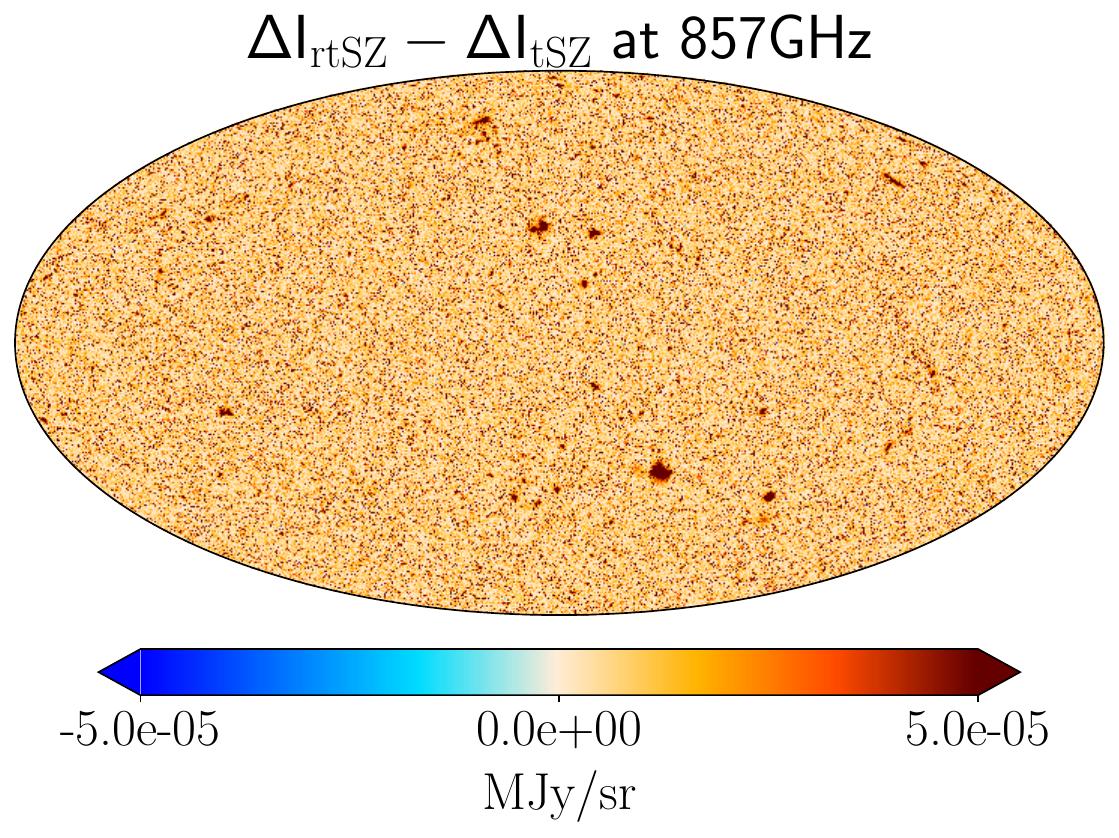}
        \includegraphics[width=0.3\linewidth]{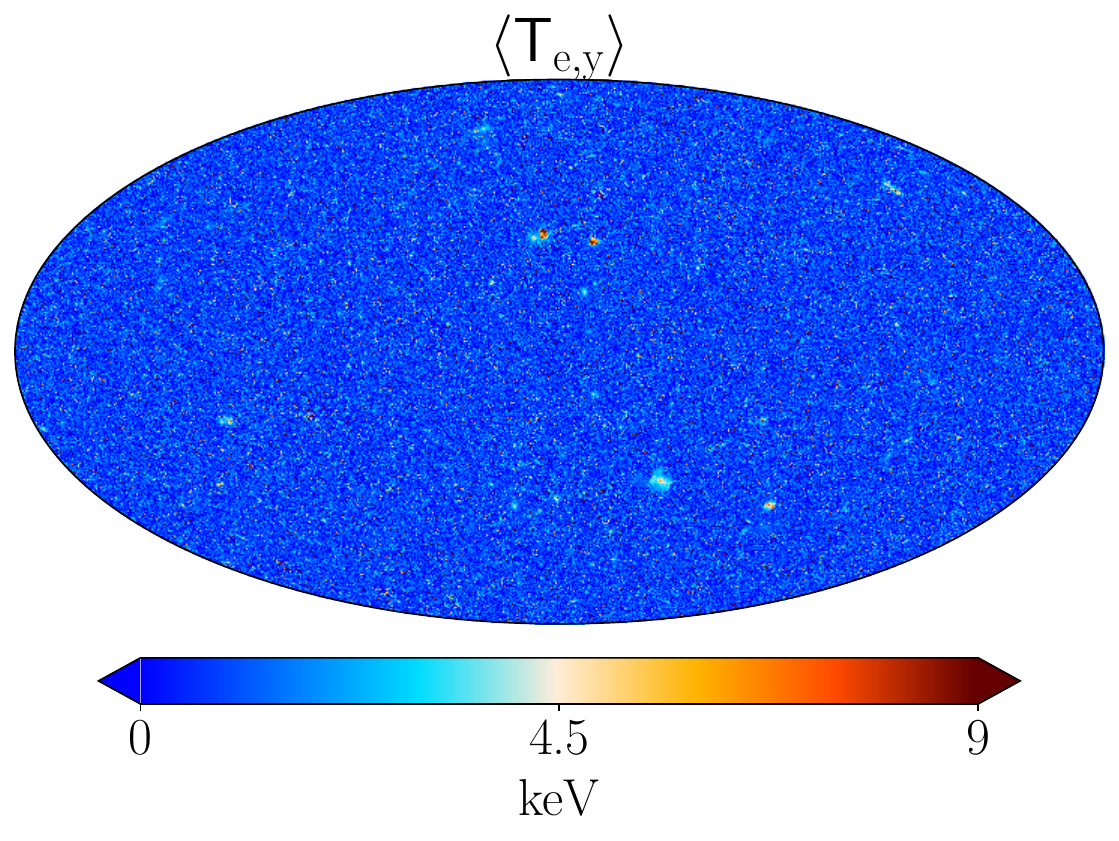}
    \end{minipage}%
    \vspace{+0.1cm}
\caption{Full-sky mock intensity maps of the lensed thermal Sunyaev–Zel’dovich (tSZ) effect and its relativistic correction. These maps are generated from the lightcone outputs of the fiducial $(2.8 ~\rm Gpc)^{3}$ run integrated up to $z = 4.5$. \textit{Left}: tSZ intensity map at 857 GHz, computed using Equation \ref{eqn::delta_tSZ} (i.e., non-relativistic tSZ). \textit{Middle}: Difference map between the relativistically corrected and non-relativistic tSZ intensity maps at the same frequency, with the relativistic corrections applied using  \textsc{SZpack}. \textit{Right}: the corresponding $y$-weighted temperature map. Note the large differences in the color bar ranges across the three maps.}
\label{intensity_map}
\end{figure*}

The left panel of Figure \ref{intensity_map} shows an illustration of the mock, non-relativistic, lensed tSZ intensity map at 857 GHz. This map is created by stacking lightcone data from the fiducial $(2.8 ~\rm Gpc)^{3}$ run up to redshift $z = 4.5$. The middle panel is the corresponding difference map between the relativistically corrected tSZ intensities and the non-relativistic anisotropy map. These relativistic corrections are computed using the \textsc{SZpack}, as described in Section \ref{ssec: SZ_effect}. The rightmost panel shows the $y$-weighted temperature map for comparison. It is visually evident that the relativistic corrections to the tSZ anisotropies are significantly enhanced in regions containing massive clusters (red regions in the rightmost panel) owing to their high temperatures.  This suggests that neglecting relativistic effects in these regions could introduce appreciable biases, especially when extracting the tSZ auto-power spectrum from the observed temperature maps \citep[see e.g.][]{relativistic_tSZ_auto_powersp}. 

\begin{figure}
\includegraphics[width=\columnwidth]{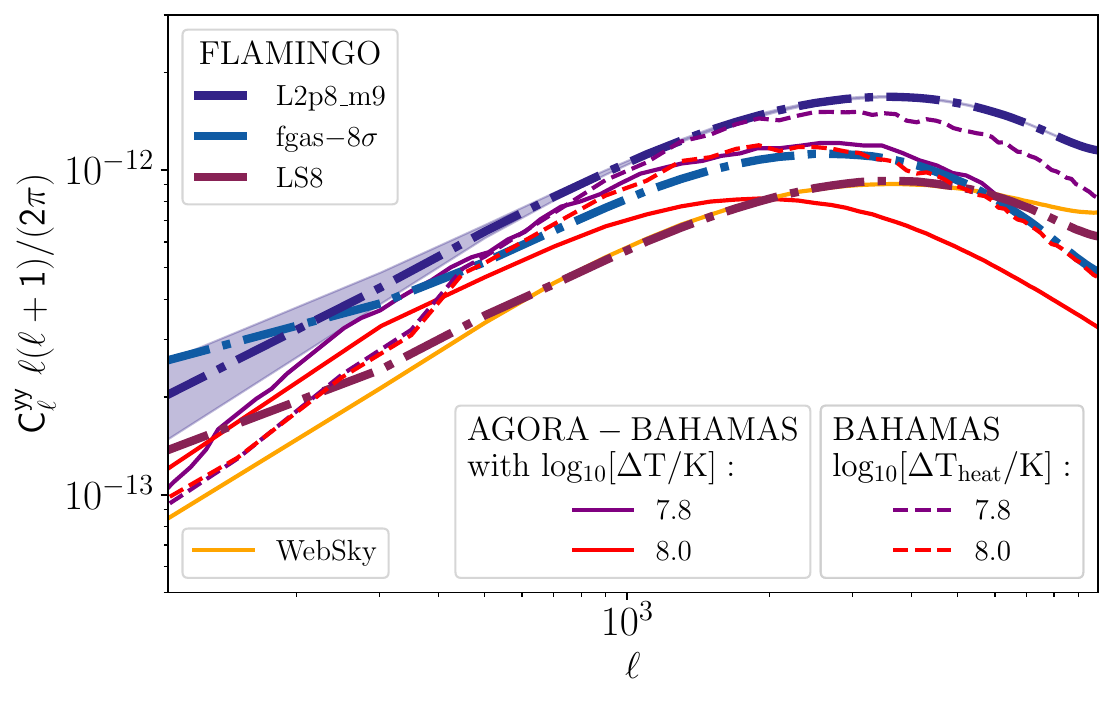}
   \caption{Comparison of the tSZ angular power spectrum from different simulations. Thick dash-dot lines in different colors show results from the FLAMINGO $(2.8 ~\rm Gpc)^{3}$ fiducial run and a subset of its $(1 ~\rm Gpc)^{3}$ model variants, integrated up to $z = 3.0$. The shaded region shows the spread obtained by averaging over 8 independent lightcones. The orange line shows the spectrum from the WebSky simulation. Solid lines are the spectra computed from the AGORA tSZ mock maps, calibrated against two BAHAMAS AGN models with heating temperatures of $10^{7.8}$ and $10^{8.0}~\rm K$ (the latter is the default AGORA gas model) using the HMx code of \citet{Mead_2020}. Note that AGORA does not model the full gas distribution outside of haloes. For comparison, dashed lines show the corresponding results from the BAHAMAS simulations using the same AGN models, where the BAHAMAS maps are constructed in a lightcone with a field of view of 5 square degrees.}
\label{yy_auto_sims_comp}
\end{figure}

In Figure \ref{yy_auto_sims_comp}, we present the comparison of tSZ ($y-y$) auto-power spectra computed from our mock maps and other simulations. For this comparison the power spectra are derived from unlensed maps without relativistic corrections, for consistency with the previous studies we compare to. Note that a detailed comparison between the FLAMINGO tSZ auto-power spectrum and observational data can be found in \citet{Ian_low_S8_FMG_paper,Ian_kSZ_feedback_FLAMINGO}; here, we mainly focus on comparing different simulation results.

Among the FLAMINGO model variants, the LS8 run yields the lowest power amplitude, due to the strong $\sigma_{8}$ dependence of the tSZ power spectrum. And, as expected, the strongest AGN model (fgas-$8\sigma$) shows the greatest suppression of power on small angular scales, due to gas explusion from groups and low-mass clusters. The two BAHAMAS curves show a similar trend and amplitude to the FLAMINGO fiducial and fgas-$8\sigma$ curves, which is perhaps unsurprising given the similar approaches to feedback calibration of the two suites of simulations.  Note that the differences between the FLAMINGO fiducial and BAHAMAS curves on large-scales ($\ell \la 300$) are likely due to a combination of factors, including cosmic variance, differences in cosmology, and box size limitations for BAHAMAS \citep[also see discussions in][]{Matthieu_24}. The shaded region shows the spread in power spectra for the large FLAMINGO 2.8 Gpc using 8 independent lightcones. 

The WebSky simulation shows a trend similar to that in the fiducial case but with a lower power amplitude, in spite of the WebSky cosmology being much more similar to the fiducial FLAMINGO cosmology rather than LS8.  It is unclear what is driving this difference.  An interesting comparison comes from the AGORA and BAHAMAS results. In principle, AGORA adopted the bound gas profile from the HMx code \citep{Mead_2020}, which was calibrated on the BAHAMAS simulations. However, AGORA produces significantly less tSZ power than BAHAMAS at small scales.  \citet{Mead_2020} cautioned against using HMx for tSZ effect power spectrum analyses given the relatively poor fit of the model to the pressure power spectrum of BAHAMAS. 
In addition, AGORA does not model gas outside of haloes, apart from gas that was ejected from the haloes (the typical effective range of the feedback effect is $\approx2-3~r_{200}$). The diffuse electron distribution in the intergalactic medium is therefore likely to be underestimated. Together, these factors plausibly explain the lower tSZ power predicted by AGORA compared to BAHAMAS and FLAMINGO. 

\begin{table}
    \centering
    \caption{tSZ monopole measured from observations and from different FLAMINGO model variations. For each model, the predicted monopole is estimated by averaging the mock full-sky Compton $y$ map integrated up to $z=3.0$.}
    \begin{tabular}{cc}
    \toprule
      Observations   & $\left<y\right>$ \\
      \midrule
        COBE-FIRAS \citep{COBE_FIRAS} & < $15\times10^{-6}$\\
        COBE-FIRAS re-analysis \citep{new_FIRAS_measure} & < $5.2\times10^{-6}$\\
        \textit{Planck} \citep{tSZ_tomo_Chiang} & $1.22^{+0.23}_{-0.17} \times10^{-6}$\\
    \midrule
       FLAMINGO models   & $\left<y\right>$ \\
       \midrule
       Fiducial&$(1.52\pm0.01)\times10^{-6}$\\
       $M_{\ast}-\sigma$&$1.59\times10^{-6}$\\
       $f_{\rm gas}+ 2\sigma$&$1.46\times10^{-6}$\\
       $f_{\rm gas}- 8\sigma$&$1.75\times10^{-6}$\\
       Jet&$1.40\times10^{-6}$\\
       Planck&$1.57\times10^{-6}$\\
       LS8&$1.17\times10^{-6}$\\
     \bottomrule
    \end{tabular}
    \label{tab:y_monopole_table}
\end{table}

In principle the tSZ monopole can also be used as a test of cosmology and astrophysics \citep[e.g.,][]{Hill2015, Thiele2022}.  However, measurements of the monopole are challenging. In fact, the only experiment to date that has been able to place a \textit{direct} constraint on $\left<y\right>$, from spectral distortion measurements of the primary CMB, is COBE-FIRAS \citep{COBE_FIRAS}, which placed an upper limit on the mean $y$ of the Universe.  A very recent re-analysis of the COBE-FIRAS data using an improved astrophysical foreground cleaning technique has reduced this upper limit by approximately a factor of three \citep{new_FIRAS_measure}.  Future CMB spectrometers, such as BISOU \citep{BISOU}, TMS \citep{TMS}, COSMO \citep{COSMO}, and FOSSIL \citep{fossil_link}, will be able to place much stronger direct constraints on the mean $y$-distortion. Note that the monopole from haloes (which is expected to dominate the overall $y$ signal) can be indirectly inferred from experiments such as \textit{Planck}, by cross-correlating Compton $y$ maps with tracers of large-scale structure (e.g., galaxies, quasars) in a tomographic way, allowing one to constrain a bias-weighted estimate of $dy/dz$.  Modelling the bias of the tracers and integrating over redshift allows one to infer $\left<y\right>$ (e.g., \citealt{tSZ_tomo_Chiang}).

Table \ref{tab:y_monopole_table} compares the observed tSZ monopole constraints from COBE-FIRAS and \textit{Planck} with predictions from the various FLAMINGO model variations. The $\left<y\right>$ value for each model is obtained by averaging the Compton $y$ map integrated up to $z = 3.0$. The uncertainty value quoted after the fiducial run result is obtained by averaging over 8 different observers from the $(2.8~\rm Gpc)^{3}$ box. All runs are easily compatible with the 95$\%$ upper limit provided by COBE-FIRAS, including the more stringent re-analysis by \citet{new_FIRAS_measure}. For the fiducial run, the result is slightly higher than the model-dependent tomographic measurement from \textit{Planck}. The strongest AGN model ($f_{\rm gas}- 8\sigma$ run) yields the largest $y$-monopole due to the enhanced thermal energy of the expelled gas, yielding a $\left<y\right>$ that is approximately $2.3\sigma$ larger than derived by \citet{tSZ_tomo_Chiang}. The LS8 model predicts the lowest value and, interestingly, is in excellent agreement with the \textit{Planck} mean value.  This is consistent with the findings of \citet{Ian_low_S8_FMG_paper}, who shown that the measured tSZ power spectrum (\citealt{Bolliet2018}, but see \citealt{CIB_deproj_ref3} for a discussion of the uncertainties in the power spectrum measurements) has a lower amplitude than expected for a \textit{Planck} primary CMB cosmology, while a LS8 cosmology was in relatively good agreement with the measurements on large scales.

\begin{figure}
\includegraphics[width=\columnwidth]{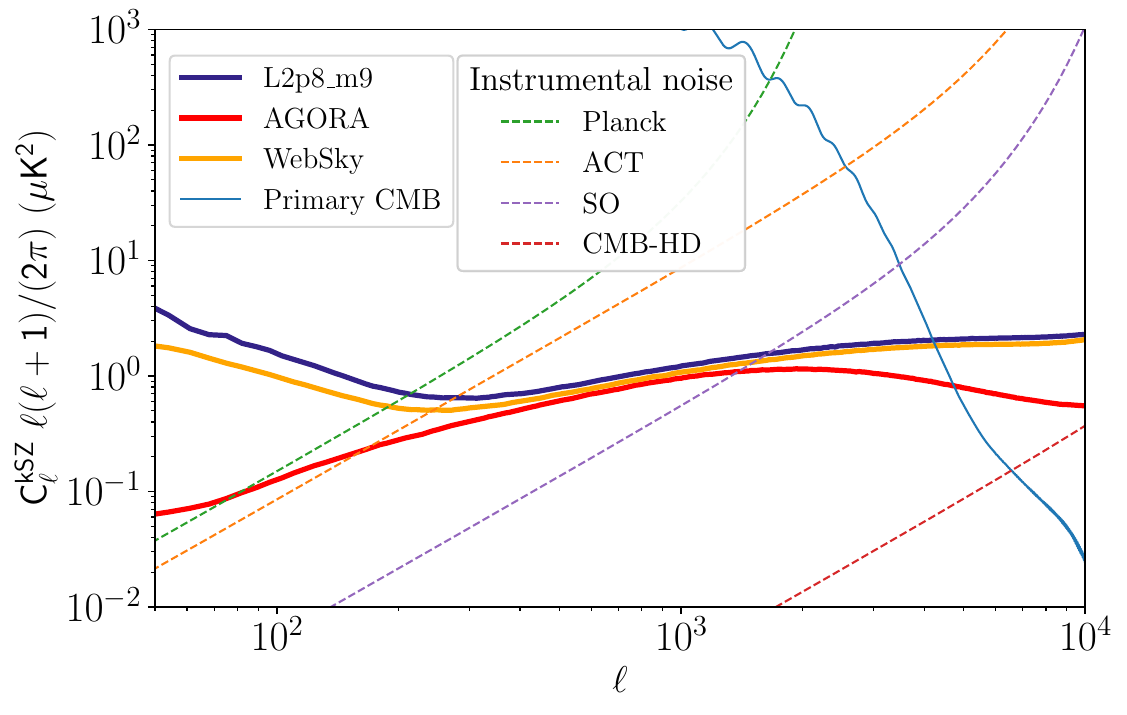}
   \caption{Comparison of the kSZ auto-power spectra from different simulations. The thick dark line shows the result from the FLAMINGO fiducial $(2.8~\rm Gpc)^3$ run, integrated up to $z = 4.5$. The thick red line shows the AGORA result up to $z = 3.0$, and the thick orange line shows the WebSky result up to $z = 4.5$. For reference, the primary CMB power spectrum (thin blue line) and the CMB instrumental noise curves for some current and future CMB surveys (dashed lines) are over-plotted.}
\label{kSZ_kSZ_auto}
\end{figure}

We now move on to the kSZ effect.  Figure \ref{kSZ_kSZ_auto} shows the kSZ auto-power spectrum computed from the fiducial $(2.8~\rm Gpc)^{3}$ run, compared with kSZ power spectra from previous simulations. Here we see that the fiducial FLAMINGO model yields results that are fairly consistent with the WebSky simulation. Although the AGORA result is integrated only up to $z = 3.0$, its amplitude is comparable to ours at intermediate scales ($\ell \approx 500$–1000). However, it shows a significant suppression of power at large scales ($\ell \lesssim 300$), which is plausibly due to the lack of a diffuse IGM component in their model. The difference at small scales ($\ell \gtrsim 2000$) is likely because their model is based on AGN 8.0 BAHAMAS, which has stronger feedback effects on the gas distribution than fiducial FLAMINGO.  We comment on the cosmology and feedback dependence of the kSZ power spectra in Section \ref{sec::discussions}.

For comparison, we also include the primary CMB power spectrum and the instrumental noise power spectra of several current and upcoming CMB surveys in the same figure. The primary CMB spectrum is generated using the CAMB \citep{CAMB_paper}, while the instrumental noise power spectrum is modelled as $C^{\ell}_{\rm noise}/W^{2}_{\rm beam}(\ell)$, where:
\begin{equation}\label{eqn::white_noise_ps}
C^{\ell}_{\rm noise} = \left(\frac{\Delta{T}}{T_{\rm CMB}}\right)^{2},
\end{equation}
and
\begin{equation}\label{eqn::beam_function}
W_{\rm beam}(\ell) = \textrm{exp}(-\ell(\ell + 1)/\ell^{2}_{\rm beam}/2).
\end{equation}
Here, $\ell_{\rm beam}$ is defined as $\sqrt{8\ln2}/\theta_{\rm FWHM}$, and $\Delta_{T}$ is the white noise level. We adopted the values of $\theta_{\rm FWHM}=1.3^{\prime}$ and $\Delta_{T}=25~\mu \rm K$-arcmin for the ACT survey at 150 GHz \citep{ACT_beam_size,ACTDR6_lensing_ref}, $\theta_{\rm FWHM}=7.2^{\prime}$ and $\Delta_{T}=33~\mu \rm K$-arcmin for the \textit{Planck} survey at 143 GHz \citep{Planck_beam_noise_ref}, $\theta_{\rm FWHM}=0.25^{\prime}$ and $\Delta_{T}=0.5~\mu \rm K$-arcmin for the CMB-HD survey at 150 GHz \citep{CMB_HD}, and $\theta_{\rm FWHM}=1.4^{\prime}$ and $\Delta_{T}=6.3~\mu \rm K$-arcmin for the SO-LAT survey at 145 GHz \citep{Sehgal_radio_SO_forecast}. In general, the kSZ power spectrum is subdominant to the combined signal from the primary CMB and instrumental noise. However, detection seems to be achievable at small angular scales ($\ell \gtrsim 5000$), particularly with next-generation experiments such as the CMB-HD. As we will show later, this is also the regime where the kSZ signal becomes most sensitive to feedback and cosmological effects. This highlights the importance of accurately modelling the kSZ signal for future CMB surveys.

\subsection{CIB statistics: auto-spectra and cross correlations with other probes}\label{ssec:CIB_statistics}

\begin{figure*}
    \begin{minipage}[b]{1.0\textwidth}
        \centering
        \includegraphics[width=0.45\linewidth]{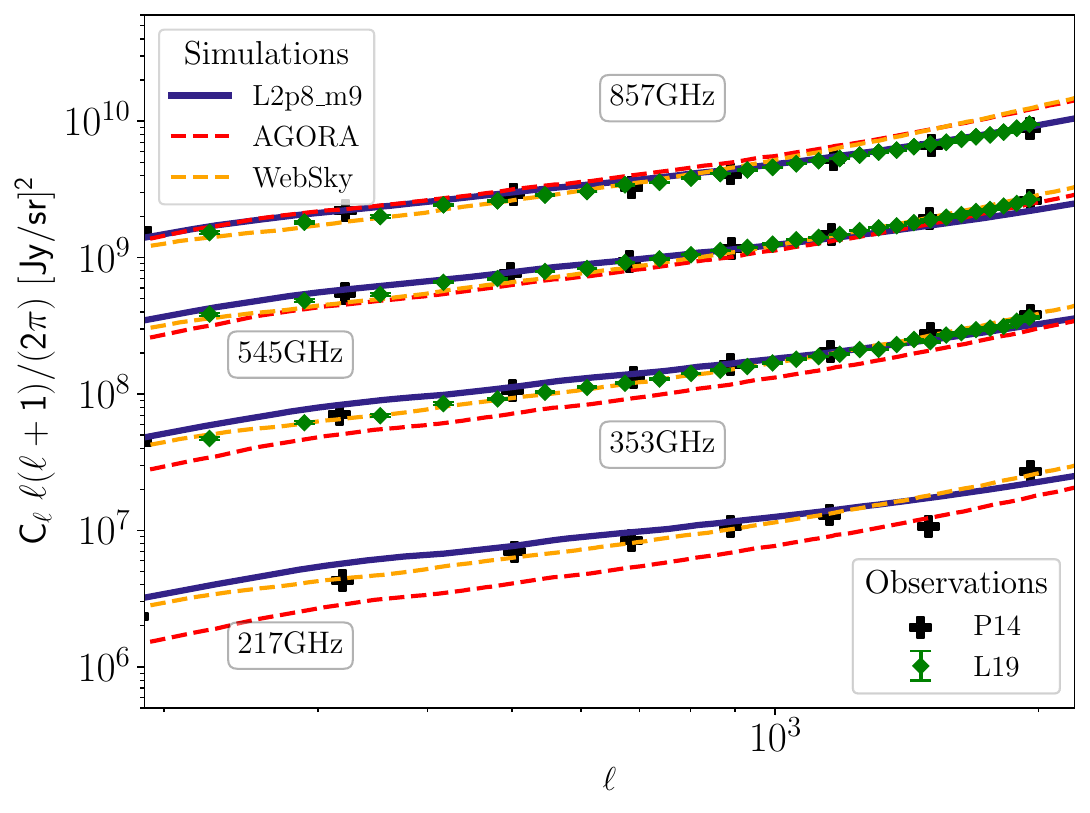}
        \includegraphics[width=0.45\linewidth]{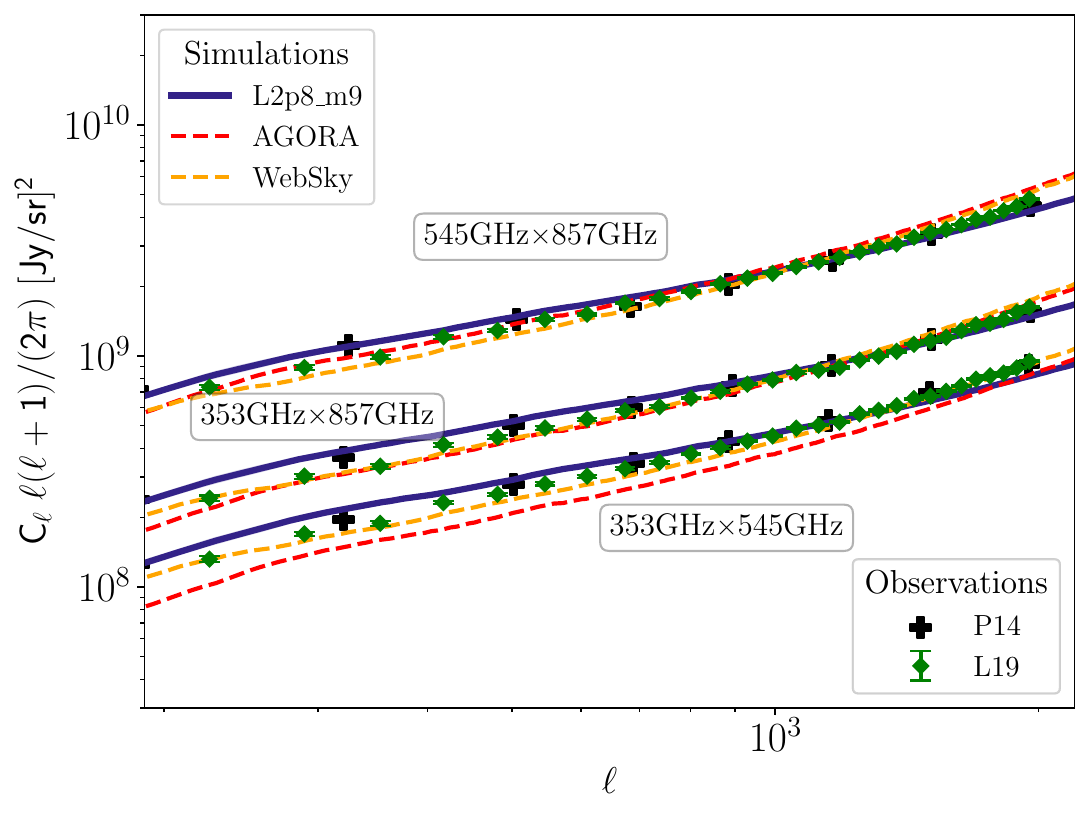}
    \end{minipage}%
    \vspace{+0.1cm}
\caption{The auto- (\textit{left}) and cross- (\textit{right}) power spectra measured from the CIB maps generated from the FLAMINGO fiducial $(2.8~\rm Gpc)^3$ run at the \textit{Planck} HFI frequencies. All curves are obtained by averaging over 8 independent lightcones. Results from the AGORA (red dashed) and the WebSky (orange dashed) simulations are displayed for comparison. Observational data points from \citet{P14_CIB} and \citet{L19_CIB} are shown as black crosses and green diamonds, respectively.}
\label{CIB_CIB_stats}
\end{figure*}

\begin{figure}
\includegraphics[width=\columnwidth]{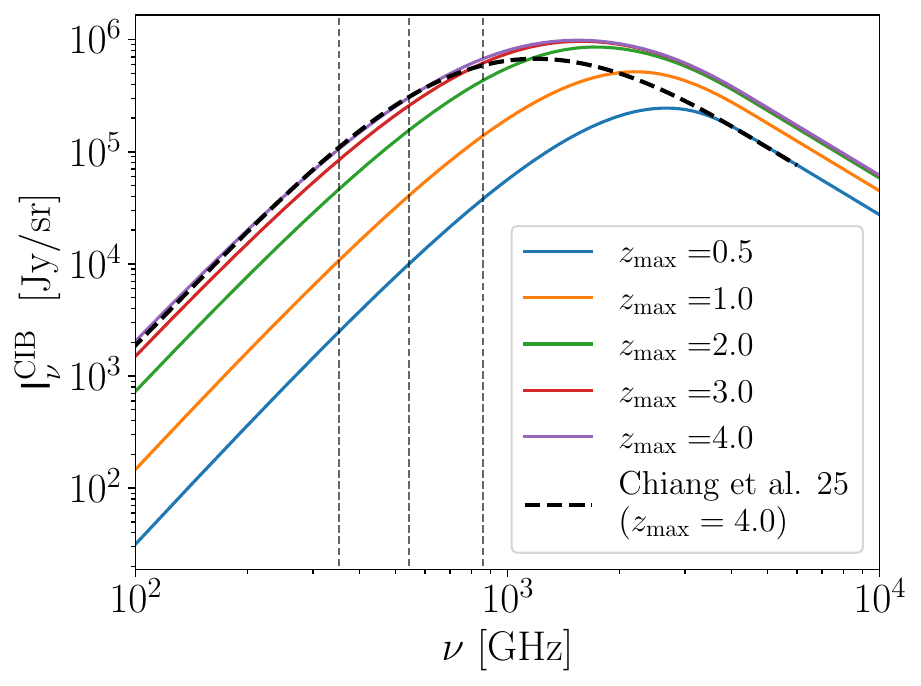}
   \caption{CIB monopole computed using the best-fitting CIB SED from the FLAMINGO fiducial $(2.8~\rm Gpc)^3$ run, integrated up to different maximum redshifts. For comparison, the measurements provided by \citet{CIB_tomo_ref1} are overplotted as a black dashed line. For reference, the vertical dashed lines are the \textit{Planck} HFI frequencies — 353 GHz, 545 GHz, and 857 GHz — which are the frequencies used in the SED fitting to the CIB power spectra.}
\label{CIB_monopole}
\end{figure}

The best-fitting CIB auto- and cross-power spectra are shown in Figure \ref{CIB_CIB_stats}, where we compare the 353/545/857 GHz power spectra with observational measurements from \citet{L19_CIB} and \citet{P14_CIB}. The model provides a qualitatively good fit to the observed data across a broad range of frequencies and multipoles.  In Appendix \ref{appen:C}, we show the CIB power spectra as $ \ell C_\ell$ and plotted on a linear y-axis scale. The curves are also qualitatively similar to the AGORA prediction, for which the free parameters of its adopted SED model were also constrained by fitting to the 353/545/857 GHz data. Note, however, that neither our model nor AGORA were fitted to the 217 GHz measurements, which is where we see the largest discrepancy between the observations and the AGORA model.  Although we did not explicitly fit the 217 GHz data, our best-fitting SED model reproduces the observational measurements at this frequency reasonably well. This provides a reassuring check on the accuracy of the best-fitting SED. 

We also find broad agreement between the FLAMINGO results and those from WebSky.  Note that the WebSky model was not fitted to the \citet{L19_CIB} data, but was adjusted to approximately match the 545 GHz measurement at $\ell = 500$ from \citet{P14_CIB}. For this comparison, we use the \textit{masked} WebSky CIB maps, in which all pixels with flux density greater than 400 mJy are masked and refilled. As already noted, such bright-source masking has only a very minor effect on our model (as well as on AGORA), but it has a very large impact on the WebSky CIB power spectrum at high $\ell$ (see Appendix \ref{appen:D} for a detailed discussion). Therefore, we use the masked WebSky CIB maps for the following comparisons.

\begin{table}
    \centering
    \caption{Frequency decoherence of the CIB measured by averaging $C_\ell^{\nu\nu'}/\sqrt{C_\ell^{\nu\nu}C_\ell^{\nu'\nu'}}$ over the range $150 < \ell < 1000$. Error bars correspond to the standard deviation in this range. The results from \citet{P14_CIB}, \citet{L19_CIB}, the WebSky and AGORA simulations are also listed for comparison.}
    \label{tab:CIB_decoherence}
    \resizebox{\columnwidth}{!}{
    \begin{tabular}{lccccccc}
        \toprule
        & \multicolumn{5}{c}{Frequency (GHz)} \\ 
        \cmidrule{2-6}
        & & 857 & 545 & 353 & 217 &\\ 
        \midrule
        \multirow{3}{*}{857} &
        \textit{Planck} & 1 & $0.949 \pm 0.005$ & $0.911 \pm 0.003$ & $ 0.85 \pm 0.05$\\
        & Lenz et al. & 1 & $0.96 \pm 0.01$ & $0.91 \pm 0.01$ & -\\
        & \textbf{FLAMINGO} & 1 & $0.959 \pm 0.004$ & $0.895 \pm 0.009$ & $0.841\pm 0.012$\\
        & WebSky & 1 & $0.933 \pm 0.017$ & $0.882 \pm 0.021$ & $0.838 \pm 0.026$\\
        & AGORA & 1 & $0.970 \pm 0.003$ & $0.907 \pm 0.008$ & $0.840 \pm 0.013$ \\
        \midrule
        \multirow{3}{*}{545} &
        \textit{Planck} & ... & 1 & $0.983 \pm 0.007$ & $0.90 \pm 0.05$\\
        & Lenz et al. & ... & 1 & $0.98 \pm 0.01$ & -\\
        & \textbf{FLAMINGO} & ... & 1 & $0.983 \pm 0.001$ & $0.956\pm 0.003$\\
        & WebSky & ... & 1 & $0.960 \pm 0.014$ &  $0.935 \pm 0.018$\\
        & AGORA & ... & 1 & $0.981 \pm 0.002$ & $0.941\pm 0.005$\\
        \midrule
        \multirow{3}{*}{353} & 
        \textit{Planck} & ... & ... & 1 & $ 0.91 \pm 0.05$\\
        & Lenz et al. & ... & ... & 1 &  -\\
        & \textbf{FLAMINGO} & ... & ... & 1 & $0.993\pm 0.001$\\
        & WebSky & ... & ... & 1 & $0.968 \pm 0.014$ \\
        & AGORA & ... & ... & 1 & $0.989\pm 0.001$\\
        \bottomrule
    \end{tabular}
    }
\end{table}

Another interesting property of the CIB map is its frequency decoherence, which quantifies how similar the spatial structures of CIB anisotropies are across different frequency channels. This is characterised by the decorrelation coefficient, defined as $C_\ell^{\nu\nu'}/\sqrt{C_\ell^{\nu\nu}C_\ell^{\nu'\nu'}}$. This quantity captures the degree of overlap in the redshift contributions of CIB signals across frequencies: adjacent frequency channels tend to be more correlated (with value closer to 1) because their redshift-dependent emission peaks overlap more. To estimate this, we compute the mean and standard deviation of this decorrelation coefficient within the multipole range $150 < \ell < 1000$, where the CIB clustering signal dominates. The results are summarised in Table \ref{tab:CIB_decoherence}. Compared with other simulations and observations, our predictions show good agreement with previous measurements, where the decorrelation level increases as the frequency separation between channels gets larger. 

Similar to the case for the tSZ effect, the CIB monopole can be inferred via cross correlation between CIB maps with other tracers of large-scale structure.  Figure \ref{CIB_monopole} shows the CIB monopole integrated up to different maximum redshifts, using our best-fitting CIB SED as defined in Equation \ref{eqn::SED_galaxies}. For each frequency
\begin{equation}
    I_{\nu} = \int^{z_{\rm max}}_{0} dz \frac{dI_{\nu}}{dz},
\end{equation}
with $dI_{\nu}/dz$ defined as
\begin{equation}
    \frac{dI_{\nu}}{dz} = \frac{c}{4\pi}\frac{\epsilon({\nu_{\rm emitted}})}{H(z)(1+z)},
\end{equation}
where the emissivity $\epsilon({\nu_{\rm emitted}})$ is derived from our best-fitting CIB SED. From this figure, we see that the CIB signal converges by $z\approx 3$, and the power-law break happens at $\nu\gtrsim1000~\rm GHz$. We find that our predictions are in reasonably good agreement with the \textit{Planck}-based measurements of \citet{CIB_tomo_ref1} (we show their best-fitting parametric model to their measurements).  It would also be interesting to compare to tomographic measurements of CIB intensity using similar observational techniques (e.g. galaxy–CIB cross-correlations as applied by \citealt{CIB_tomo_ref2} and \citealt{CIB_tomo_ref1}). We plan to explore this comparison in detail in future work.    

\begin{figure*}
    \begin{minipage}[b]{1.0\textwidth}
        \centering
        \includegraphics[width=0.45\linewidth]{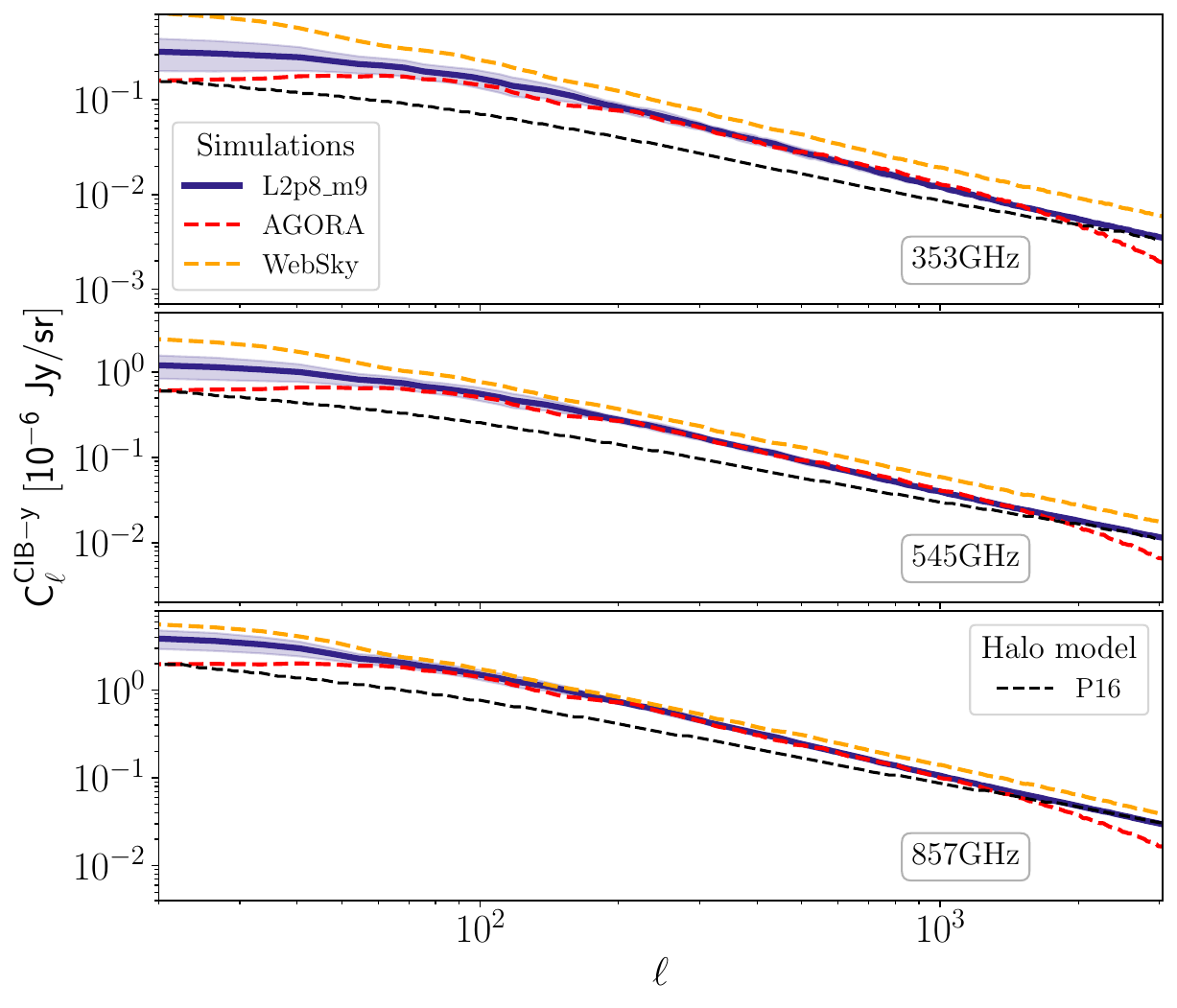}
        \includegraphics[width=0.54\linewidth]{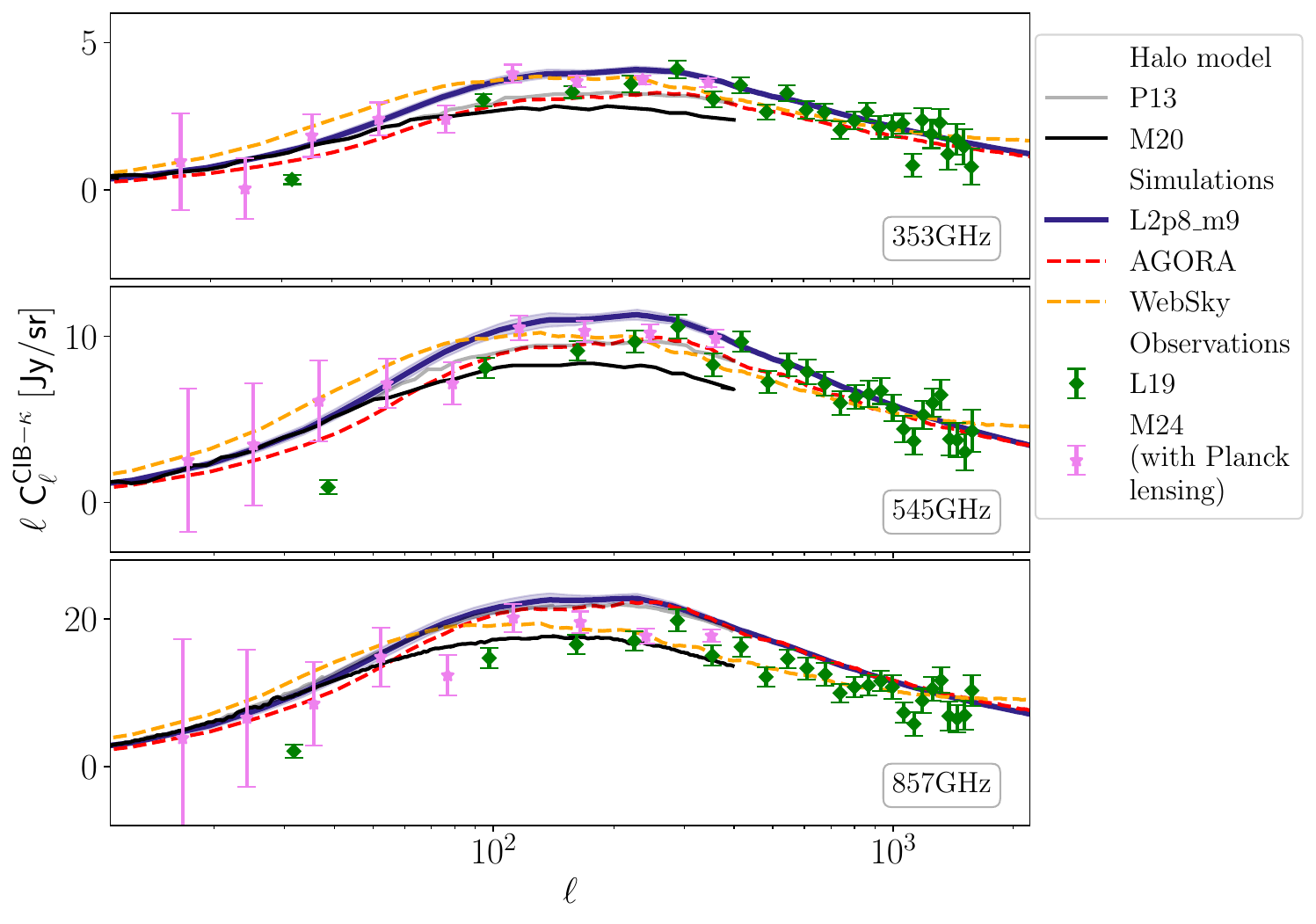}
    \end{minipage}%
    \vspace{+0.1cm}
\caption{CIB-LSS tracers cross-power spectra at different \textit{Planck} frequencies. \textit{Left}: Comparison of the CIB-tSZ cross-power spectrum between simulations (dark solid: fiducial $(2.8~\rm Gpc)^{3}$ FLAMINGO run, with shaded regions estimated by averaging the results from eight different lightcones; red dashed: AGORA; orange dashed: WebSky) and halo model predictions from \citet{P16_CIB_tSZ} (black dashed). \textit{Right}: Comparison of the CIB-CMB lensing ($\kappa$) cross-power spectrum between simulations, halo model predictions (black solid: \citet{P14_CIB}; gray solid: \citet{M2020_CIB_kappa}), and observations (green diamonds: \citet{L19_CIB}, violet stars: \citet{Fiona_CIB23}, taken from the Galactic dust-cleaning case of $N_{\rm HI}<2.5 \times10^{20}~\rm cm^{-2}$ and real ILC on $N_{\rm side}=1$ super pixels).}
\label{CIB_LSS_stats}
\end{figure*}

We now explore the cross-power spectra between the CIB and other LSS tracers. Here, we present the results for the CIB–tSZ and CIB–CMB lensing ($\kappa$) cross-power spectra, as shown in Figure \ref{CIB_LSS_stats}. 

For the CIB–tSZ cross-correlation, our predicted curves are qualitatively consistent with those from the AGORA. The simulation uncertainties on the FLAMINGO curves, which are estimated by averaging the results from 8 different lightcones in the $(2.8 ~\rm Gpc)^{3}$ fiducial run, are relatively large at large angular scales (small $\ell$s). This is expected since the CIB–tSZ signal mainly traces warm–hot gas surrounding star-forming regions in intermediate-mass haloes. On large scales, the contribution is dominated by more massive and rarer structures, which introduces significant sample variance and hence larger uncertainties in the measurement. The WebSky results are higher at all three frequencies compared to our predicted curves (as well as AGORA’s; see Appendix \ref{appen:C} for the power spectra plotted with a linear y-axis scale, where the differences are more noticeable). This may be due to their adopted CIB modelling, which could result in higher spatial correlations between the CIB and tSZ signals for low-redshift haloes \citep[see e.g.][]{Inigo_CIB_tSZ_lowz_corr_ref1,Inigo_CIB_tSZ_lowz_corr_ref2}. The halo model-based prediction from \citet{P16_CIB_tSZ} is systematically lower at all frequencies compared to the simulation-based results.

In practice, detecting the CIB–tSZ cross-correlation in observations can be challenging because the two signals trace quite distinct physical processes: the CIB is more sensitive to star-forming galaxies, while the tSZ effect traces hot electron pressure, preferentially in massive galaxy groups and clusters. Their overlap is limited due to environmental processes (e.g., ram pressure stripping), which tends to quench star formation in high density regions. This leads to a relatively weak correlation signal that is challenging to measure, particularly in the face of limited sensitivity and foreground contaminants. Due to the limited observational constraints for this statistic, we simply highlight the potential non-negligible contribution of this correlation and leave a more detailed discussion for future work. 

The right panel of Figure \ref{CIB_LSS_stats} shows the CIB–$\kappa$ cross-power spectrum, which is a more robustly measured quantity compared to the CIB–tSZ cross-correlation. Note that the uncertainties in this statistic are much smaller across all multipole scales. This is because $\kappa$ traces the total matter distribution and has a broad redshift kernel similar to that of the CIB.

Overall, the shape and amplitude predicted by our models are consistent with those from other simulations and observations. However, caution must be taken when interpreting the observational measurements: the data from \citet{L19_CIB} are known to be biased at $\ell \lesssim 100$, due to a non-optimal method of Galactic dust foreground subtraction (see e.g. \citealt{Fiona_CIB23}). Instead, the measurements from \citet{Fiona_CIB23}, which adopt an improved dust-cleaning approach, are considered more reliable on these large angular scales. At smaller scales, the simulations match the \citet{L19_CIB} data relatively well. Although we have a slightly higher amplitude at $\ell \sim 100$–1000 when compared to the \citet{Fiona_CIB23} measurements, the predictions are still reasonably close to the measurements. A small adjustment in cosmology (see Fig.~\ref{CIB_LSS_cross_feed_models_SED_fixed} below) might completely remove this offset.  On larger scales, all curves are consistent with the measurements of \citet{L19_CIB} and halo model predictions.

\subsection{Radio point source statistics}\label{ssec:radio_ps_statistics}

\begin{figure*}
    \begin{minipage}[b]{1.0\textwidth}
        \centering
        \includegraphics[width=0.9\linewidth]{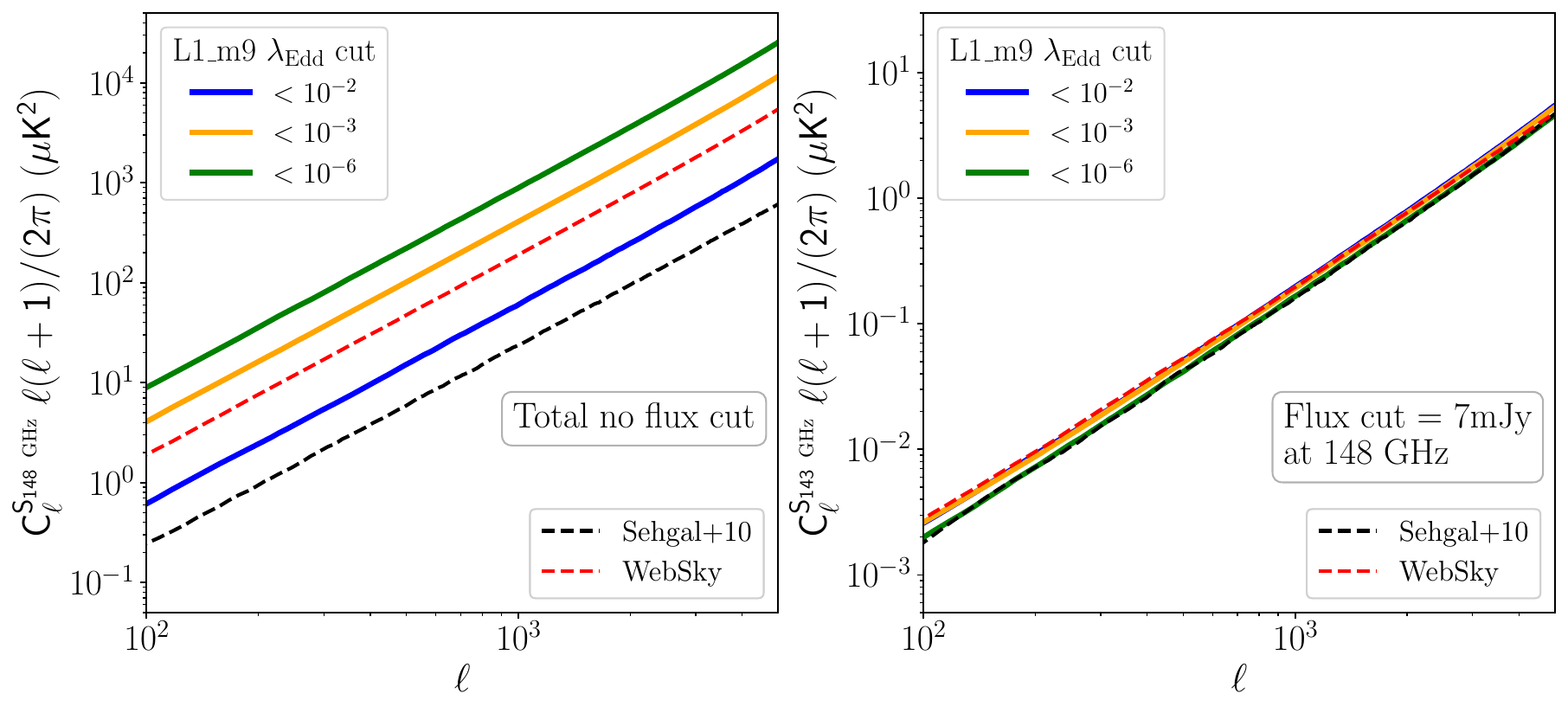}
    \end{minipage}%
    \vspace{+0.1cm}
\caption{Auto-power spectrum of radio point sources for the three $\lambda_{\rm Edd}$ cuts considered in our mock catalogue, where the radio luminosities are assigned by abundance matching to the observed radio luminosity function of low-excitation radio galaxies, measured at 150 MHz from the LOFAR survey (see Section \ref{ssec::Radio_Point_Sources} for details). \textit{Left}: power spectrum computed from the full sample without masking bright sources. For comparison, the corresponding full power spectra at the same frequency from other simulations are shown as dashed lines. \textit{Right}: power spectrum after removing bright sources with a 7 mJy flux cut at 148 GHz. Dashed lines are the curves from other simulations with the same flux cut applied.}
\label{fluxCl_radio_143GHz}
\end{figure*}

\begin{figure*}
    \begin{minipage}[b]{1.0\textwidth}
        \centering
        \includegraphics[width=0.45\linewidth]{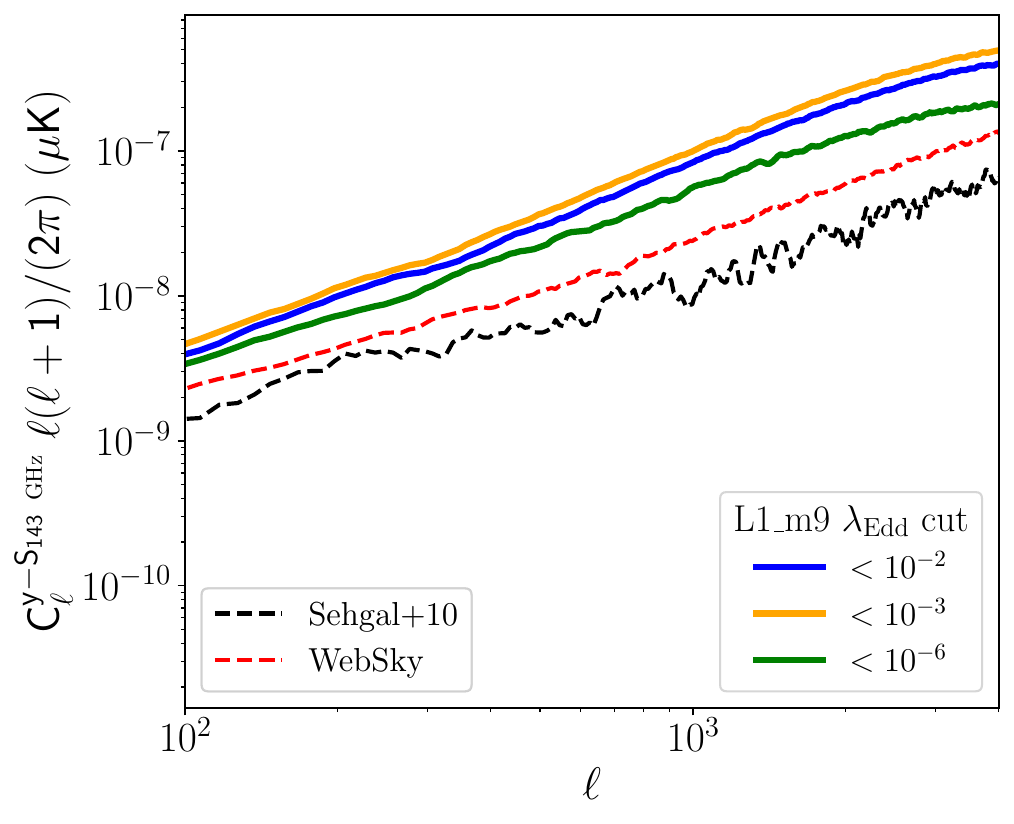}
        \includegraphics[width=0.45\linewidth]{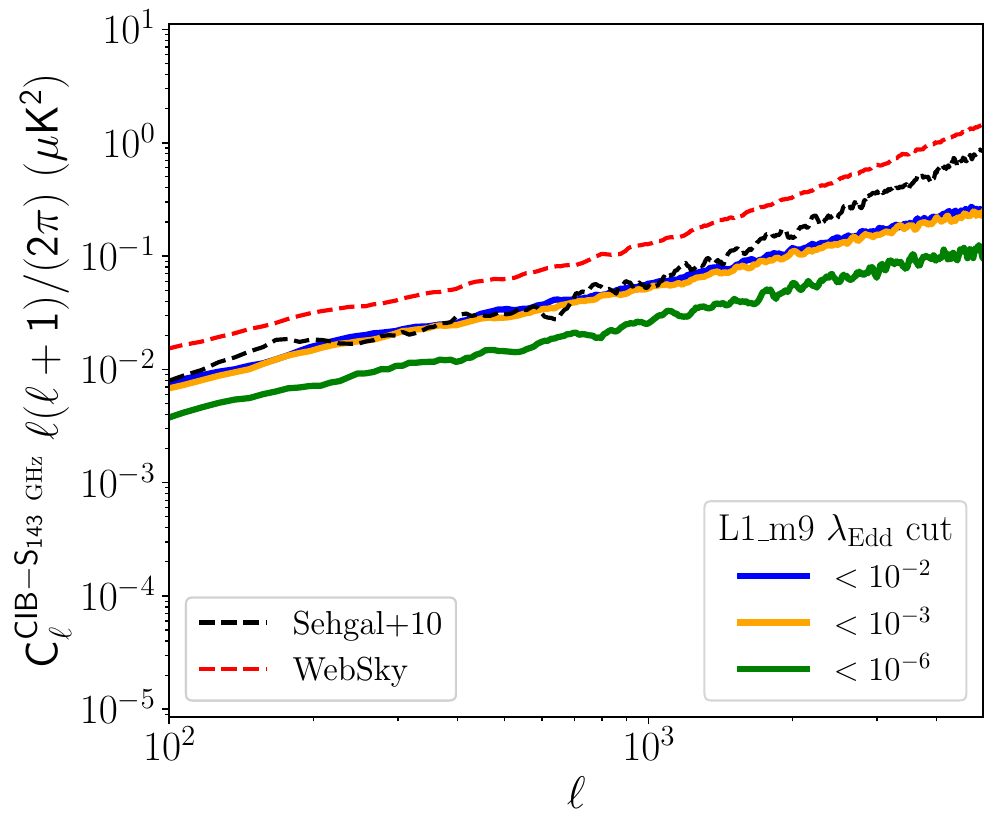}
    \end{minipage}%
    \vspace{+0.1cm}
\caption{Cross-power spectrum between radio flux density (after bright source removal) and other LSS tracers, where the radio luminosities are assigned by abundance matching to the observed radio luminosity function of low-excitation radio galaxies, measured at 150 MHz from the LOFAR survey (see Section \ref{ssec::Radio_Point_Sources} for details). \textit{Left}: cross-correlation with the Compton $y$ field for three different $\lambda_{\rm Edd}$ cuts. Right: cross-correlation with the CIB map at 143 GHz. Predictions from \citet{Sehgal_radio_SO_forecast} (based on the \citealt{Sehgal_2010} mocks) and WebSky are shown as black and red dashed lines.}
\label{Clxxx_radio_143GHz_different_lambda_edd}
\end{figure*}

The left panel of Figure \ref{fluxCl_radio_143GHz} shows the auto-power spectrum of the radio point sources for three different $\lambda_{\rm Edd}$ cases, measured on our stacked HEALP\textsc{ix} maps integrated to $z = 2.5$. The power spectra for all cases generally follow a Poisson-like distribution, which is consistent with both model predictions and other simulation results. In particular, our predictions show that the population with the lowest accretion rate threshold ($\lambda_{\rm Edd} < 10^{-6}$) has the highest power. By construction, we have the same space density of bright sources regardless of the applied Eddington cuts. However, in the low Eddington-cut case, we select sources that typically reside in massive clusters (as shown in Figure \ref{Clxxx_radio_143GHz_different_lambda_edd}), which are more highly biased tracers with respect to the overall matter distribution, and therefore appear more strongly clustered in the resulting power spectrum.

Observational analyses often mask individually-detected bright sources.  We demonstrate the impact this has by excluding sources with flux densities greater than 7 mJy at 148 GHz. This is the same flux cut as adopted by \citet{Sehgal_radio_SO_forecast} and in the WebSky simulation \citep{Li_websky_radio}. As done in \citet{Sehgal_radio_SO_forecast}, we populate the remaining sources onto a HEALP\textsc{ix} map and recompute the power spectrum. While in the WebSky simulation, an apodized mask was applied with a 15-arcminute C2 apodization around pixels with sources brighter than the flux cut. The results are shown in the right panel of Figure \ref{fluxCl_radio_143GHz}. After applying the flux cut, the recovered power spectra show excellent agreement among the three cases, as well as with the predictions from \citet{Sehgal_radio_SO_forecast} (based on the \citealt{Sehgal_2010} mocks) and WebSky. This is expected because all the cases considered in our study and other simulations are calibrated to reproduce the fainter end of the differential source count curves at CMB frequencies. Given this consistency, we then use the maps with bright sources removed to further study the cross-correlation between radio point sources and other secondary CMB anisotropies. 

To test the influence of bright-source masking on the recovered statistics, we perform a test using the Compton $y-y$ power spectrum. Specifically, we replace pixels with radio source flux densities greater than 7 mJy by the average of their nearest neighbors (as used by e.g. \citealt{Pearly_CIB_results, Websky_ref}). We repeat this test for maps of different resolutions, from $n_{\rm side}=512$ up to $n_{\rm side}=4096$. We find that for maps with a resolution higher than $n_{\rm side}=1024$ (corresponding to an angular resolution of 3.4 arcmin), the true $y-y$ power spectrum can be accurately recovered. We leave the investigation of the effects of different masking schemes on various statistics and their observational implications to a future study.

Figure \ref{Clxxx_radio_143GHz_different_lambda_edd} shows the cross-correlation between radio point sources (after bright source removal) and other LSS tracers. The left panel shows the cross-correlation between tSZ and radio flux density. 
The FLAMINGO predictions broadly agree in shape with those based on the maps from \citet{Sehgal_radio_SO_forecast} and Websky, although with a higher amplitude compared to the other curves.
The differences among the three $\lambda_{\rm Edd}$ cuts are relatively small, which is likely due to the removal of bright and massive sources which typically trace massive clusters that also dominate the tSZ signal. The predicted correlation coefficient\footnote{See its definition in Section \ref{sssec::feedback_dependencies_CIB_SED_varied}} values at $\ell=3000$ for radio-tSZ, $\rho^{y-\rm S_{143 ~GHz}}_{\ell=3000}$, are  0.19, 0.24, 0.11 for the $\lambda_{\rm Edd}<10^{-2}, 10^{-3} ~\textrm{and}~10^{-6}$ cases respectively, compared to 0.056 and 0.089 as predicted by \citet{Sehgal_2010} and the WebSky simulations.

The right panel shows the cross-correlation between flux density and the CIB at 143 GHz (after bright source removal in the radio catalogues). Compared to the left panel, these curves are much smoother, as the CIB is more sensitive to more common, lower mass haloes, where the bright source cut has less of an impact.  The lowest-$\lambda_{\rm Edd}$ case yields the lowest cross-correlation amplitude. This trend is likely driven by the fact that imposing a lower $\lambda_{\rm Edd}$ limit preferentially picks out relatively higher mass haloes which tend to have lower star formation activity due to environmental quenching and enhanced AGN feedback. correlation coefficient values at $\ell=3000$ for radio-CIB, $\rho^{\textrm{CIB}-\rm S_{143 ~GHz}}_{\ell=3000}$, are 0.056, 0.052, 0.023 for the $\lambda_{\rm Edd}<10^{-2}, 10^{-3} ~\textrm{and}~10^{-6}$ cases respectively, compared to 0.08 and 0.074 as predicted by \citet{Sehgal_2010} and the WebSky simulations.

\section{Discussion: Feedback and cosmology dependencies}\label{sec::discussions}

A major advantage of the FLAMINGO suite is that it has a number of variations in feedback modelling and background cosmology. This enables us to explore the dependencies of the key summary statistics on feedback processes and cosmology. In this section, we focus mainly on the CIB statistics (Section \ref{ssec::feedback_dependencies_CIB}) and the kSZ auto-power spectrum (Section \ref{ssec::feedback_dependencies_kSZ}). For a discussion of such dependencies in other statistics (such as the tSZ effect and its cross correlations), see \citet{Ian_low_S8_FMG_paper,Ian_kSZ_feedback_FLAMINGO}.

\subsection{Variations in the CIB power spectrum}\label{ssec::feedback_dependencies_CIB}

In previous studies modelling the CIB \citep[e.g. in ][]{P14_CIB,Websky_ref,M2020_CIB_kappa,agora_ref}, the analysis is typically performed under the assumption of a fixed cosmological model. In addition, the nature of the SED has no direct link to the physics of galaxy formation and feedback making the derived best-fitting parameters difficult to interpret.  While this indeed simplifies the modelling, both the CIB power spectrum and the SED modelling encode information about feedback physics and cosmology, as we will demonstrate below. 

We adopt two approaches to explore the dependencies of the CIB power spectrum on feedback and cosmological models.  In the first approach, we fix the SED parameters to the best-fitting values obtained from the fiducial $(2.8 ~\rm Gpc)^{3}$ run in Section \ref{sssec::CIB_default_model} and use the same set of parameters to compute the CIB statistics for other model variants. As a result, any variations in the resulting CIB-CIB auto- and cross-power spectra are solely driven by differences in the spatial clustering of star-forming galaxies predicted by each FLAMINGO model. This is discussed in Section \ref{sssec::feedback_dependencies_CIB_SED_fixed}.

In the second approach, we refit the CIB SED model from each FLAMINGO variant to the measurements from \citet{L19_CIB}. This method ensures that the CIB auto- and cross-power spectra are well matched across all models by construction, but it results in a distinct set of best-fitting SED parameters for each case. We further examine how much variation remains in the CIB–LSS cross-correlations under consistent CIB power constraints. Results from this approach are presented in Section \ref{sssec::feedback_dependencies_CIB_SED_varied}.

\subsubsection{Fixed SED template}\label{sssec::feedback_dependencies_CIB_SED_fixed}

\begin{figure*}
    \begin{minipage}[b]{1.0\textwidth}
        \centering
        \includegraphics[width=0.45\linewidth]{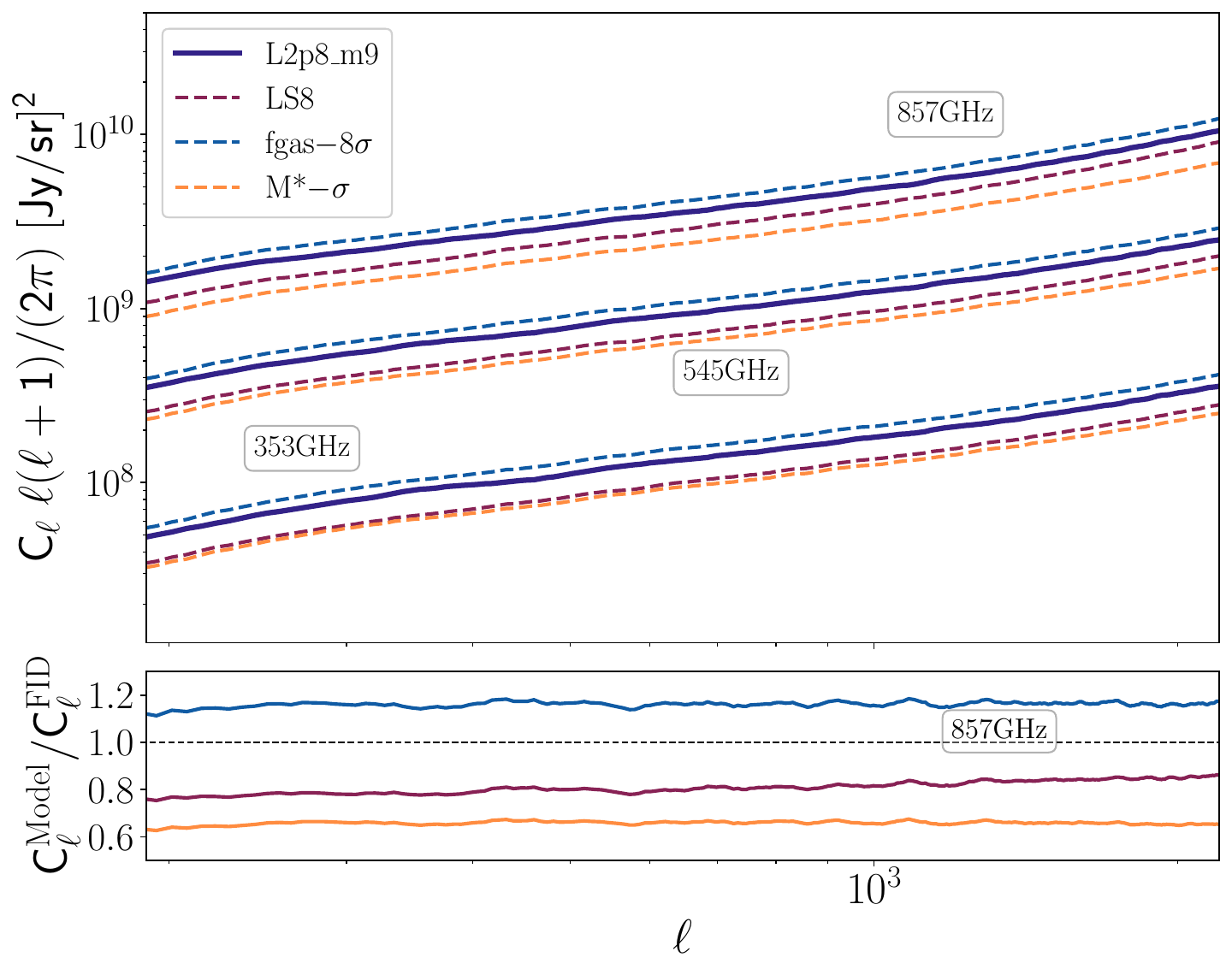}
        \includegraphics[width=0.45\linewidth]{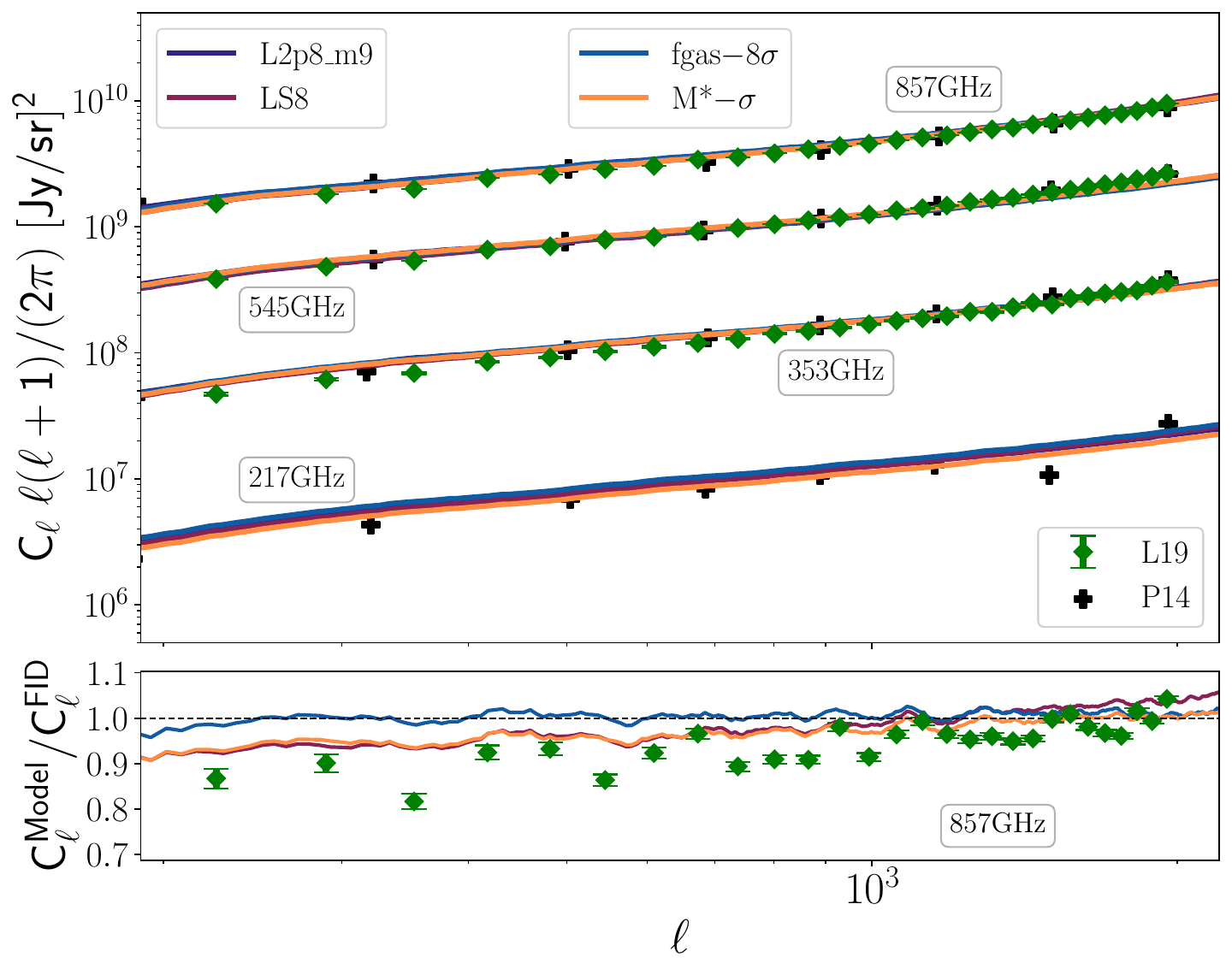}
    \end{minipage}%
    \vspace{+0.1cm}
\caption{Feedback and cosmology dependencies of the CIB auto-power spectra for the case when SED parameters are fixed to the values from fitting the fiducial $(2.8 ~\rm Gpc)^{3}$ run to the \citet{L19_CIB} data (see Section \ref{sssec::CIB_default_model}, \textit{left}), and for the case when SEDs are refitted to the \citet{L19_CIB} data for different models (\textit{right}). Green data points are the measurements from \citet{L19_CIB}. The ratio curves between different models and the fiducial curves for 857 GHz are shown on the bottom, along with the ratios between the \citet{L19_CIB} measurements and the fiducial curves for comparison.}
\label{CIB_CIB_auto_feed_models}
\end{figure*}

\begin{figure*}
    \begin{minipage}[b]{1.0\textwidth}
        \centering
        \includegraphics[width=0.45\linewidth]{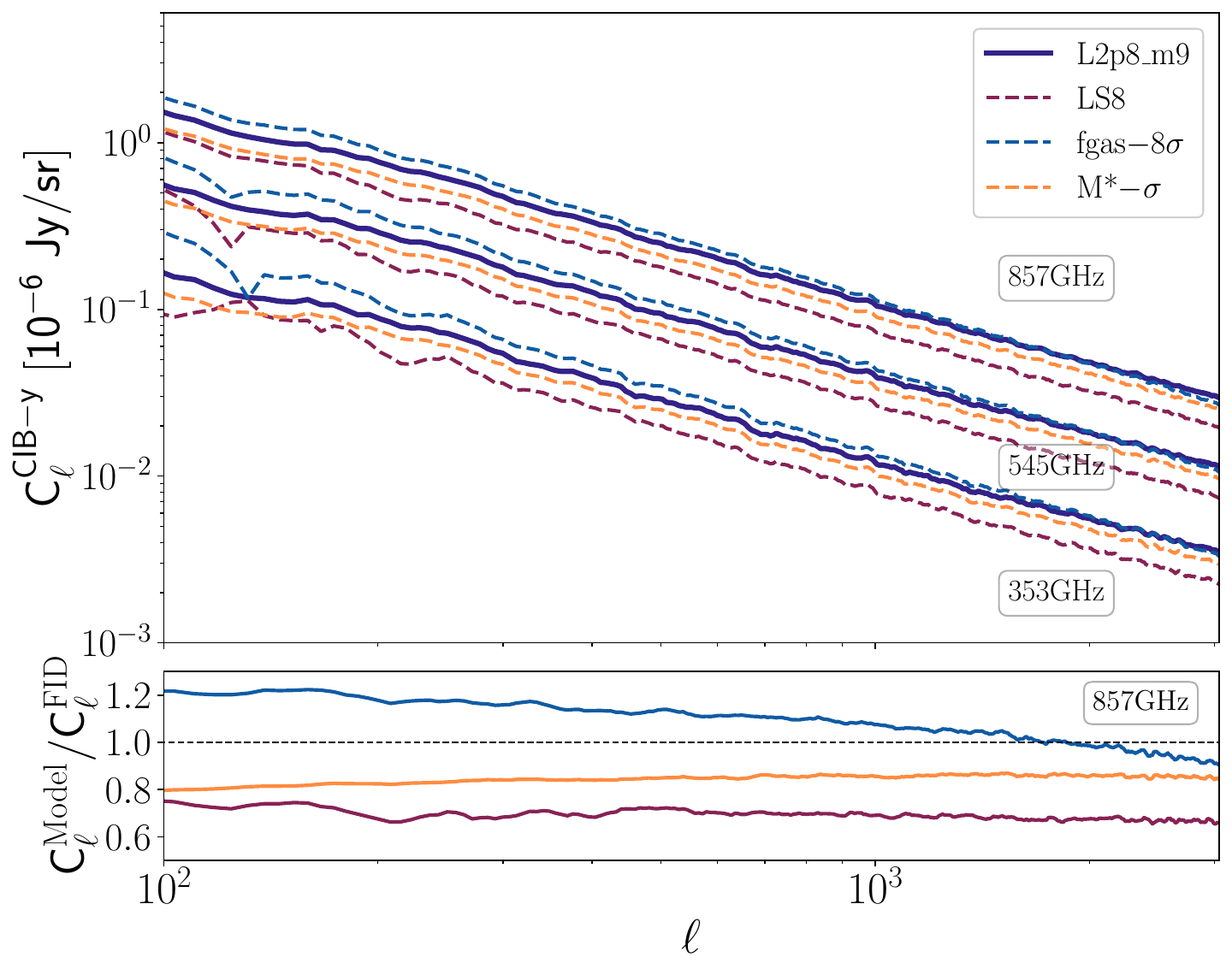}
        \includegraphics[width=0.45\linewidth]{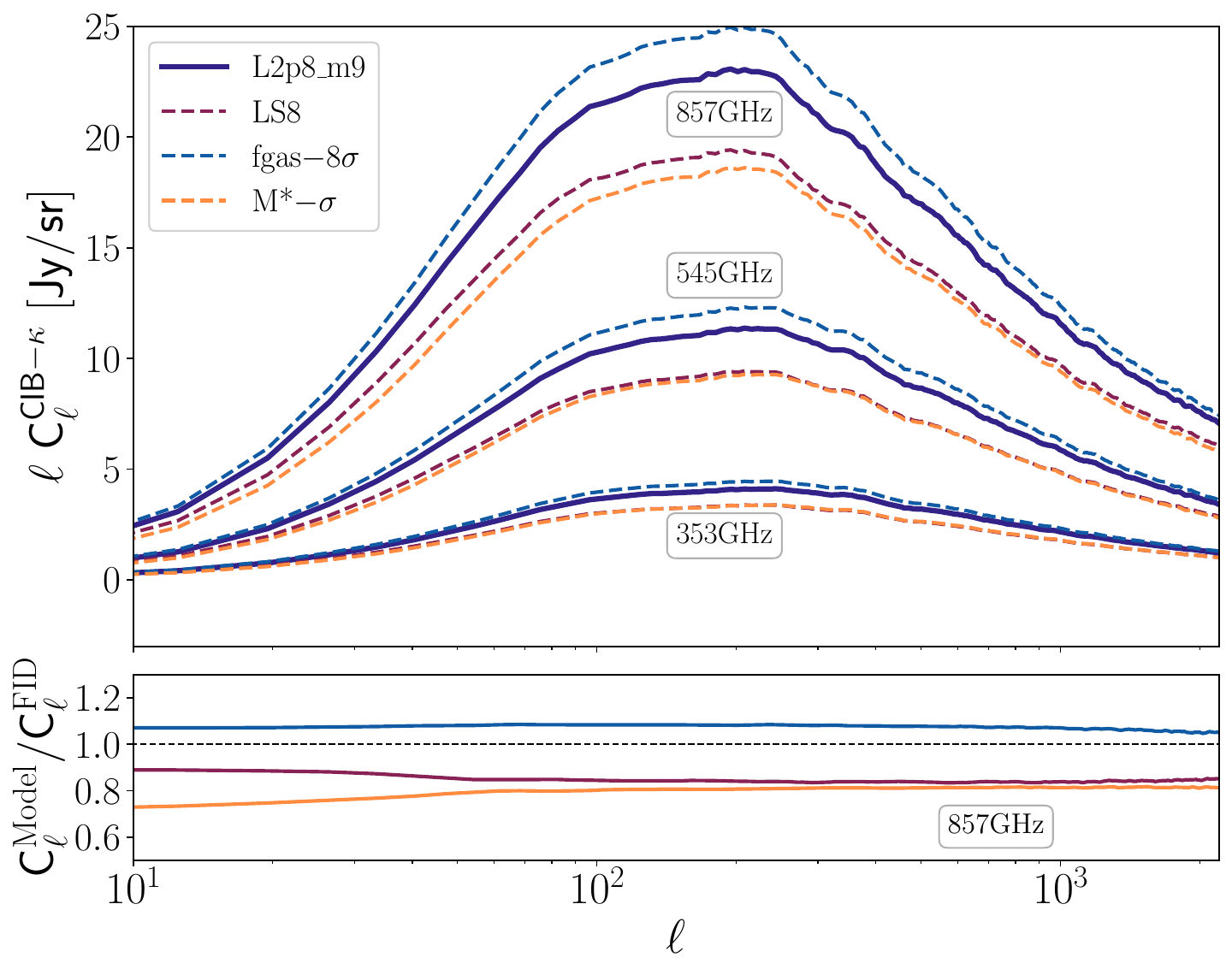}
    \end{minipage}%
    \vspace{+0.1cm}
\caption{Feedback and cosmology dependencies of the CIB-tSZ (\textit{left}) and CIB-$\kappa$ (\textit{right}) cross-power spectra for the case when CIB SED parameters are fixed to the values obtained from fitting the fiducial $(2.8 ~\rm Gpc)^{3}$ run to the data from \citet{L19_CIB} (see Section \ref{sssec::CIB_default_model}). Ratios between different models and the fiducial curves for 857 GHz are shown on the bottom panel.}
\label{CIB_LSS_cross_feed_models_SED_fixed}
\end{figure*}

The left panel of Figure \ref{CIB_CIB_auto_feed_models} shows the CIB auto-power spectrum for different feedback and cosmology models, adopting the best-fitting SED parameters derived from the fiducial run. Here, we show only the curves from the strong stellar feedback ($M_{\ast}-\sigma$), strongest AGN (fgas-$8\sigma$), and LS8 models for illustration. In the bottom panel, we show the ratio between different models and the fiducial curve. Only the results at 857 GHz are shown here for illustration, as the other frequencies display the same general trends across models. This comparison clearly shows the sensitivity of the CIB power spectrum to feedback and cosmological variations: the $M_{\ast}-\sigma$ model predicts the lowest power amplitude ($\approx30-40\%$ reduction at all scales) due to its strong suppression of star formation activity. The fgas-$8\sigma$ model yields a moderately higher power amplitude ($\approx 20\%$) compared to the fiducial model.  Here we point out that, while the energy per AGN feedback event was increased to yield a lower gas fraction in galaxy groups in the fgas-$8\sigma$ model, the other subgrid feedback parameters (particularly controlling stellar feedback) were also varied to attempt to rematch the observed galaxy stellar mass function.  From figure 8 of \citet{FLAMINGO_ref_Joop}, we see that the fgas-$8\sigma$ has a slightly higher mass function compared to the fiducial model (see also the SFR density curve in the right panel of Figure \ref{CIB_SEDfit_SFR_feed_models}), which is consistent with the difference between models in the CIB power spectrum.  For the LS8 model, although it uses same subgrid physics as the fiducial model, it also predicts a lower power amplitude, primarily due to an overall suppression of matter clustering. However, this suppression is generally less significant than that seen in the strong stellar feedback model.  We discuss the right panel of Figure \ref{CIB_CIB_auto_feed_models} below.

Figure \ref{CIB_LSS_cross_feed_models_SED_fixed} shows the feedback and cosmology dependencies of the CIB–tSZ and CIB–$\kappa$ cross-power spectra. On large scales, the CIB–tSZ cross spectrum (left panel) is noisy due to the limited number of massive structures that dominate at these scales\footnote{It is possible to explore these fluctuations further by, e.g., decomposing the CIB–tSZ cross-correlation by halo mass. We leave this for future work.}. 
The LS8 model predicts the lowest power amplitude, which is expected due to the strong $\sigma_{8}$ dependence of tSZ statistics. On small scales, the strongest AGN model shows a suppression of power due to gas depletion within the one-halo regime. 

For the CIB–$\kappa$ cross-power spectrum (right panel), the feedback dependencies follow similar trends as shown in the CIB auto-spectrum though with generally reduced offset amplitudes, which is expected given that the $\kappa$ statistics are not strongly sensitive to feedback processes. For the LS8 model, however, a similar amplitude offset with respect to the fiducial run is seen in the CIB–$\kappa$ cross spectrum and the CIB auto-spectrum, which is due to the suppressed overall matter clustering predicted by this low-$\sigma_{8}$ cosmology. 

In summary, if the SED is fixed (or precisely known), the CIB auto-power spectrum and the cross-correlation of the CIB with other LSS tracers are sensitive probes of both cosmology and feedback.  In reality, however, the SED is not precisely known, and in general, one often constrains a model of the SED to match observations, an approach which we consider immediately below.

\subsubsection{Varying the SED template}\label{sssec::feedback_dependencies_CIB_SED_varied}

\begin{figure*}
\includegraphics[width=0.9\textwidth]{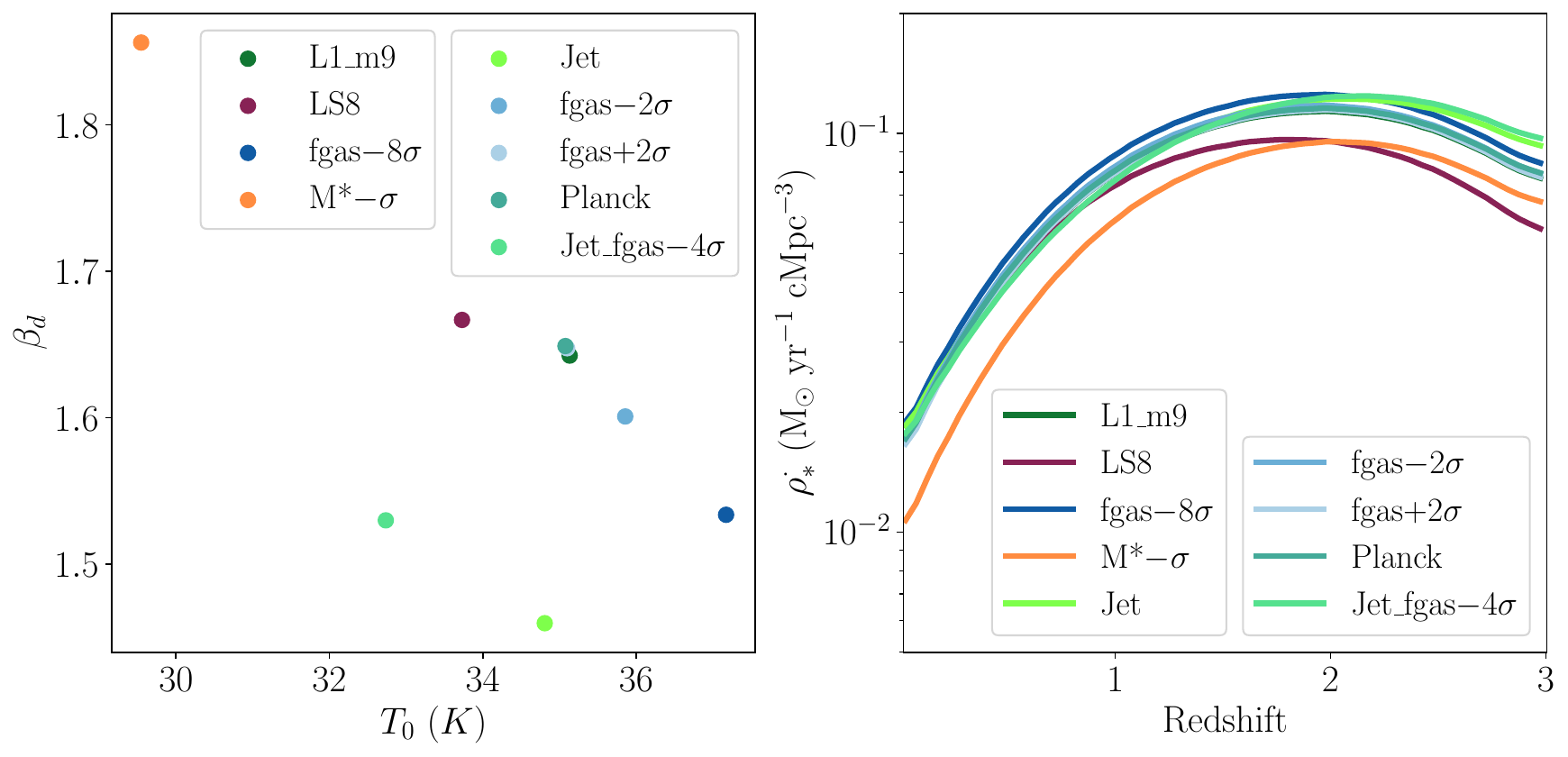}
   \caption{\textit{Left}: The best-fitting rest-frame dust temperature ($T_{0}$) and spectral index ($\beta_{\rm d}$) for all FLAMINGO model variants considered in this study, with best-fitting values obtained by refitting the modelled CIB power spectra to \citet{L19_CIB} data. \textit{Right}: Cosmic star formation rate density as a function of redshift for the different model variants.}
\label{CIB_SEDfit_SFR_feed_models}
\end{figure*}

\begin{figure*}
    \begin{minipage}[b]{1.0\textwidth}
        \centering
        \includegraphics[width=0.45\linewidth]{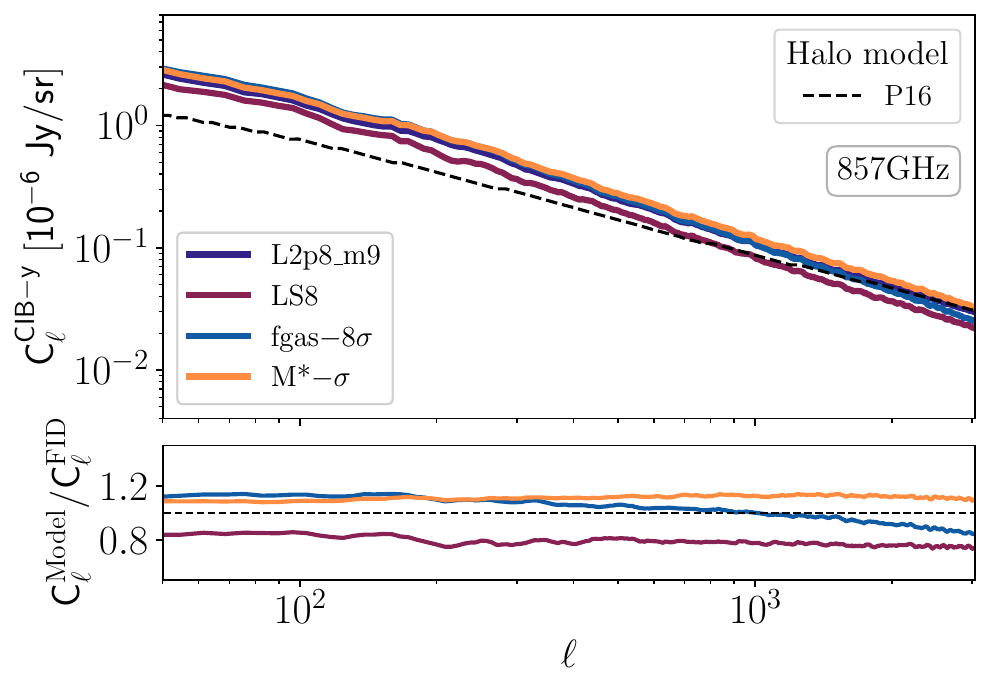}
        \includegraphics[width=0.45\linewidth]{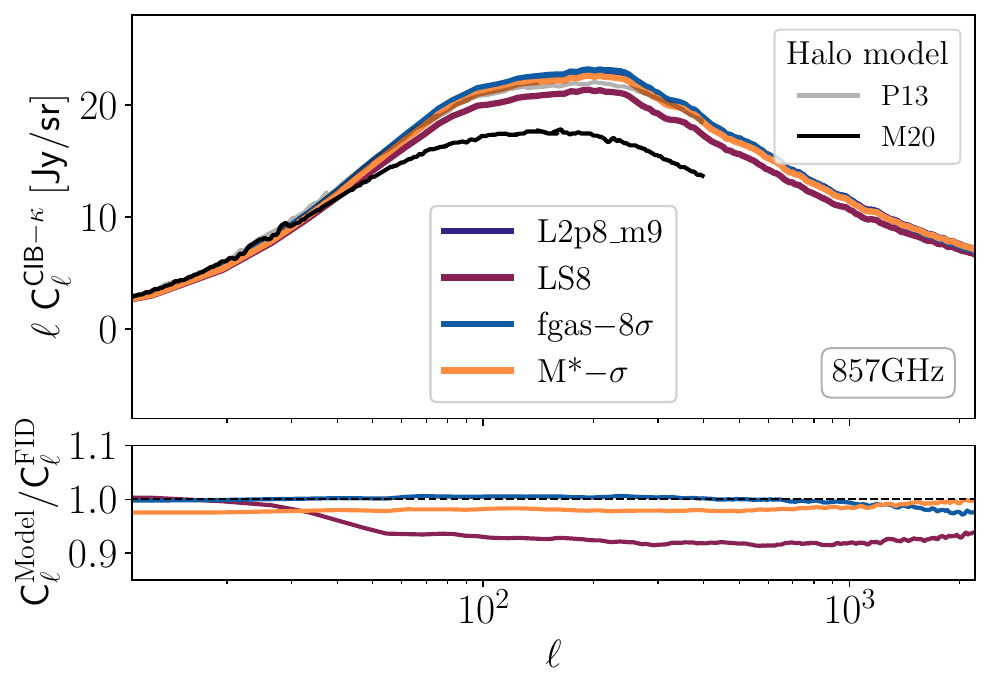}
    \end{minipage}%
    \vspace{+0.1cm}
\caption{Feedback and cosmology dependencies of the CIB-LSS cross-power spectrum at 857 GHz for the case when SEDs are refitted to the \citet{L19_CIB} data for different models. Ratios between different models and the fiducial curves are shown in the bottom subpanels. The left panel shows the results for the CIB-tSZ cross-power spectrum, and the right panel shows the  CIB-$\kappa$ curves.}
\label{CIB_LSS_feed_models}
\end{figure*}

\begin{figure*}
    \begin{minipage}[b]{1.0\textwidth}
        \centering
        \includegraphics[width=0.45\linewidth]{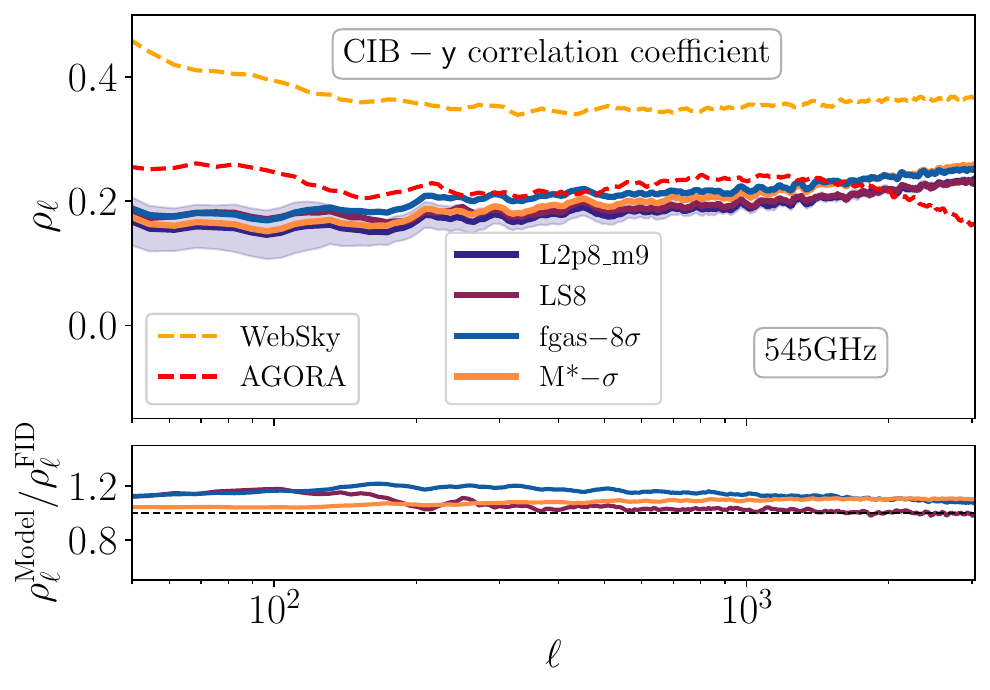}
        \includegraphics[width=0.45\linewidth]{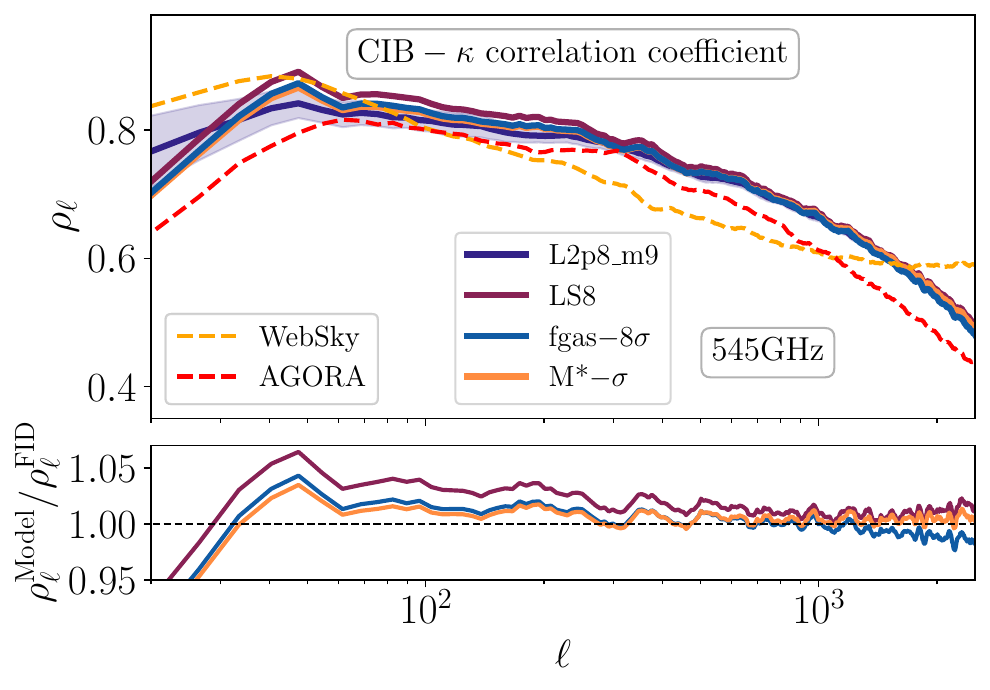}
    \end{minipage}%
    \vspace{+0.1cm}
\caption{Feedback and cosmology dependencies of the CIB-LSS correlation coefficient at 545 GHz for the case when SEDs are refitted to the \citet{L19_CIB} data for different models, with shaded regions estimated by averaging the results from eight different lightcones. Ratios between different models and the fiducial curves are shown in the bottom subpanels. The left panel shows the results for the CIB-tSZ correlation coefficient, and the right panel shows the  CIB-$\kappa$ correlation coefficient.}
\label{CIB_LSS_feed_models_correlation coefficient}
\end{figure*}

The right panel of Figure \ref{CIB_CIB_auto_feed_models} presents the CIB auto-power spectra for different FLAMINGO feedback and cosmology models, where the SED parameters have now been refitted for each model to match observational data of \citet{L19_CIB}. As an illustration, the ratio between each model and the fiducial run for 857 GHz is shown in the bottom panel. In general all the models match the CIB auto-power spectra similarly well. Green data points in the bottom panel are the ratios between the \citet{L19_CIB} measurements and the fiducial curves. The ratio plot suggests that the fiducial model provides a slightly worse fit to the data compared to other model variations. A better fit, however, can be achieved with the \textit{four-parameter} model (see the right panel of Figure \ref{CIB_CIB_auto_feed_models_four_params}).

The best-fitting SED parameters for all considered model variations, specifically the present-day dust temperature ($T_{0}$) and the spectral index ($\beta_{\rm d}$), are shown in the left panel of Figure \ref{CIB_SEDfit_SFR_feed_models}. This plot highlights how these parameters vary across different physical and cosmological assumptions (see Equation \ref{eqn::SED_galaxies} for the SED definition). For reference, the right panel of Figure \ref{CIB_SEDfit_SFR_feed_models} shows the cosmic SFR density as a function of redshift, computed from all gas cells in each lightcone shell for every model variant.

Interestingly, an anti-correlation is observed between $T_{0}$ and $\beta_{\rm d}$, which is consistent with the parameter degeneracies revealed in the MCMC corner plot (Figure \ref{lensed_CIB_mcmc_L2800N5040_three_params}). Moreover, the $M_{\ast}-\sigma$ and LS8 models, which predict the lowest SFRs across redshifts, are also the most extreme outliers in the $T_{0}-\beta_{\rm d}$ parameter space, for models with thermal (non-jet) AGN feedback. These findings demonstrate that the SED fitting procedure is flexible enough to absorb most of the feedback and cosmology dependencies into the SED parameters. This implies that, if one constrains the SED parameters to match the CIB auto-power spectrum, the sensitivity to feedback and cosmology variations is mostly lost.

Importantly, however, this does \textit{not} imply that cross-correlations between the CIB and others tracers of LSS will be insensitive to feedback and cosmology variations.
To test this, Figure \ref{CIB_LSS_feed_models} shows the resulting CIB–tSZ and CIB–$\kappa$ cross-power spectra for this analysis at 857 GHz. Compared to Figure \ref{CIB_LSS_cross_feed_models_SED_fixed}, where SED parameters were fixed, we see there is still tangible (if somewhat reduced) sensitivity to variations in the feedback model and background cosmology. The LS8 model shows a $\approx 10\%$ suppression in the CIB–$\kappa$ cross-spectrum and up to $\approx 20\%$ in the CIB–tSZ case. Also, the small-scale suppression in the CIB–tSZ cross spectrum is still significant for the fgas-$8\sigma$ run relative to the fiducial model. 

Figure \ref{CIB_LSS_feed_models_correlation coefficient} shows the cross-correlation coefficient for the CIB–tSZ (left) and CIB–$\kappa$ (right) cross correlations, computed as $\rho_{\ell}$  = $C^{\rm CIB-XXX}_{\ell}/\sqrt{C^{\rm CIB-CIB}_{\ell}C^{\rm XXX-XXX}_{\ell}}$, where XXX refers to either the tSZ or $\kappa$ field. Note that the CIB and tSZ statistics are well converged up to redshift $z \approx 3$ (see Figure \ref{CIB_monopole} and discussions in \citealt{Ian_low_S8_FMG_paper}). However, the $\kappa$–$\kappa$ auto-power spectrum still has non-negligible contributions from higher redshifts. Since our $\kappa$ maps are integrated only up to $z = 4.5$, we correct for the missing high-redshift power using the Limber approximation \citep[e.g. see][]{limber_approx,Ian_low_S8_FMG_paper} and the matter power spectrum from different FLAMINGO model variants, evaluated at particle snapshots up to $z = 30$. For illustration, we only show results for the 545 GHz channel. 

The predicted correlation coefficients $\rho^{\mathrm{CIB-tSZ}}_{\ell}$ are qualitatively consistent with those from the AGORA simulations. The values from WebSky are however higher than those from FLAMINGO and AGORA, mainly due to WebSky’s lower tSZ auto power (see Figure \ref{yy_auto_sims_comp}). The $\rho^{\mathrm{CIB-\kappa}}_{\ell}$ predictions are quite consistent across different models.

Compared with Figure \ref{CIB_LSS_feed_models}, once the model dependence on the auto-power spectra is divided out, the differences of the CIB statistics become weaker than in the cross-power spectra alone. This is expected since the auto-power spectra themselves also carry model-dependent information. Nevertheless, as seen in the $\rho^{\mathrm{CIB-tSZ}}_{\ell}$ curves, differences of $\sim 10-20\%$ still remain among the various FLAMINGO models, and there are larger discrepancies among CMB simulations when different gas-physics prescriptions and cosmology are adopted.

These results suggest that, even after SED-based matching, CIB–LSS cross-correlations are still sensitive to the underlying baryonic physics and cosmology. Although statistics such as the CIB–tSZ cross-correlation are still difficult to measure with current survey setups, here we simply note that future CIB observations, especially when they are combined with other LSS tracers, may be able to provide meaningful, independent constraints on stellar (CIB auto) and AGN (CIB-LSS cross) feedback models, as well as cosmological parameters.

Note that an alternative approach to constraining the SED of the CIB could be to instead match the CIB cross-correlations and predict the auto-correlations.  The optimal strategy for calibration likely depends on the scientific goals (e.g., to best constrain cosmological parameters as opposed to test feedback models).

\subsection{Variations in kSZ power spectrum}\label{ssec::feedback_dependencies_kSZ}

\begin{figure}
\includegraphics[width=\columnwidth]{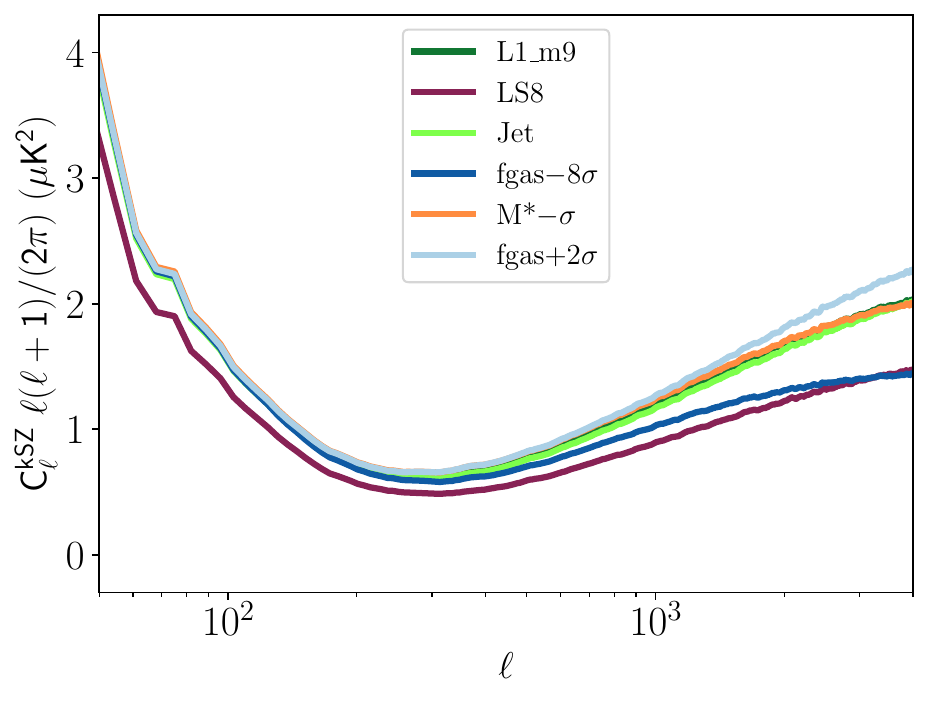}
   \caption{The kSZ auto-power spectrum for different FLAMINGO model variants.}
\label{kSZ_kSZ_auto_feed_models}
\end{figure}

Figure \ref{kSZ_kSZ_auto_feed_models} shows the impact of baryonic feedback and cosmology on the kSZ auto power spectrum. This summary statistic has several interesting trends at intermediate to small angular scales across the different FLAMINGO models. Notably, the fiducial, $M_{\ast}-\sigma$, and jet models yield very similar predictions, particularly at high multipoles. This is not unexpected, since all three of these models have been calibrated to the same gas mass fraction data and the kSZ effect scales in proportion to the integrated gas mass.
In contrast, models which systematically vary the gas mass fractions, such as fgas$+2\sigma$ and fgas$-8\sigma$, lead to very different behaviours on small angular scales.

In the LS8 cosmology, fewer massive haloes form. This results in a globally lower abundance of dense, coherently moving ionised gas in this model, which is reflected as a significant suppression of the kSZ power spectrum across all scales. As shown in Figure \ref{kSZ_kSZ_auto}, upcoming high-resolution CMB experiments may be sensitive to the kSZ signal at multipoles $l\gtrsim5000$. Our calculations highlight the potential the kSZ effect power spectrum has as a probe of both feedback and cosmology.

\section{Conclusion}\label{sec::conclusion}

Secondary anisotropies in the cosmic microwave background (CMB) contain a wealth of cosmological and astrophysical information. In this work, we have used the state-of-the-art FLAMINGO cosmological hydrodynamical simulations to generate a set of realistic mock CMB anisotropy maps, which are derived directly from the simulated properties of matter, gas, and accreting black holes. To our knowledge, this is the first time in which a broad range of anisotropies on full-sky lightcones has been self-consistently produced from cosmological hydrodynamical simulations. 

Using the full-sky HEALP\textsc{ix} lightcone maps, we have generated mock CMB lensing maps, Sunyaev–Zel’dovich (SZ) effect maps (with and without relativistic corrections), anisotropic screening maps (or equivalently, optical depth $\tau$ maps), and cosmic infrared background (CIB) maps, all integrated up to $z = 4.5$. In addition, we have produced radio point source maps from the black hole lightcone catalogues, integrated up to $z = 2.5$. Descriptions of how each component are modelled and derived from the lightcone-based catalogues are outlined in Section \ref{sec::Map_Generation_and_Power_Spectrum_Modelling}, and comparisons between the power spectra from our mock maps and observational data are presented in Section \ref{sec::Maps_and_power_spectra}. To fully utilise the model variations within the FLAMINGO simulations, we further explored the impact of feedback effects and cosmology on the auto- and cross-correlations of various anisotropies, with a particular focus on CMB lensing, the SZ effect, CIB, and radio point sources. The dependence of different statistics on feedback and cosmology is discussed in Section \ref{sec::discussions}. Our main findings are summarised as follows:

\begin{itemize}
   
    \item From the $(2.8~\rm Gpc)^{3}$ intermediate-resolution run (L2p8$\_$m9), we recover the general shape of the thermal SZ and kinetic SZ auto-power spectra, which are consistent with the predictions from other hydrodynamical and CMB mock simulations. However, the amplitudes differ due to varying treatments of subgrid physics and intergalactic medium (IGM) modelling (Figures \ref{yy_auto_sims_comp} and \ref{kSZ_kSZ_auto}).
    
    \item Using a simplified \textit{three-parameter} model, which includes a linear SFR$-L_{\rm bol, IR}$ conversion law, a modified blackbody spectral energy distribution (SED) template for infrared sources, and a power law redshift evolution of the dust temperature, we obtain good fits to the CIB auto- and cross-power spectra for the \textit{Planck} HFI channels (Figure \ref{CIB_CIB_stats}). From this fitting, we successfully constrain the SED and dust properties to values that are broadly consistent with those adopted in \citet{P16_CIB_tSZ}, although our best-fitting model favours a weaker redshift evolution of the dust temperature (Figure \ref{lensed_CIB_mcmc_L2800N5040_three_params}).

    \item Additional complexity can be introduced into this simplified CIB model, such as varying the proportionality constant in the SFR$-L_{\rm bol, IR}$ conversion law, or incorporating a dust-mass dependence with a more sophisticated SFR scaling. In these extended frameworks, we are still able to recover both the shape and amplitude of the measured curves (Figure \ref{CIB_CIB_auto_new_LIR_cal}). Other improvements in the SED, for example, treating the dust temperature as a function of various galaxy properties, are left for future work.
 
    \item The predicted CIB–Compton $y$ and CIB–$\kappa$ cross-correlations are in good agreement with both observations and previous halo model predictions (Figure \ref{CIB_LSS_stats}).
    
    \item We construct radio point-source catalogues from the black hole lightcone data, with luminosities assigned via abundance matching to the 150 MHz radio luminosity function from the LOFAR survey \citep{LOFAR_RLF}. A frequency scaling is then applied to convert $L_{150 ~\rm MHz}$ to CMB frequencies, with the scaling factor calibrated against the observed differential source counts (Figure \ref{radio_source_num_count}). Based on this, we successfully reproduce the radio point-source auto-power spectrum at 143 GHz (Figure \ref{fluxCl_radio_143GHz}). Furthermore, we find that the cross-correlations between radio sources and other anisotropies depend on the properties of the radio population (characterised by the Eddington accretion ratio $\lambda_{\rm Edd}$ of black holes, Figure \ref{Clxxx_radio_143GHz_different_lambda_edd}).
    
    \item The CIB power spectrum and its cross-correlations with other large-scale structure (LSS) tracers show strong dependence on feedback processes and cosmology. When keeping the CIB modelling parameters fixed across all model variations, we find that the strong stellar feedback model produces the largest suppression of both the CIB auto-correlation and the CIB–$\kappa$ cross-correlation, which is consistent with the suppression of star formation. AGN feedback suppresses small-scale power in the CIB–Compton $y$ cross-correlation. Finally, the low-$\sigma_{8}$ (LS8) model consistently yields the lowest amplitudes across all CIB statistics (Figures \ref{CIB_CIB_auto_feed_models} and \ref{CIB_LSS_cross_feed_models_SED_fixed}).
    
    \item When refitting the CIB-predicted curves from each model variation to the same set of measured power spectra (and noting that the various runs can reproduce the CIB power spectra similarly well when the SED is refitted), we find that the best-fitting SED parameters show strong model dependence (Figure \ref{CIB_SEDfit_SFR_feed_models}). In other words, the flexibility/uncertainty in the SED can absorb the dependence of the CIB power spectrum on feedback and cosmology. However, cross-correlations between the CIB and other LSS tracers retain sensitivity to the feedback modelling and background cosmology (Figures \ref{CIB_LSS_feed_models} and \ref{CIB_LSS_feed_models_correlation coefficient}) even after the SED parameters have been refitted to the CIB auto power spectra.  An alternative approach to SED calibration could be to fit the CIB-LSS cross-correlations and predict the CIB auto power spectra.  
     
    \item The kinetic SZ power spectrum also shows a strong dependence on baryonic feedback and cosmology, especially at small scales where feedback effects dominate. Increasing the strength of AGN feedback results in the greatest suppression of power on small scales, while the LS8 model yields the lowest amplitude across all scales (Figure \ref{kSZ_kSZ_auto_feed_models}). 

    \item We also present predictions for the tSZ monopole (Table \ref{tab:y_monopole_table}) and the CIB monopole (Figure \ref{CIB_monopole}). All predicted $y$ monopoles from the different FLAMINGO models lie well below the direct COBE–FIRAS 95$\%$ upper limit.  The fiducial model sits slightly above the indirect \textit{Planck} tomographic estimate of \citet{tSZ_tomo_Chiang}. The strongest AGN model ($f_{\rm gas}-8\sigma$ run) yields the largest value, which is in mild tension with the tomographic measurements, while the LS8 cosmology (with fiducial feedback modelling) gives the lowest monopole and is in excellent agreement with the \textit{Planck} measurement from \citet{tSZ_tomo_Chiang}. For the CIB monopole, using the best-fitting SED from the \textit{three-parameter} model, we show that the CIB intensity converges by $z\approx 3$ with a spectral break above $\nu\gtrsim1000~\rm GHz$. The predicted CIB monopoles broadly match the \textit{Planck}-based constraints of \citet{CIB_tomo_ref1}.
    
\end{itemize}

Our simulations reproduce a wide range of observational constraints with good accuracy and can be used to make forecasts for upcoming CMB surveys such as the Simons Observatory \citep{Sehgal_radio_SO_forecast} and CMB-HD \citep{CMB_HD}. In particular, we highlight the importance of CIB and radio point-source statistics, as well as their dependence on feedback models and cosmology. We expect these mock CMB maps to be valuable for testing cross-correlations between different secondary anisotropies, and for validating component-separation pipelines used to separate their individual contributions in real observations.

\section*{Acknowledgements}

The authors thank the anonymous referee for helpful suggestions which improved the paper. This work was supported by the Science and Technology Facilities Council (grant number ST/Y002733/1). This project has received funding from the European Research Council (ERC) under the European Union’s Horizon 2020 research and innovation programme (grant agreement No 769130). This work used the DiRAC@Durham facility managed by the Institute for Computational Cosmology on behalf of the STFC DiRAC HPC Facility (\url{https://dirac.ac.uk/}). The equipment was funded by BEIS capital funding via STFC capital grants ST/K00042X/1, ST/P002293/1, ST/R002371/1 and ST/S002502/1, Durham University and STFC operations grant ST/R000832/1. DiRAC is part of the National e-Infrastructure. JCH is supported by STFC consolidated grant ST/X001075/1.

\section*{Data availability}
The maps and catalogues produced as part of this study will be made publicly available on the FLAMINGO data release website upon acceptance of the paper. Additional derived data products can be provided upon reasonable request to the corresponding author.

\bibliographystyle{mnras}
\bibliography{draft}

@ARTICLE{Braspenning2024,
       author = {{Braspenning}, Joey and {Schaye}, Joop and {Schaller}, Matthieu and {McCarthy}, Ian G. and {Kay}, Scott T. and {Helly}, John C. and {Kugel}, Roi and {Elbers}, Willem and {Frenk}, Carlos S. and {Kwan}, Juliana and {Salcido}, Jaime and {van Daalen}, Marcel P. and {Vandenbroucke}, Bert},
        title = "{The FLAMINGO project: galaxy clusters in comparison to X-ray observations}",
      journal = {\mnras},
         year = 2024,
        month = sep,
       volume = {533},
       number = {3},
        pages = {2656-2676},
          doi = {10.1093/mnras/stae1436},
archivePrefix = {arXiv},
       eprint = {2312.08277},
 primaryClass = {astro-ph.GA},
       adsurl = {https://ui.adsabs.harvard.edu/abs/2024MNRAS.533.2656B}
}

@ARTICLE{referee_ref_Ian_lowsigma8,
       author = {{McCarthy}, Ian G. and {Salcido}, Jaime and {Schaye}, Joop and {Kwan}, Juliana and {Elbers}, Willem and {Kugel}, Roi and {Schaller}, Matthieu and {Helly}, John C. and {Braspenning}, Joey and {Frenk}, Carlos S. and {van Daalen}, Marcel P. and {Vandenbroucke}, Bert and {Conley}, Jonah T. and {Font}, Andreea S. and {Upadhye}, Amol},
        title = "{The FLAMINGO project: revisiting the S$_{8}$ tension and the role of baryonic physics}",
      journal = {MNRAS},
         year = 2023,
        month = dec,
       volume = {526},
       number = {4},
        pages = {5494-5519},
          doi = {10.1093/mnras/stad3107},
archivePrefix = {arXiv},
       eprint = {2309.07959},
 primaryClass = {astro-ph.CO},
       adsurl = {https://ui.adsabs.harvard.edu/abs/2023MNRAS.526.5494M}
}

@ARTICLE{referee_ref_BA_ref4,
       author = {{Broxterman}, Jeger C. and {Schaller}, Matthieu and {Schaye}, Joop and {Hoekstra}, Henk and {Kuijken}, Konrad and {Helly}, John C. and {Kugel}, Roi and {Braspenning}, Joey and {Elbers}, Willem and {Frenk}, Carlos S. and {Kwan}, Juliana and {McCarthy}, Ian G. and {Salcido}, Jaime and {van Daalen}, Marcel P. and {Vandenbroucke}, Bert},
        title = "{The FLAMINGO project: baryonic impact on weak gravitational lensing convergence peak counts}",
      journal = {MNRAS},
         year = 2024,
        month = apr,
       volume = {529},
       number = {3},
        pages = {2309-2326},
          doi = {10.1093/mnras/stae698},
archivePrefix = {arXiv},
       eprint = {2312.08450},
 primaryClass = {astro-ph.CO},
       adsurl = {https://ui.adsabs.harvard.edu/abs/2024MNRAS.529.2309B}
}

@ARTICLE{referee_ref_BA_ref3,
       author = {{Hilbert}, Stefan and {Barreira}, Alexandre and {Fabbian}, Giulio and {Fosalba}, Pablo and {Giocoli}, Carlo and {Bose}, Sownak and {Calabrese}, Matteo and {Carbone}, Carmelita and {Davies}, Christopher T. and {Li}, Baojiu and {Llinares}, Claudio and {Monaco}, Pierluigi},
        title = "{The accuracy of weak lensing simulations}",
      journal = {MNRAS},
         year = 2020,
        month = mar,
       volume = {493},
       number = {1},
        pages = {305-319},
          doi = {10.1093/mnras/staa281},
archivePrefix = {arXiv},
       eprint = {1910.10625},
 primaryClass = {astro-ph.CO},
       adsurl = {https://ui.adsabs.harvard.edu/abs/2020MNRAS.493..305H}
}

@ARTICLE{referee_ref_BA_ref2,
       author = {{Fabbian}, Giulio and {Lewis}, Antony and {Beck}, Dominic},
        title = "{CMB lensing reconstruction biases in cross-correlation with large-scale structure probes}",
      journal = {JCAP},
         year = 2019,
        month = oct,
       volume = {2019},
       number = {10},
          eid = {057},
        pages = {057},
          doi = {10.1088/1475-7516/2019/10/057},
archivePrefix = {arXiv},
       eprint = {1906.08760},
 primaryClass = {astro-ph.CO},
       adsurl = {https://ui.adsabs.harvard.edu/abs/2019JCAP...10..057F}
}

@ARTICLE{referee_ref_BA_ref1,
       author = {{Giocoli}, Carlo and {Jullo}, Eric and {Metcalf}, R. Benton and {de la Torre}, Sylvain and {Yepes}, Gustavo and {Prada}, Francisco and {Comparat}, Johan and {G{\"o}ttlober}, Stefan and {Kyplin}, Anatoly and {Kneib}, Jean-Paul and {Petkova}, Margarita and {Shan}, Huan Yuan and {Tessore}, Nicolas},
        title = "{Multi Dark Lens Simulations: weak lensing light-cones and data base presentation}",
      journal = {MNRAS},
         year = 2016,
        month = sep,
       volume = {461},
       number = {1},
        pages = {209-223},
          doi = {10.1093/mnras/stw1336},
archivePrefix = {arXiv},
       eprint = {1511.08211},
 primaryClass = {astro-ph.CO},
       adsurl = {https://ui.adsabs.harvard.edu/abs/2016MNRAS.461..209G}
}

@ARTICLE{Akino_fgas,
       author = {{Akino}, Daichi and {Eckert}, Dominique and {Okabe}, Nobuhiro and {Sereno}, Mauro and {Umetsu}, Keiichi and {Oguri}, Masamune and {Gastaldello}, Fabio and {Chiu}, I.-Non and {Ettori}, Stefano and {Evrard}, August E. and {Farahi}, Arya and {Maughan}, Ben and {Pierre}, Marguerite and {Ricci}, Marina and {Valtchanov}, Ivan and {McCarthy}, Ian and {McGee}, Sean and {Miyazaki}, Satoshi and {Nishizawa}, Atsushi J. and {Tanaka}, Masayuki},
        title = "{HSC-XXL: Baryon budget of the 136 XXL groups and clusters}",
      journal = {\pasj},
         year = 2022,
        month = feb,
       volume = {74},
       number = {1},
        pages = {175-208},
          doi = {10.1093/pasj/psab115},
archivePrefix = {arXiv},
       eprint = {2111.10080},
 primaryClass = {astro-ph.CO},
       adsurl = {https://ui.adsabs.harvard.edu/abs/2022PASJ...74..175A}
}

@ARTICLE{Driver_SMF,
       author = {{Driver}, Simon P. and {Bellstedt}, Sabine and {Robotham}, Aaron S.~G. and {Baldry}, Ivan K. and {Davies}, Luke J. and {Liske}, Jochen and {Obreschkow}, Danail and {Taylor}, Edward N. and {Wright}, Angus H. and {Alpaslan}, Mehmet and {Bamford}, Steven P. and {Bauer}, Amanda E. and {Bland-Hawthorn}, Joss and {Bilicki}, Maciej and {Bravo}, Mat{\'\i}as and {Brough}, Sarah and {Casura}, Sarah and {Cluver}, Michelle E. and {Colless}, Matthew and {Conselice}, Christopher J. and {Croom}, Scott M. and {de Jong}, Jelte and {D'Eugenio}, Franceso and {De Propris}, Roberto and {Dogruel}, Burak and {Drinkwater}, Michael J. and {Dvornik}, Andrej and {Farrow}, Daniel J. and {Frenk}, Carlos S. and {Giblin}, Benjamin and {Graham}, Alister W. and {Grootes}, Meiert W. and {Gunawardhana}, Madusha L.~P. and {Hashemizadeh}, Abdolhosein and {H{\"a}u{\ss}ler}, Boris and {Heymans}, Catherine and {Hildebrandt}, Hendrik and {Holwerda}, Benne W. and {Hopkins}, Andrew M. and {Jarrett}, Tom H. and {Heath Jones}, D. and {Kelvin}, Lee S. and {Koushan}, Soheil and {Kuijken}, Konrad and {Lara-L{\'o}pez}, Maritza A. and {Lange}, Rebecca and {L{\'o}pez-S{\'a}nchez}, {\'A}ngel R. and {Loveday}, Jon and {Mahajan}, Smriti and {Meyer}, Martin and {Moffett}, Amanda J. and {Napolitano}, Nicola R. and {Norberg}, Peder and {Owers}, Matt S. and {Radovich}, Mario and {Raouf}, Mojtaba and {Peacock}, John A. and {Phillipps}, Steven and {Pimbblet}, Kevin A. and {Popescu}, Cristina and {Said}, Khaled and {Sansom}, Anne E. and {Seibert}, Mark and {Sutherland}, Will J. and {Thorne}, Jessica E. and {Tuffs}, Richard J. and {Turner}, Ryan and {van der Wel}, Arjen and {van Kampen}, Eelco and {Wilkins}, Steve M.},
        title = "{Galaxy And Mass Assembly (GAMA): Data Release 4 and the z < 0.1 total and z < 0.08 morphological galaxy stellar mass functions}",
      journal = {\mnras},
         year = 2022,
        month = jun,
       volume = {513},
       number = {1},
        pages = {439-467},
          doi = {10.1093/mnras/stac472},
archivePrefix = {arXiv},
       eprint = {2203.08539},
 primaryClass = {astro-ph.GA},
       adsurl = {https://ui.adsabs.harvard.edu/abs/2022MNRAS.513..439D}
}

@ARTICLE{Matthieu_24,
       author = {{Schaller}, Matthieu and {Schaye}, Joop and {Kugel}, Roi and {Broxterman}, Jeger C. and {van Daalen}, Marcel P.},
        title = "{The FLAMINGO project: baryon effects on the matter power spectrum}",
      journal = {\mnras},
         year = 2025,
        month = may,
       volume = {539},
       number = {2},
        pages = {1337-1351},
          doi = {10.1093/mnras/staf569},
archivePrefix = {arXiv},
       eprint = {2410.17109},
 primaryClass = {astro-ph.CO},
       adsurl = {https://ui.adsabs.harvard.edu/abs/2025MNRAS.539.1337S}
}

@ARTICLE{QLF_shortage_of_bright_quasar,
       author = {{Ding}, Boyi and {Pizzati}, Elia and {Schaye}, Joop and {Hennawi}, Joseph F. and {McDonald}, William and {Schaller}, Matthieu},
        title = "{The luminosity function and clustering of bright quasars in the FLAMINGO cosmological simulations}",
      journal = {arXiv e-prints},
         year = 2025,
        month = oct,
          eid = {arXiv:2510.24283},
        pages = {arXiv:2510.24283},
          doi = {10.48550/arXiv.2510.24283},
archivePrefix = {arXiv},
       eprint = {2510.24283},
 primaryClass = {astro-ph.GA},
       adsurl = {https://ui.adsabs.harvard.edu/abs/2025arXiv251024283D}
}

@ARTICLE{Hill2015,
       author = {{Hill}, J. Colin and {Battaglia}, Nick and {Chluba}, Jens and {Ferraro}, Simone and {Schaan}, Emmanuel and {Spergel}, David N.},
        title = "{Taking the Universe's Temperature with Spectral Distortions of the Cosmic Microwave Background}",
      journal = {\prl},
         year = 2015,
        month = dec,
       volume = {115},
       number = {26},
          eid = {261301},
        pages = {261301},
          doi = {10.1103/PhysRevLett.115.261301},
archivePrefix = {arXiv},
       eprint = {1507.01583},
 primaryClass = {astro-ph.CO},
       adsurl = {https://ui.adsabs.harvard.edu/abs/2015PhRvL.115z1301H}
}

@ARTICLE{Thiele2022,
       author = {{Thiele}, Leander and {Wadekar}, Digvijay and {Hill}, J. Colin and {Battaglia}, Nicholas and {Chluba}, Jens and {Villaescusa-Navarro}, Francisco and {Hernquist}, Lars and {Vogelsberger}, Mark and {Angl{\'e}s-Alc{\'a}zar}, Daniel and {Marinacci}, Federico},
        title = "{Percent-level constraints on baryonic feedback with spectral distortion measurements}",
      journal = {\prd},
         year = 2022,
        month = apr,
       volume = {105},
       number = {8},
          eid = {083505},
        pages = {083505},
          doi = {10.1103/PhysRevD.105.083505},
archivePrefix = {arXiv},
       eprint = {2201.01663},
 primaryClass = {astro-ph.CO},
       adsurl = {https://ui.adsabs.harvard.edu/abs/2022PhRvD.105h3505T}
}

@ARTICLE{Heckman_Best_14,
       author = {{Heckman}, Timothy M. and {Best}, Philip N.},
        title = "{The Coevolution of Galaxies and Supermassive Black Holes: Insights from Surveys of the Contemporary Universe}",
      journal = {\araa},
         year = 2014,
        month = aug,
       volume = {52},
        pages = {589-660},
          doi = {10.1146/annurev-astro-081913-035722},
archivePrefix = {arXiv},
       eprint = {1403.4620},
 primaryClass = {astro-ph.GA},
       adsurl = {https://ui.adsabs.harvard.edu/abs/2014ARA&A..52..589H}
}

@ARTICLE{Fabian_2012,
       author = {{Fabian}, A.~C.},
        title = "{Observational Evidence of Active Galactic Nuclei Feedback}",
      journal = {\araa},
         year = 2012,
        month = sep,
       volume = {50},
        pages = {455-489},
          doi = {10.1146/annurev-astro-081811-125521},
archivePrefix = {arXiv},
       eprint = {1204.4114},
 primaryClass = {astro-ph.CO},
       adsurl = {https://ui.adsabs.harvard.edu/abs/2012ARA&A..50..455F}
}

@ARTICLE{Best_Heckman_12,
       author = {{Best}, P.~N. and {Heckman}, T.~M.},
        title = "{On the fundamental dichotomy in the local radio-AGN population: accretion, evolution and host galaxy properties}",
      journal = {\mnras},
         year = 2012,
        month = apr,
       volume = {421},
       number = {2},
        pages = {1569-1582},
          doi = {10.1111/j.1365-2966.2012.20414.x},
archivePrefix = {arXiv},
       eprint = {1201.2397},
 primaryClass = {astro-ph.CO},
       adsurl = {https://ui.adsabs.harvard.edu/abs/2012MNRAS.421.1569B}
}

@ARTICLE{Bolliet2018,
       author = {{Bolliet}, Boris and {Comis}, Barbara and {Komatsu}, Eiichiro and {Mac{\'\i}as-P{\'e}rez}, Juan Francisco},
        title = "{Dark energy constraints from the thermal Sunyaev-Zeldovich power spectrum}",
      journal = {\mnras},
         year = 2018,
        month = jul,
       volume = {477},
       number = {4},
        pages = {4957-4967},
          doi = {10.1093/mnras/sty823},
archivePrefix = {arXiv},
       eprint = {1712.00788},
 primaryClass = {astro-ph.CO},
       adsurl = {https://ui.adsabs.harvard.edu/abs/2018MNRAS.477.4957B}
}

@ARTICLE{antilles,
       author = {{Salcido}, Jaime and {McCarthy}, Ian G. and {Kwan}, Juliana and {Upadhye}, Amol and {Font}, Andreea S.},
        title = "{SP(k) - a hydrodynamical simulation-based model for the impact of baryon physics on the non-linear matter power spectrum}",
      journal = {\mnras},
         year = 2023,
        month = aug,
       volume = {523},
       number = {2},
        pages = {2247-2262},
          doi = {10.1093/mnras/stad1474},
archivePrefix = {arXiv},
       eprint = {2305.09710},
 primaryClass = {astro-ph.CO},
       adsurl = {https://ui.adsabs.harvard.edu/abs/2023MNRAS.523.2247S}
}

@INPROCEEDINGS{class_sz,
       author = {{Bolliet}, B. and {Kusiak}, A. and {McCarthy}, F. and {Sabyr}, A. and {Surrao}, K. and {Hill}, J.~C. and {Chluba}, J. and {Ferraro}, S. and {Hadzhiyska}, B. and {Han}, D. and {Mac{\'\i}as-P{\'e}rez}, J.~F. and {Madhavacheril}, M. and {Maniyar}, A. and {Mehta}, Y. and {Pandey}, S. and {Schaan}, E. and {Sherwin}, B. and {Mancini}, A. Spurio and {Zubeldia}, {\'I}.},
        title = "{class\_sz I: Overview}",
    booktitle = {mm Universe 2023 - Observing the Universe at mm Wavelengths},
         year = 2024,
       series = {European Physical Journal Web of Conferences},
       volume = {293},
        month = jun,
    publisher = {EDP},
          eid = {00008},
        pages = {00008},
          doi = {10.1051/epjconf/202429300008},
       adsurl = {https://ui.adsabs.harvard.edu/abs/2024EPJWC.29300008B}
}

@ARTICLE{agora_ref,
       author = {{Omori}, Yuuki},
        title = "{AGORA: Multicomponent simulation for cross-survey science}",
      journal = {\mnras},
         year = 2024,
        month = jun,
       volume = {530},
       number = {4},
        pages = {5030-5068},
          doi = {10.1093/mnras/stae1031},
archivePrefix = {arXiv},
       eprint = {2212.07420},
 primaryClass = {astro-ph.CO},
       adsurl = {https://ui.adsabs.harvard.edu/abs/2024MNRAS.530.5030O}
}

@ARTICLE{P14_CIB,
       author = {{Planck Collaboration} and {Ade}, P.~A.~R. and {Aghanim}, N. and {Armitage-Caplan}, C. and {Arnaud}, M. and {Ashdown}, M. and {Atrio-Barandela}, F. and {Aumont}, J. and {Baccigalupi}, C. and {Banday}, A.~J. and {Barreiro}, R.~B. and {Bartlett}, J.~G. and {Battaner}, E. and {Benabed}, K. and {Beno{\^\i}t}, A. and {Benoit-L{\'e}vy}, A. and {Bernard}, J. -P. and {Bersanelli}, M. and {Bethermin}, M. and {Bielewicz}, P. and {Blagrave}, K. and {Bobin}, J. and {Bock}, J.~J. and {Bonaldi}, A. and {Bond}, J.~R. and {Borrill}, J. and {Bouchet}, F.~R. and {Boulanger}, F. and {Bridges}, M. and {Bucher}, M. and {Burigana}, C. and {Butler}, R.~C. and {Cardoso}, J. -F. and {Catalano}, A. and {Challinor}, A. and {Chamballu}, A. and {Chen}, X. and {Chiang}, H.~C. and {Chiang}, L. -Y. and {Christensen}, P.~R. and {Church}, S. and {Clements}, D.~L. and {Colombi}, S. and {Colombo}, L.~P.~L. and {Couchot}, F. and {Coulais}, A. and {Crill}, B.~P. and {Curto}, A. and {Cuttaia}, F. and {Danese}, L. and {Davies}, R.~D. and {Davis}, R.~J. and {de Bernardis}, P. and {de Rosa}, A. and {de Zotti}, G. and {Delabrouille}, J. and {Delouis}, J. -M. and {D{\'e}sert}, F. -X. and {Dickinson}, C. and {Diego}, J.~M. and {Dole}, H. and {Donzelli}, S. and {Dor{\'e}}, O. and {Douspis}, M. and {Dupac}, X. and {Efstathiou}, G. and {En{\ss}lin}, T.~A. and {Eriksen}, H.~K. and {Finelli}, F. and {Forni}, O. and {Frailis}, M. and {Franceschi}, E. and {Galeotta}, S. and {Ganga}, K. and {Ghosh}, T. and {Giard}, M. and {Giraud-H{\'e}raud}, Y. and {Gonz{\'a}lez-Nuevo}, J. and {G{\'o}rski}, K.~M. and {Gratton}, S. and {Gregorio}, A. and {Gruppuso}, A. and {Hansen}, F.~K. and {Hanson}, D. and {Harrison}, D. and {Helou}, G. and {Henrot-Versill{\'e}}, S. and {Hern{\'a}ndez-Monteagudo}, C. and {Herranz}, D. and {Hildebrandt}, S.~R. and {Hivon}, E. and {Hobson}, M. and {Holmes}, W.~A. and {Hornstrup}, A. and {Hovest}, W. and {Huffenberger}, K.~M. and {Jaffe}, A.~H. and {Jaffe}, T.~R. and {Jones}, W.~C. and {Juvela}, M. and {Kalberla}, P. and {Keih{\"a}nen}, E. and {Kerp}, J. and {Keskitalo}, R. and {Kisner}, T.~S. and {Kneissl}, R. and {Knoche}, J. and {Knox}, L. and {Kunz}, M. and {Kurki-Suonio}, H. and {Lacasa}, F. and {Lagache}, G. and {L{\"a}hteenm{\"a}ki}, A. and {Lamarre}, J. -M. and {Langer}, M. and {Lasenby}, A. and {Laureijs}, R.~J. and {Lawrence}, C.~R. and {Leonardi}, R. and {Le{\'o}n-Tavares}, J. and {Lesgourgues}, J. and {Liguori}, M. and {Lilje}, P.~B. and {Linden-V{\o}rnle}, M. and {L{\'o}pez-Caniego}, M. and {Lubin}, P.~M. and {Mac{\'\i}as-P{\'e}rez}, J.~F. and {Maffei}, B. and {Maino}, D. and {Mandolesi}, N. and {Maris}, M. and {Marshall}, D.~J. and {Martin}, P.~G. and {Mart{\'\i}nez-Gonz{\'a}lez}, E. and {Masi}, S. and {Massardi}, M. and {Matarrese}, S. and {Matthai}, F. and {Mazzotta}, P. and {Melchiorri}, A. and {Mendes}, L. and {Mennella}, A. and {Migliaccio}, M. and {Mitra}, S. and {Miville-Desch{\^e}nes}, M. -A. and {Moneti}, A. and {Montier}, L. and {Morgante}, G. and {Mortlock}, D. and {Munshi}, D. and {Murphy}, J.~A. and {Naselsky}, P. and {Nati}, F. and {Natoli}, P. and {Netterfield}, C.~B. and {N{\o}rgaard-Nielsen}, H.~U. and {Noviello}, F. and {Novikov}, D. and {Novikov}, I. and {Osborne}, S. and {Oxborrow}, C.~A. and {Paci}, F. and {Pagano}, L. and {Pajot}, F. and {Paladini}, R. and {Paoletti}, D. and {Partridge}, B. and {Pasian}, F. and {Patanchon}, G. and {Perdereau}, O. and {Perotto}, L. and {Perrotta}, F. and {Piacentini}, F. and {Piat}, M. and {Pierpaoli}, E. and {Pietrobon}, D. and {Plaszczynski}, S. and {Pointecouteau}, E. and {Polenta}, G. and {Ponthieu}, N. and {Popa}, L. and {Poutanen}, T. and {Pratt}, G.~W. and {Pr{\'e}zeau}, G. and {Prunet}, S. and {Puget}, J. -L. and {Rachen}, J.~P. and {Reach}, W.~T. and {Rebolo}, R. and {Reinecke}, M. and {Remazeilles}, M. and {Renault}, C. and {Ricciardi}, S. and {Riller}, T. and {Ristorcelli}, I. and {Rocha}, G. and {Rosset}, C. and {Roudier}, G. and {Rowan-Robinson}, M. and {Rubi{\~n}o-Mart{\'\i}n}, J.~A. and {Rusholme}, B. and {Sandri}, M. and {Santos}, D. and {Savini}, G. and {Scott}, D. and {Seiffert}, M.~D. and {Serra}, P. and {Shellard}, E.~P.~S. and {Spencer}, L.~D. and {Starck}, J. -L. and {Stolyarov}, V. and {Stompor}, R. and {Sudiwala}, R. and {Sunyaev}, R. and {Sureau}, F. and {Sutton}, D. and {Suur-Uski}, A. -S. and {Sygnet}, J. -F. and {Tauber}, J.~A. and {Tavagnacco}, D. and {Terenzi}, L. and {Toffolatti}, L. and {Tomasi}, M. and {Tristram}, M. and {Tucci}, M. and {Tuovinen}, J. and {T{\"u}rler}, M. and {Valenziano}, L. and {Valiviita}, J. and {Van Tent}, B. and {Vielva}, P. and {Villa}, F. and {Vittorio}, N. and {Wade}, L.~A. and {Wandelt}, B.~D. and {Welikala}, N. and {White}, M. and {White}, S.~D.~M. and {Winkel}, B. and {Yvon}, D. and {Zacchei}, A. and {Zonca}, A.},
        title = "{Planck 2013 results. XXX. Cosmic infrared background measurements and implications for star formation}",
      journal = {\aap},
         year = 2014,
        month = nov,
       volume = {571},
          eid = {A30},
        pages = {A30},
          doi = {10.1051/0004-6361/201322093},
archivePrefix = {arXiv},
       eprint = {1309.0382},
 primaryClass = {astro-ph.CO},
       adsurl = {https://ui.adsabs.harvard.edu/abs/2014A&A...571A..30P}
}

@ARTICLE{L19_CIB,
       author = {{Lenz}, Daniel and {Dor{\'e}}, Olivier and {Lagache}, Guilaine},
        title = "{Large-scale Maps of the Cosmic Infrared Background from Planck}",
      journal = {\apj},
         year = 2019,
        month = sep,
       volume = {883},
       number = {1},
          eid = {75},
        pages = {75},
          doi = {10.3847/1538-4357/ab3c2b},
archivePrefix = {arXiv},
       eprint = {1905.00426},
 primaryClass = {astro-ph.CO},
       adsurl = {https://ui.adsabs.harvard.edu/abs/2019ApJ...883...75L}
}

@ARTICLE{M2020_CIB_kappa,
       author = {{Maniyar}, A. and {B{\'e}thermin}, M. and {Lagache}, G.},
        title = "{Simple halo model formalism for the cosmic infrared background and its correlation with the thermal Sunyaev-Zel'dovich effect}",
      journal = {\aap},
         year = 2021,
        month = jan,
       volume = {645},
          eid = {A40},
        pages = {A40},
          doi = {10.1051/0004-6361/202038790},
archivePrefix = {arXiv},
       eprint = {2006.16329},
 primaryClass = {astro-ph.CO},
       adsurl = {https://ui.adsabs.harvard.edu/abs/2021A&A...645A..40M}
}

@ARTICLE{P16_CIB_tSZ,
       author = {{Planck Collaboration} and {Ade}, P.~A.~R. and {Aghanim}, N. and {Arnaud}, M. and {Aumont}, J. and {Baccigalupi}, C. and {Banday}, A.~J. and {Barreiro}, R.~B. and {Bartlett}, J.~G. and {Bartolo}, N. and {Battaner}, E. and {Benabed}, K. and {Benoit-L{\'e}vy}, A. and {Bernard}, J. -P. and {Bersanelli}, M. and {Bielewicz}, P. and {Bock}, J.~J. and {Bonaldi}, A. and {Bonavera}, L. and {Bond}, J.~R. and {Borrill}, J. and {Bouchet}, F.~R. and {Burigana}, C. and {Butler}, R.~C. and {Calabrese}, E. and {Catalano}, A. and {Chamballu}, A. and {Chiang}, H.~C. and {Christensen}, P.~R. and {Churazov}, E. and {Clements}, D.~L. and {Colombo}, L.~P.~L. and {Combet}, C. and {Comis}, B. and {Couchot}, F. and {Coulais}, A. and {Crill}, B.~P. and {Curto}, A. and {Cuttaia}, F. and {Danese}, L. and {Davies}, R.~D. and {Davis}, R.~J. and {de Bernardis}, P. and {de Rosa}, A. and {de Zotti}, G. and {Delabrouille}, J. and {Dickinson}, C. and {Diego}, J.~M. and {Dole}, H. and {Donzelli}, S. and {Dor{\'e}}, O. and {Douspis}, M. and {Ducout}, A. and {Dupac}, X. and {Efstathiou}, G. and {Elsner}, F. and {En{\ss}lin}, T.~A. and {Eriksen}, H.~K. and {Finelli}, F. and {Flores-Cacho}, I. and {Forni}, O. and {Frailis}, M. and {Fraisse}, A.~A. and {Franceschi}, E. and {Galeotta}, S. and {Galli}, S. and {Ganga}, K. and {G{\'e}nova-Santos}, R.~T. and {Giard}, M. and {Giraud-H{\'e}raud}, Y. and {Gjerl{\o}w}, E. and {Gonz{\'a}lez-Nuevo}, J. and {G{\'o}rski}, K.~M. and {Gregorio}, A. and {Gruppuso}, A. and {Gudmundsson}, J.~E. and {Hansen}, F.~K. and {Harrison}, D.~L. and {Helou}, G. and {Hern{\'a}ndez-Monteagudo}, C. and {Herranz}, D. and {Hildebrandt}, S.~R. and {Hivon}, E. and {Hobson}, M. and {Hornstrup}, A. and {Hovest}, W. and {Huffenberger}, K.~M. and {Hurier}, G. and {Jaffe}, A.~H. and {Jaffe}, T.~R. and {Jones}, W.~C. and {Keih{\"a}nen}, E. and {Keskitalo}, R. and {Kisner}, T.~S. and {Kneissl}, R. and {Knoche}, J. and {Kunz}, M. and {Kurki-Suonio}, H. and {Lagache}, G. and {Lamarre}, J. -M. and {Langer}, M. and {Lasenby}, A. and {Lattanzi}, M. and {Lawrence}, C.~R. and {Leonardi}, R. and {Levrier}, F. and {Lilje}, P.~B. and {Linden-V{\o}rnle}, M. and {L{\'o}pez-Caniego}, M. and {Lubin}, P.~M. and {Mac{\'\i}as-P{\'e}rez}, J.~F. and {Maffei}, B. and {Maggio}, G. and {Maino}, D. and {Mak}, D.~S.~Y. and {Mandolesi}, N. and {Mangilli}, A. and {Maris}, M. and {Martin}, P.~G. and {Mart{\'\i}nez-Gonz{\'a}lez}, E. and {Masi}, S. and {Matarrese}, S. and {Melchiorri}, A. and {Mennella}, A. and {Migliaccio}, M. and {Mitra}, S. and {Miville-Desch{\^e}nes}, M. -A. and {Moneti}, A. and {Montier}, L. and {Morgante}, G. and {Mortlock}, D. and {Munshi}, D. and {Murphy}, J.~A. and {Nati}, F. and {Natoli}, P. and {Noviello}, F. and {Novikov}, D. and {Novikov}, I. and {Oxborrow}, C.~A. and {Paci}, F. and {Pagano}, L. and {Pajot}, F. and {Paoletti}, D. and {Partridge}, B. and {Pasian}, F. and {Pearson}, T.~J. and {Perdereau}, O. and {Perotto}, L. and {Pettorino}, V. and {Piacentini}, F. and {Piat}, M. and {Pierpaoli}, E. and {Plaszczynski}, S. and {Pointecouteau}, E. and {Polenta}, G. and {Ponthieu}, N. and {Pratt}, G.~W. and {Prunet}, S. and {Puget}, J. -L. and {Rachen}, J.~P. and {Reinecke}, M. and {Remazeilles}, M. and {Renault}, C. and {Renzi}, A. and {Ristorcelli}, I. and {Rocha}, G. and {Rosset}, C. and {Rossetti}, M. and {Roudier}, G. and {Rubi{\~n}o-Mart{\'\i}n}, J.~A. and {Rusholme}, B. and {Sandri}, M. and {Santos}, D. and {Savelainen}, M. and {Savini}, G. and {Scott}, D. and {Spencer}, L.~D. and {Stolyarov}, V. and {Stompor}, R. and {Sunyaev}, R. and {Sutton}, D. and {Suur-Uski}, A. -S. and {Sygnet}, J. -F. and {Tauber}, J.~A. and {Terenzi}, L. and {Toffolatti}, L. and {Tomasi}, M. and {Tristram}, M. and {Tucci}, M. and {Umana}, G. and {Valenziano}, L. and {Valiviita}, J. and {Van Tent}, B. and {Vielva}, P. and {Villa}, F. and {Wade}, L.~A. and {Wandelt}, B.~D. and {Wehus}, I.~K. and {Welikala}, N. and {Yvon}, D. and {Zacchei}, A. and {Zonca}, A.},
        title = "{Planck 2015 results. XXIII. The thermal Sunyaev-Zeldovich effect-cosmic infrared background correlation}",
      journal = {\aap},
         year = 2016,
        month = sep,
       volume = {594},
          eid = {A23},
        pages = {A23},
          doi = {10.1051/0004-6361/201527418},
archivePrefix = {arXiv},
       eprint = {1509.06555},
 primaryClass = {astro-ph.CO},
       adsurl = {https://ui.adsabs.harvard.edu/abs/2016A&A...594A..23P}
}

@ARTICLE{P16_CIB_SED_template,
       author = {{Planck Collaboration} and {Ade}, P.~A.~R. and {Aghanim}, N. and {Arnaud}, M. and {Ashdown}, M. and {Aumont}, J. and {Baccigalupi}, C. and {Banday}, A.~J. and {Barreiro}, R.~B. and {Bartlett}, J.~G. and {Bartolo}, N. and {Basak}, S. and {Battaner}, E. and {Benabed}, K. and {Beno{\^\i}t}, A. and {Benoit-L{\'e}vy}, A. and {Bernard}, J. -P. and {Bersanelli}, M. and {Bielewicz}, P. and {Bock}, J.~J. and {Bonaldi}, A. and {Bonavera}, L. and {Bond}, J.~R. and {Borrill}, J. and {Bouchet}, F.~R. and {Boulanger}, F. and {Bucher}, M. and {Burigana}, C. and {Butler}, R.~C. and {Calabrese}, E. and {Cardoso}, J. -F. and {Catalano}, A. and {Challinor}, A. and {Chamballu}, A. and {Chiang}, H.~C. and {Christensen}, P.~R. and {Church}, S. and {Clements}, D.~L. and {Colombi}, S. and {Colombo}, L.~P.~L. and {Combet}, C. and {Couchot}, F. and {Coulais}, A. and {Crill}, B.~P. and {Curto}, A. and {Cuttaia}, F. and {Danese}, L. and {Davies}, R.~D. and {Davis}, R.~J. and {de Bernardis}, P. and {de Rosa}, A. and {de Zotti}, G. and {Delabrouille}, J. and {D{\'e}sert}, F. -X. and {Diego}, J.~M. and {Dole}, H. and {Donzelli}, S. and {Dor{\'e}}, O. and {Douspis}, M. and {Ducout}, A. and {Dunkley}, J. and {Dupac}, X. and {Efstathiou}, G. and {Elsner}, F. and {En{\ss}lin}, T.~A. and {Eriksen}, H.~K. and {Fergusson}, J. and {Finelli}, F. and {Forni}, O. and {Frailis}, M. and {Fraisse}, A.~A. and {Franceschi}, E. and {Frejsel}, A. and {Galeotta}, S. and {Galli}, S. and {Ganga}, K. and {Giard}, M. and {Giraud-H{\'e}raud}, Y. and {Gjerl{\o}w}, E. and {Gonz{\'a}lez-Nuevo}, J. and {G{\'o}rski}, K.~M. and {Gratton}, S. and {Gregorio}, A. and {Gruppuso}, A. and {Gudmundsson}, J.~E. and {Hansen}, F.~K. and {Hanson}, D. and {Harrison}, D.~L. and {Henrot-Versill{\'e}}, S. and {Hern{\'a}ndez-Monteagudo}, C. and {Herranz}, D. and {Hildebrandt}, S.~R. and {Hivon}, E. and {Hobson}, M. and {Holmes}, W.~A. and {Hornstrup}, A. and {Hovest}, W. and {Huffenberger}, K.~M. and {Hurier}, G. and {Jaffe}, A.~H. and {Jaffe}, T.~R. and {Jones}, W.~C. and {Juvela}, M. and {Keih{\"a}nen}, E. and {Keskitalo}, R. and {Kisner}, T.~S. and {Kneissl}, R. and {Knoche}, J. and {Kunz}, M. and {Kurki-Suonio}, H. and {Lagache}, G. and {L{\"a}hteenm{\"a}ki}, A. and {Lamarre}, J. -M. and {Lasenby}, A. and {Lattanzi}, M. and {Lawrence}, C.~R. and {Leonardi}, R. and {Lesgourgues}, J. and {Levrier}, F. and {Lewis}, A. and {Liguori}, M. and {Lilje}, P.~B. and {Linden-V{\o}rnle}, M. and {L{\'o}pez-Caniego}, M. and {Lubin}, P.~M. and {Mac{\'\i}as-P{\'e}rez}, J.~F. and {Maggio}, G. and {Maino}, D. and {Mandolesi}, N. and {Mangilli}, A. and {Maris}, M. and {Martin}, P.~G. and {Mart{\'\i}nez-Gonz{\'a}lez}, E. and {Masi}, S. and {Matarrese}, S. and {McGehee}, P. and {Meinhold}, P.~R. and {Melchiorri}, A. and {Mendes}, L. and {Mennella}, A. and {Migliaccio}, M. and {Mitra}, S. and {Miville-Desch{\^e}nes}, M. -A. and {Moneti}, A. and {Montier}, L. and {Morgante}, G. and {Mortlock}, D. and {Moss}, A. and {Munshi}, D. and {Murphy}, J.~A. and {Naselsky}, P. and {Nati}, F. and {Natoli}, P. and {Netterfield}, C.~B. and {N{\o}rgaard-Nielsen}, H.~U. and {Noviello}, F. and {Novikov}, D. and {Novikov}, I. and {Oxborrow}, C.~A. and {Paci}, F. and {Pagano}, L. and {Pajot}, F. and {Paoletti}, D. and {Pasian}, F. and {Patanchon}, G. and {Perdereau}, O. and {Perotto}, L. and {Perrotta}, F. and {Pettorino}, V. and {Piacentini}, F. and {Piat}, M. and {Pierpaoli}, E. and {Pietrobon}, D. and {Plaszczynski}, S. and {Pointecouteau}, E. and {Polenta}, G. and {Popa}, L. and {Pratt}, G.~W. and {Pr{\'e}zeau}, G. and {Prunet}, S. and {Puget}, J. -L. and {Rachen}, J.~P. and {Reach}, W.~T. and {Rebolo}, R. and {Reinecke}, M. and {Remazeilles}, M. and {Renault}, C. and {Renzi}, A. and {Ristorcelli}, I. and {Rocha}, G. and {Rosset}, C. and {Rossetti}, M. and {Roudier}, G. and {Rowan-Robinson}, M. and {Rubi{\~n}o-Mart{\'\i}n}, J.~A. and {Rusholme}, B. and {Sandri}, M. and {Santos}, D. and {Savelainen}, M. and {Savini}, G. and {Scott}, D. and {Seiffert}, M.~D. and {Shellard}, E.~P.~S. and {Spencer}, L.~D. and {Stolyarov}, V. and {Stompor}, R. and {Sudiwala}, R. and {Sunyaev}, R. and {Sutton}, D. and {Suur-Uski}, A. -S. and {Sygnet}, J. -F. and {Tauber}, J.~A. and {Terenzi}, L. and {Toffolatti}, L. and {Tomasi}, M. and {Tristram}, M. and {Tucci}, M. and {Tuovinen}, J. and {Valenziano}, L. and {Valiviita}, J. and {Van Tent}, B. and {Vielva}, P. and {Villa}, F. and {Wade}, L.~A. and {Wandelt}, B.~D. and {Wehus}, I.~K. and {White}, M. and {Yvon}, D. and {Zacchei}, A. and {Zonca}, A.},
        title = "{Planck 2015 results. XV. Gravitational lensing}",
      journal = {\aap},
         year = 2016,
        month = sep,
       volume = {594},
          eid = {A15},
        pages = {A15},
          doi = {10.1051/0004-6361/201525941},
archivePrefix = {arXiv},
       eprint = {1502.01591},
 primaryClass = {astro-ph.CO},
       adsurl = {https://ui.adsabs.harvard.edu/abs/2016A&A...594A..15P}
}

@ARTICLE{LIR_SFR_relation,
       author = {{Kennicutt}, Robert C., Jr.},
        title = "{The Global Schmidt Law in Star-forming Galaxies}",
      journal = {\apj},
         year = 1998,
        month = may,
       volume = {498},
       number = {2},
        pages = {541-552},
          doi = {10.1086/305588},
archivePrefix = {arXiv},
       eprint = {astro-ph/9712213},
 primaryClass = {astro-ph},
       adsurl = {https://ui.adsabs.harvard.edu/abs/1998ApJ...498..541K}
}

@ARTICLE{SZpack_ref1,
       author = {{Chluba}, Jens and {Nagai}, Daisuke and {Sazonov}, Sergey and {Nelson}, Kaylea},
        title = "{A fast and accurate method for computing the Sunyaev-Zel'dovich signal of hot galaxy clusters}",
      journal = {\mnras},
         year = 2012,
        month = oct,
       volume = {426},
       number = {1},
        pages = {510-530},
          doi = {10.1111/j.1365-2966.2012.21741.x},
archivePrefix = {arXiv},
       eprint = {1205.5778},
 primaryClass = {astro-ph.CO},
       adsurl = {https://ui.adsabs.harvard.edu/abs/2012MNRAS.426..510C}
}

@ARTICLE{Szpack_ref2,
       author = {{Chluba}, Jens and {Switzer}, Eric and {Nelson}, Kaylea and {Nagai}, Daisuke},
        title = "{Sunyaev-Zeldovich signal processing and temperature-velocity moment method for individual clusters}",
      journal = {\mnras},
         year = 2013,
        month = apr,
       volume = {430},
       number = {4},
        pages = {3054-3069},
          doi = {10.1093/mnras/stt110},
archivePrefix = {arXiv},
       eprint = {1211.3206},
 primaryClass = {astro-ph.CO},
       adsurl = {https://ui.adsabs.harvard.edu/abs/2013MNRAS.430.3054C}
}

@ARTICLE{Websky_ref,
       author = {{Stein}, George and {Alvarez}, Marcelo A. and {Bond}, J. Richard and {van Engelen}, Alexander and {Battaglia}, Nicholas},
        title = "{The Websky extragalactic CMB simulations}",
      journal = {\jcap},
         year = 2020,
        month = oct,
       volume = {2020},
       number = {10},
          eid = {012},
        pages = {012},
          doi = {10.1088/1475-7516/2020/10/012},
archivePrefix = {arXiv},
       eprint = {2001.08787},
 primaryClass = {astro-ph.CO},
       adsurl = {https://ui.adsabs.harvard.edu/abs/2020JCAP...10..012S}
}

@ARTICLE{FLAMINGO_ref_Joop,
       author = {{Schaye}, Joop and {Kugel}, Roi and {Schaller}, Matthieu and {Helly}, John C. and {Braspenning}, Joey and {Elbers}, Willem and {McCarthy}, Ian G. and {van Daalen}, Marcel P. and {Vandenbroucke}, Bert and {Frenk}, Carlos S. and {Kwan}, Juliana and {Salcido}, Jaime and {Bah{\'e}}, Yannick M. and {Borrow}, Josh and {Chaikin}, Evgenii and {Hahn}, Oliver and {Hu{\v{s}}ko}, Filip and {Jenkins}, Adrian and {Lacey}, Cedric G. and {Nobels}, Folkert S.~J.},
        title = "{The FLAMINGO project: cosmological hydrodynamical simulations for large-scale structure and galaxy cluster surveys}",
      journal = {\mnras},
         year = 2023,
        month = dec,
       volume = {526},
       number = {4},
        pages = {4978-5020},
          doi = {10.1093/mnras/stad2419},
archivePrefix = {arXiv},
       eprint = {2306.04024},
 primaryClass = {astro-ph.CO},
       adsurl = {https://ui.adsabs.harvard.edu/abs/2023MNRAS.526.4978S}
}

@ARTICLE{C13,
       author = {{Hayward}, Christopher C. and {Narayanan}, Desika and {Kere{\v{s}}}, Du{\v{s}}an and {Jonsson}, Patrik and {Hopkins}, Philip F. and {Cox}, T.~J. and {Hernquist}, Lars},
        title = "{Submillimetre galaxies in a hierarchical universe: number counts, redshift distribution and implications for the IMF}",
      journal = {\mnras},
         year = 2013,
        month = jan,
       volume = {428},
       number = {3},
        pages = {2529-2547},
          doi = {10.1093/mnras/sts222},
archivePrefix = {arXiv},
       eprint = {1209.2413},
 primaryClass = {astro-ph.CO},
       adsurl = {https://ui.adsabs.harvard.edu/abs/2013MNRAS.428.2529H}
}

@ARTICLE{L21,
       author = {{Lovell}, Christopher C. and {Geach}, James E. and {Dav{\'e}}, Romeel and {Narayanan}, Desika and {Li}, Qi},
        title = "{Reproducing submillimetre galaxy number counts with cosmological hydrodynamic simulations}",
      journal = {\mnras},
         year = 2021,
        month = mar,
       volume = {502},
       number = {1},
        pages = {772-793},
          doi = {10.1093/mnras/staa4043},
archivePrefix = {arXiv},
       eprint = {2006.15156},
 primaryClass = {astro-ph.GA},
       adsurl = {https://ui.adsabs.harvard.edu/abs/2021MNRAS.502..772L}
}

@ARTICLE{C23,
       author = {{Cochrane}, R.~K. and {Hayward}, C.~C. and {Angl{\'e}s-Alc{\'a}zar}, D. and {Somerville}, R.~S.},
        title = "{Predicting sub-millimetre flux densities from global galaxy properties}",
      journal = {\mnras},
         year = 2023,
        month = feb,
       volume = {518},
       number = {4},
        pages = {5522-5535},
          doi = {10.1093/mnras/stac3451},
archivePrefix = {arXiv},
       eprint = {2211.11702},
 primaryClass = {astro-ph.GA},
       adsurl = {https://ui.adsabs.harvard.edu/abs/2023MNRAS.518.5522C}
}

@ARTICLE{K25,
       author = {{Kumar}, Ankit and {Artale}, M. Celeste and {Montero-Dorta}, Antonio D. and {Guaita}, Lucia and {Lee}, Kyoung-Soo and {Pope}, Alexandra and {Schaye}, Joop and {Schaller}, Matthieu and {Gawiser}, Eric and {Hwang}, Ho Seong and {Jeong}, Woong-Seob and {Lee}, Jaehyun and {Padilla}, Nelson and {Park}, Changbom and {Ramakrishnan}, Vandana and {Singh}, Akriti and {Yang}, Yujin},
        title = "{Modeling submillimeter galaxies in cosmological simulations: Contribution to the cosmic star formation density and predictions for future surveys}",
      journal = {arXiv e-prints},
         year = 2025,
        month = jan,
          eid = {arXiv:2501.19327},
        pages = {arXiv:2501.19327},
          doi = {10.48550/arXiv.2501.19327},
archivePrefix = {arXiv},
       eprint = {2501.19327},
 primaryClass = {astro-ph.CO},
       adsurl = {https://ui.adsabs.harvard.edu/abs/2025arXiv250119327K}
}

@ARTICLE{Fiona_CIB23,
       author = {{McCarthy}, Fiona},
        title = "{Large-scale galactic-dust-cleaned cosmic infrared background maps from \textbackslashtextit\{Planck\} PR4 and HI4PI with \textbackslashtexttt\{pyilc\}}",
      journal = {arXiv e-prints},
         year = 2024,
        month = may,
          eid = {arXiv:2405.13470},
        pages = {arXiv:2405.13470},
          doi = {10.48550/arXiv.2405.13470},
archivePrefix = {arXiv},
       eprint = {2405.13470},
 primaryClass = {astro-ph.CO},
       adsurl = {https://ui.adsabs.harvard.edu/abs/2024arXiv240513470M}
}

@ARTICLE{LOFAR_RLF,
       author = {{Kondapally}, Rohit and {Best}, Philip N. and {Cochrane}, Rachel K. and {Sabater}, Jos{\'e} and {Duncan}, Kenneth J. and {Hardcastle}, Martin J. and {Haskell}, Paul and {Mingo}, Beatriz and {R{\"o}ttgering}, Huub J.~A. and {Smith}, Daniel J.~B. and {Williams}, Wendy L. and {Bonato}, Matteo and {Calistro Rivera}, Gabriela and {Gao}, Fangyou and {Hale}, Catherine L. and {Ma{\l}ek}, Katarzyna and {Miley}, George K. and {Prandoni}, Isabella and {Wang}, Lingyu},
        title = "{Cosmic evolution of low-excitation radio galaxies in the LOFAR two-metre sky survey deep fields}",
      journal = {\mnras},
         year = 2022,
        month = jul,
       volume = {513},
       number = {3},
        pages = {3742-3767},
          doi = {10.1093/mnras/stac1128},
archivePrefix = {arXiv},
       eprint = {2204.07588},
 primaryClass = {astro-ph.GA},
       adsurl = {https://ui.adsabs.harvard.edu/abs/2022MNRAS.513.3742K}
}

@ARTICLE{Li_websky_radio,
       author = {{Li}, Zack and {Puglisi}, Giuseppe and {Madhavacheril}, Mathew S. and {Alvarez}, Marcelo A.},
        title = "{Simulated catalogs and maps of radio galaxies at millimeter wavelengths in Websky}",
      journal = {\jcap},
         year = 2022,
        month = aug,
       volume = {2022},
       number = {8},
          eid = {029},
        pages = {029},
          doi = {10.1088/1475-7516/2022/08/029},
archivePrefix = {arXiv},
       eprint = {2110.15357},
 primaryClass = {astro-ph.GA},
       adsurl = {https://ui.adsabs.harvard.edu/abs/2022JCAP...08..029L}
}

@ARTICLE{Sehgal_radio_SO_forecast,
       author = {{Ade}, Peter and {Aguirre}, James and {Ahmed}, Zeeshan and {Aiola}, Simone and {Ali}, Aamir and {Alonso}, David and {Alvarez}, Marcelo A. and {Arnold}, Kam and {Ashton}, Peter and {Austermann}, Jason and {Awan}, Humna and {Baccigalupi}, Carlo and {Baildon}, Taylor and {Barron}, Darcy and {Battaglia}, Nick and {Battye}, Richard and {Baxter}, Eric and {Bazarko}, Andrew and {Beall}, James A. and {Bean}, Rachel and {Beck}, Dominic and {Beckman}, Shawn and {Beringue}, Benjamin and {Bianchini}, Federico and {Boada}, Steven and {Boettger}, David and {Bond}, J. Richard and {Borrill}, Julian and {Brown}, Michael L. and {Bruno}, Sarah Marie and {Bryan}, Sean and {Calabrese}, Erminia and {Calafut}, Victoria and {Calisse}, Paolo and {Carron}, Julien and {Challinor}, Anthony and {Chesmore}, Grace and {Chinone}, Yuji and {Chluba}, Jens and {Cho}, Hsiao-Mei Sherry and {Choi}, Steve and {Coppi}, Gabriele and {Cothard}, Nicholas F. and {Coughlin}, Kevin and {Crichton}, Devin and {Crowley}, Kevin D. and {Crowley}, Kevin T. and {Cukierman}, Ari and {D'Ewart}, John M. and {D{\"u}nner}, Rolando and {de Haan}, Tijmen and {Devlin}, Mark and {Dicker}, Simon and {Didier}, Joy and {Dobbs}, Matt and {Dober}, Bradley and {Duell}, Cody J. and {Duff}, Shannon and {Duivenvoorden}, Adri and {Dunkley}, Jo and {Dusatko}, John and {Errard}, Josquin and {Fabbian}, Giulio and {Feeney}, Stephen and {Ferraro}, Simone and {Flux{\`a}}, Pedro and {Freese}, Katherine and {Frisch}, Josef C. and {Frolov}, Andrei and {Fuller}, George and {Fuzia}, Brittany and {Galitzki}, Nicholas and {Gallardo}, Patricio A. and {Tomas Galvez Ghersi}, Jose and {Gao}, Jiansong and {Gawiser}, Eric and {Gerbino}, Martina and {Gluscevic}, Vera and {Goeckner-Wald}, Neil and {Golec}, Joseph and {Gordon}, Sam and {Gralla}, Megan and {Green}, Daniel and {Grigorian}, Arpi and {Groh}, John and {Groppi}, Chris and {Guan}, Yilun and {Gudmundsson}, Jon E. and {Han}, Dongwon and {Hargrave}, Peter and {Hasegawa}, Masaya and {Hasselfield}, Matthew and {Hattori}, Makoto and {Haynes}, Victor and {Hazumi}, Masashi and {He}, Yizhou and {Healy}, Erin and {Henderson}, Shawn W. and {Hervias-Caimapo}, Carlos and {Hill}, Charles A. and {Hill}, J. Colin and {Hilton}, Gene and {Hilton}, Matt and {Hincks}, Adam D. and {Hinshaw}, Gary and {Hlo{\v{z}}ek}, Ren{\'e}e and {Ho}, Shirley and {Ho}, Shuay-Pwu Patty and {Howe}, Logan and {Huang}, Zhiqi and {Hubmayr}, Johannes and {Huffenberger}, Kevin and {Hughes}, John P. and {Ijjas}, Anna and {Ikape}, Margaret and {Irwin}, Kent and {Jaffe}, Andrew H. and {Jain}, Bhuvnesh and {Jeong}, Oliver and {Kaneko}, Daisuke and {Karpel}, Ethan D. and {Katayama}, Nobuhiko and {Keating}, Brian and {Kernasovskiy}, Sarah S. and {Keskitalo}, Reijo and {Kisner}, Theodore and {Kiuchi}, Kenji and {Klein}, Jeff and {Knowles}, Kenda and {Koopman}, Brian and {Kosowsky}, Arthur and {Krachmalnicoff}, Nicoletta and {Kuenstner}, Stephen E. and {Kuo}, Chao-Lin and {Kusaka}, Akito and {Lashner}, Jacob and {Lee}, Adrian and {Lee}, Eunseong and {Leon}, David and {Leung}, Jason S. -Y. and {Lewis}, Antony and {Li}, Yaqiong and {Li}, Zack and {Limon}, Michele and {Linder}, Eric and {Lopez-Caraballo}, Carlos and {Louis}, Thibaut and {Lowry}, Lindsay and {Lungu}, Marius and {Madhavacheril}, Mathew and {Mak}, Daisy and {Maldonado}, Felipe and {Mani}, Hamdi and {Mates}, Ben and {Matsuda}, Frederick and {Maurin}, Lo{\"\i}c and {Mauskopf}, Phil and {May}, Andrew and {McCallum}, Nialh and {McKenney}, Chris and {McMahon}, Jeff and {Meerburg}, P. Daniel and {Meyers}, Joel and {Miller}, Amber and {Mirmelstein}, Mark and {Moodley}, Kavilan and {Munchmeyer}, Moritz and {Munson}, Charles and {Naess}, Sigurd and {Nati}, Federico and {Navaroli}, Martin and {Newburgh}, Laura and {Nguyen}, Ho Nam and {Niemack}, Michael and {Nishino}, Haruki and {Orlowski-Scherer}, John and {Page}, Lyman and {Partridge}, Bruce and {Peloton}, Julien and {Perrotta}, Francesca and {Piccirillo}, Lucio and {Pisano}, Giampaolo and {Poletti}, Davide and {Puddu}, Roberto and {Puglisi}, Giuseppe and {Raum}, Chris and {Reichardt}, Christian L. and {Remazeilles}, Mathieu and {Rephaeli}, Yoel and {Riechers}, Dominik and {Rojas}, Felipe and {Roy}, Anirban and {Sadeh}, Sharon and {Sakurai}, Yuki and {Salatino}, Maria and {Sathyanarayana Rao}, Mayuri and {Schaan}, Emmanuel and {Schmittfull}, Marcel and {Sehgal}, Neelima and {Seibert}, Joseph},
        title = "{The Simons Observatory: science goals and forecasts}",
      journal = {\jcap},
         year = 2019,
        month = feb,
       volume = {2019},
       number = {2},
          eid = {056},
        pages = {056},
          doi = {10.1088/1475-7516/2019/02/056},
archivePrefix = {arXiv},
       eprint = {1808.07445},
 primaryClass = {astro-ph.CO},
       adsurl = {https://ui.adsabs.harvard.edu/abs/2019JCAP...02..056A}
}

@ARTICLE{Sehgal_2010,
       author = {{Sehgal}, Neelima and {Bode}, Paul and {Das}, Sudeep and {Hernandez-Monteagudo}, Carlos and {Huffenberger}, Kevin and {Lin}, Yen-Ting and {Ostriker}, Jeremiah P. and {Trac}, Hy},
        title = "{Simulations of the Microwave Sky}",
      journal = {\apj},
         year = 2010,
        month = feb,
       volume = {709},
       number = {2},
        pages = {920-936},
          doi = {10.1088/0004-637X/709/2/920},
archivePrefix = {arXiv},
       eprint = {0908.0540},
 primaryClass = {astro-ph.CO},
       adsurl = {https://ui.adsabs.harvard.edu/abs/2010ApJ...709..920S}
}

@ARTICLE{Everett_SPT,
       author = {{Everett}, W.~B. and {Zhang}, L. and {Crawford}, T.~M. and {Vieira}, J.~D. and {Aravena}, M. and {Archipley}, M.~A. and {Austermann}, J.~E. and {Benson}, B.~A. and {Bleem}, L.~E. and {Carlstrom}, J.~E. and {Chang}, C.~L. and {Chapman}, S. and {Crites}, A.~T. and {de Haan}, T. and {Dobbs}, M.~A. and {George}, E.~M. and {Halverson}, N.~W. and {Harrington}, N. and {Holder}, G.~P. and {Holzapfel}, W.~L. and {Hrubes}, J.~D. and {Knox}, L. and {Lee}, A.~T. and {Luong-Van}, D. and {Mangian}, A.~C. and {Marrone}, D.~P. and {McMahon}, J.~J. and {Meyer}, S.~S. and {Mocanu}, L.~M. and {Mohr}, J.~J. and {Natoli}, T. and {Padin}, S. and {Pryke}, C. and {Reichardt}, C.~L. and {Reuter}, C.~A. and {Ruhl}, J.~E. and {Sayre}, J.~T. and {Schaffer}, K.~K. and {Shirokoff}, E. and {Spilker}, J.~S. and {Stalder}, B. and {Staniszewski}, Z. and {Stark}, A.~A. and {Story}, K.~T. and {Switzer}, E.~R. and {Vanderlinde}, K. and {Wei{\ss}}, A. and {Williamson}, R.},
        title = "{Millimeter-wave Point Sources from the 2500 Square Degree SPT-SZ Survey: Catalog and Population Statistics}",
      journal = {\apj},
         year = 2020,
        month = sep,
       volume = {900},
       number = {1},
          eid = {55},
        pages = {55},
          doi = {10.3847/1538-4357/ab9df7},
archivePrefix = {arXiv},
       eprint = {2003.03431},
 primaryClass = {astro-ph.IM},
       adsurl = {https://ui.adsabs.harvard.edu/abs/2020ApJ...900...55E}
}

@ARTICLE{Vargas_ACT,
       author = {{Vargas}, Cristian and {L{\'o}pez-Caraballo}, Carlos H. and {Battistelli}, Elia S. and {Dunner}, Rolando and {Farren}, Gerrit and {Gralla}, Megan and {Hall}, Kirsten R. and {Herv{\'\i}as-Caimapo}, Carlos and {Hilton}, Matt and {Hincks}, Adam D. and {Huffenberger}, Kevin and {Marriage}, Tobias and {Mroczkowski}, Tony and {Niemack}, Michael D. and {Page}, Lyman and {Partridge}, Bruce and {Rojas}, Felipe and {Rizzo}, Francesca and {Sif{\'o}n}, Crist{\'o}bal and {Staggs}, Suzanne and {Wollack}, Edward J.},
        title = "{The Atacama Cosmology Telescope: Extragalactic Point Sources in the Southern Surveys at 150, 220 and 280 GHz observed between 2008-2010}",
      journal = {arXiv e-prints},
         year = 2023,
        month = oct,
          eid = {arXiv:2310.17535},
        pages = {arXiv:2310.17535},
          doi = {10.48550/arXiv.2310.17535},
archivePrefix = {arXiv},
       eprint = {2310.17535},
 primaryClass = {astro-ph.GA},
       adsurl = {https://ui.adsabs.harvard.edu/abs/2023arXiv231017535V}
}

@ARTICLE{P18_radio,
       author = {{Planck Collaboration} and {Akrami}, Y. and {Arg{\"u}eso}, F. and {Ashdown}, M. and {Aumont}, J. and {Baccigalupi}, C. and {Ballardini}, M. and {Banday}, A.~J. and {Barreiro}, R.~B. and {Bartolo}, N. and {Basak}, S. and {Benabed}, K. and {Bernard}, J. -P. and {Bersanelli}, M. and {Bielewicz}, P. and {Bonavera}, L. and {Bond}, J.~R. and {Borrill}, J. and {Bouchet}, F.~R. and {Burigana}, C. and {Butler}, R.~C. and {Calabrese}, E. and {Carron}, J. and {Chiang}, H.~C. and {Combet}, C. and {Crill}, B.~P. and {Cuttaia}, F. and {de Bernardis}, P. and {de Rosa}, A. and {de Zotti}, G. and {Delabrouille}, J. and {Delouis}, J. -M. and {Di Valentino}, E. and {Dickinson}, C. and {Diego}, J.~M. and {Ducout}, A. and {Dupac}, X. and {Efstathiou}, G. and {Elsner}, F. and {En{\ss}lin}, T.~A. and {Eriksen}, H.~K. and {Fantaye}, Y. and {Finelli}, F. and {Frailis}, M. and {Fraisse}, A.~A. and {Franceschi}, E. and {Frolov}, A. and {Galeotta}, S. and {Galli}, S. and {Ganga}, K. and {G{\'e}nova-Santos}, R.~T. and {Gerbino}, M. and {Ghosh}, T. and {Gonz{\'a}lez-Nuevo}, J. and {G{\'o}rski}, K.~M. and {Gratton}, S. and {Gruppuso}, A. and {Gudmundsson}, J.~E. and {Handley}, W. and {Hansen}, F.~K. and {Herranz}, D. and {Hivon}, E. and {Huang}, Z. and {Jaffe}, A.~H. and {Jones}, W.~C. and {Keih{\"a}nen}, E. and {Keskitalo}, R. and {Kiiveri}, K. and {Kim}, J. and {Kisner}, T.~S. and {Krachmalnicoff}, N. and {Kunz}, M. and {Kurki-Suonio}, H. and {L{\"a}hteenm{\"a}ki}, A. and {Lamarre}, J. -M. and {Lasenby}, A. and {Lattanzi}, M. and {Lawrence}, C.~R. and {Levrier}, F. and {Liguori}, M. and {Lilje}, P.~B. and {Lindholm}, V. and {L{\'o}pez-Caniego}, M. and {Ma}, Y. -Z. and {Mac{\'\i}as-P{\'e}rez}, J.~F. and {Maggio}, G. and {Maino}, D. and {Mandolesi}, N. and {Mangilli}, A. and {Maris}, M. and {Martin}, P.~G. and {Mart{\'\i}nez-Gonz{\'a}lez}, E. and {Matarrese}, S. and {McEwen}, J.~D. and {Meinhold}, P.~R. and {Melchiorri}, A. and {Mennella}, A. and {Migliaccio}, M. and {Miville-Desch{\^e}nes}, M. -A. and {Molinari}, D. and {Moneti}, A. and {Montier}, L. and {Morgante}, G. and {Natoli}, P. and {Oxborrow}, C.~A. and {Pagano}, L. and {Paoletti}, D. and {Partridge}, B. and {Patanchon}, G. and {Pearson}, T.~J. and {Pettorino}, V. and {Piacentini}, F. and {Polenta}, G. and {Puget}, J. -L. and {Rachen}, J.~P. and {Racine}, B. and {Reinecke}, M. and {Remazeilles}, M. and {Renzi}, A. and {Rocha}, G. and {Roudier}, G. and {Rubi{\~n}o-Mart{\'\i}n}, J.~A. and {Salvati}, L. and {Sandri}, M. and {Savelainen}, M. and {Scott}, D. and {Suur-Uski}, A. -S. and {Tauber}, J.~A. and {Tavagnacco}, D. and {Toffolatti}, L. and {Tomasi}, M. and {Trombetti}, T. and {Tucci}, M. and {Valiviita}, J. and {Van Tent}, B. and {Vielva}, P. and {Villa}, F. and {Vittorio}, N. and {Wehus}, I.~K. and {Zacchei}, A. and {Zonca}, A.},
        title = "{Planck intermediate results. LIV. The Planck multi-frequency catalogue of non-thermal sources}",
      journal = {\aap},
         year = 2018,
        month = nov,
       volume = {619},
          eid = {A94},
        pages = {A94},
          doi = {10.1051/0004-6361/201832888},
archivePrefix = {arXiv},
       eprint = {1802.08649},
 primaryClass = {astro-ph.CO},
       adsurl = {https://ui.adsabs.harvard.edu/abs/2018A&A...619A..94P}
}

@ARTICLE{kappa_phi_gamma_alm_conversion,
       author = {{Hu}, Wayne},
        title = "{Weak lensing of the CMB: A harmonic approach}",
      journal = {\prd},
         year = 2000,
        month = aug,
       volume = {62},
       number = {4},
          eid = {043007},
        pages = {043007},
          doi = {10.1103/PhysRevD.62.043007},
archivePrefix = {arXiv},
       eprint = {astro-ph/0001303},
 primaryClass = {astro-ph},
       adsurl = {https://ui.adsabs.harvard.edu/abs/2000PhRvD..62d3007H}
}

@ARTICLE{FLAMINGO_kugel,
       author = {{Kugel}, Roi and {Schaye}, Joop and {Schaller}, Matthieu and {Helly}, John C. and {Braspenning}, Joey and {Elbers}, Willem and {Frenk}, Carlos S. and {McCarthy}, Ian G. and {Kwan}, Juliana and {Salcido}, Jaime and {van Daalen}, Marcel P. and {Vandenbroucke}, Bert and {Bah{\'e}}, Yannick M. and {Borrow}, Josh and {Chaikin}, Evgenii and {Hu{\v{s}}ko}, Filip and {Jenkins}, Adrian and {Lacey}, Cedric G. and {Nobels}, Folkert S.~J. and {Vernon}, Ian},
        title = "{FLAMINGO: calibrating large cosmological hydrodynamical simulations with machine learning}",
      journal = {\mnras},
         year = 2023,
        month = dec,
       volume = {526},
       number = {4},
        pages = {6103-6127},
          doi = {10.1093/mnras/stad2540},
archivePrefix = {arXiv},
       eprint = {2306.05492},
 primaryClass = {astro-ph.CO},
       adsurl = {https://ui.adsabs.harvard.edu/abs/2023MNRAS.526.6103K}
}

@ARTICLE{SWIFT_ref,
       author = {{Schaller}, Matthieu and {Borrow}, Josh and {Draper}, Peter W. and {Ivkovic}, Mladen and {McAlpine}, Stuart and {Vandenbroucke}, Bert and {Bah{\'e}}, Yannick and {Chaikin}, Evgenii and {Chalk}, Aidan B.~G. and {Chan}, Tsang Keung and {Correa}, Camila and {van Daalen}, Marcel and {Elbers}, Willem and {Gonnet}, Pedro and {Hausammann}, Lo{\"\i}c and {Helly}, John and {Hu{\v{s}}ko}, Filip and {Kegerreis}, Jacob A. and {Nobels}, Folkert S.~J. and {Ploeckinger}, Sylvia and {Revaz}, Yves and {Roper}, William J. and {Ruiz-Bonilla}, Sergio and {Sandnes}, Thomas D. and {Uyttenhove}, Yolan and {Willis}, James S. and {Xiang}, Zhen},
        title = "{SWIFT: A modern highly-parallel gravity and smoothed particle hydrodynamics solver for astrophysical and cosmological applications}",
      journal = {\mnras},
         year = 2024,
        month = may,
       volume = {530},
       number = {2},
        pages = {2378-2419},
          doi = {10.1093/mnras/stae922},
archivePrefix = {arXiv},
       eprint = {2305.13380},
 primaryClass = {astro-ph.IM},
       adsurl = {https://ui.adsabs.harvard.edu/abs/2024MNRAS.530.2378S}
}

@ARTICLE{force_cal_ref1,
       author = {{Greengard}, L. and {Rokhlin}, V.},
        title = "{A Fast Algorithm for Particle Simulations}",
      journal = {Journal of Computational Physics},
         year = 1987,
        month = dec,
       volume = {73},
       number = {2},
        pages = {325-348},
          doi = {10.1016/0021-9991(87)90140-9},
       adsurl = {https://ui.adsabs.harvard.edu/abs/1987JCoPh..73..325G}
}

@ARTICLE{force_cal_ref2,
       author = {{Cheng}, H. and {Greengard}, L. and {Rokhlin}, V.},
        title = "{A Fast Adaptive Multipole Algorithm in Three Dimensions}",
      journal = {Journal of Computational Physics},
         year = 1999,
        month = nov,
       volume = {155},
       number = {2},
        pages = {468-498},
          doi = {10.1006/jcph.1999.6355},
       adsurl = {https://ui.adsabs.harvard.edu/abs/1999JCoPh.155..468C}
}

@ARTICLE{force_cal_ref3,
       author = {{Dehnen}, Walter},
        title = "{A fast multipole method for stellar dynamics}",
      journal = {Computational Astrophysics and Cosmology},
         year = 2014,
        month = sep,
       volume = {1},
          eid = {1},
        pages = {1},
          doi = {10.1186/s40668-014-0001-7},
archivePrefix = {arXiv},
       eprint = {1405.2255},
 primaryClass = {astro-ph.IM},
       adsurl = {https://ui.adsabs.harvard.edu/abs/2014ComAC...1....1D}
}

@ARTICLE{force_cal_ref4,
       author = {{Bagla}, J.~S. and {Ray}, Suryadeep},
        title = "{Performance characteristics of TreePM codes}",
      journal = {\na},
         year = 2003,
        month = sep,
       volume = {8},
       number = {7},
        pages = {665-677},
          doi = {10.1016/S1384-1076(03)00056-3},
archivePrefix = {arXiv},
       eprint = {astro-ph/0212129},
 primaryClass = {astro-ph},
       adsurl = {https://ui.adsabs.harvard.edu/abs/2003NewA....8..665B}
}

@ARTICLE{SPH_review_price,
       author = {{Price}, Daniel J.},
        title = "{Smoothed particle hydrodynamics and magnetohydrodynamics}",
      journal = {Journal of Computational Physics},
         year = 2012,
        month = feb,
       volume = {231},
       number = {3},
        pages = {759-794},
          doi = {10.1016/j.jcp.2010.12.011},
archivePrefix = {arXiv},
       eprint = {1012.1885},
 primaryClass = {astro-ph.IM},
       adsurl = {https://ui.adsabs.harvard.edu/abs/2012JCoPh.231..759P}
}

@ARTICLE{DES_Y3_Abbott,
       author = {{Abbott}, T.~M.~C. and {Aguena}, M. and {Alarcon}, A. and {Allam}, S. and {Alves}, O. and {Amon}, A. and {Andrade-Oliveira}, F. and {Annis}, J. and {Avila}, S. and {Bacon}, D. and {Baxter}, E. and {Bechtol}, K. and {Becker}, M.~R. and {Bernstein}, G.~M. and {Bhargava}, S. and {Birrer}, S. and {Blazek}, J. and {Brandao-Souza}, A. and {Bridle}, S.~L. and {Brooks}, D. and {Buckley-Geer}, E. and {Burke}, D.~L. and {Camacho}, H. and {Campos}, A. and {Carnero Rosell}, A. and {Carrasco Kind}, M. and {Carretero}, J. and {Castander}, F.~J. and {Cawthon}, R. and {Chang}, C. and {Chen}, A. and {Chen}, R. and {Choi}, A. and {Conselice}, C. and {Cordero}, J. and {Costanzi}, M. and {Crocce}, M. and {da Costa}, L.~N. and {da Silva Pereira}, M.~E. and {Davis}, C. and {Davis}, T.~M. and {De Vicente}, J. and {DeRose}, J. and {Desai}, S. and {Di Valentino}, E. and {Diehl}, H.~T. and {Dietrich}, J.~P. and {Dodelson}, S. and {Doel}, P. and {Doux}, C. and {Drlica-Wagner}, A. and {Eckert}, K. and {Eifler}, T.~F. and {Elsner}, F. and {Elvin-Poole}, J. and {Everett}, S. and {Evrard}, A.~E. and {Fang}, X. and {Farahi}, A. and {Fernandez}, E. and {Ferrero}, I. and {Fert{\'e}}, A. and {Fosalba}, P. and {Friedrich}, O. and {Frieman}, J. and {Garc{\'\i}a-Bellido}, J. and {Gatti}, M. and {Gaztanaga}, E. and {Gerdes}, D.~W. and {Giannantonio}, T. and {Giannini}, G. and {Gruen}, D. and {Gruendl}, R.~A. and {Gschwend}, J. and {Gutierrez}, G. and {Harrison}, I. and {Hartley}, W.~G. and {Herner}, K. and {Hinton}, S.~R. and {Hollowood}, D.~L. and {Honscheid}, K. and {Hoyle}, B. and {Huff}, E.~M. and {Huterer}, D. and {Jain}, B. and {James}, D.~J. and {Jarvis}, M. and {Jeffrey}, N. and {Jeltema}, T. and {Kovacs}, A. and {Krause}, E. and {Kron}, R. and {Kuehn}, K. and {Kuropatkin}, N. and {Lahav}, O. and {Leget}, P. -F. and {Lemos}, P. and {Liddle}, A.~R. and {Lidman}, C. and {Lima}, M. and {Lin}, H. and {MacCrann}, N. and {Maia}, M.~A.~G. and {Marshall}, J.~L. and {Martini}, P. and {McCullough}, J. and {Melchior}, P. and {Mena-Fern{\'a}ndez}, J. and {Menanteau}, F. and {Miquel}, R. and {Mohr}, J.~J. and {Morgan}, R. and {Muir}, J. and {Myles}, J. and {Nadathur}, S. and {Navarro-Alsina}, A. and {Nichol}, R.~C. and {Ogando}, R.~L.~C. and {Omori}, Y. and {Palmese}, A. and {Pandey}, S. and {Park}, Y. and {Paz-Chinch{\'o}n}, F. and {Petravick}, D. and {Pieres}, A. and {Plazas Malag{\'o}n}, A.~A. and {Porredon}, A. and {Prat}, J. and {Raveri}, M. and {Rodriguez-Monroy}, M. and {Rollins}, R.~P. and {Romer}, A.~K. and {Roodman}, A. and {Rosenfeld}, R. and {Ross}, A.~J. and {Rykoff}, E.~S. and {Samuroff}, S. and {S{\'a}nchez}, C. and {Sanchez}, E. and {Sanchez}, J. and {Sanchez Cid}, D. and {Scarpine}, V. and {Schubnell}, M. and {Scolnic}, D. and {Secco}, L.~F. and {Serrano}, S. and {Sevilla-Noarbe}, I. and {Sheldon}, E. and {Shin}, T. and {Smith}, M. and {Soares-Santos}, M. and {Suchyta}, E. and {Swanson}, M.~E.~C. and {Tabbutt}, M. and {Tarle}, G. and {Thomas}, D. and {To}, C. and {Troja}, A. and {Troxel}, M.~A. and {Tucker}, D.~L. and {Tutusaus}, I. and {Varga}, T.~N. and {Walker}, A.~R. and {Weaverdyck}, N. and {Wechsler}, R. and {Weller}, J. and {Yanny}, B. and {Yin}, B. and {Zhang}, Y. and {Zuntz}, J. and {DES Collaboration}},
        title = "{Dark Energy Survey Year 3 results: Cosmological constraints from galaxy clustering and weak lensing}",
      journal = {\prd},
         year = 2022,
        month = jan,
       volume = {105},
       number = {2},
          eid = {023520},
        pages = {023520},
          doi = {10.1103/PhysRevD.105.023520},
archivePrefix = {arXiv},
       eprint = {2105.13549},
 primaryClass = {astro-ph.CO},
       adsurl = {https://ui.adsabs.harvard.edu/abs/2022PhRvD.105b3520A}
}

@ARTICLE{OWLS_ref,
       author = {{Schaye}, Joop and {Dalla Vecchia}, Claudio and {Booth}, C.~M. and {Wiersma}, Robert P.~C. and {Theuns}, Tom and {Haas}, Marcel R. and {Bertone}, Serena and {Duffy}, Alan R. and {McCarthy}, I.~G. and {van de Voort}, Freeke},
        title = "{The physics driving the cosmic star formation history}",
      journal = {\mnras},
         year = 2010,
        month = mar,
       volume = {402},
       number = {3},
        pages = {1536-1560},
          doi = {10.1111/j.1365-2966.2009.16029.x},
archivePrefix = {arXiv},
       eprint = {0909.5196},
 primaryClass = {astro-ph.CO},
       adsurl = {https://ui.adsabs.harvard.edu/abs/2010MNRAS.402.1536S}
}

@ARTICLE{Planck2020_cosmology,
       author = {{Planck Collaboration} and {Aghanim}, N. and {Akrami}, Y. and {Ashdown}, M. and {Aumont}, J. and {Baccigalupi}, C. and {Ballardini}, M. and {Banday}, A.~J. and {Barreiro}, R.~B. and {Bartolo}, N. and {Basak}, S. and {Battye}, R. and {Benabed}, K. and {Bernard}, J. -P. and {Bersanelli}, M. and {Bielewicz}, P. and {Bock}, J.~J. and {Bond}, J.~R. and {Borrill}, J. and {Bouchet}, F.~R. and {Boulanger}, F. and {Bucher}, M. and {Burigana}, C. and {Butler}, R.~C. and {Calabrese}, E. and {Cardoso}, J. -F. and {Carron}, J. and {Challinor}, A. and {Chiang}, H.~C. and {Chluba}, J. and {Colombo}, L.~P.~L. and {Combet}, C. and {Contreras}, D. and {Crill}, B.~P. and {Cuttaia}, F. and {de Bernardis}, P. and {de Zotti}, G. and {Delabrouille}, J. and {Delouis}, J. -M. and {Di Valentino}, E. and {Diego}, J.~M. and {Dor{\'e}}, O. and {Douspis}, M. and {Ducout}, A. and {Dupac}, X. and {Dusini}, S. and {Efstathiou}, G. and {Elsner}, F. and {En{\ss}lin}, T.~A. and {Eriksen}, H.~K. and {Fantaye}, Y. and {Farhang}, M. and {Fergusson}, J. and {Fernandez-Cobos}, R. and {Finelli}, F. and {Forastieri}, F. and {Frailis}, M. and {Fraisse}, A.~A. and {Franceschi}, E. and {Frolov}, A. and {Galeotta}, S. and {Galli}, S. and {Ganga}, K. and {G{\'e}nova-Santos}, R.~T. and {Gerbino}, M. and {Ghosh}, T. and {Gonz{\'a}lez-Nuevo}, J. and {G{\'o}rski}, K.~M. and {Gratton}, S. and {Gruppuso}, A. and {Gudmundsson}, J.~E. and {Hamann}, J. and {Handley}, W. and {Hansen}, F.~K. and {Herranz}, D. and {Hildebrandt}, S.~R. and {Hivon}, E. and {Huang}, Z. and {Jaffe}, A.~H. and {Jones}, W.~C. and {Karakci}, A. and {Keih{\"a}nen}, E. and {Keskitalo}, R. and {Kiiveri}, K. and {Kim}, J. and {Kisner}, T.~S. and {Knox}, L. and {Krachmalnicoff}, N. and {Kunz}, M. and {Kurki-Suonio}, H. and {Lagache}, G. and {Lamarre}, J. -M. and {Lasenby}, A. and {Lattanzi}, M. and {Lawrence}, C.~R. and {Le Jeune}, M. and {Lemos}, P. and {Lesgourgues}, J. and {Levrier}, F. and {Lewis}, A. and {Liguori}, M. and {Lilje}, P.~B. and {Lilley}, M. and {Lindholm}, V. and {L{\'o}pez-Caniego}, M. and {Lubin}, P.~M. and {Ma}, Y. -Z. and {Mac{\'\i}as-P{\'e}rez}, J.~F. and {Maggio}, G. and {Maino}, D. and {Mandolesi}, N. and {Mangilli}, A. and {Marcos-Caballero}, A. and {Maris}, M. and {Martin}, P.~G. and {Martinelli}, M. and {Mart{\'\i}nez-Gonz{\'a}lez}, E. and {Matarrese}, S. and {Mauri}, N. and {McEwen}, J.~D. and {Meinhold}, P.~R. and {Melchiorri}, A. and {Mennella}, A. and {Migliaccio}, M. and {Millea}, M. and {Mitra}, S. and {Miville-Desch{\^e}nes}, M. -A. and {Molinari}, D. and {Montier}, L. and {Morgante}, G. and {Moss}, A. and {Natoli}, P. and {N{\o}rgaard-Nielsen}, H.~U. and {Pagano}, L. and {Paoletti}, D. and {Partridge}, B. and {Patanchon}, G. and {Peiris}, H.~V. and {Perrotta}, F. and {Pettorino}, V. and {Piacentini}, F. and {Polastri}, L. and {Polenta}, G. and {Puget}, J. -L. and {Rachen}, J.~P. and {Reinecke}, M. and {Remazeilles}, M. and {Renzi}, A. and {Rocha}, G. and {Rosset}, C. and {Roudier}, G. and {Rubi{\~n}o-Mart{\'\i}n}, J.~A. and {Ruiz-Granados}, B. and {Salvati}, L. and {Sandri}, M. and {Savelainen}, M. and {Scott}, D. and {Shellard}, E.~P.~S. and {Sirignano}, C. and {Sirri}, G. and {Spencer}, L.~D. and {Sunyaev}, R. and {Suur-Uski}, A. -S. and {Tauber}, J.~A. and {Tavagnacco}, D. and {Tenti}, M. and {Toffolatti}, L. and {Tomasi}, M. and {Trombetti}, T. and {Valenziano}, L. and {Valiviita}, J. and {Van Tent}, B. and {Vibert}, L. and {Vielva}, P. and {Villa}, F. and {Vittorio}, N. and {Wandelt}, B.~D. and {Wehus}, I.~K. and {White}, M. and {White}, S.~D.~M. and {Zacchei}, A. and {Zonca}, A.},
        title = "{Planck 2018 results. VI. Cosmological parameters}",
      journal = {\aap},
         year = 2020,
        month = sep,
       volume = {641},
          eid = {A6},
        pages = {A6},
          doi = {10.1051/0004-6361/201833910},
archivePrefix = {arXiv},
       eprint = {1807.06209},
 primaryClass = {astro-ph.CO},
       adsurl = {https://ui.adsabs.harvard.edu/abs/2020A&A...641A...6P}
}

@ARTICLE{amon_lensing_cosmo,
       author = {{Amon}, A. and {Robertson}, N.~C. and {Miyatake}, H. and {Heymans}, C. and {White}, M. and {DeRose}, J. and {Yuan}, S. and {Wechsler}, R.~H. and {Varga}, T.~N. and {Bocquet}, S. and {Dvornik}, A. and {More}, S. and {Ross}, A.~J. and {Hoekstra}, H. and {Alarcon}, A. and {Asgari}, M. and {Blazek}, J. and {Campos}, A. and {Chen}, R. and {Choi}, A. and {Crocce}, M. and {Diehl}, H.~T. and {Doux}, C. and {Eckert}, K. and {Elvin-Poole}, J. and {Everett}, S. and {Fert{\'e}}, A. and {Gatti}, M. and {Giannini}, G. and {Gruen}, D. and {Gruendl}, R.~A. and {Hartley}, W.~G. and {Herner}, K. and {Hildebrandt}, H. and {Huang}, S. and {Huff}, E.~M. and {Joachimi}, B. and {Lee}, S. and {MacCrann}, N. and {Myles}, J. and {Navarro-Alsina}, A. and {Nishimichi}, T. and {Prat}, J. and {Secco}, L.~F. and {Sevilla-Noarbe}, I. and {Sheldon}, E. and {Shin}, T. and {Tr{\"o}ster}, T. and {Troxel}, M.~A. and {Tutusaus}, I. and {Wright}, A.~H. and {Yin}, B. and {Aguena}, M. and {Allam}, S. and {Annis}, J. and {Bacon}, D. and {Bilicki}, M. and {Brooks}, D. and {Burke}, D.~L. and {Carnero Rosell}, A. and {Carretero}, J. and {Castander}, F.~J. and {Cawthon}, R. and {Costanzi}, M. and {da Costa}, L.~N. and {Pereira}, M.~E.~S. and {de Jong}, J. and {De Vicente}, J. and {Desai}, S. and {Dietrich}, J.~P. and {Doel}, P. and {Ferrero}, I. and {Frieman}, J. and {Garc{\'\i}a-Bellido}, J. and {Gerdes}, D.~W. and {Gschwend}, J. and {Gutierrez}, G. and {Hinton}, S.~R. and {Hollowood}, D.~L. and {Honscheid}, K. and {Huterer}, D. and {Kannawadi}, A. and {Kuehn}, K. and {Kuropatkin}, N. and {Lahav}, O. and {Lima}, M. and {Maia}, M.~A.~G. and {Marshall}, J.~L. and {Menanteau}, F. and {Miquel}, R. and {Mohr}, J.~J. and {Morgan}, R. and {Muir}, J. and {Paz-Chinch{\'o}n}, F. and {Pieres}, A. and {Plazas Malag{\'o}n}, A.~A. and {Porredon}, A. and {Rodriguez-Monroy}, M. and {Roodman}, A. and {Sanchez}, E. and {Serrano}, S. and {Shan}, H. and {Suchyta}, E. and {Swanson}, M.~E.~C. and {Tarle}, G. and {Thomas}, D. and {To}, C. and {Zhang}, Y.},
        title = "{Consistent lensing and clustering in a low-S$_{8}$ Universe with BOSS, DES Year 3, HSC Year 1, and KiDS-1000}",
      journal = {\mnras},
         year = 2023,
        month = jan,
       volume = {518},
       number = {1},
        pages = {477-503},
          doi = {10.1093/mnras/stac2938},
archivePrefix = {arXiv},
       eprint = {2202.07440},
 primaryClass = {astro-ph.CO},
       adsurl = {https://ui.adsabs.harvard.edu/abs/2023MNRAS.518..477A}
}

@ARTICLE{Siegel2026,
       author = {{Siegel}, Jared and {Amon}, Alexandra and {McCarthy}, Ian G. and {Bigwood}, Leah and {Yamamoto}, Masaya and {Bulbul}, Esra and {Greene}, Jenny E. and {McCullough}, Jamie and {Schaller}, Matthieu and {Schaye}, Joop},
        title = "{Joint X-ray, kinetic Sunyaev-Zeldovich, and weak lensing measurements: toward a consensus picture of efficient gas expulsion from groups and clusters}",
      journal = {arXiv e-prints},
         year = 2026,
        month = sep,
          eid = {arXiv:2509.10455},
        pages = {arXiv:2509.10455},
          doi = {10.48550/arXiv.2509.10455},
archivePrefix = {arXiv},
       eprint = {2509.10455},
 primaryClass = {astro-ph.CO},
       adsurl = {https://ui.adsabs.harvard.edu/abs/2025arXiv250910455S}
}

@ARTICLE{Bigwood2026,
       author = {{Bigwood}, Leah and {Yamamoto}, Masaya and {Siegel}, Jared and {Amon}, Alexandra and {McCarthy}, Ian G. and {Dave}, Romeel and {Salcido}, Jaime and {Schaller}, Matthieu and {Schaye}, Joop and {Yang}, Tianyi},
        title = "{The kinetic Sunyaev Zeldovich effect as a benchmark for AGN feedback models in hydrodynamical simulations: insights from DESI + ACT}",
      journal = {arXiv e-prints},
         year = 2026,
        month = oct,
          eid = {arXiv:2510.15822},
        pages = {arXiv:2510.15822},
          doi = {10.48550/arXiv.2510.15822},
archivePrefix = {arXiv},
       eprint = {2510.15822},
 primaryClass = {astro-ph.CO},
       adsurl = {https://ui.adsabs.harvard.edu/abs/2025arXiv251015822B}
}

@ARTICLE{McDonald2026,
       author = {{McDonald}, William and {Schaye}, Joop and {Kuijken}, Konrad and {Helly}, John and {Braspenning}, Joey and {Schaller}, Matthieu},
        title = "{The FLAMINGO Project: Exploring the X-ray--cosmic-shear cross-correlation as a probe of large-scale structure}",
      journal = {arXiv e-prints},
         year = 2026,
        month = feb,
          eid = {arXiv:2602.02484},
        pages = {arXiv:2602.02484},
          doi = {10.48550/arXiv.2602.02484},
archivePrefix = {arXiv},
       eprint = {2602.02484},
 primaryClass = {astro-ph.CO},
       adsurl = {https://ui.adsabs.harvard.edu/abs/2026arXiv260202484M}
}

@ARTICLE{Aihara_HSC_Y1,
       author = {{Aihara}, Hiroaki and {Armstrong}, Robert and {Bickerton}, Steven and {Bosch}, James and {Coupon}, Jean and {Furusawa}, Hisanori and {Hayashi}, Yusuke and {Ikeda}, Hiroyuki and {Kamata}, Yukiko and {Karoji}, Hiroshi and {Kawanomoto}, Satoshi and {Koike}, Michitaro and {Komiyama}, Yutaka and {Lang}, Dustin and {Lupton}, Robert H. and {Mineo}, Sogo and {Miyatake}, Hironao and {Miyazaki}, Satoshi and {Morokuma}, Tomoki and {Obuchi}, Yoshiyuki and {Oishi}, Yukie and {Okura}, Yuki and {Price}, Paul A. and {Takata}, Tadafumi and {Tanaka}, Manobu M. and {Tanaka}, Masayuki and {Tanaka}, Yoko and {Uchida}, Tomohisa and {Uraguchi}, Fumihiro and {Utsumi}, Yousuke and {Wang}, Shiang-Yu and {Yamada}, Yoshihiko and {Yamanoi}, Hitomi and {Yasuda}, Naoki and {Arimoto}, Nobuo and {Chiba}, Masashi and {Finet}, Francois and {Fujimori}, Hiroki and {Fujimoto}, Seiji and {Furusawa}, Junko and {Goto}, Tomotsugu and {Goulding}, Andy and {Gunn}, James E. and {Harikane}, Yuichi and {Hattori}, Takashi and {Hayashi}, Masao and {He{\l}miniak}, Krzysztof G. and {Higuchi}, Ryo and {Hikage}, Chiaki and {Ho}, Paul T.~P. and {Hsieh}, Bau-Ching and {Huang}, Kuiyun and {Huang}, Song and {Imanishi}, Masatoshi and {Iwata}, Ikuru and {Jaelani}, Anton T. and {Jian}, Hung-Yu and {Kashikawa}, Nobunari and {Katayama}, Nobuhiko and {Kojima}, Takashi and {Konno}, Akira and {Koshida}, Shintaro and {Kusakabe}, Haruka and {Leauthaud}, Alexie and {Lee}, Chien-Hsiu and {Lin}, Lihwai and {Lin}, Yen-Ting and {Mandelbaum}, Rachel and {Matsuoka}, Yoshiki and {Medezinski}, Elinor and {Miyama}, Shoken and {Momose}, Rieko and {More}, Anupreeta and {More}, Surhud and {Mukae}, Shiro and {Murata}, Ryoma and {Murayama}, Hitoshi and {Nagao}, Tohru and {Nakata}, Fumiaki and {Niida}, Mana and {Niikura}, Hiroko and {Nishizawa}, Atsushi J. and {Oguri}, Masamune and {Okabe}, Nobuhiro and {Ono}, Yoshiaki and {Onodera}, Masato and {Onoue}, Masafusa and {Ouchi}, Masami and {Pyo}, Tae-Soo and {Shibuya}, Takatoshi and {Shimasaku}, Kazuhiro and {Simet}, Melanie and {Speagle}, Joshua and {Spergel}, David N. and {Strauss}, Michael A. and {Sugahara}, Yuma and {Sugiyama}, Naoshi and {Suto}, Yasushi and {Suzuki}, Nao and {Tait}, Philip J. and {Takada}, Masahiro and {Terai}, Tsuyoshi and {Toba}, Yoshiki and {Turner}, Edwin L. and {Uchiyama}, Hisakazu and {Umetsu}, Keiichi and {Urata}, Yuji and {Usuda}, Tomonori and {Yeh}, Sherry and {Yuma}, Suraphong},
        title = "{First data release of the Hyper Suprime-Cam Subaru Strategic Program}",
      journal = {\pasj},
         year = 2018,
        month = jan,
       volume = {70},
          eid = {S8},
        pages = {S8},
          doi = {10.1093/pasj/psx081},
archivePrefix = {arXiv},
       eprint = {1702.08449},
 primaryClass = {astro-ph.IM},
       adsurl = {https://ui.adsabs.harvard.edu/abs/2018PASJ...70S...8A}
}

@ARTICLE{Kuijken_KIDS,
       author = {{Kuijken}, K. and {Heymans}, C. and {Dvornik}, A. and {Hildebrandt}, H. and {de Jong}, J.~T.~A. and {Wright}, A.~H. and {Erben}, T. and {Bilicki}, M. and {Giblin}, B. and {Shan}, H. -Y. and {Getman}, F. and {Grado}, A. and {Hoekstra}, H. and {Miller}, L. and {Napolitano}, N. and {Paolilo}, M. and {Radovich}, M. and {Schneider}, P. and {Sutherland}, W. and {Tewes}, M. and {Tortora}, C. and {Valentijn}, E.~A. and {Verdoes Kleijn}, G.~A.},
        title = "{The fourth data release of the Kilo-Degree Survey: ugri imaging and nine-band optical-IR photometry over 1000 square degrees}",
      journal = {\aap},
         year = 2019,
        month = may,
       volume = {625},
          eid = {A2},
        pages = {A2},
          doi = {10.1051/0004-6361/201834918},
archivePrefix = {arXiv},
       eprint = {1902.11265},
 primaryClass = {astro-ph.GA},
       adsurl = {https://ui.adsabs.harvard.edu/abs/2019A&A...625A...2K}
}

@ARTICLE{Borrow_SPHENIX,
       author = {{Borrow}, Josh and {Schaller}, Matthieu and {Bower}, Richard G. and {Schaye}, Joop},
        title = "{SPHENIX: smoothed particle hydrodynamics for the next generation of galaxy formation simulations}",
      journal = {\mnras},
         year = 2022,
        month = apr,
       volume = {511},
       number = {2},
        pages = {2367-2389},
          doi = {10.1093/mnras/stab3166},
archivePrefix = {arXiv},
       eprint = {2012.03974},
 primaryClass = {astro-ph.GA},
       adsurl = {https://ui.adsabs.harvard.edu/abs/2022MNRAS.511.2367B}
}

@ARTICLE{IC_ref1,
       author = {{Hahn}, Oliver and {Rampf}, Cornelius and {Uhlemann}, Cora},
        title = "{Higher order initial conditions for mixed baryon-CDM simulations}",
      journal = {\mnras},
         year = 2021,
        month = may,
       volume = {503},
       number = {1},
        pages = {426-445},
          doi = {10.1093/mnras/staa3773},
archivePrefix = {arXiv},
       eprint = {2008.09124},
 primaryClass = {astro-ph.CO},
       adsurl = {https://ui.adsabs.harvard.edu/abs/2021MNRAS.503..426H}
}

@ARTICLE{IC_ref2,
       author = {{Elbers}, Willem and {Frenk}, Carlos S. and {Jenkins}, Adrian and {Li}, Baojiu and {Pascoli}, Silvia},
        title = "{Higher order initial conditions with massive neutrinos}",
      journal = {\mnras},
         year = 2022,
        month = nov,
       volume = {516},
       number = {3},
        pages = {3821-3836},
          doi = {10.1093/mnras/stac2365},
archivePrefix = {arXiv},
       eprint = {2202.00670},
 primaryClass = {astro-ph.CO},
       adsurl = {https://ui.adsabs.harvard.edu/abs/2022MNRAS.516.3821E}
}

@ARTICLE{FLAMINGO_neutrino,
       author = {{Elbers}, Willem and {Frenk}, Carlos S. and {Jenkins}, Adrian and {Li}, Baojiu and {Pascoli}, Silvia},
        title = "{An optimal non-linear method for simulating relic neutrinos}",
      journal = {\mnras},
         year = 2021,
        month = oct,
       volume = {507},
       number = {2},
        pages = {2614-2631},
          doi = {10.1093/mnras/stab2260},
archivePrefix = {arXiv},
       eprint = {2010.07321},
 primaryClass = {astro-ph.CO},
       adsurl = {https://ui.adsabs.harvard.edu/abs/2021MNRAS.507.2614E}
}

@ARTICLE{Healpix_ref,
       author = {{G{\'o}rski}, K.~M. and {Hivon}, E. and {Banday}, A.~J. and {Wandelt}, B.~D. and {Hansen}, F.~K. and {Reinecke}, M. and {Bartelmann}, M.},
        title = "{HEALPix: A Framework for High-Resolution Discretization and Fast Analysis of Data Distributed on the Sphere}",
      journal = {\apj},
         year = 2005,
        month = apr,
       volume = {622},
       number = {2},
        pages = {759-771},
          doi = {10.1086/427976},
archivePrefix = {arXiv},
       eprint = {astro-ph/0409513},
 primaryClass = {astro-ph},
       adsurl = {https://ui.adsabs.harvard.edu/abs/2005ApJ...622..759G}
}

@ARTICLE{halo_lc_ref1,
       author = {{Han}, Jiaxin and {Jing}, Y.~P. and {Wang}, Huiyuan and {Wang}, Wenting},
        title = "{Resolving subhaloes' lives with the Hierarchical Bound-Tracing algorithm}",
      journal = {\mnras},
         year = 2012,
        month = dec,
       volume = {427},
       number = {3},
        pages = {2437-2449},
          doi = {10.1111/j.1365-2966.2012.22111.x},
archivePrefix = {arXiv},
       eprint = {1103.2099},
 primaryClass = {astro-ph.CO},
       adsurl = {https://ui.adsabs.harvard.edu/abs/2012MNRAS.427.2437H}
}

@ARTICLE{halo_lc_ref2,
       author = {{Han}, Jiaxin and {Cole}, Shaun and {Frenk}, Carlos S. and {Benitez-Llambay}, Alejandro and {Helly}, John},
        title = "{HBT+: an improved code for finding subhaloes and building merger trees in cosmological simulations}",
      journal = {\mnras},
         year = 2018,
        month = feb,
       volume = {474},
       number = {1},
        pages = {604-617},
          doi = {10.1093/mnras/stx2792},
archivePrefix = {arXiv},
       eprint = {1708.03646},
 primaryClass = {astro-ph.CO},
       adsurl = {https://ui.adsabs.harvard.edu/abs/2018MNRAS.474..604H}
}

@ARTICLE{halo_lc_ref3,
       author = {{Forouhar Moreno}, Victor J. and {Helly}, John and {McGibbon}, Rob and {Schaye}, Joop and {Schaller}, Matthieu and {Han}, Jiaxin and {Kugel}, Roi},
        title = "{Assessing subhalo finders in cosmological hydrodynamical simulations}",
      journal = {arXiv e-prints},
         year = 2025,
        month = feb,
          eid = {arXiv:2502.06932},
        pages = {arXiv:2502.06932},
          doi = {10.48550/arXiv.2502.06932},
archivePrefix = {arXiv},
       eprint = {2502.06932},
 primaryClass = {astro-ph.CO},
       adsurl = {https://ui.adsabs.harvard.edu/abs/2025arXiv250206932F}
}

@ARTICLE{Ian_low_S8_FMG_paper,
       author = {{McCarthy}, Ian G. and {Salcido}, Jaime and {Schaye}, Joop and {Kwan}, Juliana and {Elbers}, Willem and {Kugel}, Roi and {Schaller}, Matthieu and {Helly}, John C. and {Braspenning}, Joey and {Frenk}, Carlos S. and {van Daalen}, Marcel P. and {Vandenbroucke}, Bert and {Conley}, Jonah T. and {Font}, Andreea S. and {Upadhye}, Amol},
        title = "{The FLAMINGO project: revisiting the S$_{8}$ tension and the role of baryonic physics}",
      journal = {\mnras},
         year = 2023,
        month = dec,
       volume = {526},
       number = {4},
        pages = {5494-5519},
          doi = {10.1093/mnras/stad3107},
archivePrefix = {arXiv},
       eprint = {2309.07959},
 primaryClass = {astro-ph.CO},
       adsurl = {https://ui.adsabs.harvard.edu/abs/2023MNRAS.526.5494M}
}

@ARTICLE{SPTpol_lensing_ref,
       author = {{Wu}, W.~L.~K. and {Mocanu}, L.~M. and {Ade}, P.~A.~R. and {Anderson}, A.~J. and {Austermann}, J.~E. and {Avva}, J.~S. and {Beall}, J.~A. and {Bender}, A.~N. and {Benson}, B.~A. and {Bianchini}, F. and {Bleem}, L.~E. and {Carlstrom}, J.~E. and {Chang}, C.~L. and {Chiang}, H.~C. and {Citron}, R. and {Corbett Moran}, C. and {Crawford}, T.~M. and {Crites}, A.~T. and {de Haan}, T. and {Dobbs}, M.~A. and {Everett}, W. and {Gallicchio}, J. and {George}, E.~M. and {Gilbert}, A. and {Gupta}, N. and {Halverson}, N.~W. and {Harrington}, N. and {Henning}, J.~W. and {Hilton}, G.~C. and {Holder}, G.~P. and {Holzapfel}, W.~L. and {Hou}, Z. and {Hrubes}, J.~D. and {Huang}, N. and {Hubmayr}, J. and {Irwin}, K.~D. and {Knox}, L. and {Lee}, A.~T. and {Li}, D. and {Lowitz}, A. and {Manzotti}, A. and {McMahon}, J.~J. and {Meyer}, S.~S. and {Millea}, M. and {Montgomery}, J. and {Nadolski}, A. and {Natoli}, T. and {Nibarger}, J.~P. and {Noble}, G.~I. and {Novosad}, V. and {Omori}, Y. and {Padin}, S. and {Patil}, S. and {Pryke}, C. and {Reichardt}, C.~L. and {Ruhl}, J.~E. and {Saliwanchik}, B.~R. and {Sayre}, J.~T. and {Schaffer}, K.~K. and {Sievers}, C. and {Simard}, G. and {Smecher}, G. and {Stark}, A.~A. and {Story}, K.~T. and {Tucker}, C. and {Vanderlinde}, K. and {Veach}, T. and {Vieira}, J.~D. and {Wang}, G. and {Whitehorn}, N. and {Yefremenko}, V.},
        title = "{A Measurement of the Cosmic Microwave Background Lensing Potential and Power Spectrum from 500 deg$^{2}$ of SPTpol Temperature and Polarization Data}",
      journal = {\apj},
         year = 2019,
        month = oct,
       volume = {884},
       number = {1},
          eid = {70},
        pages = {70},
          doi = {10.3847/1538-4357/ab4186},
archivePrefix = {arXiv},
       eprint = {1905.05777},
 primaryClass = {astro-ph.CO},
       adsurl = {https://ui.adsabs.harvard.edu/abs/2019ApJ...884...70W}
}

@ARTICLE{ACTDR6_lensing_ref,
       author = {{Qu}, Frank J. and {Sherwin}, Blake D. and {Madhavacheril}, Mathew S. and {Han}, Dongwon and {Crowley}, Kevin T. and {Abril-Cabezas}, Irene and {Ade}, Peter A.~R. and {Aiola}, Simone and {Alford}, Tommy and {Amiri}, Mandana and {Amodeo}, Stefania and {An}, Rui and {Atkins}, Zachary and {Austermann}, Jason E. and {Battaglia}, Nicholas and {Battistelli}, Elia Stefano and {Beall}, James A. and {Bean}, Rachel and {Beringue}, Benjamin and {Bhandarkar}, Tanay and {Biermann}, Emily and {Bolliet}, Boris and {Bond}, J. Richard and {Cai}, Hongbo and {Calabrese}, Erminia and {Calafut}, Victoria and {Capalbo}, Valentina and {Carrero}, Felipe and {Carron}, Julien and {Challinor}, Anthony and {Chesmore}, Grace E. and {Cho}, Hsiao-mei and {Choi}, Steve K. and {Clark}, Susan E. and {C{\'o}rdova Rosado}, Rodrigo and {Cothard}, Nicholas F. and {Coughlin}, Kevin and {Coulton}, William and {Dalal}, Roohi and {Darwish}, Omar and {Devlin}, Mark J. and {Dicker}, Simon and {Doze}, Peter and {Duell}, Cody J. and {Duff}, Shannon M. and {Duivenvoorden}, Adriaan J. and {Dunkley}, Jo and {D{\"u}nner}, Rolando and {Fanfani}, Valentina and {Fankhanel}, Max and {Farren}, Gerrit and {Ferraro}, Simone and {Freundt}, Rodrigo and {Fuzia}, Brittany and {Gallardo}, Patricio A. and {Garrido}, Xavier and {Gluscevic}, Vera and {Golec}, Joseph E. and {Guan}, Yilun and {Halpern}, Mark and {Harrison}, Ian and {Hasselfield}, Matthew and {Healy}, Erin and {Henderson}, Shawn and {Hensley}, Brandon and {Herv{\'\i}as-Caimapo}, Carlos and {Hill}, J. Colin and {Hilton}, Gene C. and {Hilton}, Matt and {Hincks}, Adam D. and {Hlo{\v{z}}ek}, Ren{\'e}e and {Ho}, Shuay-Pwu Patty and {Huber}, Zachary B. and {Hubmayr}, Johannes and {Huffenberger}, Kevin M. and {Hughes}, John P. and {Irwin}, Kent and {Isopi}, Giovanni and {Jense}, Hidde T. and {Keller}, Ben and {Kim}, Joshua and {Knowles}, Kenda and {Koopman}, Brian J. and {Kosowsky}, Arthur and {Kramer}, Darby and {Kusiak}, Aleksandra and {La Posta}, Adrien and {Lague}, Alex and {Lakey}, Victoria and {Lee}, Eunseong and {Li}, Zack and {Li}, Yaqiong and {Limon}, Michele and {Lokken}, Martine and {Louis}, Thibaut and {Lungu}, Marius and {MacCrann}, Niall and {MacInnis}, Amanda and {Maldonado}, Diego and {Maldonado}, Felipe and {Mallaby-Kay}, Maya and {Marques}, Gabriela A. and {McMahon}, Jeff and {Mehta}, Yogesh and {Menanteau}, Felipe and {Moodley}, Kavilan and {Morris}, Thomas W. and {Mroczkowski}, Tony and {Naess}, Sigurd and {Namikawa}, Toshiya and {Nati}, Federico and {Newburgh}, Laura and {Nicola}, Andrina and {Niemack}, Michael D. and {Nolta}, Michael R. and {Orlowski-Scherer}, John and {Page}, Lyman A. and {Pandey}, Shivam and {Partridge}, Bruce and {Prince}, Heather and {Puddu}, Roberto and {Radiconi}, Federico and {Robertson}, Naomi and {Rojas}, Felipe and {Sakuma}, Tai and {Salatino}, Maria and {Schaan}, Emmanuel and {Schmitt}, Benjamin L. and {Sehgal}, Neelima and {Shaikh}, Shabbir and {Sierra}, Carlos and {Sievers}, Jon and {Sif{\'o}n}, Crist{\'o}bal and {Simon}, Sara and {Sonka}, Rita and {Spergel}, David N. and {Staggs}, Suzanne T. and {Storer}, Emilie and {Switzer}, Eric R. and {Tampier}, Niklas and {Thornton}, Robert and {Trac}, Hy and {Treu}, Jesse and {Tucker}, Carole and {Ullom}, Joel and {Vale}, Leila R. and {Van Engelen}, Alexander and {Van Lanen}, Jeff and {van Marrewijk}, Joshiwa and {Vargas}, Cristian and {Vavagiakis}, Eve M. and {Wagoner}, Kasey and {Wang}, Yuhan and {Wenzl}, Lukas and {Wollack}, Edward J. and {Xu}, Zhilei and {Zago}, Fernando and {Zheng}, Kaiwen},
        title = "{The Atacama Cosmology Telescope: A Measurement of the DR6 CMB Lensing Power Spectrum and Its Implications for Structure Growth}",
      journal = {\apj},
         year = 2024,
        month = feb,
       volume = {962},
       number = {2},
          eid = {112},
        pages = {112},
          doi = {10.3847/1538-4357/acfe06},
archivePrefix = {arXiv},
       eprint = {2304.05202},
 primaryClass = {astro-ph.CO},
       adsurl = {https://ui.adsabs.harvard.edu/abs/2024ApJ...962..112Q}
}

@ARTICLE{P18_lensing_ref,
       author = {{Planck Collaboration} and {Aghanim}, N. and {Akrami}, Y. and {Ashdown}, M. and {Aumont}, J. and {Baccigalupi}, C. and {Ballardini}, M. and {Banday}, A.~J. and {Barreiro}, R.~B. and {Bartolo}, N. and {Basak}, S. and {Benabed}, K. and {Bernard}, J. -P. and {Bersanelli}, M. and {Bielewicz}, P. and {Bock}, J.~J. and {Bond}, J.~R. and {Borrill}, J. and {Bouchet}, F.~R. and {Boulanger}, F. and {Bucher}, M. and {Burigana}, C. and {Calabrese}, E. and {Cardoso}, J. -F. and {Carron}, J. and {Challinor}, A. and {Chiang}, H.~C. and {Colombo}, L.~P.~L. and {Combet}, C. and {Crill}, B.~P. and {Cuttaia}, F. and {de Bernardis}, P. and {de Zotti}, G. and {Delabrouille}, J. and {Di Valentino}, E. and {Diego}, J.~M. and {Dor{\'e}}, O. and {Douspis}, M. and {Ducout}, A. and {Dupac}, X. and {Efstathiou}, G. and {Elsner}, F. and {En{\ss}lin}, T.~A. and {Eriksen}, H.~K. and {Fantaye}, Y. and {Fernandez-Cobos}, R. and {Finelli}, F. and {Forastieri}, F. and {Frailis}, M. and {Fraisse}, A.~A. and {Franceschi}, E. and {Frolov}, A. and {Galeotta}, S. and {Galli}, S. and {Ganga}, K. and {G{\'e}nova-Santos}, R.~T. and {Gerbino}, M. and {Ghosh}, T. and {Gonz{\'a}lez-Nuevo}, J. and {G{\'o}rski}, K.~M. and {Gratton}, S. and {Gruppuso}, A. and {Gudmundsson}, J.~E. and {Hamann}, J. and {Handley}, W. and {Hansen}, F.~K. and {Herranz}, D. and {Hivon}, E. and {Huang}, Z. and {Jaffe}, A.~H. and {Jones}, W.~C. and {Karakci}, A. and {Keih{\"a}nen}, E. and {Keskitalo}, R. and {Kiiveri}, K. and {Kim}, J. and {Knox}, L. and {Krachmalnicoff}, N. and {Kunz}, M. and {Kurki-Suonio}, H. and {Lagache}, G. and {Lamarre}, J. -M. and {Lasenby}, A. and {Lattanzi}, M. and {Lawrence}, C.~R. and {Le Jeune}, M. and {Levrier}, F. and {Lewis}, A. and {Liguori}, M. and {Lilje}, P.~B. and {Lindholm}, V. and {L{\'o}pez-Caniego}, M. and {Lubin}, P.~M. and {Ma}, Y. -Z. and {Mac{\'\i}as-P{\'e}rez}, J.~F. and {Maggio}, G. and {Maino}, D. and {Mandolesi}, N. and {Mangilli}, A. and {Marcos-Caballero}, A. and {Maris}, M. and {Martin}, P.~G. and {Mart{\'\i}nez-Gonz{\'a}lez}, E. and {Matarrese}, S. and {Mauri}, N. and {McEwen}, J.~D. and {Melchiorri}, A. and {Mennella}, A. and {Migliaccio}, M. and {Miville-Desch{\^e}nes}, M. -A. and {Molinari}, D. and {Moneti}, A. and {Montier}, L. and {Morgante}, G. and {Moss}, A. and {Natoli}, P. and {Pagano}, L. and {Paoletti}, D. and {Partridge}, B. and {Patanchon}, G. and {Perrotta}, F. and {Pettorino}, V. and {Piacentini}, F. and {Polastri}, L. and {Polenta}, G. and {Puget}, J. -L. and {Rachen}, J.~P. and {Reinecke}, M. and {Remazeilles}, M. and {Renzi}, A. and {Rocha}, G. and {Rosset}, C. and {Roudier}, G. and {Rubi{\~n}o-Mart{\'\i}n}, J.~A. and {Ruiz-Granados}, B. and {Salvati}, L. and {Sandri}, M. and {Savelainen}, M. and {Scott}, D. and {Sirignano}, C. and {Sunyaev}, R. and {Suur-Uski}, A. -S. and {Tauber}, J.~A. and {Tavagnacco}, D. and {Tenti}, M. and {Toffolatti}, L. and {Tomasi}, M. and {Trombetti}, T. and {Valiviita}, J. and {Van Tent}, B. and {Vielva}, P. and {Villa}, F. and {Vittorio}, N. and {Wandelt}, B.~D. and {Wehus}, I.~K. and {White}, M. and {White}, S.~D.~M. and {Zacchei}, A. and {Zonca}, A.},
        title = "{Planck 2018 results. VIII. Gravitational lensing}",
      journal = {\aap},
         year = 2020,
        month = sep,
       volume = {641},
          eid = {A8},
        pages = {A8},
          doi = {10.1051/0004-6361/201833886},
archivePrefix = {arXiv},
       eprint = {1807.06210},
 primaryClass = {astro-ph.CO},
       adsurl = {https://ui.adsabs.harvard.edu/abs/2020A&A...641A...8P}
}

@ARTICLE{KIDS_Planck_ACT_shear_cross,
       author = {{Robertson}, Naomi Clare and {Alonso}, David and {Harnois-D{\'e}raps}, Joachim and {Darwish}, Omar and {Kannawadi}, Arun and {Amon}, Alexandra and {Asgari}, Marika and {Bilicki}, Maciej and {Calabrese}, Erminia and {Choi}, Steve K. and {Devlin}, Mark J. and {Dunkley}, Jo and {Dvornik}, Andrej and {Erben}, Thomas and {Ferraro}, Simone and {Fortuna}, Maria Cristina and {Giblin}, Benjamin and {Han}, Dongwon and {Heymans}, Catherine and {Hildebrandt}, Hendrik and {Hill}, J. Colin and {Hilton}, Matt and {Ho}, Shuay-Pwu P. and {Hoekstra}, Henk and {Hubmayr}, Johannes and {Hughes}, John P. and {Joachimi}, Benjamin and {Joudaki}, Shahab and {Knowles}, Kenda and {Kuijken}, Konrad and {Madhavacheril}, Mathew S. and {Moodley}, Kavilan and {Miller}, Lance and {Namikawa}, Toshiya and {Nati}, Federico and {Niemack}, Michael D. and {Page}, Lyman A. and {Partridge}, Bruce and {Schaan}, Emmanuel and {Schillaci}, Alessandro and {Schneider}, Peter and {Sehgal}, Neelima and {Sherwin}, Blake D. and {Sif{\'o}n}, Crist{\'o}bal and {Staggs}, Suzanne T. and {Tr{\"o}ster}, Tilman and {van Engelen}, Alexander and {Valentijn}, Edwin and {Wollack}, Edward J. and {Wright}, Angus H. and {Xu}, Zhilei},
        title = "{Strong detection of the CMB lensing and galaxy weak lensing cross-correlation from ACT-DR4, Planck Legacy, and KiDS-1000}",
      journal = {\aap},
         year = 2021,
        month = may,
       volume = {649},
          eid = {A146},
        pages = {A146},
          doi = {10.1051/0004-6361/202039975},
archivePrefix = {arXiv},
       eprint = {2011.11613},
 primaryClass = {astro-ph.CO},
       adsurl = {https://ui.adsabs.harvard.edu/abs/2021A&A...649A.146R}
}

@ARTICLE{NaMaster_ref,
       author = {{Alonso}, David and {Sanchez}, Javier and {Slosar}, An{\v{z}}e and {LSST Dark Energy Science Collaboration}},
        title = "{A unified pseudo-C$_{{\ensuremath{\ell}}}$ framework}",
      journal = {\mnras},
         year = 2019,
        month = apr,
       volume = {484},
       number = {3},
        pages = {4127-4151},
          doi = {10.1093/mnras/stz093},
archivePrefix = {arXiv},
       eprint = {1809.09603},
 primaryClass = {astro-ph.CO},
       adsurl = {https://ui.adsabs.harvard.edu/abs/2019MNRAS.484.4127A}
}

@ARTICLE{CMB_weak_lensing_review,
       author = {{Lewis}, Antony and {Challinor}, Anthony},
        title = "{Weak gravitational lensing of the CMB}",
      journal = {\physrep},
         year = 2006,
        month = jun,
       volume = {429},
       number = {1},
        pages = {1-65},
          doi = {10.1016/j.physrep.2006.03.002},
archivePrefix = {arXiv},
       eprint = {astro-ph/0601594},
 primaryClass = {astro-ph},
       adsurl = {https://ui.adsabs.harvard.edu/abs/2006PhR...429....1L}
}

@ARTICLE{SZ_effecr_ref1,
       author = {{Sunyaev}, R.~A. and {Zeldovich}, Ya. B.},
        title = "{The Observations of Relic Radiation as a Test of the Nature of X-Ray Radiation from the Clusters of Galaxies}",
      journal = {Comments on Astrophysics and Space Physics},
         year = 1972,
        month = nov,
       volume = {4},
        pages = {173},
       adsurl = {https://ui.adsabs.harvard.edu/abs/1972CoASP...4..173S}
}

@ARTICLE{SZ_effecr_ref2,
       author = {{Sunyaev}, R.~A. and {Zeldovich}, Ia. B.},
        title = "{Microwave background radiation as a probe of the contemporary structure and history of the universe}",
      journal = {ARAA},
         year = 1980,
        month = jan,
       volume = {18},
        pages = {537-560},
          doi = {10.1146/annurev.aa.18.090180.002541},
       adsurl = {https://ui.adsabs.harvard.edu/abs/1980ARA&A..18..537S}
}

@ARTICLE{tSZ_review_tony,
       author = {{Mroczkowski}, Tony and {Nagai}, Daisuke and {Basu}, Kaustuv and {Chluba}, Jens and {Sayers}, Jack and {Adam}, R{\'e}mi and {Churazov}, Eugene and {Crites}, Abigail and {Di Mascolo}, Luca and {Eckert}, Dominique and {Macias-Perez}, Juan and {Mayet}, Fr{\'e}d{\'e}ric and {Perotto}, Laurence and {Pointecouteau}, Etienne and {Romero}, Charles and {Ruppin}, Florian and {Scannapieco}, Evan and {ZuHone}, John},
        title = "{Astrophysics with the Spatially and Spectrally Resolved Sunyaev-Zeldovich Effects. A Millimetre/Submillimetre Probe of the Warm and Hot Universe}",
      journal = {SSR},
         year = 2019,
        month = feb,
       volume = {215},
       number = {1},
          eid = {17},
        pages = {17},
          doi = {10.1007/s11214-019-0581-2},
archivePrefix = {arXiv},
       eprint = {1811.02310},
 primaryClass = {astro-ph.CO},
       adsurl = {https://ui.adsabs.harvard.edu/abs/2019SSRv..215...17M}
}

@ARTICLE{Vikram_2017_group_SZ,
       author = {{Vikram}, Vinu and {Lidz}, Adam and {Jain}, Bhuvnesh},
        title = "{A Measurement of the Galaxy Group-Thermal Sunyaev-Zel'dovich Effect Cross-Correlation Function}",
      journal = {MNRAS},
         year = 2017,
        month = may,
       volume = {467},
       number = {2},
        pages = {2315-2330},
          doi = {10.1093/mnras/stw3311},
archivePrefix = {arXiv},
       eprint = {1608.04160},
 primaryClass = {astro-ph.CO},
       adsurl = {https://ui.adsabs.harvard.edu/abs/2017MNRAS.467.2315V}
}

@ARTICLE{Tilman_KIDS_y_WL,
       author = {{Tr{\"o}ster}, Tilman and {Mead}, Alexander J. and {Heymans}, Catherine and {Yan}, Ziang and {Alonso}, David and {Asgari}, Marika and {Bilicki}, Maciej and {Dvornik}, Andrej and {Hildebrandt}, Hendrik and {Joachimi}, Benjamin and {Kannawadi}, Arun and {Kuijken}, Konrad and {Schneider}, Peter and {Shan}, Huan Yuan and {van Waerbeke}, Ludovic and {Wright}, Angus H.},
        title = "{Joint constraints on cosmology and the impact of baryon feedback: Combining KiDS-1000 lensing with the thermal Sunyaev-Zeldovich effect from Planck and ACT}",
      journal = {AAP},
         year = 2022,
        month = apr,
       volume = {660},
          eid = {A27},
        pages = {A27},
          doi = {10.1051/0004-6361/202142197},
archivePrefix = {arXiv},
       eprint = {2109.04458},
 primaryClass = {astro-ph.CO},
       adsurl = {https://ui.adsabs.harvard.edu/abs/2022A&A...660A..27T}
}

@ARTICLE{Battaglia_2015_y_lensing,
       author = {{Battaglia}, N. and {Hill}, J.~C. and {Murray}, N.},
        title = "{Deconstructing Thermal Sunyaev-Zel{\textquoteright}dovich - Gravitational Lensing Cross-correlations: Implications for the Intracluster Medium}",
      journal = {APJ},
         year = 2015,
        month = oct,
       volume = {812},
       number = {2},
          eid = {154},
        pages = {154},
          doi = {10.1088/0004-637X/812/2/154},
archivePrefix = {arXiv},
       eprint = {1412.5593},
 primaryClass = {astro-ph.CO},
       adsurl = {https://ui.adsabs.harvard.edu/abs/2015ApJ...812..154B}
}

@ARTICLE{b_pe_constrain_most_precise,
       author = {{Koukoufilippas}, Nick and {Alonso}, David and {Bilicki}, Maciej and {Peacock}, John A.},
        title = "{Tomographic measurement of the intergalactic gas pressure through galaxy-tSZ cross-correlations}",
      journal = {MNRAS},
         year = 2020,
        month = feb,
       volume = {491},
       number = {4},
        pages = {5464-5480},
          doi = {10.1093/mnras/stz3351},
archivePrefix = {arXiv},
       eprint = {1909.09102},
 primaryClass = {astro-ph.CO},
       adsurl = {https://ui.adsabs.harvard.edu/abs/2020MNRAS.491.5464K}
}

@ARTICLE{Will_forecast_for_rSZ,
       author = {{Kuhn}, L. and {Li}, Z. and {Coulton}, William R.},
        title = "{Forecasts and Simulations for Relativistic Corrections to the Sunyaev-Zeldovich Effect}",
      journal = {arXiv e-prints},
         year = 2025,
        month = apr,
          eid = {arXiv:2504.18637},
        pages = {arXiv:2504.18637},
          doi = {10.48550/arXiv.2504.18637},
archivePrefix = {arXiv},
       eprint = {2504.18637},
 primaryClass = {astro-ph.CO},
       adsurl = {https://ui.adsabs.harvard.edu/abs/2025arXiv250418637K}
}

@ARTICLE{rSZ_cluster_temp_ref1,
       author = {{Coulton}, William R. and {Duivenvoorden}, Adriaan J. and {Atkins}, Zachary and {Battaglia}, Nicholas and {Battistelli}, Elia Stefano and {Bond}, J Richard and {Cai}, Hongbo and {Calabrese}, Erminia and {Choi}, Steve K. and {Crowley}, Kevin T. and {Devlin}, Mark J. and {Dunkley}, Jo and {Ferraro}, Simone and {Guan}, Yilun and {Herv{\'\i}as-Caimapo}, Carlos and {Hill}, J. Colin and {Hilton}, Matt and {Hincks}, Adam D. and {Kosowsky}, Arthur and {Madhavacheril}, Mathew S. and {van Marrewijk}, Joshiwa and {McCarthy}, Fiona and {Moodley}, Kavilan and {Mroczkowski}, Tony and {Niemack}, Michael D. and {Page}, Lyman A. and {Partridge}, Bruce and {Schaan}, Emmanuel and {Sehgal}, Neelima and {Sherwin}, Blake and {Sif{\'o}n}, Crist{\'o}bal and {Spergel}, David N. and {Staggs}, Suzanne T. and {Vavagiakis}, Eve M. and {Wollack}, Edward J.},
        title = "{The Atacama Cosmology Telescope: A measurement of galaxy cluster temperatures through relativistic corrections to the thermal Sunyaev-Zeldovich effect}",
      journal = {arXiv e-prints},
         year = 2024,
        month = oct,
          eid = {arXiv:2410.19046},
        pages = {arXiv:2410.19046},
          doi = {10.48550/arXiv.2410.19046},
archivePrefix = {arXiv},
       eprint = {2410.19046},
 primaryClass = {astro-ph.CO},
       adsurl = {https://ui.adsabs.harvard.edu/abs/2024arXiv241019046C}
}

@ARTICLE{Challinor_rSZ,
       author = {{Challinor}, Anthony and {Lasenby}, Anthony},
        title = "{Relativistic Corrections to the Sunyaev-Zeldovich Effect}",
      journal = {\apj},
         year = 1998,
        month = may,
       volume = {499},
       number = {1},
        pages = {1-6},
          doi = {10.1086/305623},
archivePrefix = {arXiv},
       eprint = {astro-ph/9711161},
 primaryClass = {astro-ph},
       adsurl = {https://ui.adsabs.harvard.edu/abs/1998ApJ...499....1C}
}

@ARTICLE{rSZ_cluster_temp_ref2,
       author = {{Remazeilles}, Mathieu and {Chluba}, Jens},
        title = "{Mapping the relativistic electron gas temperature across the sky}",
      journal = {\mnras},
         year = 2020,
        month = jun,
       volume = {494},
       number = {4},
        pages = {5734-5750},
          doi = {10.1093/mnras/staa1135},
archivePrefix = {arXiv},
       eprint = {1907.00916},
 primaryClass = {astro-ph.CO},
       adsurl = {https://ui.adsabs.harvard.edu/abs/2020MNRAS.494.5734R}
}

@ARTICLE{rSZ_cluster_temp_ref3,
       author = {{Remazeilles}, Mathieu and {Chluba}, Jens},
        title = "{Evidence for relativistic Sunyaev-Zeldovich effect in Planck CMB maps with an average electron-gas temperature of T$_{e}$ ≃ 5 keV}",
      journal = {\mnras},
         year = 2025,
        month = apr,
       volume = {538},
       number = {3},
        pages = {1576-1586},
          doi = {10.1093/mnras/staf384},
archivePrefix = {arXiv},
       eprint = {2410.02488},
 primaryClass = {astro-ph.CO},
       adsurl = {https://ui.adsabs.harvard.edu/abs/2025MNRAS.538.1576R}
}

@ARTICLE{Mead_2020,
       author = {{Mead}, A.~J. and {Tr{\"o}ster}, T. and {Heymans}, C. and {Van Waerbeke}, L. and {McCarthy}, I.~G.},
        title = "{A hydrodynamical halo model for weak-lensing cross correlations}",
      journal = {\aap},
         year = 2020,
        month = sep,
       volume = {641},
          eid = {A130},
        pages = {A130},
          doi = {10.1051/0004-6361/202038308},
archivePrefix = {arXiv},
       eprint = {2005.00009},
 primaryClass = {astro-ph.CO},
       adsurl = {https://ui.adsabs.harvard.edu/abs/2020A&A...641A.130M}
}

@ARTICLE{mass_peak_patch_ref1,
       author = {{Bond}, J.~R. and {Myers}, S.~T.},
        title = "{The Peak-Patch Picture of Cosmic Catalogs. I. Algorithms}",
      journal = {\apjs},
         year = 1996,
        month = mar,
       volume = {103},
        pages = {1},
          doi = {10.1086/192267},
       adsurl = {https://ui.adsabs.harvard.edu/abs/1996ApJS..103....1B}
}

@ARTICLE{2LPT_ref,
       author = {{Bouchet}, F.~R. and {Colombi}, S. and {Hivon}, E. and {Juszkiewicz}, R.},
        title = "{Perturbative Lagrangian approach to gravitational instability.}",
      journal = {\aap},
         year = 1995,
        month = apr,
       volume = {296},
        pages = {575},
          doi = {10.48550/arXiv.astro-ph/9406013},
archivePrefix = {arXiv},
       eprint = {astro-ph/9406013},
 primaryClass = {astro-ph},
       adsurl = {https://ui.adsabs.harvard.edu/abs/1995A&A...296..575B}
}

@ARTICLE{mass_peak_patch_ref2,
       author = {{Stein}, George and {Alvarez}, Marcelo A. and {Bond}, J. Richard},
        title = "{The mass-Peak Patch algorithm for fast generation of deep all-sky dark matter halo catalogues and its N-body validation}",
      journal = {\mnras},
         year = 2019,
        month = feb,
       volume = {483},
       number = {2},
        pages = {2236-2250},
          doi = {10.1093/mnras/sty3226},
archivePrefix = {arXiv},
       eprint = {1810.07727},
 primaryClass = {astro-ph.CO},
       adsurl = {https://ui.adsabs.harvard.edu/abs/2019MNRAS.483.2236S}
}

@ARTICLE{relativistic_tSZ_auto_powersp,
       author = {{Remazeilles}, Mathieu and {Bolliet}, Boris and {Rotti}, Aditya and {Chluba}, Jens},
        title = "{Can we neglect relativistic temperature corrections in the Planck thermal SZ analysis?}",
      journal = {\mnras},
         year = 2019,
        month = mar,
       volume = {483},
       number = {3},
        pages = {3459-3464},
          doi = {10.1093/mnras/sty3352},
archivePrefix = {arXiv},
       eprint = {1809.09666},
 primaryClass = {astro-ph.CO},
       adsurl = {https://ui.adsabs.harvard.edu/abs/2019MNRAS.483.3459R}
}

@ARTICLE{IRIS_CIB,
       author = {{Miville-Desch{\^e}nes}, Marc-Antoine and {Lagache}, Guilaine},
        title = "{IRIS: A New Generation of IRAS Maps}",
      journal = {\apjs},
         year = 2005,
        month = apr,
       volume = {157},
       number = {2},
        pages = {302-323},
          doi = {10.1086/427938},
archivePrefix = {arXiv},
       eprint = {astro-ph/0412216},
 primaryClass = {astro-ph},
       adsurl = {https://ui.adsabs.harvard.edu/abs/2005ApJS..157..302M}
}

@ARTICLE{Hershel_CIB_ref1,
       author = {{Berta}, S. and {Magnelli}, B. and {Nordon}, R. and {Lutz}, D. and {Wuyts}, S. and {Altieri}, B. and {Andreani}, P. and {Aussel}, H. and {Casta{\~n}eda}, H. and {Cepa}, J. and {Cimatti}, A. and {Daddi}, E. and {Elbaz}, D. and {F{\"o}rster Schreiber}, N.~M. and {Genzel}, R. and {Le Floc'h}, E. and {Maiolino}, R. and {P{\'e}rez-Fournon}, I. and {Poglitsch}, A. and {Popesso}, P. and {Pozzi}, F. and {Riguccini}, L. and {Rodighiero}, G. and {Sanchez-Portal}, M. and {Sturm}, E. and {Tacconi}, L.~J. and {Valtchanov}, I.},
        title = "{Building the cosmic infrared background brick by brick with Herschel/PEP}",
      journal = {\aap},
         year = 2011,
        month = aug,
       volume = {532},
          eid = {A49},
        pages = {A49},
          doi = {10.1051/0004-6361/201116844},
archivePrefix = {arXiv},
       eprint = {1106.3070},
 primaryClass = {astro-ph.CO},
       adsurl = {https://ui.adsabs.harvard.edu/abs/2011A&A...532A..49B}
}

@ARTICLE{Hershel_CIB_ref2,
       author = {{Viero}, M.~P. and {Wang}, L. and {Zemcov}, M. and {Addison}, G. and {Amblard}, A. and {Arumugam}, V. and {Aussel}, H. and {B{\'e}thermin}, M. and {Bock}, J. and {Boselli}, A. and {Buat}, V. and {Burgarella}, D. and {Casey}, C.~M. and {Clements}, D.~L. and {Conley}, A. and {Conversi}, L. and {Cooray}, A. and {De Zotti}, G. and {Dowell}, C.~D. and {Farrah}, D. and {Franceschini}, A. and {Glenn}, J. and {Griffin}, M. and {Hatziminaoglou}, E. and {Heinis}, S. and {Ibar}, E. and {Ivison}, R.~J. and {Lagache}, G. and {Levenson}, L. and {Marchetti}, L. and {Marsden}, G. and {Nguyen}, H.~T. and {O'Halloran}, B. and {Oliver}, S.~J. and {Omont}, A. and {Page}, M.~J. and {Papageorgiou}, A. and {Pearson}, C.~P. and {P{\'e}rez-Fournon}, I. and {Pohlen}, M. and {Rigopoulou}, D. and {Roseboom}, I.~G. and {Rowan-Robinson}, M. and {Schulz}, B. and {Scott}, D. and {Seymour}, N. and {Shupe}, D.~L. and {Smith}, A.~J. and {Symeonidis}, M. and {Vaccari}, M. and {Valtchanov}, I. and {Vieira}, J.~D. and {Wardlow}, J. and {Xu}, C.~K.},
        title = "{HerMES: Cosmic Infrared Background Anisotropies and the Clustering of Dusty Star-forming Galaxies}",
      journal = {\apj},
         year = 2013,
        month = jul,
       volume = {772},
       number = {1},
          eid = {77},
        pages = {77},
          doi = {10.1088/0004-637X/772/1/77},
archivePrefix = {arXiv},
       eprint = {1208.5049},
 primaryClass = {astro-ph.CO},
       adsurl = {https://ui.adsabs.harvard.edu/abs/2013ApJ...772...77V}
}

@ARTICLE{Pearly_CIB_results,
       author = {{Planck Collaboration} and {Ade}, P.~A.~R. and {Aghanim}, N. and {Arnaud}, M. and {Ashdown}, M. and {Aumont}, J. and {Baccigalupi}, C. and {Balbi}, A. and {Banday}, A.~J. and {Barreiro}, R.~B. and {Bartlett}, J.~G. and {Battaner}, E. and {Benabed}, K. and {Beno{\^\i}t}, A. and {Bernard}, J. -P. and {Bersanelli}, M. and {Bhatia}, R. and {Blagrave}, K. and {Bock}, J.~J. and {Bonaldi}, A. and {Bonavera}, L. and {Bond}, J.~R. and {Borrill}, J. and {Bouchet}, F.~R. and {Bucher}, M. and {Burigana}, C. and {Cabella}, P. and {Cardoso}, J. -F. and {Catalano}, A. and {Cay{\'o}n}, L. and {Challinor}, A. and {Chamballu}, A. and {Chiang}, L. -Y. and {Chiang}, C. and {Christensen}, P.~R. and {Clements}, D.~L. and {Colombi}, S. and {Couchot}, F. and {Coulais}, A. and {Crill}, B.~P. and {Cuttaia}, F. and {Danese}, L. and {Davies}, R.~D. and {Davis}, R.~J. and {de Bernardis}, P. and {de Gasperis}, G. and {de Rosa}, A. and {de Zotti}, G. and {Delabrouille}, J. and {Delouis}, J. -M. and {D{\'e}sert}, F. -X. and {Dole}, H. and {Donzelli}, S. and {Dor{\'e}}, O. and {D{\"o}rl}, U. and {Douspis}, M. and {Dupac}, X. and {Efstathiou}, G. and {En{\ss}lin}, T.~A. and {Eriksen}, H.~K. and {Finelli}, F. and {Forni}, O. and {Fosalba}, P. and {Frailis}, M. and {Franceschi}, E. and {Galeotta}, S. and {Ganga}, K. and {Giard}, M. and {Giardino}, G. and {Giraud-H{\'e}raud}, Y. and {Gonz{\'a}lez-Nuevo}, J. and {G{\'o}rski}, K.~M. and {Grain}, J. and {Gratton}, S. and {Gregorio}, A. and {Gruppuso}, A. and {Hansen}, F.~K. and {Harrison}, D. and {Helou}, G. and {Henrot-Versill{\'e}}, S. and {Herranz}, D. and {Hildebrandt}, S.~R. and {Hivon}, E. and {Hobson}, M. and {Holmes}, W.~A. and {Hovest}, W. and {Hoyland}, R.~J. and {Huffenberger}, K.~M. and {Jaffe}, A.~H. and {Jones}, W.~C. and {Juvela}, M. and {Keih{\"a}nen}, E. and {Keskitalo}, R. and {Kisner}, T.~S. and {Kneissl}, R. and {Knox}, L. and {Kurki-Suonio}, H. and {Lagache}, G. and {Lamarre}, J. -M. and {Lasenby}, A. and {Laureijs}, R.~J. and {Lawrence}, C.~R. and {Leach}, S. and {Leonardi}, R. and {Leroy}, C. and {Lilje}, P.~B. and {Linden-V{\o}rnle}, M. and {Lockman}, F.~J. and {L{\'o}pez-Caniego}, M. and {Lubin}, P.~M. and {Mac{\'\i}as-P{\'e}rez}, J.~F. and {MacTavish}, C.~J. and {Maffei}, B. and {Maino}, D. and {Mandolesi}, N. and {Mann}, R. and {Maris}, M. and {Martin}, P. and {Mart{\'\i}nez-Gonz{\'a}lez}, E. and {Masi}, S. and {Matarrese}, S. and {Matthai}, F. and {Mazzotta}, P. and {Melchiorri}, A. and {Mendes}, L. and {Mennella}, A. and {Mitra}, S. and {Miville-Desch{\^e}nes}, M. -A. and {Moneti}, A. and {Montier}, L. and {Morgante}, G. and {Mortlock}, D. and {Munshi}, D. and {Murphy}, A. and {Naselsky}, P. and {Natoli}, P. and {Netterfield}, C.~B. and {N{\o}rgaard-Nielsen}, H.~U. and {Novikov}, D. and {Novikov}, I. and {O'Dwyer}, I.~J. and {Oliver}, S. and {Osborne}, S. and {Pajot}, F. and {Pasian}, F. and {Patanchon}, G. and {Perdereau}, O. and {Perotto}, L. and {Perrotta}, F. and {Piacentini}, F. and {Piat}, M. and {Pinheiro Gon{\c{c}}alves}, D. and {Plaszczynski}, S. and {Pointecouteau}, E. and {Polenta}, G. and {Ponthieu}, N. and {Poutanen}, T. and {Pr{\'e}zeau}, G. and {Prunet}, S. and {Puget}, J. -L. and {Rachen}, J.~P. and {Reach}, W.~T. and {Reinecke}, M. and {Remazeilles}, M. and {Renault}, C. and {Ricciardi}, S. and {Riller}, T. and {Ristorcelli}, I. and {Rocha}, G. and {Rosset}, C. and {Rowan-Robinson}, M. and {Rubi{\~n}o-Mart{\'\i}n}, J.~A. and {Rusholme}, B. and {Sandri}, M. and {Santos}, D. and {Savini}, G. and {Scott}, D. and {Seiffert}, M.~D. and {Shellard}, P. and {Smoot}, G.~F. and {Starck}, J. -L. and {Stivoli}, F. and {Stolyarov}, V. and {Stompor}, R. and {Sudiwala}, R. and {Sunyaev}, R. and {Sygnet}, J. -F. and {Tauber}, J.~A. and {Terenzi}, L. and {Toffolatti}, L. and {Tomasi}, M. and {Torre}, J. -P. and {Tristram}, M. and {Tuovinen}, J. and {Umana}, G. and {Valenziano}, L. and {Vielva}, P. and {Villa}, F. and {Vittorio}, N. and {Wade}, L.~A.},
        title = "{Planck early results. XVIII. The power spectrum of cosmic infrared background anisotropies}",
      journal = {\aap},
         year = 2011,
        month = dec,
       volume = {536},
          eid = {A18},
        pages = {A18},
          doi = {10.1051/0004-6361/201116461},
archivePrefix = {arXiv},
       eprint = {1101.2028},
 primaryClass = {astro-ph.CO},
       adsurl = {https://ui.adsabs.harvard.edu/abs/2011A&A...536A..18P}
}

@ARTICLE{CIB_deproj_ref1,
       author = {{Hill}, J. Colin and {Spergel}, David N.},
        title = "{Detection of thermal SZ-CMB lensing cross-correlation in Planck nominal mission data}",
      journal = {\jcap},
         year = 2014,
        month = feb,
       volume = {2014},
       number = {2},
          eid = {030},
        pages = {030},
          doi = {10.1088/1475-7516/2014/02/030},
archivePrefix = {arXiv},
       eprint = {1312.4525},
 primaryClass = {astro-ph.CO},
       adsurl = {https://ui.adsabs.harvard.edu/abs/2014JCAP...02..030H}
}

@ARTICLE{CIB_deproj_ref2,
       author = {{McCarthy}, Fiona and {Hill}, J. Colin},
        title = "{Component-separated, CIB-cleaned thermal Sunyaev-Zel'dovich maps from Planck PR4 data with a flexible public needlet ILC pipeline}",
      journal = {\prd},
         year = 2024,
        month = jan,
       volume = {109},
       number = {2},
          eid = {023528},
        pages = {023528},
          doi = {10.1103/PhysRevD.109.023528},
archivePrefix = {arXiv},
       eprint = {2307.01043},
 primaryClass = {astro-ph.CO},
       adsurl = {https://ui.adsabs.harvard.edu/abs/2024PhRvD.109b3528M}
}

@ARTICLE{CIB_deproj_ref3,
       author = {{Efstathiou}, George and {McCarthy}, Fiona},
        title = "{The power spectrum of the thermal Sunyaev{\textendash}Zeldovich effect}",
      journal = {\mnras},
         year = 2025,
        month = jun,
       volume = {540},
       number = {1},
        pages = {1055-1068},
          doi = {10.1093/mnras/staf709},
archivePrefix = {arXiv},
       eprint = {2502.10232},
 primaryClass = {astro-ph.CO},
       adsurl = {https://ui.adsabs.harvard.edu/abs/2025MNRAS.540.1055E}
}

@ARTICLE{ACT_CIB_ref1,
       author = {{Dunkley}, J. and {Calabrese}, E. and {Sievers}, J. and {Addison}, G.~E. and {Battaglia}, N. and {Battistelli}, E.~S. and {Bond}, J.~R. and {Das}, S. and {Devlin}, M.~J. and {D{\"u}nner}, R. and {Fowler}, J.~W. and {Gralla}, M. and {Hajian}, A. and {Halpern}, M. and {Hasselfield}, M. and {Hincks}, A.~D. and {Hlozek}, R. and {Hughes}, J.~P. and {Irwin}, K.~D. and {Kosowsky}, A. and {Louis}, T. and {Marriage}, T.~A. and {Marsden}, D. and {Menanteau}, F. and {Moodley}, K. and {Niemack}, M. and {Nolta}, M.~R. and {Page}, L.~A. and {Partridge}, B. and {Sehgal}, N. and {Spergel}, D.~N. and {Staggs}, S.~T. and {Switzer}, E.~R. and {Trac}, H. and {Wollack}, E.},
        title = "{The Atacama Cosmology Telescope: likelihood for small-scale CMB data}",
      journal = {\jcap},
         year = 2013,
        month = jul,
       volume = {2013},
       number = {7},
          eid = {025},
        pages = {025},
          doi = {10.1088/1475-7516/2013/07/025},
archivePrefix = {arXiv},
       eprint = {1301.0776},
 primaryClass = {astro-ph.CO},
       adsurl = {https://ui.adsabs.harvard.edu/abs/2013JCAP...07..025D}
}

@ARTICLE{SPT_CIB_ref,
       author = {{George}, E.~M. and {Reichardt}, C.~L. and {Aird}, K.~A. and {Benson}, B.~A. and {Bleem}, L.~E. and {Carlstrom}, J.~E. and {Chang}, C.~L. and {Cho}, H. -M. and {Crawford}, T.~M. and {Crites}, A.~T. and {de Haan}, T. and {Dobbs}, M.~A. and {Dudley}, J. and {Halverson}, N.~W. and {Harrington}, N.~L. and {Holder}, G.~P. and {Holzapfel}, W.~L. and {Hou}, Z. and {Hrubes}, J.~D. and {Keisler}, R. and {Knox}, L. and {Lee}, A.~T. and {Leitch}, E.~M. and {Lueker}, M. and {Luong-Van}, D. and {McMahon}, J.~J. and {Mehl}, J. and {Meyer}, S.~S. and {Millea}, M. and {Mocanu}, L.~M. and {Mohr}, J.~J. and {Montroy}, T.~E. and {Padin}, S. and {Plagge}, T. and {Pryke}, C. and {Ruhl}, J.~E. and {Schaffer}, K.~K. and {Shaw}, L. and {Shirokoff}, E. and {Spieler}, H.~G. and {Staniszewski}, Z. and {Stark}, A.~A. and {Story}, K.~T. and {van Engelen}, A. and {Vanderlinde}, K. and {Vieira}, J.~D. and {Williamson}, R. and {Zahn}, O.},
        title = "{A Measurement of Secondary Cosmic Microwave Background Anisotropies from the 2500 Square-degree SPT-SZ Survey}",
      journal = {\apj},
         year = 2015,
        month = feb,
       volume = {799},
       number = {2},
          eid = {177},
        pages = {177},
          doi = {10.1088/0004-637X/799/2/177},
archivePrefix = {arXiv},
       eprint = {1408.3161},
 primaryClass = {astro-ph.CO},
       adsurl = {https://ui.adsabs.harvard.edu/abs/2015ApJ...799..177G}
}

@ARTICLE{ACT_DR6_CIB_ref,
       author = {{Louis}, Thibaut and {La Posta}, Adrien and {Atkins}, Zachary and {Jense}, Hidde T. and {Abril-Cabezas}, Irene and {Addison}, Graeme E. and {Ade}, Peter A.~R. and {Aiola}, Simone and {Alford}, Tommy and {Alonso}, David and {Amiri}, Mandana and {An}, Rui and {Austermann}, Jason E. and {Barbavara}, Eleonora and {Battaglia}, Nicholas and {Battistelli}, Elia Stefano and {Beall}, James A. and {Bean}, Rachel and {Beheshti}, Ali and {Beringue}, Benjamin and {Bhandarkar}, Tanay and {Biermann}, Emily and {Bolliet}, Boris and {Bond}, J Richard and {Calabrese}, Erminia and {Capalbo}, Valentina and {Carrero}, Felipe and {Chen}, Shi-Fan and {Chesmore}, Grace and {Cho}, Hsiao-mei and {Choi}, Steve K. and {Clark}, Susan E. and {Cothard}, Nicholas F. and {Coughlin}, Kevin and {Coulton}, William and {Crichton}, Devin and {Crowley}, Kevin T. and {Darwish}, Omar and {Devlin}, Mark J. and {Dicker}, Simon and {Duell}, Cody J. and {Duff}, Shannon M. and {Duivenvoorden}, Adriaan J. and {Dunkley}, Jo and {Dunner}, Rolando and {Embil Villagra}, Carmen and {Fankhanel}, Max and {Farren}, Gerrit S. and {Ferraro}, Simone and {Foster}, Allen and {Freundt}, Rodrigo and {Fuzia}, Brittany and {Gallardo}, Patricio A. and {Garrido}, Xavier and {Gerbino}, Martina and {Giardiello}, Serena and {Gill}, Ajay and {Givans}, Jahmour and {Gluscevic}, Vera and {Goldstein}, Samuel and {Golec}, Joseph E. and {Gong}, Yulin and {Guan}, Yilun and {Halpern}, Mark and {Harrison}, Ian and {Hasselfield}, Matthew and {Healy}, Erin and {Henderson}, Shawn and {Hensley}, Brandon and {Herv{\'\i}as-Caimapo}, Carlos and {Hill}, J. Colin and {Hilton}, Gene C. and {Hilton}, Matt and {Hincks}, Adam D. and {Hlo{\v{z}}ek}, Ren{\'e}e and {Ho}, Shuay-Pwu Patty and {Hood}, John and {Hornecker}, Erika and {Huber}, Zachary B. and {Hubmayr}, Johannes and {Huffenberger}, Kevin M. and {Hughes}, John P. and {Ikape}, Margaret and {Irwin}, Kent and {Isopi}, Giovanni and {Joshi}, Neha and {Keller}, Ben and {Kim}, Joshua and {Knowles}, Kenda and {Koopman}, Brian J. and {Kosowsky}, Arthur and {Kramer}, Darby and {Kusiak}, Aleksandra and {Lague}, Alex and {Lakey}, Victoria and {Lee}, Eunseong and {Li}, Yaqiong and {Li}, Zack and {Limon}, Michele and {Lokken}, Martine and {Lungu}, Marius and {MacCrann}, Niall and {MacInnis}, Amanda and {Madhavacheril}, Mathew S. and {Maldonado}, Diego and {Maldonado}, Felipe and {Mallaby-Kay}, Maya and {Marques}, Gabriela A. and {van Marrewijk}, Joshiwa and {McCarthy}, Fiona and {McMahon}, Jeff and {Mehta}, Yogesh and {Menanteau}, Felipe and {Moodley}, Kavilan and {Morris}, Thomas W. and {Mroczkowski}, Tony and {Naess}, Sigurd and {Namikawa}, Toshiya and {Nati}, Federico and {Nerval}, Simran K. and {Newburgh}, Laura and {Nicola}, Andrina and {Niemack}, Michael D. and {Nolta}, Michael R. and {Orlowski-Scherer}, John and {Pagano}, Luca and {Page}, Lyman A. and {Pandey}, Shivam and {Partridge}, Bruce and {Perez Sarmiento}, Karen and {Prince}, Heather and {Puddu}, Roberto and {Qu}, Frank J. and {Ragavan}, Damien C. and {Ried Guachalla}, Bernardita and {Rogers}, Keir K. and {Rojas}, Felipe and {Sakuma}, Tai and {Schaan}, Emmanuel and {Schmitt}, Benjamin L. and {Sehgal}, Neelima and {Shaikh}, Shabbir and {Sherwin}, Blake D. and {Sierra}, Carlos and {Sievers}, Jon and {Sif{\'o}n}, Crist{\'o}bal and {Simon}, Sara and {Sonka}, Rita and {Spergel}, David N. and {Staggs}, Suzanne T. and {Storer}, Emilie and {Surrao}, Kristen and {Switzer}, Eric R. and {Tampier}, Niklas and {Thornton}, Robert and {Trac}, Hy and {Tucker}, Carole and {Ullom}, Joel and {Vale}, Leila R. and {Van Engelen}, Alexander and {Van Lanen}, Jeff and {Vargas}, Cristian and {Vavagiakis}, Eve M. and {Wagoner}, Kasey and {Wang}, Yuhan and {Wenzl}, Lukas and {Wollack}, Edward J. and {Zheng}, Kaiwen},
        title = "{The Atacama Cosmology Telescope: DR6 Power Spectra, Likelihoods and $Λ$CDM Parameters}",
      journal = {arXiv e-prints},
         year = 2025,
        month = mar,
          eid = {arXiv:2503.14452},
        pages = {arXiv:2503.14452},
          doi = {10.48550/arXiv.2503.14452},
archivePrefix = {arXiv},
       eprint = {2503.14452},
 primaryClass = {astro-ph.CO},
       adsurl = {https://ui.adsabs.harvard.edu/abs/2025arXiv250314452L}
}

@ARTICLE{SPTpol_CIB_ref,
       author = {{Reichardt}, C.~L. and {Patil}, S. and {Ade}, P.~A.~R. and {Anderson}, A.~J. and {Austermann}, J.~E. and {Avva}, J.~S. and {Baxter}, E. and {Beall}, J.~A. and {Bender}, A.~N. and {Benson}, B.~A. and {Bianchini}, F. and {Bleem}, L.~E. and {Carlstrom}, J.~E. and {Chang}, C.~L. and {Chaubal}, P. and {Chiang}, H.~C. and {Chou}, T.~L. and {Citron}, R. and {Moran}, C. Corbett and {Crawford}, T.~M. and {Crites}, A.~T. and {de Haan}, T. and {Dobbs}, M.~A. and {Everett}, W. and {Gallicchio}, J. and {George}, E.~M. and {Gilbert}, A. and {Gupta}, N. and {Halverson}, N.~W. and {Harrington}, N. and {Henning}, J.~W. and {Hilton}, G.~C. and {Holder}, G.~P. and {Holzapfel}, W.~L. and {Hrubes}, J.~D. and {Huang}, N. and {Hubmayr}, J. and {Irwin}, K.~D. and {Knox}, L. and {Lee}, A.~T. and {Li}, D. and {Lowitz}, A. and {Luong-Van}, D. and {McMahon}, J.~J. and {Mehl}, J. and {Meyer}, S.~S. and {Millea}, M. and {Mocanu}, L.~M. and {Mohr}, J.~J. and {Montgomery}, J. and {Nadolski}, A. and {Natoli}, T. and {Nibarger}, J.~P. and {Noble}, G. and {Novosad}, V. and {Omori}, Y. and {Padin}, S. and {Pryke}, C. and {Ruhl}, J.~E. and {Saliwanchik}, B.~R. and {Sayre}, J.~T. and {Schaffer}, K.~K. and {Shirokoff}, E. and {Sievers}, C. and {Smecher}, G. and {Spieler}, H.~G. and {Staniszewski}, Z. and {Stark}, A.~A. and {Tucker}, C. and {Vanderlinde}, K. and {Veach}, T. and {Vieira}, J.~D. and {Wang}, G. and {Whitehorn}, N. and {Williamson}, R. and {Wu}, W.~L.~K. and {Yefremenko}, V.},
        title = "{An Improved Measurement of the Secondary Cosmic Microwave Background Anisotropies from the SPT-SZ + SPTpol Surveys}",
      journal = {\apj},
         year = 2021,
        month = feb,
       volume = {908},
       number = {2},
          eid = {199},
        pages = {199},
          doi = {10.3847/1538-4357/abd407},
archivePrefix = {arXiv},
       eprint = {2002.06197},
 primaryClass = {astro-ph.CO},
       adsurl = {https://ui.adsabs.harvard.edu/abs/2021ApJ...908..199R}
}

@ARTICLE{viero_CIB_galaxy_pop,
       author = {{Viero}, M.~P. and {Moncelsi}, L. and {Quadri}, R.~F. and {Arumugam}, V. and {Assef}, R.~J. and {B{\'e}thermin}, M. and {Bock}, J. and {Bridge}, C. and {Casey}, C.~M. and {Conley}, A. and {Cooray}, A. and {Farrah}, D. and {Glenn}, J. and {Heinis}, S. and {Ibar}, E. and {Ikarashi}, S. and {Ivison}, R.~J. and {Kohno}, K. and {Marsden}, G. and {Oliver}, S.~J. and {Roseboom}, I.~G. and {Schulz}, B. and {Scott}, D. and {Serra}, P. and {Vaccari}, M. and {Vieira}, J.~D. and {Wang}, L. and {Wardlow}, J. and {Wilson}, G.~W. and {Yun}, M.~S. and {Zemcov}, M.},
        title = "{HerMES: The Contribution to the Cosmic Infrared Background from Galaxies Selected by Mass and Redshift}",
      journal = {\apj},
         year = 2013,
        month = dec,
       volume = {779},
       number = {1},
          eid = {32},
        pages = {32},
          doi = {10.1088/0004-637X/779/1/32},
archivePrefix = {arXiv},
       eprint = {1304.0446},
 primaryClass = {astro-ph.CO},
       adsurl = {https://ui.adsabs.harvard.edu/abs/2013ApJ...779...32V}
}

@ARTICLE{CIB_tomo_ref1,
       author = {{Chiang}, Yi-Kuan and {Makiya}, Ryu and {M{\'e}nard}, Brice},
        title = "{Cosmic Infrared Background Tomography and a Census of Cosmic Dust and Star Formation}",
      journal = {arXiv e-prints},
         year = 2025,
        month = apr,
          eid = {arXiv:2504.05384},
        pages = {arXiv:2504.05384},
          doi = {10.48550/arXiv.2504.05384},
archivePrefix = {arXiv},
       eprint = {2504.05384},
 primaryClass = {astro-ph.CO},
       adsurl = {https://ui.adsabs.harvard.edu/abs/2025arXiv250405384C}
}

@ARTICLE{CIB_tomo_ref2,
       author = {{Yan}, Ziang and {van Waerbeke}, Ludovic and {Wright}, Angus H. and {Bilicki}, Maciej and {Gu}, Shiming and {Hildebrandt}, Hendrik and {Maniyar}, Abhishek S. and {Tr{\"o}ster}, Tilman},
        title = "{Cosmic star formation history with tomographic cosmic infrared background-galaxy cross-correlation}",
      journal = {\aap},
         year = 2022,
        month = sep,
       volume = {665},
          eid = {A52},
        pages = {A52},
          doi = {10.1051/0004-6361/202243710},
archivePrefix = {arXiv},
       eprint = {2204.01649},
 primaryClass = {astro-ph.GA},
       adsurl = {https://ui.adsabs.harvard.edu/abs/2022A&A...665A..52Y}
}

@ARTICLE{SFR_CIB_M18,
       author = {{Maniyar}, A.~S. and {B{\'e}thermin}, M. and {Lagache}, G.},
        title = "{Star formation history from the cosmic infrared background anisotropies}",
      journal = {\aap},
         year = 2018,
        month = jun,
       volume = {614},
          eid = {A39},
        pages = {A39},
          doi = {10.1051/0004-6361/201732499},
archivePrefix = {arXiv},
       eprint = {1801.10146},
 primaryClass = {astro-ph.CO},
       adsurl = {https://ui.adsabs.harvard.edu/abs/2018A&A...614A..39M}
}

@ARTICLE{Magdis_CIB_SED,
       author = {{Magdis}, Georgios E. and {Daddi}, E. and {B{\'e}thermin}, M. and {Sargent}, M. and {Elbaz}, D. and {Pannella}, M. and {Dickinson}, M. and {Dannerbauer}, H. and {da Cunha}, E. and {Walter}, F. and {Rigopoulou}, D. and {Charmandaris}, V. and {Hwang}, H.~S. and {Kartaltepe}, J.},
        title = "{The Evolving Interstellar Medium of Star-forming Galaxies since z = 2 as Probed by Their Infrared Spectral Energy Distributions}",
      journal = {\apj},
         year = 2012,
        month = nov,
       volume = {760},
       number = {1},
          eid = {6},
        pages = {6},
          doi = {10.1088/0004-637X/760/1/6},
archivePrefix = {arXiv},
       eprint = {1210.1035},
 primaryClass = {astro-ph.CO},
       adsurl = {https://ui.adsabs.harvard.edu/abs/2012ApJ...760....6M}
}

@ARTICLE{Bethermin_CIB_SED,
       author = {{B{\'e}thermin}, Matthieu and {Daddi}, Emanuele and {Magdis}, Georgios and {Sargent}, Mark T. and {Hezaveh}, Yashar and {Elbaz}, David and {Le Borgne}, Damien and {Mullaney}, James and {Pannella}, Maurilio and {Buat}, V{\'e}ronique and {Charmandaris}, Vassilis and {Lagache}, Guilaine and {Scott}, Douglas},
        title = "{A Unified Empirical Model for Infrared Galaxy Counts Based on the Observed Physical Evolution of Distant Galaxies}",
      journal = {\apjl},
         year = 2012,
        month = oct,
       volume = {757},
       number = {2},
          eid = {L23},
        pages = {L23},
          doi = {10.1088/2041-8205/757/2/L23},
archivePrefix = {arXiv},
       eprint = {1208.6512},
 primaryClass = {astro-ph.CO},
       adsurl = {https://ui.adsabs.harvard.edu/abs/2012ApJ...757L..23B}
}

@ARTICLE{Chabrier_2003_IMF,
       author = {{Chabrier}, Gilles},
        title = "{Galactic Stellar and Substellar Initial Mass Function}",
      journal = {\pasp},
         year = 2003,
        month = jul,
       volume = {115},
       number = {809},
        pages = {763-795},
          doi = {10.1086/376392},
archivePrefix = {arXiv},
       eprint = {astro-ph/0304382},
 primaryClass = {astro-ph},
       adsurl = {https://ui.adsabs.harvard.edu/abs/2003PASP..115..763C}
}

@ARTICLE{CAMB_paper,
       author = {{Lewis}, Antony and {Challinor}, Anthony and {Lasenby}, Anthony},
        title = "{Efficient Computation of Cosmic Microwave Background Anisotropies in Closed Friedmann-Robertson-Walker Models}",
      journal = {\apj},
         year = 2000,
        month = aug,
       volume = {538},
       number = {2},
        pages = {473-476},
          doi = {10.1086/309179},
archivePrefix = {arXiv},
       eprint = {astro-ph/9911177},
 primaryClass = {astro-ph},
       adsurl = {https://ui.adsabs.harvard.edu/abs/2000ApJ...538..473L}
}

@ARTICLE{Ian_kSZ_feedback_FLAMINGO,
       author = {{McCarthy}, Ian G. and {Amon}, Alexandra and {Schaye}, Joop and {Schaan}, Emmanuel and {Angulo}, Raul E. and {Salcido}, Jaime and {Schaller}, Matthieu and {Bigwood}, Leah and {Elbers}, Willem and {Kugel}, Roi and {Helly}, John C. and {Forouhar Moreno}, Victor J. and {Frenk}, Carlos S. and {McGibbon}, Robert J. and {Ondaro-Mallea}, Lurdes and {van Daalen}, Marcel P.},
        title = "{FLAMINGO: combining kinetic SZ effect and galaxy{\textendash}galaxy lensing measurements to gauge the impact of feedback on large-scale structure}",
      journal = {\mnras},
         year = 2025,
        month = jun,
       volume = {540},
       number = {1},
        pages = {143-163},
          doi = {10.1093/mnras/staf731},
archivePrefix = {arXiv},
       eprint = {2410.19905},
 primaryClass = {astro-ph.CO},
       adsurl = {https://ui.adsabs.harvard.edu/abs/2025MNRAS.540..143M}
}

@ARTICLE{P16_tSZ_map,
       author = {{Planck Collaboration} and {Aghanim}, N. and {Arnaud}, M. and {Ashdown}, M. and {Aumont}, J. and {Baccigalupi}, C. and {Banday}, A.~J. and {Barreiro}, R.~B. and {Bartlett}, J.~G. and {Bartolo}, N. and {Battaner}, E. and {Battye}, R. and {Benabed}, K. and {Beno{\^\i}t}, A. and {Benoit-L{\'e}vy}, A. and {Bernard}, J. -P. and {Bersanelli}, M. and {Bielewicz}, P. and {Bock}, J.~J. and {Bonaldi}, A. and {Bonavera}, L. and {Bond}, J.~R. and {Borrill}, J. and {Bouchet}, F.~R. and {Burigana}, C. and {Butler}, R.~C. and {Calabrese}, E. and {Cardoso}, J. -F. and {Catalano}, A. and {Challinor}, A. and {Chiang}, H.~C. and {Christensen}, P.~R. and {Churazov}, E. and {Clements}, D.~L. and {Colombo}, L.~P.~L. and {Combet}, C. and {Comis}, B. and {Coulais}, A. and {Crill}, B.~P. and {Curto}, A. and {Cuttaia}, F. and {Danese}, L. and {Davies}, R.~D. and {Davis}, R.~J. and {de Bernardis}, P. and {de Rosa}, A. and {de Zotti}, G. and {Delabrouille}, J. and {D{\'e}sert}, F. -X. and {Dickinson}, C. and {Diego}, J.~M. and {Dolag}, K. and {Dole}, H. and {Donzelli}, S. and {Dor{\'e}}, O. and {Douspis}, M. and {Ducout}, A. and {Dupac}, X. and {Efstathiou}, G. and {Elsner}, F. and {En{\ss}lin}, T.~A. and {Eriksen}, H.~K. and {Fergusson}, J. and {Finelli}, F. and {Forni}, O. and {Frailis}, M. and {Fraisse}, A.~A. and {Franceschi}, E. and {Frejsel}, A. and {Galeotta}, S. and {Galli}, S. and {Ganga}, K. and {G{\'e}nova-Santos}, R.~T. and {Giard}, M. and {Gonz{\'a}lez-Nuevo}, J. and {G{\'o}rski}, K.~M. and {Gregorio}, A. and {Gruppuso}, A. and {Gudmundsson}, J.~E. and {Hansen}, F.~K. and {Harrison}, D.~L. and {Henrot-Versill{\'e}}, S. and {Hern{\'a}ndez-Monteagudo}, C. and {Herranz}, D. and {Hildebrandt}, S.~R. and {Hivon}, E. and {Holmes}, W.~A. and {Hornstrup}, A. and {Huffenberger}, K.~M. and {Hurier}, G. and {Jaffe}, A.~H. and {Jones}, W.~C. and {Juvela}, M. and {Keih{\"a}nen}, E. and {Keskitalo}, R. and {Kneissl}, R. and {Knoche}, J. and {Kunz}, M. and {Kurki-Suonio}, H. and {Lacasa}, F. and {Lagache}, G. and {L{\"a}hteenm{\"a}ki}, A. and {Lamarre}, J. -M. and {Lasenby}, A. and {Lattanzi}, M. and {Leonardi}, R. and {Lesgourgues}, J. and {Levrier}, F. and {Liguori}, M. and {Lilje}, P.~B. and {Linden-V{\o}rnle}, M. and {L{\'o}pez-Caniego}, M. and {Mac{\'\i}as-P{\'e}rez}, J.~F. and {Maffei}, B. and {Maggio}, G. and {Maino}, D. and {Mandolesi}, N. and {Mangilli}, A. and {Maris}, M. and {Martin}, P.~G. and {Mart{\'\i}nez-Gonz{\'a}lez}, E. and {Masi}, S. and {Matarrese}, S. and {Melchiorri}, A. and {Melin}, J. -B. and {Migliaccio}, M. and {Miville-Desch{\^e}nes}, M. -A. and {Moneti}, A. and {Montier}, L. and {Morgante}, G. and {Mortlock}, D. and {Munshi}, D. and {Murphy}, J.~A. and {Naselsky}, P. and {Nati}, F. and {Natoli}, P. and {Noviello}, F. and {Novikov}, D. and {Novikov}, I. and {Paci}, F. and {Pagano}, L. and {Pajot}, F. and {Paoletti}, D. and {Pasian}, F. and {Patanchon}, G. and {Perdereau}, O. and {Perotto}, L. and {Pettorino}, V. and {Piacentini}, F. and {Piat}, M. and {Pierpaoli}, E. and {Pietrobon}, D. and {Plaszczynski}, S. and {Pointecouteau}, E. and {Polenta}, G. and {Ponthieu}, N. and {Pratt}, G.~W. and {Prunet}, S. and {Puget}, J. -L. and {Rachen}, J.~P. and {Reinecke}, M. and {Remazeilles}, M. and {Renault}, C. and {Renzi}, A. and {Ristorcelli}, I. and {Rocha}, G. and {Rossetti}, M. and {Roudier}, G. and {Rubi{\~n}o-Mart{\'\i}n}, J.~A. and {Rusholme}, B. and {Sandri}, M. and {Santos}, D. and {Sauv{\'e}}, A. and {Savelainen}, M. and {Savini}, G. and {Scott}, D. and {Spencer}, L.~D. and {Stolyarov}, V. and {Stompor}, R. and {Sunyaev}, R. and {Sutton}, D. and {Suur-Uski}, A. -S. and {Sygnet}, J. -F. and {Tauber}, J.~A. and {Terenzi}, L. and {Toffolatti}, L. and {Tomasi}, M. and {Tramonte}, D. and {Tristram}, M. and {Tucci}, M. and {Tuovinen}, J. and {Valenziano}, L. and {Valiviita}, J. and {Van Tent}, B. and {Vielva}, P. and {Villa}, F. and {Wade}, L.~A. and {Wandelt}, B.~D. and {Wehus}, I.~K. and {Yvon}, D.},
        title = "{Planck 2015 results. XXII. A map of the thermal Sunyaev-Zeldovich effect}",
      journal = {\aap},
         year = 2016,
        month = sep,
       volume = {594},
          eid = {A22},
        pages = {A22},
          doi = {10.1051/0004-6361/201525826},
archivePrefix = {arXiv},
       eprint = {1502.01596},
 primaryClass = {astro-ph.CO},
       adsurl = {https://ui.adsabs.harvard.edu/abs/2016A&A...594A..22P}
}

@ARTICLE{Fiona_2024_CMBlensing_tSZ,
       author = {{McCarthy}, Fiona and {Hill}, J. Colin},
        title = "{Cross-correlation of the thermal Sunyaev-Zel'dovich and CMB lensing signals in Planck PR4 data with robust CIB decontamination}",
      journal = {\prd},
         year = 2024,
        month = jan,
       volume = {109},
       number = {2},
          eid = {023529},
        pages = {023529},
          doi = {10.1103/PhysRevD.109.023529},
archivePrefix = {arXiv},
       eprint = {2308.16260},
 primaryClass = {astro-ph.CO},
       adsurl = {https://ui.adsabs.harvard.edu/abs/2024PhRvD.109b3529M}
}

@ARTICLE{tSZ_tomo_Chiang,
       author = {{Chiang}, Yi-Kuan and {Makiya}, Ryu and {M{\'e}nard}, Brice and {Komatsu}, Eiichiro},
        title = "{The Cosmic Thermal History Probed by Sunyaev-Zeldovich Effect Tomography}",
      journal = {\apj},
         year = 2020,
        month = oct,
       volume = {902},
       number = {1},
          eid = {56},
        pages = {56},
          doi = {10.3847/1538-4357/abb403},
archivePrefix = {arXiv},
       eprint = {2006.14650},
 primaryClass = {astro-ph.CO},
       adsurl = {https://ui.adsabs.harvard.edu/abs/2020ApJ...902...56C}
}

@ARTICLE{Hojjati_tSZ_weak_lensing,
       author = {{Hojjati}, Alireza and {Tr{\"o}ster}, Tilman and {Harnois-D{\'e}raps}, Joachim and {McCarthy}, Ian G. and {van Waerbeke}, Ludovic and {Choi}, Ami and {Erben}, Thomas and {Heymans}, Catherine and {Hildebrandt}, Hendrik and {Hinshaw}, Gary and {Ma}, Yin-Zhe and {Miller}, Lance and {Viola}, Massimo and {Tanimura}, Hideki},
        title = "{Cross-correlating Planck tSZ with RCSLenS weak lensing: implications for cosmology and AGN feedback}",
      journal = {\mnras},
         year = 2017,
        month = oct,
       volume = {471},
       number = {2},
        pages = {1565-1580},
          doi = {10.1093/mnras/stx1659},
archivePrefix = {arXiv},
       eprint = {1608.07581},
 primaryClass = {astro-ph.CO},
       adsurl = {https://ui.adsabs.harvard.edu/abs/2017MNRAS.471.1565H}
}

@ARTICLE{COMMANDER_ref,
       author = {{Eriksen}, H.~K. and {Dickinson}, C. and {Lawrence}, C.~R. and {Baccigalupi}, C. and {Banday}, A.~J. and {G{\'o}rski}, K.~M. and {Hansen}, F.~K. and {Lilje}, P.~B. and {Pierpaoli}, E. and {Seiffert}, M.~D. and {Smith}, K.~M. and {Vanderlinde}, K.},
        title = "{Cosmic Microwave Background Component Separation by Parameter Estimation}",
      journal = {\apj},
         year = 2006,
        month = apr,
       volume = {641},
       number = {2},
        pages = {665-682},
          doi = {10.1086/500499},
archivePrefix = {arXiv},
       eprint = {astro-ph/0508268},
 primaryClass = {astro-ph},
       adsurl = {https://ui.adsabs.harvard.edu/abs/2006ApJ...641..665E}
}

@ARTICLE{CMB_separation_pipeline_paper,
       author = {{Planck Collaboration} and {Akrami}, Y. and {Ashdown}, M. and {Aumont}, J. and {Baccigalupi}, C. and {Ballardini}, M. and {Banday}, A.~J. and {Barreiro}, R.~B. and {Bartolo}, N. and {Basak}, S. and {Benabed}, K. and {Bersanelli}, M. and {Bielewicz}, P. and {Bond}, J.~R. and {Borrill}, J. and {Bouchet}, F.~R. and {Boulanger}, F. and {Bucher}, M. and {Burigana}, C. and {Calabrese}, E. and {Cardoso}, J. -F. and {Carron}, J. and {Casaponsa}, B. and {Challinor}, A. and {Colombo}, L.~P.~L. and {Combet}, C. and {Crill}, B.~P. and {Cuttaia}, F. and {de Bernardis}, P. and {de Rosa}, A. and {de Zotti}, G. and {Delabrouille}, J. and {Delouis}, J. -M. and {Di Valentino}, E. and {Dickinson}, C. and {Diego}, J.~M. and {Donzelli}, S. and {Dor{\'e}}, O. and {Ducout}, A. and {Dupac}, X. and {Efstathiou}, G. and {Elsner}, F. and {En{\ss}lin}, T.~A. and {Eriksen}, H.~K. and {Falgarone}, E. and {Fernandez-Cobos}, R. and {Finelli}, F. and {Forastieri}, F. and {Frailis}, M. and {Fraisse}, A.~A. and {Franceschi}, E. and {Frolov}, A. and {Galeotta}, S. and {Galli}, S. and {Ganga}, K. and {G{\'e}nova-Santos}, R.~T. and {Gerbino}, M. and {Ghosh}, T. and {Gonz{\'a}lez-Nuevo}, J. and {G{\'o}rski}, K.~M. and {Gratton}, S. and {Gruppuso}, A. and {Gudmundsson}, J.~E. and {Handley}, W. and {Hansen}, F.~K. and {Helou}, G. and {Herranz}, D. and {Hildebrandt}, S.~R. and {Huang}, Z. and {Jaffe}, A.~H. and {Karakci}, A. and {Keih{\"a}nen}, E. and {Keskitalo}, R. and {Kiiveri}, K. and {Kim}, J. and {Kisner}, T.~S. and {Krachmalnicoff}, N. and {Kunz}, M. and {Kurki-Suonio}, H. and {Lagache}, G. and {Lamarre}, J. -M. and {Lasenby}, A. and {Lattanzi}, M. and {Lawrence}, C.~R. and {Le Jeune}, M. and {Levrier}, F. and {Liguori}, M. and {Lilje}, P.~B. and {Lindholm}, V. and {L{\'o}pez-Caniego}, M. and {Lubin}, P.~M. and {Ma}, Y. -Z. and {Mac{\'\i}as-P{\'e}rez}, J.~F. and {Maggio}, G. and {Maino}, D. and {Mandolesi}, N. and {Mangilli}, A. and {Marcos-Caballero}, A. and {Maris}, M. and {Martin}, P.~G. and {Mart{\'\i}nez-Gonz{\'a}lez}, E. and {Matarrese}, S. and {Mauri}, N. and {McEwen}, J.~D. and {Meinhold}, P.~R. and {Melchiorri}, A. and {Mennella}, A. and {Migliaccio}, M. and {Miville-Desch{\^e}nes}, M. -A. and {Molinari}, D. and {Moneti}, A. and {Montier}, L. and {Morgante}, G. and {Natoli}, P. and {Oppizzi}, F. and {Pagano}, L. and {Paoletti}, D. and {Partridge}, B. and {Peel}, M. and {Pettorino}, V. and {Piacentini}, F. and {Polenta}, G. and {Puget}, J. -L. and {Rachen}, J.~P. and {Reinecke}, M. and {Remazeilles}, M. and {Renzi}, A. and {Rocha}, G. and {Roudier}, G. and {Rubi{\~n}o-Mart{\'\i}n}, J.~A. and {Ruiz-Granados}, B. and {Salvati}, L. and {Sandri}, M. and {Savelainen}, M. and {Scott}, D. and {Seljebotn}, D.~S. and {Sirignano}, C. and {Spencer}, L.~D. and {Suur-Uski}, A. -S. and {Tauber}, J.~A. and {Tavagnacco}, D. and {Tenti}, M. and {Thommesen}, H. and {Toffolatti}, L. and {Tomasi}, M. and {Trombetti}, T. and {Valiviita}, J. and {Van Tent}, B. and {Vielva}, P. and {Villa}, F. and {Vittorio}, N. and {Wandelt}, B.~D. and {Wehus}, I.~K. and {Zacchei}, A. and {Zonca}, A.},
        title = "{Planck 2018 results. IV. Diffuse component separation}",
      journal = {\aap},
         year = 2020,
        month = sep,
       volume = {641},
          eid = {A4},
        pages = {A4},
          doi = {10.1051/0004-6361/201833881},
archivePrefix = {arXiv},
       eprint = {1807.06208},
 primaryClass = {astro-ph.CO},
       adsurl = {https://ui.adsabs.harvard.edu/abs/2020A&A...641A...4P}
}

@ARTICLE{MILCA_ref,
       author = {{Hurier}, G. and {Mac{\'\i}as-P{\'e}rez}, J.~F. and {Hildebrandt}, S.},
        title = "{MILCA, a modified internal linear combination algorithm to extract astrophysical emissions from multifrequency sky maps}",
      journal = {\aap},
         year = 2013,
        month = oct,
       volume = {558},
          eid = {A118},
        pages = {A118},
          doi = {10.1051/0004-6361/201321891},
archivePrefix = {arXiv},
       eprint = {1007.1149},
 primaryClass = {astro-ph.IM},
       adsurl = {https://ui.adsabs.harvard.edu/abs/2013A&A...558A.118H}
}

@ARTICLE{MDPL2_ref,
       author = {{Klypin}, Anatoly and {Yepes}, Gustavo and {Gottl{\"o}ber}, Stefan and {Prada}, Francisco and {He{\ss}}, Steffen},
        title = "{MultiDark simulations: the story of dark matter halo concentrations and density profiles}",
      journal = {\mnras},
         year = 2016,
        month = apr,
       volume = {457},
       number = {4},
        pages = {4340-4359},
          doi = {10.1093/mnras/stw248},
archivePrefix = {arXiv},
       eprint = {1411.4001},
 primaryClass = {astro-ph.CO},
       adsurl = {https://ui.adsabs.harvard.edu/abs/2016MNRAS.457.4340K}
}

@ARTICLE{SPTpol_intro,
       author = {{Bleem}, L. and {Ade}, P. and {Aird}, K. and {Austermann}, J. and {Beall}, J. and {Becker}, D. and {Benson}, B. and {Britton}, J. and {Carlstrom}, J. and {Chang}, C.~L. and {Cho}, H. and {de Haan}, T. and {Crawford}, T. and {Crites}, A. and {Datesman}, A. and {Dobbs}, M. and {Everett}, W. and {Ewall-Wice}, A. and {George}, E. and {Halverson}, N. and {Harrington}, N. and {Henning}, J. and {Hilton}, G. and {Holzapfel}, W. and {Hoover}, S. and {Hubmayr}, J. and {Irwin}, K. and {Keisler}, R. and {Kennedy}, J. and {Lee}, A. and {Leitch}, E. and {Li}, D. and {Lueker}, M. and {Marrone}, D.~P. and {McMahon}, J. and {Mehl}, J. and {Meyer}, S. and {Montgomery}, J. and {Montroy}, T. and {Natoli}, T. and {Nibarger}, J. and {Niemack}, M. and {Novosad}, V. and {Padin}, S. and {Pryke}, C. and {Reichardt}, C. and {Ruhl}, J. and {Saliwanchik}, B. and {Sayre}, J. and {Schafer}, K. and {Shirokoff}, E. and {Story}, K. and {Vanderlinde}, K. and {Vieira}, J. and {Wang}, G. and {Williamson}, R. and {Yefremenko}, V. and {Yoon}, K.~W. and {Young}, E.},
        title = "{An Overview of the SPTpol Experiment}",
      journal = {Journal of Low Temperature Physics},
         year = 2012,
        month = jun,
       volume = {167},
       number = {5-6},
        pages = {859-864},
          doi = {10.1007/s10909-012-0505-y},
       adsurl = {https://ui.adsabs.harvard.edu/abs/2012JLTP..167..859B}
}

@ARTICLE{Planck_intro,
       author = {{The Planck Collaboration}},
        title = "{The Scientific Programme of Planck}",
      journal = {arXiv e-prints},
         year = 2006,
        month = apr,
          eid = {astro-ph/0604069},
        pages = {astro-ph/0604069},
          doi = {10.48550/arXiv.astro-ph/0604069},
archivePrefix = {arXiv},
       eprint = {astro-ph/0604069},
 primaryClass = {astro-ph},
       adsurl = {https://ui.adsabs.harvard.edu/abs/2006astro.ph..4069T}
}

@ARTICLE{SPT_intro,
       author = {{Carlstrom}, J.~E. and {Ade}, P.~A.~R. and {Aird}, K.~A. and {Benson}, B.~A. and {Bleem}, L.~E. and {Busetti}, S. and {Chang}, C.~L. and {Chauvin}, E. and {Cho}, H. -M. and {Crawford}, T.~M. and {Crites}, A.~T. and {Dobbs}, M.~A. and {Halverson}, N.~W. and {Heimsath}, S. and {Holzapfel}, W.~L. and {Hrubes}, J.~D. and {Joy}, M. and {Keisler}, R. and {Lanting}, T.~M. and {Lee}, A.~T. and {Leitch}, E.~M. and {Leong}, J. and {Lu}, W. and {Lueker}, M. and {Luong-Van}, D. and {McMahon}, J.~J. and {Mehl}, J. and {Meyer}, S.~S. and {Mohr}, J.~J. and {Montroy}, T.~E. and {Padin}, S. and {Plagge}, T. and {Pryke}, C. and {Ruhl}, J.~E. and {Schaffer}, K.~K. and {Schwan}, D. and {Shirokoff}, E. and {Spieler}, H.~G. and {Staniszewski}, Z. and {Stark}, A.~A. and {Tucker}, C. and {Vanderlinde}, K. and {Vieira}, J.~D. and {Williamson}, R.},
        title = "{The 10 Meter South Pole Telescope}",
      journal = {\pasp},
         year = 2011,
        month = may,
       volume = {123},
       number = {903},
        pages = {568},
          doi = {10.1086/659879},
archivePrefix = {arXiv},
       eprint = {0907.4445},
 primaryClass = {astro-ph.IM},
       adsurl = {https://ui.adsabs.harvard.edu/abs/2011PASP..123..568C}
}

@ARTICLE{SZ_simba,
       author = {{Yang}, Tianyi and {Cai}, Yan-Chuan and {Cui}, Weiguang and {Dav{\'e}}, Romeel and {Peacock}, John A. and {Sorini}, Daniele},
        title = "{Understanding the relation between thermal Sunyaev-Zeldovich decrement and halo mass using the SIMBA and TNG simulations}",
      journal = {\mnras},
         year = 2022,
        month = nov,
       volume = {516},
       number = {3},
        pages = {4084-4096},
          doi = {10.1093/mnras/stac2505},
archivePrefix = {arXiv},
       eprint = {2202.11430},
 primaryClass = {astro-ph.CO},
       adsurl = {https://ui.adsabs.harvard.edu/abs/2022MNRAS.516.4084Y}
}

@ARTICLE{limber_approx,
       author = {{Kaiser}, Nick},
        title = "{Weak Gravitational Lensing of Distant Galaxies}",
      journal = {\apj},
         year = 1992,
        month = apr,
       volume = {388},
        pages = {272},
          doi = {10.1086/171151},
       adsurl = {https://ui.adsabs.harvard.edu/abs/1992ApJ...388..272K}
}

\appendix

\section{Power spectrum fits for the CIB four-parameter model and the extended model}\label{appen:A}

\begin{figure*}
    \begin{minipage}[b]{1.0\textwidth}
        \centering

        \includegraphics[width=0.33\linewidth]{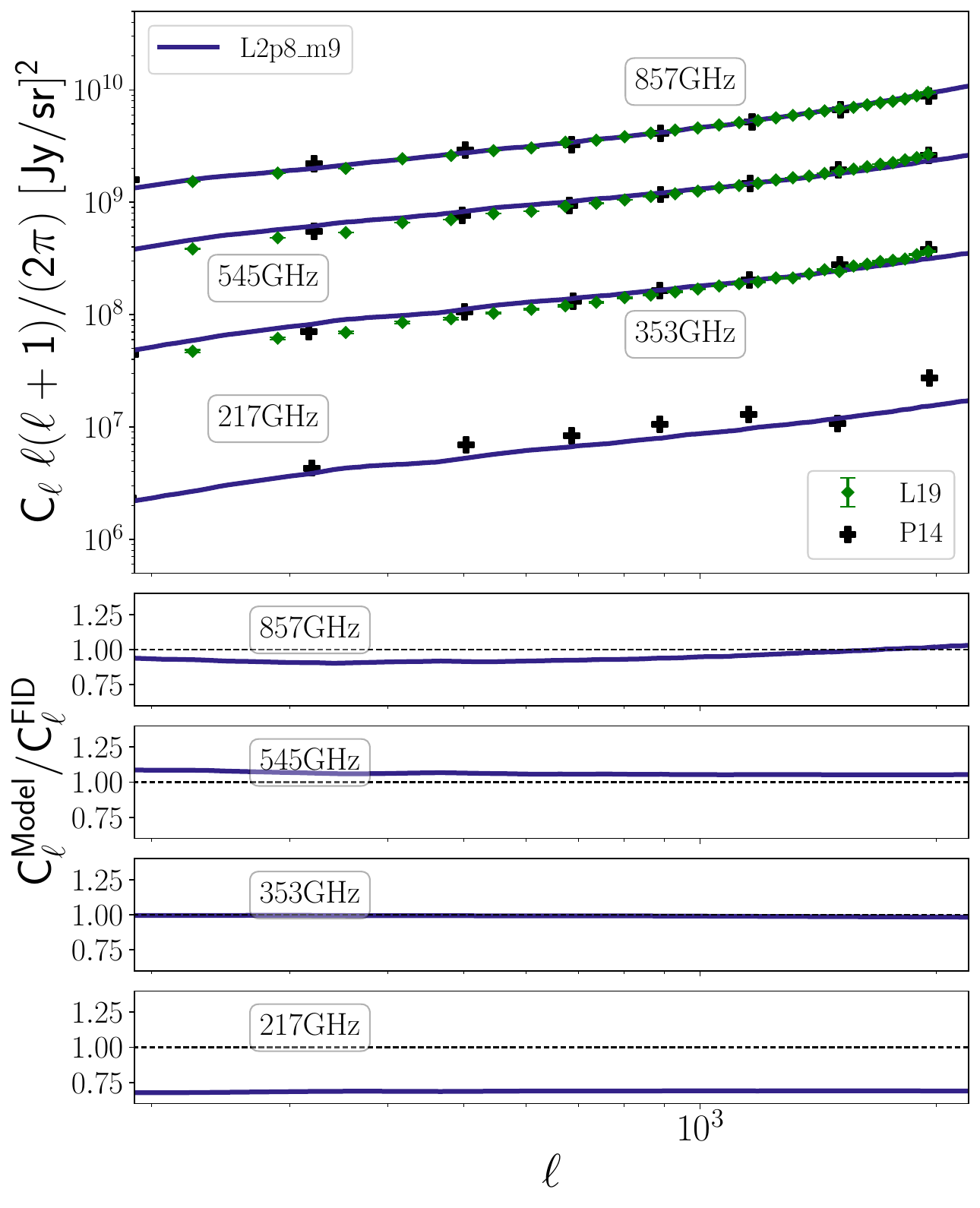}
        \includegraphics[width=0.33\linewidth]{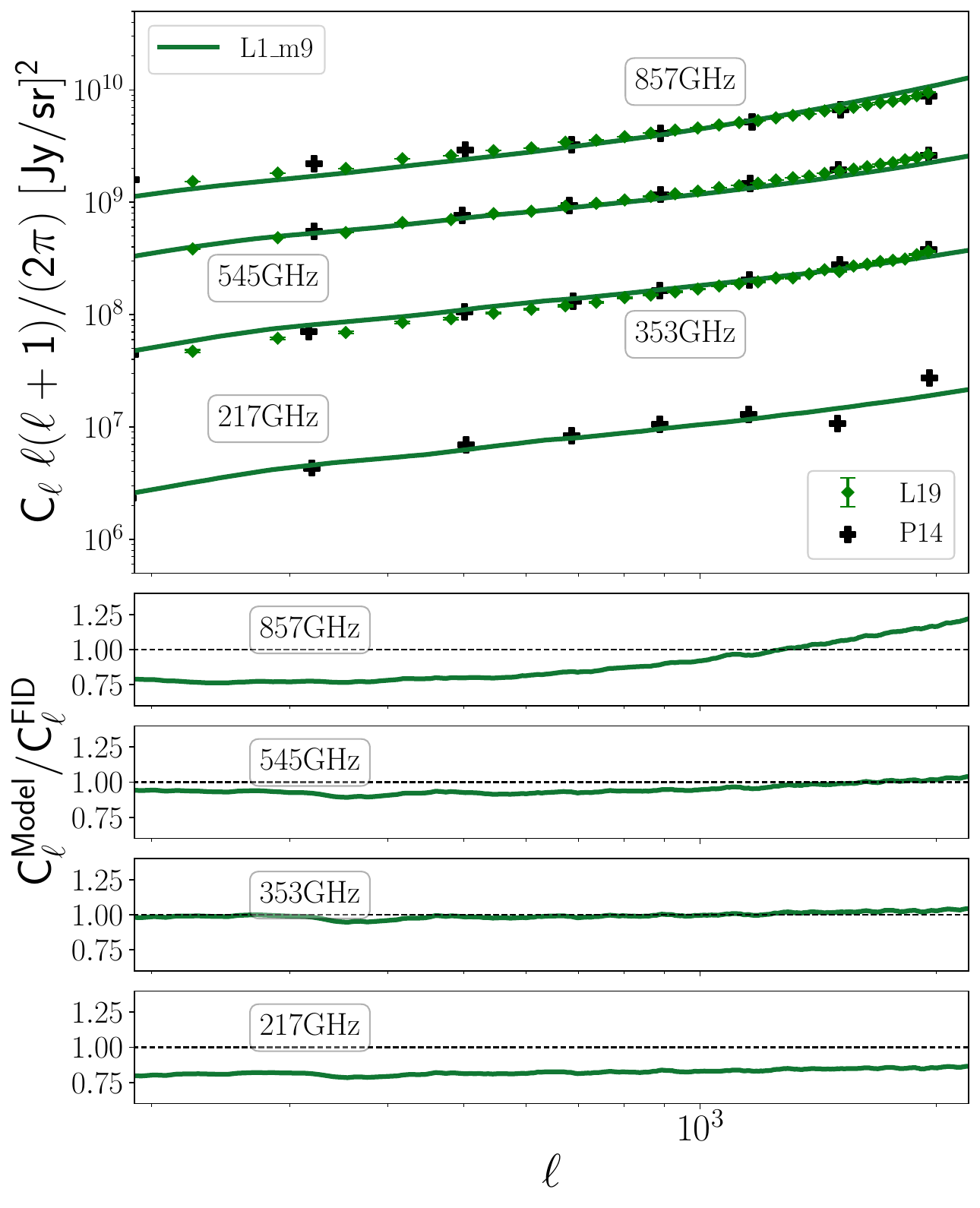}
        \includegraphics[width=0.33\linewidth]{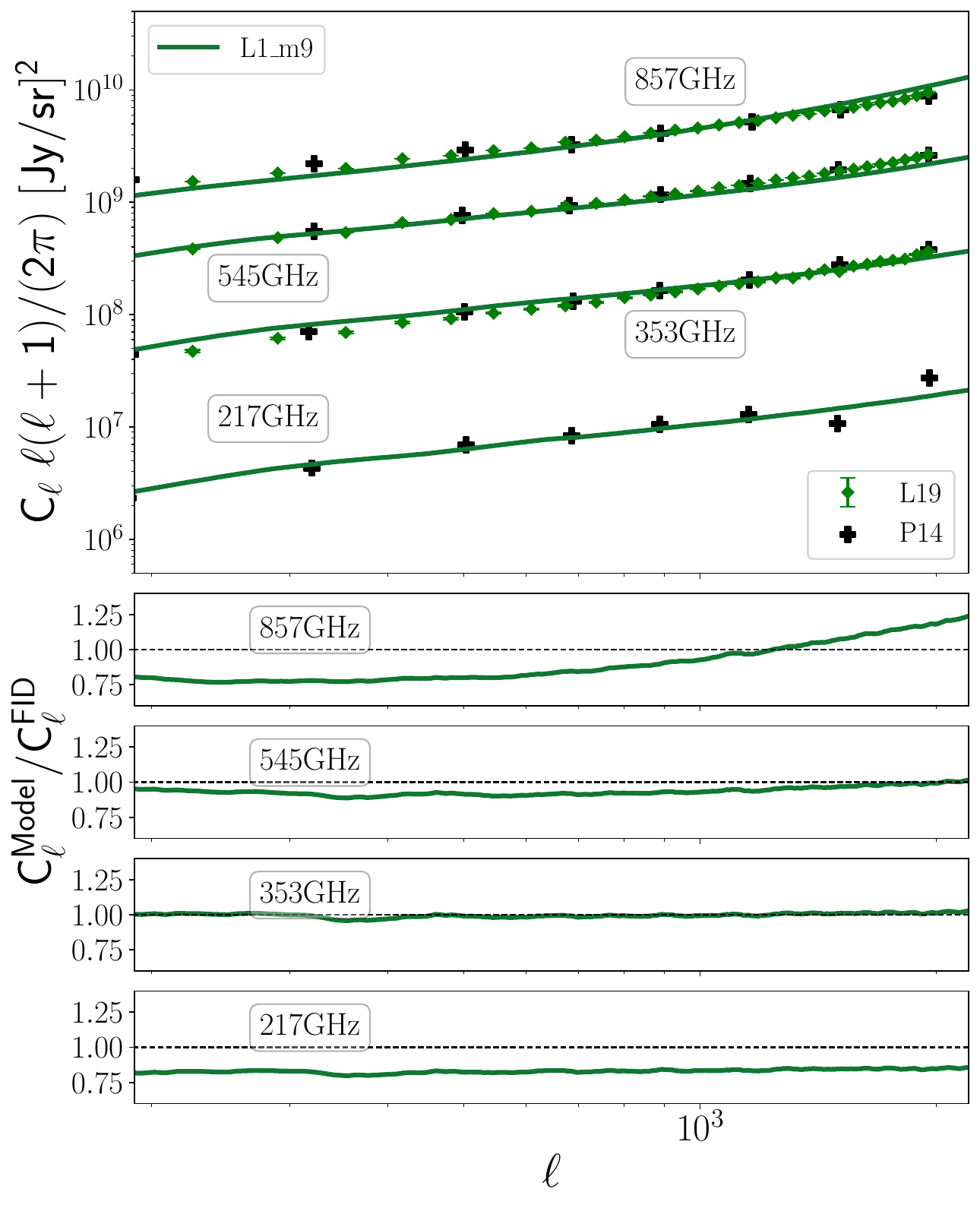}
    \end{minipage}%
    \vspace{+0.1cm}
\caption{CIB auto-power spectrum fit using  the \textit{four-parameter} model (see Equation \ref{eqn::kenicutt_law_propor_varied}) and an extended model using both SFR and dust mass (see Equation \ref{eqn::new_L_IR_eq}). Black data points show the measurements from \citet{L19_CIB} and \citet{P14_CIB} for comparison. The left panel shows the best-fitting curves from the \textit{four-parameter} model. The middle panel corresponds to the case where only star-forming gas within each halo (an aperture size of 30 kpc) is used when computing the dust mass, while the right panel shows the result when all gas is included. For reference, the bottom subpanels show the ratios between the predicted best-fitting curves at each frequency from the varied model and the default \textit{three-parameter} curves.}
\label{CIB_CIB_auto_new_LIR_cal}
\end{figure*}

In this appendix, we present the power spectrum fitting results for the two alternative CIB models: the \textit{four-parameter} model (including a free normalisation parameter in the Kennicutt relation) and a model that depends on both SFR and dust mass, as defined in equations \ref{eqn::kenicutt_law_propor_varied} and \ref{eqn::new_L_IR_eq}, respectively. Figure \ref{CIB_CIB_auto_new_LIR_cal} shows the CIB auto-power spectrum fits from these two models. The leftmost panel shows the results from the
\textit{four-parameter} case. The middle panel shows the results when only star-forming gas is used to compute the dust mass within each halo, while the rightmost panel shows the results when all gas is considered. For comparison, we also show the ratio between the best-fitting curves from different models and those obtained from the default \textit{three-parameter} case at each frequency. 

For each model, we are able to recover approximately the same shape and amplitude of the CIB power spectrum. The associated uncertainties can be largely absorbed into the SED modelling. For the \textit{four-parameter} model, we have best-fitting SED parameters of $\beta_{\rm d} = 2.65\pm0.11$, $T_{0} = 21.00\pm 0.95$ K, $\alpha = 0.07\pm 0.01$, and $N_{\rm s}=-0.26\pm0.03$. For the \textit{extended} model, we have $\beta_{\rm d} = 2.49\pm0.05~(2.56\pm0.04)$, $T_{0} = 21.55\pm 0.24 ~\textrm{K}~(20.35\pm 0.25~\textrm{K})$, $a = 1.23\pm 0.01~(1.10\pm 0.02)$, and a weak $T_{\rm dust}$ redshift evolution constraint for the star-forming gas (all gas) case, respectively.

\begin{figure*}
    \begin{minipage}[b]{1.0\textwidth}
        \centering
        \includegraphics[width=0.45\linewidth]{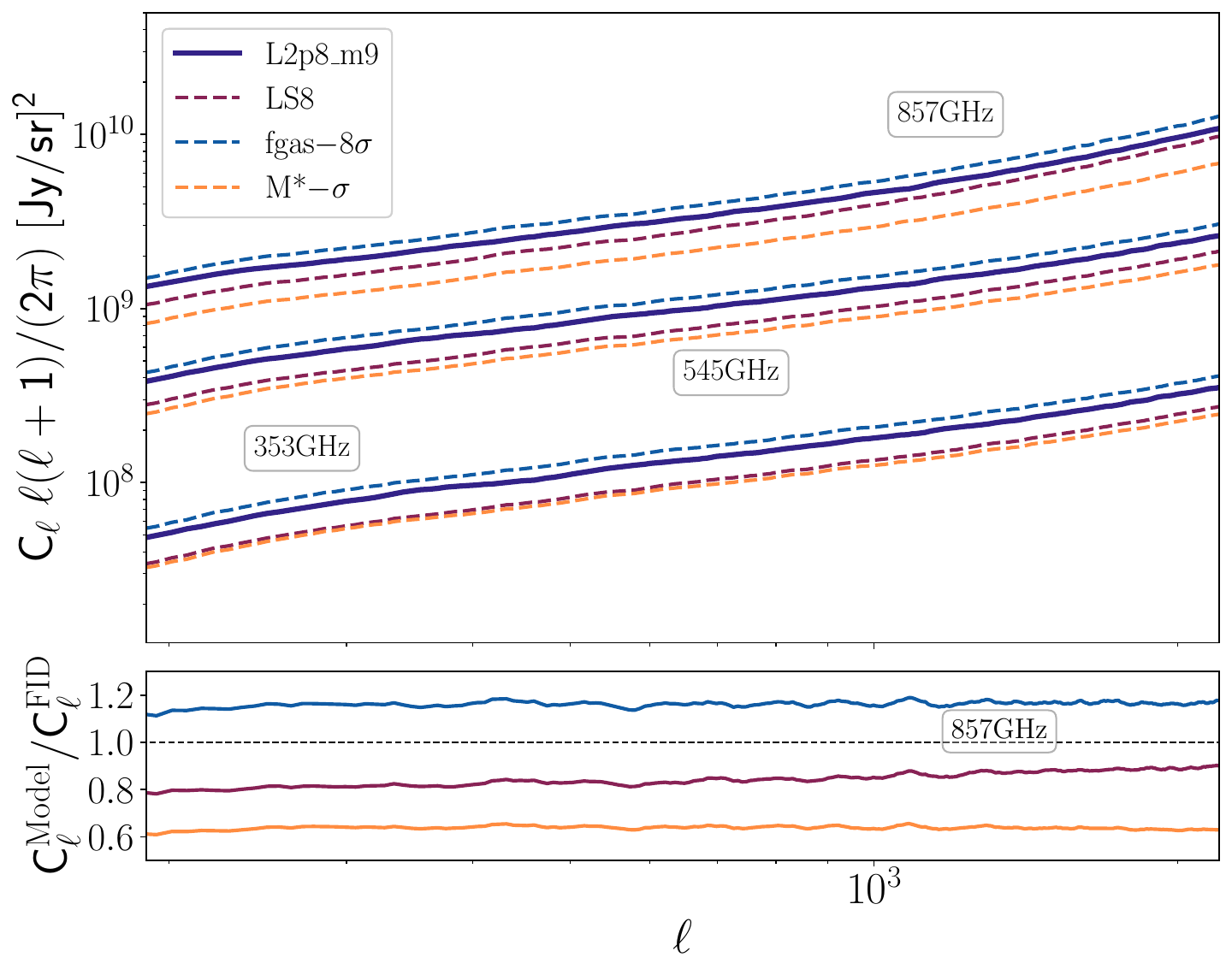}
        \includegraphics[width=0.45\linewidth]{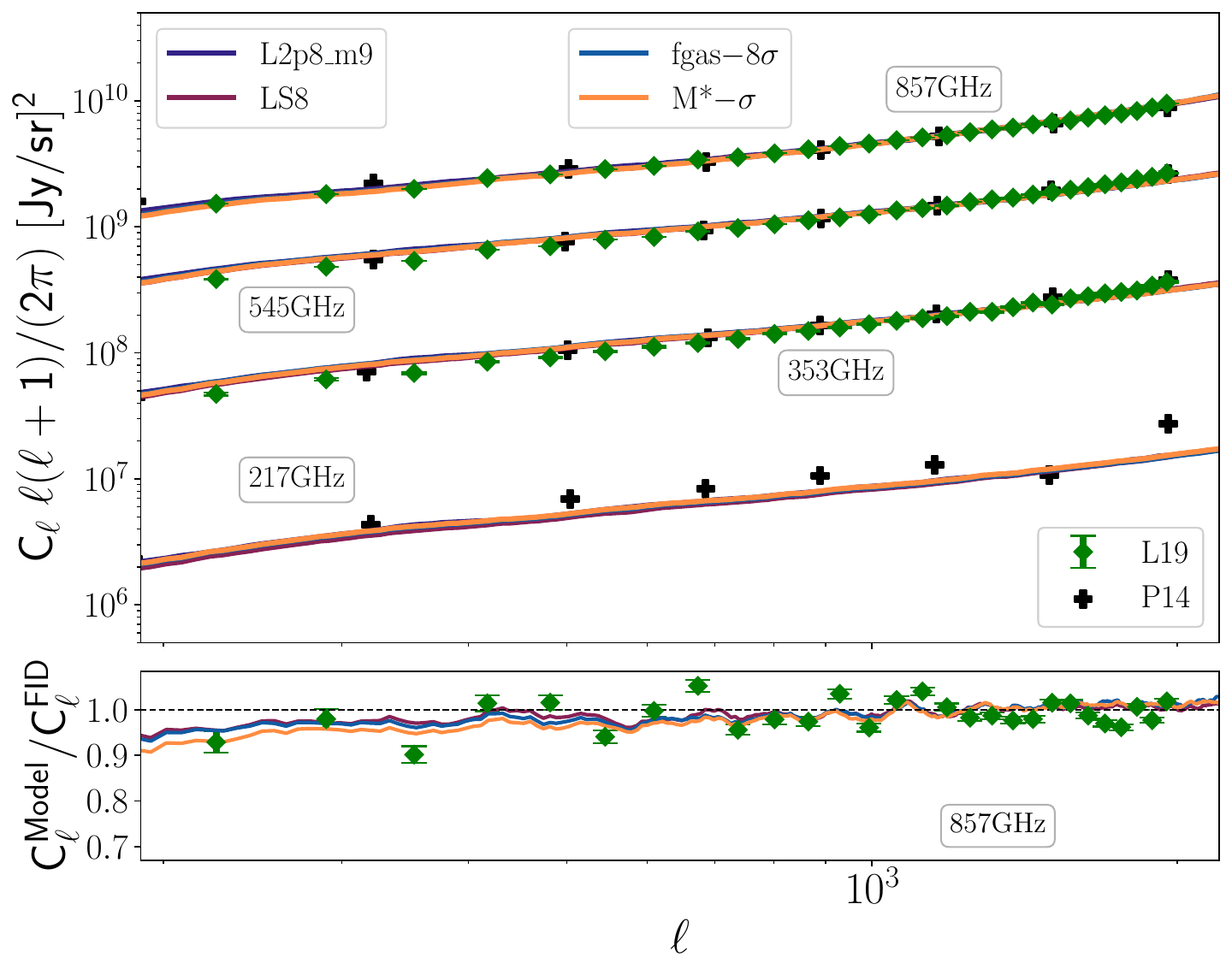}
    \end{minipage}%
    \vspace{+0.1cm}
\caption{As Figure \ref{CIB_CIB_auto_feed_models}, but for the \textit{four-parameter} case as defined in Equation \ref{eqn::kenicutt_law_propor_varied}.}
\label{CIB_CIB_auto_feed_models_four_params}
\end{figure*}

Figure \ref{CIB_CIB_auto_feed_models_four_params} shows the feedback dependence of the CIB auto-power spectrum for the \textit{four-parameter} case. By comparing this plot with Figure \ref{CIB_CIB_auto_feed_models} (for the default \textit{three-parameter} fit), we find that in general, introducing additional degrees of freedom into the CIB model does not lead to a significant improvement in the quality of the best-fitting power spectra when compared to the measured data, and the feedback dependencies of the CIB statistics are comparable to the results obtained from the \textit{three-parameter} case.

\section{Linear scale of the CIB statistics}\label{appen:C}

\begin{figure}
\includegraphics[width=\columnwidth]{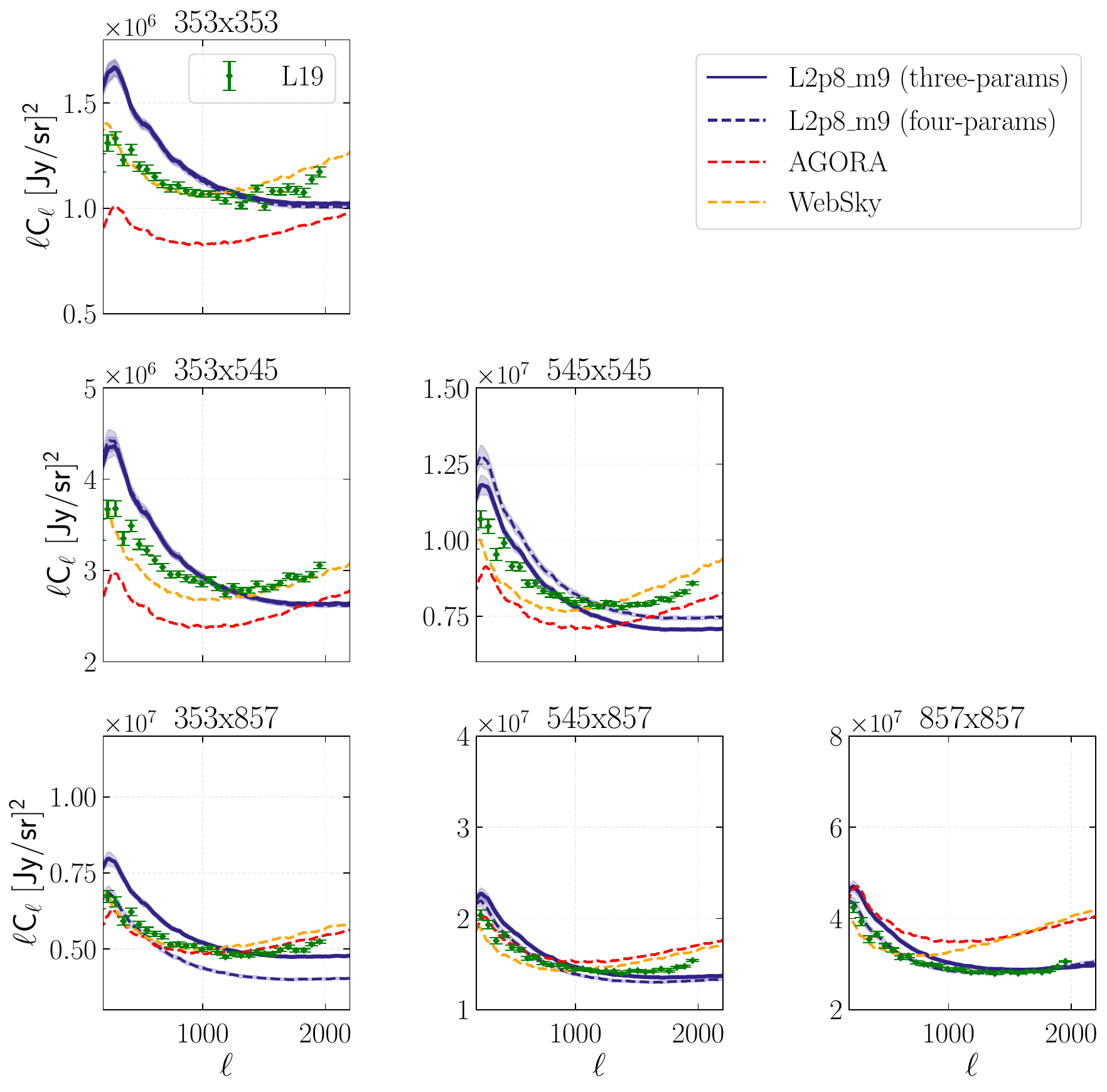}
   \caption{$C^{\rm CIB}_{\ell}$ plotted on a linear scale (solid: best-fitting curves from the \textit{three-parameter} model; dashed: best-fitting curves from the \textit{four-parameter} model). For the logarithmic-scale versions, please see Figures \ref{CIB_CIB_stats}. Shaded regions are the cosmic variance estimated by averaging the results from eight different lightcones. Green data with error bars are measurements from \citet{L19_CIB}.}
\label{CIB_CIB_linear_scale}
\end{figure}

In this appendix, we show the $\ell C^{\rm CIB}_{\ell}$ and $\ell C^{\rm CIB\text{-}y}_{\ell}$ with a linear y-axis scale to highlight the model differences more clearly. While the FLAMINGO simulations provide a qualitatively good fit to the observational measurements that improves on previous models, it is clear from visual inspection of Fig.~\ref{CIB_CIB_linear_scale} that the best-fit model does not provide a formally good fit from a statistical point of view, given the statistical uncertainties on the measurements. Note, however, that no account has been taken for any possible sources of systematic uncertainty.

\begin{figure}
\includegraphics[width=\columnwidth]{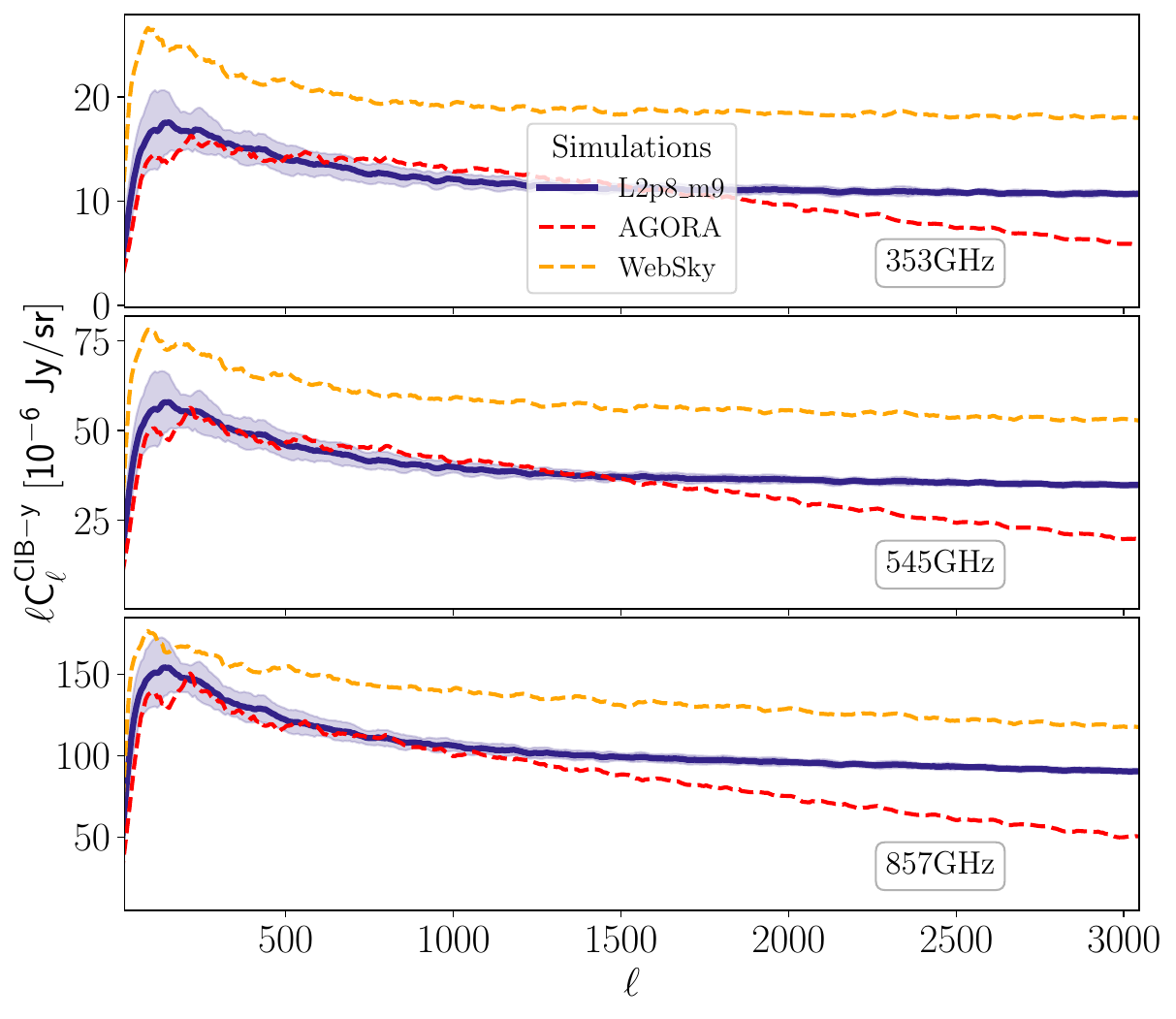}
   \caption{As Figure \ref{CIB_CIB_linear_scale}, but for $C^{\rm CIB\text{-}y}_{\ell}$. For the logarithmic-scale versions, please see Figures \ref{CIB_LSS_stats}.}
\label{CIB_tsz_linear_scale}
\end{figure}


\section{Effect of flux density cut on the CIB statistics}\label{appen:D}

\begin{figure*}
    \begin{minipage}[t]{1.0\textwidth}
        \centering
        \includegraphics[width=0.45\linewidth]{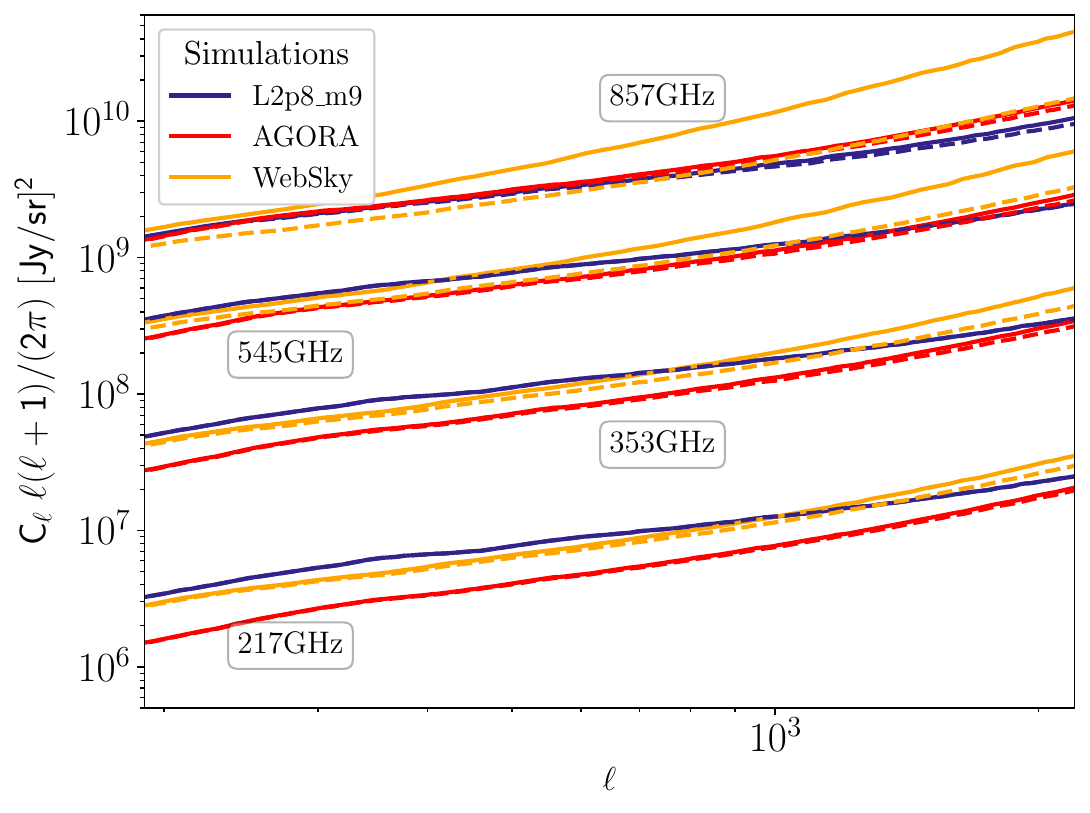}
        \includegraphics[width=0.45\linewidth]{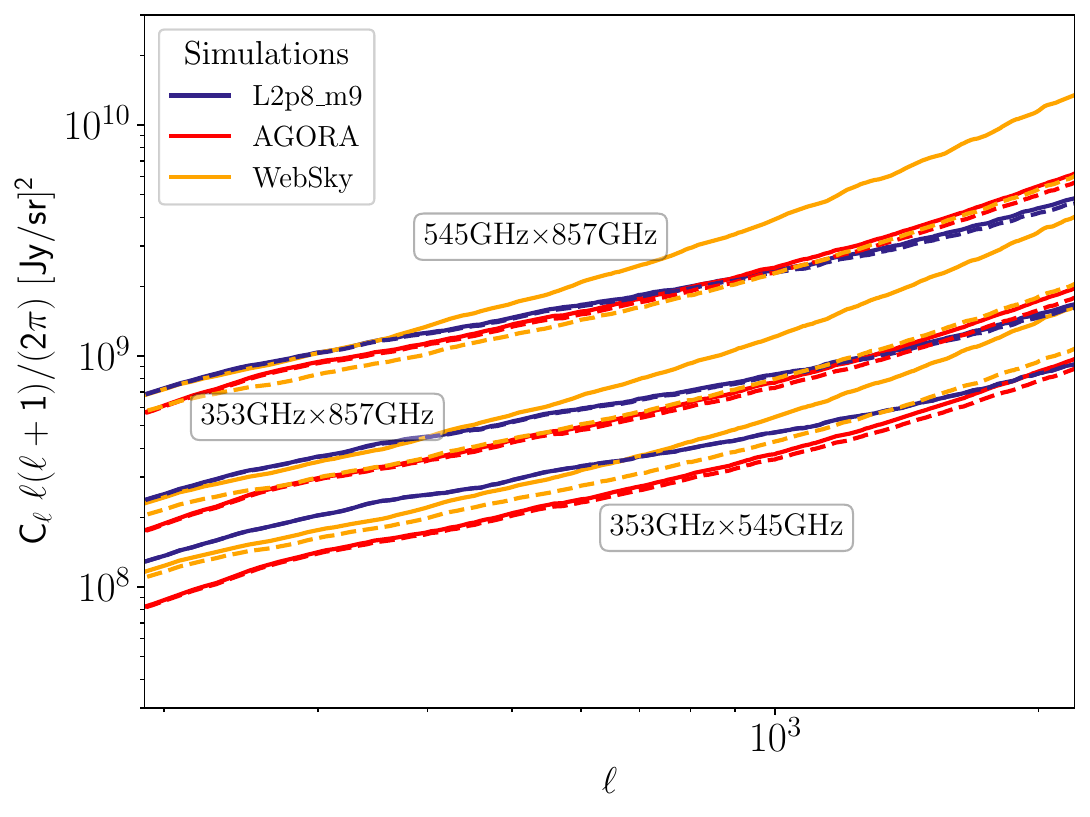}
    \end{minipage}%
    \vspace{+0.1cm}
\caption{The CIB auto- (\textit{left}) and cross- (\textit{right}) power spectra measured from the masked (dashed) and unmasked (solid) CIB maps from the FLAMINGO fiducial $(2.8~\rm Gpc)^3$ run, the AGORA (red) and the WebSky (orange) simulations.}
\label{CIB_CIB_flux_cut_test}
\end{figure*}

\begin{figure*}
    \begin{minipage}[t]{1.0\textwidth}
        \centering
        \includegraphics[width=0.45\linewidth]{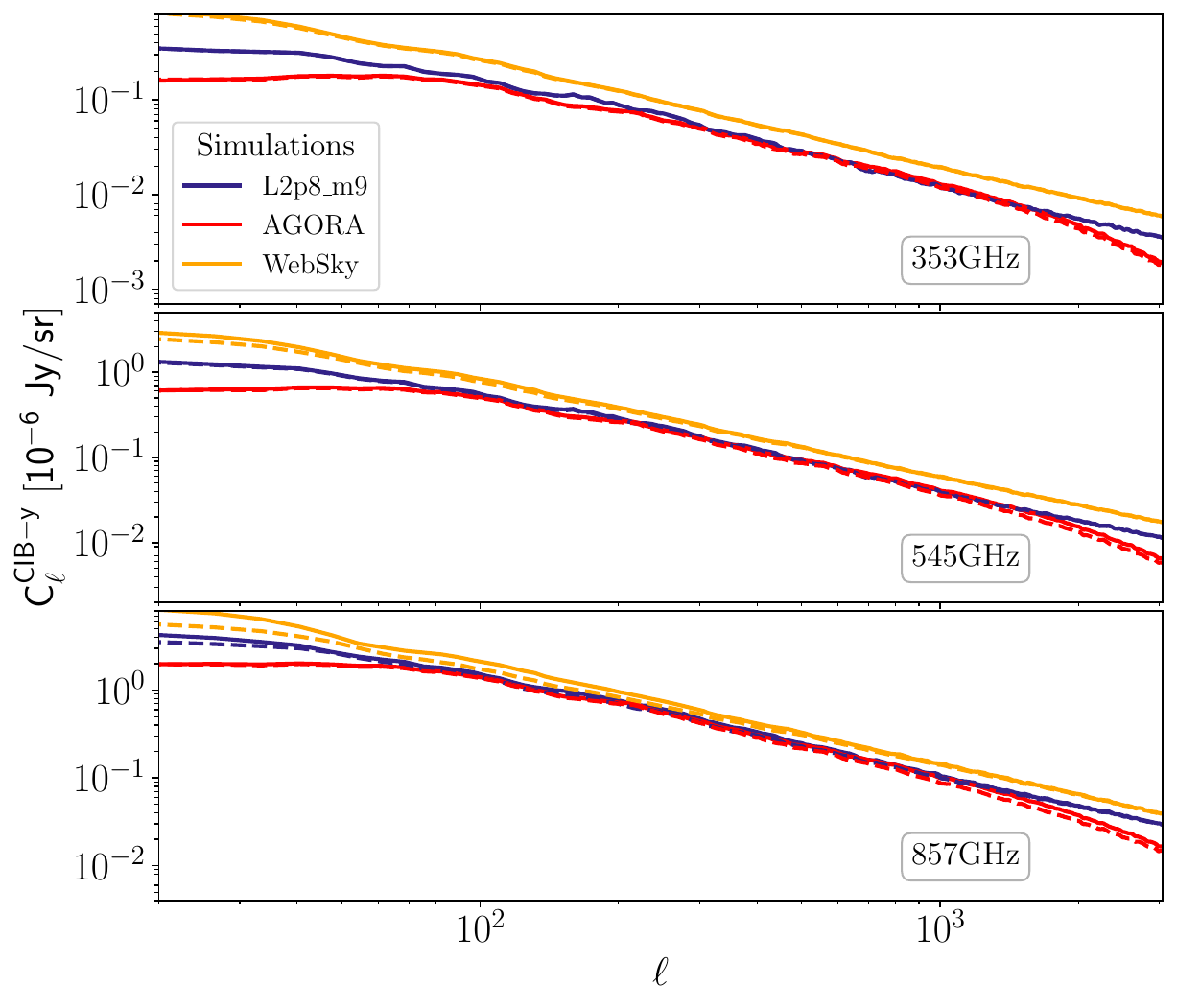}
        \includegraphics[width=0.509\linewidth]{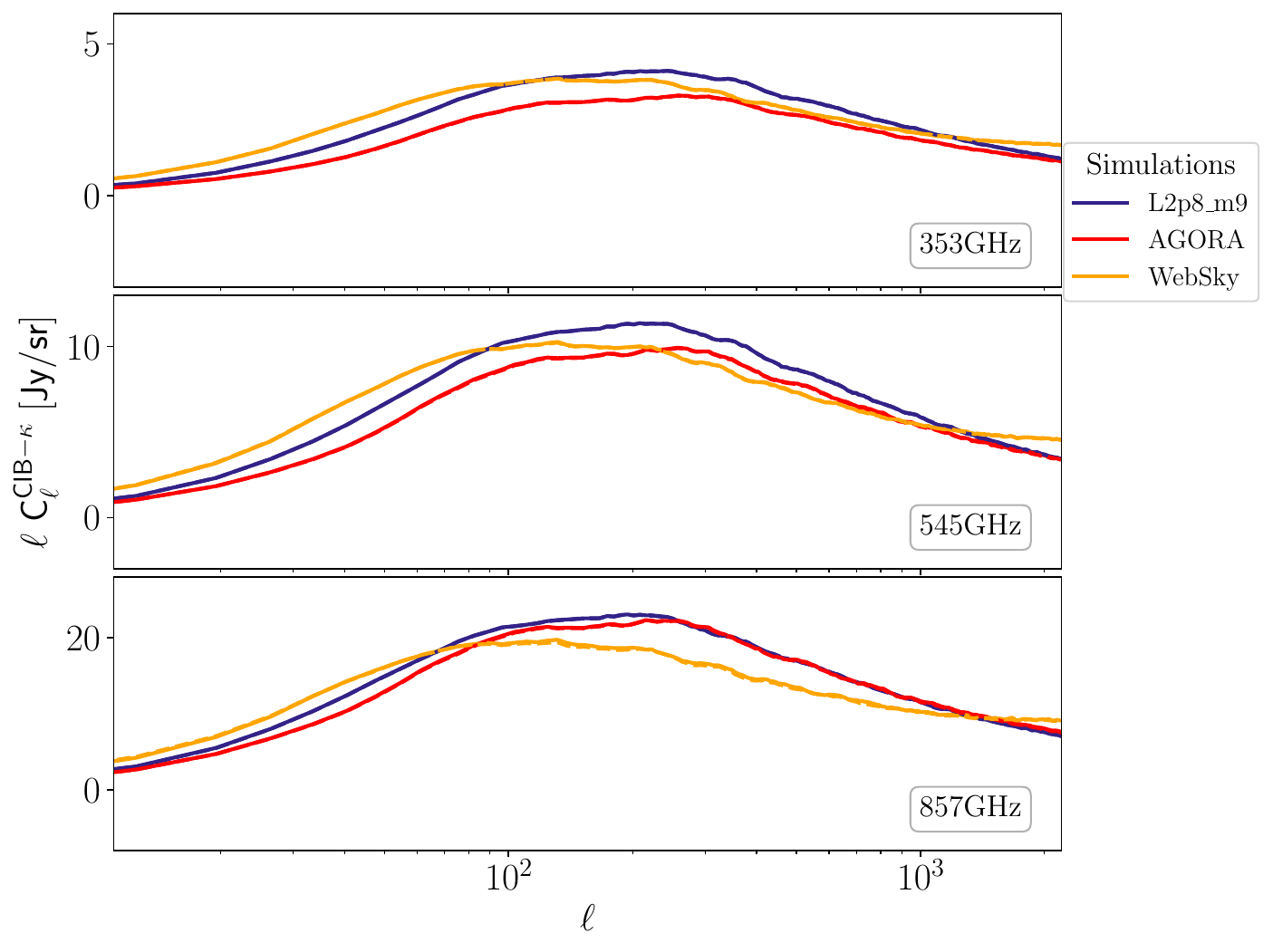}
    \end{minipage}%
    \vspace{+0.1cm}
\caption{CIB-tSZ (\textit{left}) and CIB-$\kappa$ (\textit{right}) cross-power spectra at different frequencies, generated from the masked (dashed) and unmasked (solid) CIB maps.}
\label{CIB_LSS_flux_cut_test}
\end{figure*}

In this appendix, we evaluate the impact of applying a simple flux-density cut on the CIB maps used in our analysis. The goal is to test whether bright infrared sources have a significant effect on the recovered CIB auto- and cross-power spectra for the three models considered in this work: FLAMINGO (blue), WebSky (orange), and AGORA (red), as shown in Figures \ref{CIB_CIB_flux_cut_test} and \ref{CIB_LSS_flux_cut_test}. These figures demonstrate the comparison between the unmasked and masked cases for the CIB power spectrum, the CIB–$\kappa$ cross-spectrum, and the CIB–tSZ cross-spectrum, respectively.

For simplicity, we adopt a single flux-density threshold of 400 mJy for all frequencies and all models (unlike the frequency-dependent cuts as used in the \textit{Planck} analysis, see \citealt{P14_CIB}). Pixels with values exceeding the 400 mJy threshold are replaced with the mean of the remaining pixels below the cut, which is the same refilling strategy as implemented in WebSky. For a fair comparison, the masks and refilling procedure are applied consistently across all simulations. For both the FLAMINGO and AGORA models, we find that the flux-density cut only has minor changes in the recovered CIB power spectra. The masked and unmasked curves remain nearly identical across all frequencies and for all cross-correlations with $\kappa$ and tSZ. This suggests that our model fitting is robust to such masking choices, and that flux-density cuts are not a major systematic uncertainty for the FLAMINGO-based CIB predictions.

However, the flux-density cut produces a substantial suppression in the WebSky CIB auto-spectrum, as well as noticeable changes in some of the cross-spectra. As discussed in Section \ref{ssec:CIB_statistics}, the CIB modelling adopted by WebSky appears to assign higher infrared signals to low-redshift haloes. When these bright resolved low-$z$ structures are masked and refilled, the WebSky CIB power is significantly reduced, as shown in Figure \ref{CIB_CIB_flux_cut_test}.

Despite this strong effect on the CIB auto-/cross-spectrum, the WebSky CIB–$\kappa$ cross-spectrum remains largely unaffected. This is expected because this statistic is more sensitive to the overall matter distribution and a wider range of redshifts, which is not strongly altered by removing bright, low-redshift sources. For the CIB–tSZ cross-spectrum, however, the most visible difference appears at low multipoles in 857 GHz. This is again consistent with the removal of bright, low-redshift structures, which contributes to both CIB and tSZ signals. Masking these objects removes correlated large-scale structure, leading to a slightly lower masked spectrum compared to the unmasked case.

Overall, this test shows that our FLAMINGO-based CIB modeling is stable under this simplistic flux-density cuts, whereas the WebSky predictions are more sensitive to masking due to their bright low-$z$ population. These results indicate that our model fitting is not strongly impacted by point-source masking choices.

\end{document}